\pdfoutput=1
\documentclass[vecphys]{svmult}
\usepackage{makeidx}
\usepackage{graphicx}
\usepackage{subfigure}
\usepackage{mathptmx,helvet,courier}
\usepackage{multicol}
\usepackage[bottom]{footmisc}
\usepackage{cite}
\usepackage{amsmath}
\usepackage{amssymb}

\begin{document}

\title*{Solitary waves in the Nonlinear Dirac Equation}
\titlerunning{Solitary waves in the Nonlinear Dirac Equation}
\author{
Jes\'us Cuevas-Maraver
\and
Nabile Boussa\"{\i}d
\and
Andrew Comech
\and
Ruomeng Lan
\and
Panayotis G. Kevrekidis
\and
Avadh  Saxena
}
\authorrunning{J. Cuevas-Maraver {\em et al}}
\tocauthor{J. Cuevas-Maraver, N. Boussa\"{\i}d, A. Comech, R. Lan, P.G. Kevrekidis, and A. Saxena}

\institute{
Jes\'us Cuevas-Maraver
\at Grupo de F\'{\i}sica No Lineal, Universidad de
Sevilla, Departamento de F\'{i}sica Aplicada I,
Escuela Polit\'ecnica Superior.
C/ Virgen de \'{A}frica, 7, 41011-Sevilla, Spain,
\email{jcuevas@us.es}
\and
Nabile Boussa\"{\i}d
\at Universit\'e de Franche-Comt\'e, 25030 Besan\c{c}on CEDEX, France
\and
Andrew Comech
\at
St. Petersburg State University,
St. Petersburg 199178, Russia
\at Department of Mathematics, Texas A\&M University, College Station, TX 77843-3368, USA
\at IITP,
Moscow 127994, Russia
\and
Roumeng Lan
\at Department of Mathematics, Texas A\&M University, College Station, TX 77843-3368, USA
\and
Panayotis G. Kevrekidis
\at Department of Mathematics and Statistics, University of Massachusetts, Amherst, MA 01003-4515, USA
\and
Avadh Saxena
\at Center for Nonlinear Studies and Theoretical Division, Los Alamos National Laboratory, Los Alamos, New Mexico 87545, USA
 }

\maketitle

\abstract{In the present work, we consider the existence, stability, and
dynamics of solitary waves in the nonlinear Dirac equation. We start
by introducing the Soler model of self-interacting spinors, and discuss
its localized waveforms in one, two, and three spatial dimensions and
the equations they satisfy. We present the associated explicit solutions
in one dimension and numerically obtain their analogues in higher
dimensions. The stability is subsequently discussed from a theoretical
perspective and then complemented with numerical computations. Finally,
the dynamics of the solutions is explored and compared
to its non-relativistic analogue, which is the nonlinear Schr{\"o}dinger
equation. A few special topics are also explored, including the
discrete variant of the nonlinear Dirac equation and its
solitary wave properties, as well as the {$\mathcal{P}\mathcal{T}$}-symmetric
variant of the model.}


\section{Introduction}
\label{sec:cuevas-intro}

In the last three decades, there has been an enormous interest in the study of waves in nonlinear dispersive media. Arguably, two of the most paradigmatic equations that describe such waves are the nonlinear Schr{\"o}dinger equation (NLS) and the sine--Gordon equation. The first among these equations covers a broad range of settings including atomic physics \cite{cuevas-PS03,cuevas-PS02}, nonlinear optics \cite{cuevas-KFC15,cuevas-KA03}, condensed matter physics, and mathematical physics \cite{cuevas-APT04,cuevas-SS99}. The sine--Gordon equation also covers settings in condensed matter physics and mathematical physics apart from high-energy physics models \cite{cuevas-BK04,cuevas-CKW14}. A principal focus of the relevant properties of these equations has been the study of the existence, stability, and dynamics of solitary waves (i.e. spatially localized waves supported by the nonlinearity and dispersion), both in lower-dimensional settings (such as one-dimensional solitons and multi-solitons) and in higher dimensional settings (vortices, vortex rings, and related structures) \cite{cuevas-DP06,cuevas-KFC15}.

By comparison, far less attention has been paid to the nonlinear Dirac equation (NLD), despite its presence for almost 90 years in the realm of high energy physics. The nonlinear Dirac equation with scalar-type self-interaction was initially introduced by Ivanenko in 1938 \cite{cuevas-Iva38}.
Following the ideas of Finkelstein \cite{cuevas-FLR51}, Heisenberg in 1957 \cite{cuevas-Hei57}
used this NLD model in an attempt to formulate a unified theory of elementary particles.
In 1958, a completely integrable one-dimensional model known as the Massive Thirring Model (MTM) \cite{cuevas-Thi58}, based on vector-type self-interaction of spinor field, was introduced. This model possesses solitary wave solutions. Curiously, fundamental solutions of the MTM can be transformed into solitons of the sine--Gordon equation by means of a bosonization process \cite{cuevas-Col75}.
In 1970, Soler re-introduced Ivanenko's model
with scalar-type self-interaction
in the context of extended nucleons \cite{cuevas-Sol70} and also provided
the numerical analysis of solitary wave solutions.
The one-dimensional version of the Soler model, known as the Gross--Neveu model \cite{cuevas-GN74}, was introduced in 1974 as a toy model of quark confinement in quantum chromodynamics, and explicit solitary wave solutions in the corresponding massive model were found by Lee \emph{et al.} in 1975 \cite{cuevas-LKG75}.
We can not complete this quick review of NLD models in high-energy physics without mentioning the recent work of~\cite{cuevas-NP09} (see also~\cite{cuevas-MLB14}),
where a variant of the NLD is applied to the study of neutrino oscillations.
Related systems are
the Dirac--Maxwell system~\cite{cuevas-Gro66,cuevas-Wak66,cuevas-Bou96,cuevas-EGS96,cuevas-CS12},
the Einstein--Dirac system \cite{cuevas-Rot10a,cuevas-Stu10},
and Einstein--Dirac--Maxwell system
\cite{cuevas-Rot10}.
In quantum chemistry,
the Dirac--Hartree--Fock model
\cite{cuevas-LM82,cuevas-ES99,cuevas-ES01}
takes into account the fermionic properties of electrons
(describing the exchange interaction,
which is a fundamental effect of purely quantum nature) and
is used for accurate computation
of the electronic energy
\cite{cuevas-VD97,cuevas-QGW04};
this model
has also started to receive mathematical attention \cite{cuevas-ES99,cuevas-ES01,cuevas-ES05}.

Recent years have seen a gradual increase of interest in the study of near-relativistic settings, arguably, for three principal reasons. Firstly, significant steps have been taken in the nonlinear analysis of stability of such models~\cite{cuevas-BC12,cuevas-PS14,cuevas-BC16a}, especially in one-dimensional~\cite{cuevas-PS12,cuevas-CPS17,cuevas-BC12a,cuevas-BC16a} and two-dimensional settings~\cite{cuevas-CKS+16a}. Secondly, computational advances have enabled a better understanding of the associated solitary wave solutions and their dynamics~\cite{cuevas-XST13,cuevas-CKMS10,cuevas-SQM+14,cuevas-CKS15} also in the presence of external fields~\cite{cuevas-MQC+12}. Thirdly, and perhaps most importantly, NLD starts emerging in physical systems which arise in a diverse set of contexts of considerable interest. These contexts include, in particular, bosonic evolution in honeycomb lattices~\cite{cuevas-HOC15,cuevas-HC15a} and a growing class of atomically thin 2D Dirac materials~\cite{cuevas-WBB14,cuevas-FSL+17} such as graphene, silicene, germanene, borophene, and transition metal dichalcogenides \cite{cuevas-MLH+10}. Recently, the physical aspects of nonlinear optics, such as light propagation in honeycomb photorefractive lattices (the so-called photonic graphene)~\cite{cuevas-ANZ09,cuevas-AZ10} have prompted the consideration of intriguing dynamical features, e.g. conical diffraction in 2D honeycomb lattices~\cite{cuevas-PBF+07}. Inclusion of nonlinearity is then quite natural in these models, although in a number of them (e.g., in atomic and optical physics) the nonlinearity does not couple the spinor components and breaks the Lorentz symmetry (that is, such models are not invariant under Lorentz transformations; for the explicit form of the Lorentz transformations of the spinor fields see e.g. \cite{cuevas-BD64,cuevas-Tha92}).

It would be relevant to mention one more framework where Dirac-type equations have received significant attention in recent years, that is in the context of spin-orbit coupled Bose--Einstein condensates~\cite{cuevas-DGJO11}. There, admittedly, the setup is somewhat different, as both the Dirac type operator and the Schr{\"o}dinger one co-exist, but it is relevant to point out that such settings have already been realized experimentally~\cite{cuevas-LBJ+13,cuevas-LJS11,cuevas-QHG+13}. Moreover, a wide range of coherent structures has been already proposed in them including vortices~\cite{cuevas-RSSG11,cuevas-ROL+12,cuevas-XH11}, Skyrmions \cite{cuevas-KMNM12}, Dirac monopoles \cite{cuevas-Con12}, and dark solitons~\cite{cuevas-ASK+13,cuevas-FBZ12}, as well as self-trapped states~\cite{cuevas-MJZ+10}, bright solitons~\cite{cuevas-AFKP13,cuevas-XZW13}, and gap-solitons~\cite{cuevas-KKA13}. It has also been demonstrated that in such systems it is possible to create stable vortex solitons in free space, which until recently was considered impossible due to the presence of collapse, driven by the self-attractive cubic nonlinearity~\cite{cuevas-SLM14}.

   From a mathematical perspective, Dirac models
are described by systems
(rather than by scalar equations)
that correspond to the Hamiltonian functionals
unbounded from below.
This unboundedness makes all the aspects of the
analysis of these models
(well-posedness, existence of localized solutions,
stability, numerical simulations)
much more challenging.
This has fueled an increasing interest
in the nonlinear Dirac equation and
more general models of self-interacting spinor fields,
with many results
on the existence of solitary waves
\cite{cuevas-CV86,cuevas-Mer88,cuevas-ES95} and
well-posedness in (3+1)D \cite{cuevas-EV97,cuevas-MNNO05} and
in (1+1)D
\cite{cuevas-ST10,cuevas-MNT10,cuevas-Can11,cuevas-Pel11a,cuevas-Huh13} \footnote{With the notation $(N+1)D$ we want to denote that the system possesses $N+1$ dimensions, with $N$ spatial ones plus time}.
The stability of solitary wave solutions
of the nonlinear Dirac equation
was approached via numerical simulations
\cite{cuevas-RRSV74,cuevas-AS83,cuevas-AS86,cuevas-BPZ98,cuevas-CP06,cuevas-BC12a,cuevas-MQC+12,cuevas-XST13} and
via heuristic arguments
\cite{cuevas-Bog79,cuevas-MM86,cuevas-SV86,cuevas-BSV87,cuevas-CKMS10},
but it is still not settled.
Recently, the first stability results
in the context of self-interacting spinor fields
started appearing
\cite{cuevas-Bou06,cuevas-Bou08,cuevas-BC12a,cuevas-PS14,cuevas-BC16a,cuevas-CPS17,cuevas-BC17}.

The NLD can also be viewed as a relativistic generalization (or extension) of the NLS, or, alternatively, the NLS (with additional terms) can be seen as a special case limit of the NLD at the low-energy limit. Nevertheless, it has turned out that the Dirac equation as a result of its matrix nature and the fact that it is only first order in spatial derivatives (as opposed to second order in the NLS) has proven far more computationally (and theoretically) challenging, on a number of grounds, than its NLS counterpart. This difficulty has
hindered the progress in the study of solitary waves, particularly in two-dimensional and three-dimensional settings.
However, recent developments are gradually enabling
the study of the stability and dynamical properties of solitary waves in
two-dimensional and even three-dimensional Soler models; see for
a relevant example~\cite{cuevas-CKS+16a}.
Clearly, however, this process requires numerous additional steps
that will present several challenges over the coming years.

The aim of this chapter is to give a review of recent results developed by the authors and their collaborators in the last few years, as well as to
present a basic framework of the NLD theory, mainly focused on the Soler model and its variants; this is our principal workhorse model.
The content of the chapter covers a wide spectrum of results ranging from existence and stability of solitary waves to numerical methods and
dynamics of unstable solutions. Apart from this, we also introduce
both a discrete variant of the model, as well as an
NLD model with {$\mathcal{P}\mathcal{T}$} symmetry and
analyze their principal characteristics.

This Chapter is organized as follows: in Section \ref{sec:cuevas-NLDmodels} we start with an introduction to the main nonlinear Dirac equation, namely the Soler model, and tractable expressions for the determination of solitary waves and linearizations at solitary waves in one, two, and three spatial dimensions. Section \ref{sec:cuevas-existence} is devoted to the existence properties of solitary waves and numerical methods for their calculation. Stability analysis from a theoretical and numerical point of view is the topic of Sections \ref{sec:cuevas-stability_theory} and \ref{sec:cuevas-stability}, respectively. The dynamics of solitary waves is analyzed in Section \ref{sec:cuevas-dynamics}.
The discrete version of NLD is discussed in Section \ref{sec:cuevas-DNLD}.
A {$\mathcal{P}\mathcal{T}$}-symmetric modification of the Soler model is presented in Section \ref{sec:cuevas-PT}.
We finalize the paper with a summary of the considered results and an outlook on future directions on solitary waves in nonlinear Dirac equations.

\section{The Soler model of self-interacting spinors}
\label{sec:cuevas-NLDmodels}

In this section we start with the linear Dirac equation and move on to the Soler model as a principal, Lorentz-invariant variant of the model with scalar self-interactions.
We give explicit expressions of linearization at solitary waves in one-, two-, and three-dimensional cases.

\subsection{The Dirac equation}

In December 1927, Paul Dirac
arrived at the idea of the first-order relativistically invariant equation
\cite{cuevas-Dir28}
that describes massive spin-1/2 relativistic fermions in $(3+1)$
space-time dimensions:
\begin{equation*}\label{eq:cuevas-Dirac}
i\hbar\frac{\partial}{\partial t}\psi(t,x)
=
\left(-i\hbar c \mathbf{\alpha}\cdot\mathbf{\nabla}+m c^2\beta\right)\psi(t,x),
\qquad
\psi(t,x)\in\mathbb{C}^4,
\qquad
x\in\mathbb{R}^3,
\end{equation*}
with $\psi$ being the spinor-valued wavefunction,
$\mathbf{\alpha}\cdot\mathbf{\nabla}=\sum_{j=1}^3\alpha^j\frac{\partial}{\partial x^j}$,
and $m\ge 0$ the mass of the particle.
As usual, we choose in what follows the units so that
Planck's constant $\hbar$ and the speed of light $c$
are both equal to one.
The self-adjoint $4\times 4$ matrices $\alpha^j$,
$1\le j\le 3$, and $\beta$ satisfy
\[
\{\alpha^j,\alpha^k\}=2\delta_{j k}I_4,
\qquad
\{\alpha^j,\beta\}=0,
\qquad
\beta^2=I_4,
\]
with $I_N$ being the $N\times N$ identity matrix and $\{A,B\}=AB+BA$ the anticommutator.
According to the Dirac--Pauli theorem
(see~\cite{cuevas-Dir28,cuevas-PAU36,cuevas-VDWB74},
\cite[Lemma 2.25]{cuevas-Tha92},
and also~\cite[Theorem 7]{cuevas-KES61} for general version
in odd spatial dimensions),
different choices
of the matrices $\alpha^j$ and $\beta$ are equivalent.
The most common choice, known as the Dirac--Pauli representation,
is
\begin{equation*}
 \alpha^j=\left(
  \begin{array}{cc}
  0 & \sigma_j \\
  \sigma_j & 0 \\
  \end{array}
  \right),
\qquad
 \beta=\left(
  \begin{array}{cc}
  I_2 & 0 \\
  0 & -I_2 \\
  \end{array}
  \right),
\end{equation*}
with the Pauli matrices given by
\begin{equation}\label{eq:cuevas-Pauli}
\sigma_1=\left(
  \begin{array}{cc}
  0 & 1 \\
  1 & 0 \\
  \end{array}
  \right),
\qquad
\sigma_2=\left(
  \begin{array}{cc}
  0 & -i \\
  i & 0 \\
  \end{array}
  \right),
\qquad
\sigma_3=\left(
  \begin{array}{cc}
  1 & 0 \\
  0 & -1 \\
  \end{array}
  \right).
\end{equation}
In the covariant form, the Dirac equation is written as
\begin{equation*}\label{eq:cuevas-Dirac_original2}
 i\gamma\sp\mu\partial\sb\mu\psi=m\psi,
\end{equation*}
where
$\gamma\sp\mu\partial\sb\mu=\sum_{\mu=0}^3 \gamma^\mu\partial_\mu$,
$\partial_0\equiv\partial_t$, with $\gamma^\mu$ being the Dirac $\gamma$-matrices
\begin{equation*}
 \gamma^0=\beta,
\qquad
 \gamma^j=\beta\alpha^j=\left(
  \begin{array}{cc}
  0 & \sigma_j \\
  -\sigma_j & 0 \\
  \end{array}
  \right),\ \ \
 j=1,2,3.
\end{equation*}
Matrices $\gamma^\mu$ fulfill the anticommutation relation $\{\gamma^\mu,\gamma^\nu\}=2\eta^{\mu\nu}I_4$, with $\eta^{\mu\nu}$ being the Minkowski tensor \cite{cuevas-DWS86}. In other words, $(\gamma^0)^2=I_4$ and $(\gamma^1)^2=(\gamma^2)^2=(\gamma^3)^2=-I_4$.
There exists another matrix
which anticommutes with
$\gamma^0$ and $\gamma^j$, $1\le j\le 3$,
which plays an important role in the parity transformation.
It is the $\gamma^5$ matrix, defined by

\begin{equation*}
 \gamma^5=i\gamma^0\gamma^1\gamma^2\gamma^3=
 \left(
  \begin{array}{cc}
  0 & I_2 \\
  I_2 & 0 \\
  \end{array}
  \right).
\end{equation*}
This matrix is self-adjoint and satisfies $(\gamma^5)^2=I_4$.

One can immediately generalize the ideas of Dirac to an arbitrary
spatial dimension $n\ge 1$,
writing the Dirac equation
\[
i\partial\sb t\psi=D_m\psi\equiv-i\sum\sb{j=1}\sp{n}\alpha^j\partial\sb j\psi+\beta m\psi,
\qquad
\psi(t,x)\in\mathbb{C}^N,
\quad
x\in\mathbb{R}^n,
\]
with $\alpha^j$, $1\le j\le n$, and $\beta$ being
selfadjoint matrices
satisfying the
relations
\[
\{\alpha^j,\alpha^k\}=2\delta_{j k}I_N,
\quad
\{\alpha^j,\beta\}=0,
\quad
(\alpha^j)^2=\beta^2=I_N;
\qquad
1\le j,k\le n.
\]
The smallest
number of spinor components $N$ for the spatial dimension $n\ge 1$
is obtained in the Clifford algebra theory (see e.g. \cite[Chapter 1, \S5.3]{cuevas-Fed96})
and is given by
\begin{equation}\label{eq:cuevas-Clifford}
N=2^{\lfloor(n+1)/2\rfloor}.
\end{equation}
Notice that this relation implies that in the three-dimensional case ($n=3$), the number of spinor components $N$ must be at least four.

Equation (\ref{eq:cuevas-Dirac_original2}) is derived from the following Lagrangian density:

\begin{equation*}\label{eq:cuevas-Dirac_Lagrangian}
 \mathcal{L}_{\mathrm{Dirac}}=\bar\psi(i\gamma\sp\mu\partial\sb\mu-m)\psi,
\end{equation*}
where the so-called \emph{Dirac conjugate} $\bar\psi$ is defined by
\begin{equation*}
 \bar\psi\equiv\psi^*\gamma^0,
\end{equation*}
with $\psi\sp\ast$
the Hermitian conjugate of $\psi$.

\subsection{The Soler model}

In 1938, Russian physicist Dmitri Ivanenko proposed a nonlinear model of self-interacting electrons, introducing the nonlinear term
$(\bar\psi\psi)\psi$ to the Dirac equation \cite{cuevas-Iva38}. This self-interaction term is based on the quantity $\bar\psi\psi=\psi\sp\ast\beta\psi$
which transforms as a scalar under Lorentz transformations. In 1970, Spanish physicist Mario Soler re-introduced this model in order to study, from a classical point of view, extended nucleons interacting with their own electromagnetic field  \cite{cuevas-Sol70,cuevas-Sol73}. Now this equation (or, rather, its version with an arbitrary function
of $\bar\psi\psi$)
is known as the Soler model \cite{cuevas-CV86,cuevas-Mer88,cuevas-ES95}:
\begin{equation}\label{eq:cuevas-Soler}
i\partial_t\psi=D_m\psi-f(\bar\psi\psi)\beta\psi,
\qquad
\psi(t,x)\in\mathbb{C}^N,
\quad
x\in\mathbb{R}^n,
\end{equation}
or, in the covariant form,
\begin{eqnarray*}\label{eq:cuevas-Soler2}
i\gamma\sp\mu\partial\sb\mu\psi=(m-f(\bar\psi\psi))\psi,
\end{eqnarray*}
where
$f\in C(\mathbb{R})$, $f(0)=0$.
Equation (\ref{eq:cuevas-Soler}) admits solitary wave solutions
of the form $\psi(t,x)=\phi_\omega(x)e^{-i\omega t}$,
with $\phi_\omega(x)$ exponentially localized in space
\cite{cuevas-Sol70,cuevas-VAZ77,cuevas-CV86,cuevas-Mer88,cuevas-ES95,cuevas-BC16}.
In addition, the equation is a $\mathbf{U}(1)$-invariant, relativistically invariant hamiltonian system, with the Hamiltonian represented by the density
\begin{equation}\label{eq:cuevas-hamiltonian}
\mathcal{H}_{\mathrm{Soler}}(\psi)=\psi\sp\ast D_m\psi-F(\psi\sp\ast\beta\psi),
\end{equation}
with
\begin{equation*}\label{eq:cuevas-def-F}
F(s)=\int_0^s f(t)\,dt
\end{equation*}
the antiderivative of $f$. Because of the $\psi\sp\ast D_m\psi$-term, this Hamiltonian functional is unbounded from below.
The Soler model \eqref{eq:cuevas-Soler} is also characterized by the Lagrangian density
\begin{equation*}\label{eq:cuevas-Soler_Lagrangian}
 \mathcal{L}_{\mathrm{Soler}}=\bar\psi(i\gamma\sp\mu\partial\sb\mu-m)\psi+F(\bar\psi\psi).
\end{equation*}
The $\mathbf{U}(1)$-symmetry of the Soler equation
leads to the conservation of the value of the charge functional,
given by
\begin{equation*}\label{eq:cuevas-charge}
Q(\psi(t))=\int\sb{\mathbb{R}^n}
\psi(t,x)\sp\ast\psi(t,x)\,dx,
\end{equation*}
%
which is conserved in time
(one needs to assume that the solution is smooth enough,
allowing the integration by parts).
If $\psi(t,x)$ is a solution to \eqref{eq:cuevas-Soler},
then both the charge $Q(\psi(t))$ and
the energy $E(\psi(t))=\int\mathcal{H}_{\mathrm{Soler}}(\psi(t,x))\,dx$
are conserved in time (formally; that is, as long as $\psi$ is sufficiently smooth).

A common choice of the nonlinearity is $f(s)=\vert s\vert^k$, $k>0$; this leads to $F(s)=s\vert s\vert^k/(k+1)$. We note that the absolute value is needed
when $k$ is not an integer
since the quantity $s=\bar\psi\psi$ could be negative.
Let us mention that for $k\in(0,1)$, the function $f(s)=\vert{s}\vert^k$ is not differentiable at $s=0$, which leads to certain difficulties in the construction of the solitary waves; see \cite{cuevas-BC16}.

We want to remark that the cubic Soler model
\begin{eqnarray}\label{eq:cuevas-nld-cubic}
i\partial\sb t\psi=D_m\psi-\bar\psi\psi\beta\psi,
\end{eqnarray}
which appeared in \cite{cuevas-Iva38,cuevas-Sol70}, differs from \eqref{eq:cuevas-Soler} with $f(s)=\vert{s}\vert^k$, $k=1$:

\begin{eqnarray}\label{eq:cuevas-nld-pseudocubic}
i\partial\sb t\psi=D_m\psi-\vert{\bar\psi\psi}\vert\beta\psi.
\end{eqnarray}
Both equations \eqref{eq:cuevas-nld-cubic} and \eqref{eq:cuevas-nld-pseudocubic}
are relativistically invariant Hamiltonian systems.
In particular, both equations are invariant under the time reversal
and parity transformation,
which are elements of the complete Lorentz group,
given respectively by
(see e.g. \cite{cuevas-BD64})
\[
\psi_T(t,x)
=i\gamma^1\gamma^3 K\psi(-t,x),
\]
with $K:\,\mathbb{C}^4\to \mathbb{C}^4$ the complex conjugation, and
\[
\psi_P(t,x)=\gamma^0\psi(t,-x).
\]
At the same time,
since $\bar\psi_C\psi_C=-\bar\psi\psi$, where the charge conjugation is given by
\cite{cuevas-BD64}
\[
\psi_C(t,x)=-i\gamma^2 K\psi(t,x),
\]
equation \eqref{eq:cuevas-nld-pseudocubic} is invariant under the charge conjugation,
while equation \eqref{eq:cuevas-nld-cubic} is not.
Let us mention that
the choice of unitary factors
in all these three transformations is not important.

We also point out that
the stationary waves $\phi_\omega e^{-i\omega t}$ constructed in \cite{cuevas-CV86}
in the three-dimensional case
satisfy $\bar\phi_\omega\phi_\omega>0$ for all $x\in\mathbb{R}^3$,
thus being solutions to both \eqref{eq:cuevas-nld-cubic} and \eqref{eq:cuevas-nld-pseudocubic}.


\subsection{One-dimensional Soler model}
\label{subsec:cuevas-1DSoler}

The Soler model in one spatial dimension,
Eq. (\ref{eq:cuevas-Soler}) with $n=1$, is also known as the Gross--Neveu model \cite{cuevas-GN74}.
According to relation (\ref{eq:cuevas-Clifford}),
one can take $N=2$, so that the wavefunction is represented by a bi-spinor (i.e. a spinor with only two complex components).
We will choose $\alpha^1=-\sigma_2$, $\beta=\sigma_3$.
In this case, the nonlinear Dirac equation (\ref{eq:cuevas-Soler}) can be written as a system of coupled partial differential equations of the form
\begin{equation}\label{eq:cuevas-Soler1D}
 \begin{split}
 i\partial_t\psi_1 &=\,\, \partial_x\psi_2
+(m-f(|\psi_1|^2-|\psi_2|^2))\psi_1,
\\
 i\partial_t\psi_2 &= -\partial_x\psi_1
-(m-f(|\psi_1|^2-|\psi_2|^2))\psi_2,
 \end{split}
\end{equation}
where $\psi_1,\,\psi_2\in\mathbb{C}$ denote the two components of $\psi(t,x)\in\mathbb{C}^2$.
Notice that in this equation, the spinor components are coupled both in the dispersive term and within the nonlinearity.

The focus of the present chapter is on solitary wave solutions. To this aim, we will search for standing waves of the form
\begin{equation*}
 \psi(t,x)=\phi_\omega(x)e^{-i\omega t}, \qquad
 \phi_\omega(x)=\left[\begin{array}{c}
 v(x,\omega) \\
 u(x,\omega)
 \end{array}\right] \in \mathbb{R}^2,
\end{equation*}
with $v(x,\omega)$ and $u(x,\omega)$ satisfying
\begin{equation}\label{eq:cuevas-stat1D}
 \begin{split}
 \omega v &= \,\,\partial_x u+[m-f(v^2-u^2)]v,\\
 \omega u &= -\partial_x v-[m-f(v^2-u^2)]u.
 \end{split}
\end{equation}

Once such standing wave solutions are calculated using the methods explained in Subsection \ref{subsec:cuevas-Soler2D}, their linear stability is considered by means of a Bogoliubov--de Gennes (BdG) linearized stability analysis. That is, given a solitary wave solution $\phi_\omega(x)e^{-i\omega t}$ with $\phi_\omega(x)\in\mathbb{R}^2$, we consider its perturbation in the form $\psi(t,x)=(\phi_\omega(x)+\rho(t,x))e^{-i\omega t}$, with $\rho(t,x)\in\mathbb{C}^2$. Then, the linearized equations on $R(t,x)=[\mathrm{Re}(\rho),\mathrm{Im}(\rho)]^T\in\mathbb{R}^4$ can be written as (see e.g. \cite{cuevas-BC12a})
\begin{equation}\label{eq:cuevas-JL}
 \partial_t R=\mathcal{A}\sb\omega\,R,
\end{equation}
with
\begin{equation}\label{eq:cuevas-stab1D}
\mathcal{A}\sb\omega=
\left[
\begin{array}{cc}
0 & L_{-}(\omega)
\\[2ex]
-L_{+}(\omega) & 0
\end{array}
\right],
\end{equation}
where $L_{+}(\omega)$ and $L_{-}(\omega)$ are the following self-adjoint operators:
\begin{eqnarray}
&&
L_{-}(\omega)=
\left(
\begin{array}{cc}
m-f(\tau)-\omega & \partial_x
\\[1ex]
-\partial_x & -m+f(\tau)-\omega
\end{array}
\right),
\nonumber
\\[2ex]
&&
L_{+}(\omega)=L_{-}(\omega)
-
2f'(\tau)\left(
\begin{array}{cc}
v^2&-v u
\\[1ex]
-v u&u^2
\end{array}
\right),
\nonumber
\end{eqnarray}
with $f(\tau)$ and $f'(\tau)$ evaluated at $\tau\equiv v^2-u^2$.

The potential presence of an eigenvalue with non-zero real part in the spectrum of $\mathcal{A}\sb\omega$
suggests the dynamical instability; the corresponding solitary wave is called linearly unstable.
If all the eigenvalues are purely imaginary, then the solitary wave is called spectrally (neutrally) stable.

\subsection{Two-dimensional Soler model}

Taking into account the relation given by expression (\ref{eq:cuevas-Clifford}), in two spatial dimensions one can again consider two-component spinors.
Following \cite{cuevas-CGG14}, a convenient choice for $\alpha$ and $\beta$ matrices is $\alpha^1=\sigma_1$, $\alpha^2=\sigma_2$, $\beta=\sigma_3$. With this in mind, equation (\ref{eq:cuevas-Soler}) is expressed as

\begin{equation}\label{eq:cuevas-Soler2Dcartesian}
 \begin{split}
 i\partial_t\psi_1 =& -(i\partial_x+\partial_y)\psi_2+[m-f(|\psi_1|^2-|\psi_2|^2)]\psi_1, \\
 i\partial_t\psi_2 =& -(i\partial_x-\partial_y)\psi_1-[m-f(|\psi_1|^2-|\psi_2|^2)]\psi_2.
 \end{split}
\end{equation}

In order to simplify further analysis, we use the polar coordinates
$r=\vert{\mathbf{r}}\vert$ and $\theta$;
then equation~(\ref{eq:cuevas-Soler2Dcartesian}) takes the form
\begin{equation}\label{eq:cuevas-Soler2D}
 \begin{split}
 i\partial_t\psi_1 =& - e^{-i\theta}\left(i\partial_r+\frac{\partial_\theta}{r}\right)\psi_2+[m-f(|\psi_1|^2-|\psi_2|^2)]\psi_1,
\\
 i\partial_t\psi_2 =& \, - e^{i\theta}\left(i\partial_r-\frac{\partial_\theta}{r}\right)\psi_1-[m-f(|\psi_1|^2-|\psi_2|^2)]\psi_2,
 \end{split}
\end{equation}
with $r\in(0,\infty)$ and $\theta\in[0,2\pi)$. The form of this equation suggests
to search for stationary (standing wave) solutions in the form
\begin{equation}\label{eq:cuevas-spinor2D}
 \psi(t,\mathbf{r})=\phi_\omega(\mathbf{r})e^{-i\omega t}, \qquad
 \phi_\omega(\mathbf{r})=\left[\begin{array}{c}
 v(r,\omega) e^{i S\theta} \\
 i\,u(r,\omega) e^{i(S+1)\theta}
 \end{array}\right] ,
\end{equation}
with $v(r,\omega)$ and $u(r,\omega)$ real-valued. The value $S\in\mathbb{Z}$ can be cast as the vorticity of the first spinor component. Thus, according to equations (\ref{eq:cuevas-Soler2D}) and (\ref{eq:cuevas-spinor2D}),
the equations for the stationary solutions read as follows:

\begin{equation}\label{eq:cuevas-stat2D}
 \begin{split}
 \omega v =& \,\,\, \left(\partial_r+\frac{S+1}{r}\right)u+[m-f(v^2-u^2)]v \,,\\
 \omega u =& -  \left(\partial_r-\frac{\phantom{+}S\phantom{1}}{r}\right) v-[m-f(v^2-u^2)]u.
 \end{split}
\end{equation}
This set of equations only depends on the radial coordinate $r$. The absence of angular coordinates turns the determination of stationary solutions into a one-dimensional problem, substantially simplifying the numerics.

To examine the spectral stability of a solitary wave, we consider a solution $\psi$ in the form of a perturbed solitary wave:
\begin{equation*}
\psi(t,\mathbf{r})
=\left[
 \begin{array}{c}
 \big(v(r,\omega)+\xi_1(t,r,\theta)+i\eta_1(t,r,\theta)\big)e^{i S\theta} \\[2ex]
\! i\big(u(r,\omega)+\xi_2(t,r,\theta)+i\eta_2(t,r,\theta)\big)e^{i(S+1)\theta}
 \end{array}
\!\right]
 e^{-i\omega t},
\end{equation*}
with small perturbations $\xi(t,r,\theta)=[\xi_1,  \xi_2]^T\in\mathbb{R}^2$,
$\eta(t,r,\theta)=[\eta_1,  \eta_2]^T \in \mathbb{R}^2$.

The linearized equation on $R(t,r,\theta)=[\xi_1,\xi_2,\eta_1,\eta_2]^T\in\mathbb{R}^4$ has the form
\begin{equation}\label{eq:cuevas-stab2D0}
 \partial_t R=\mathcal{A}\sb\omega R,
\end{equation}
with $\mathcal{A}\sb\omega(r,\theta,\partial_r,\partial_\theta)$ a matrix-valued first order differential operator
\begin{equation}\label{eq:cuevas-stab2D}
\mathcal{A}\sb\omega(r,\theta,\partial_r,\partial_\theta)
=
\left[
\begin{array}{cc}
-\sigma_1\frac{\partial_\theta}{r} &L_{-}(\omega)
\\[2ex]
-L_{+}(\omega)&-\sigma_1\frac{\partial_\theta}{r}
\end{array}
\right],
\end{equation}
where
\begin{eqnarray}
&&
L_{-}(\omega)=
\left(
\begin{array}{cc}
m-f(\tau)-\omega & \partial_r+\frac{S+1}{r}
\\[1ex]
-(\partial_r-\frac{S}{r}) & -m+f(\tau)-\omega
\end{array}
\right),
\nonumber
\\[2ex]
&&
L_{+}(\omega)=L_{-}(\omega)
-
2f'(\tau)\left(
\begin{array}{cc}
v^2&-v u
\\[1ex]
-v u&u^2
\end{array}
\right),
\nonumber
\end{eqnarray}
with $f(\tau)$ and $f'(\tau)$ evaluated at $\tau\equiv v^2-u^2$.

To find the spectrum of the operator $\mathcal{A}\sb\omega$,
we consider it in the space of $\mathbb{C}^4$-valued functions. The key observation which facilitates
a computation of the spectrum is that the explicit form (\ref{eq:cuevas-stab2D}) of $\mathcal{A}\sb\omega$ contains $r$, $\partial_r$, $\partial_\theta$, but not $\theta$. As a consequence, $\mathcal{A}\sb\omega$ is invariant in the spaces which correspond to the Fourier decomposition with respect to $\theta$,
\begin{equation*}
 \mathcal{X}_q=\left\{\left[a_1(r);a_2(r);b_1(r);b_2(r)\right] e^{i q\theta}\right\} \qquad
 q\in\mathbb{Z}.
\end{equation*}

The restriction of $\mathcal{A}\sb\omega$ to each such subspace is given by

\begin{equation}\label{eq:cuevas-stab2D_partial}
\mathcal{A}_{\omega,q}(r,\partial_r)=\mathcal{A}\sb\omega\vert_{\mathcal{X}_q}
=
\left[
\begin{array}{cc}
-\sigma_1\frac{i q}{r} & L_{-}(\omega)
\\[2ex]
-L_{+}(\omega) & -\sigma_1\frac{i q}{r}
\end{array}
\right],
\qquad
q\in\mathbb{Z},
\end{equation}
and this allows to compute the spectrum of $\mathcal{A}\sb\omega$ as the union of spectra of the one-dimensional spectral problems
\begin{equation*}
 \sigma\left(\mathcal{A}\sb\omega\right)
=\mathop{\bigcup}_{q\in\mathbb{Z}}\sigma\left(\mathcal{A}_{\omega,q}\right),
\end{equation*}
where the operators $\mathcal{A}_{\omega,q}$ do not contain the angular variable.

\subsection{Three-dimensional Soler model}\label{subsec:cuevas-Soler3D}

In three spatial dimensions, it is convenient
to consider equation (\ref{eq:cuevas-Soler}) in spherical coordinates.
We consider the 4-spinor solitary waves in the form of the Wakano Ansatz \cite{cuevas-Wak66}:
\begin{equation*}\label{eq:cuevas-spinor3D}
 \psi(t,\mathbf{r})=\phi_\omega(\mathbf{r})e^{-i\omega t}, \qquad
 \phi_\omega(\mathbf{r})=\left[
 \begin{array}{c}
 v(r,\omega)
 \left(\begin{array}{c}1\\0\end{array}\right)
 \\
 i u(r,\omega)
 \left(\begin{array}{c}\cos\theta
\\
e^{i\varphi}\sin\theta\end{array}\right)
 \end{array}
 \right] ,
\end{equation*}
with real-valued $v(r,\omega)$, $u(r,\omega)$ satisfying
\begin{equation}\label{eq:cuevas-stat3D}
 \begin{split}
 \omega v=& \,\, \left(\partial_r+\frac{2}{r}\right)u+[m-f(v^2-u^2)^k]v\,,\\
 \omega u =& -  \partial_r v-[m-f(v^2-u^2)^k]u.
 \end{split}
\end{equation}

To study the linearization operator in the invariant space which has the same angular structure
as the solitary waves, we consider the perturbed solutions in the form

\begin{equation*}
\psi(t,\mathbf{r})
=
\left[
\begin{array}{c}
(v(r,\omega)+\xi_1(t,r)+i\eta_1(t,r))
\left(\begin{array}{c}1\\0\end{array}\right)
\\
i(u(r,\omega)+\xi_2(t,r)+i\eta_2(t,r))
\left(\begin{array}{c}\cos\theta
\\
e^{i\varphi}\sin\theta\end{array}\right)
\end{array}
\right]e^{-i\omega t},
\end{equation*}
with real-valued $\xi=[\xi_1, \xi_2]^T\in\mathbb{R}^2$, $\eta=[\eta_1, \eta_2]^T\in\mathbb{R}^2$
(note that the considered perturbation only depends on $r$ but not on the angular variables).
The linearized equation on $R(t,r)=(\xi_1,\xi_2,\eta_1,\eta_2)^T$ is similar to equations (\ref{eq:cuevas-stab2D0}), (\ref{eq:cuevas-stab2D}):

\begin{equation*}
 \partial_t R=\mathcal{A}\sb\omega R
 \quad
\mathrm{with}
 \quad
\mathcal{A}\sb\omega=
 \left[
 \begin{array}{cc}
 0 & L_{-}(\omega) \\[2ex]
 -L_{+}(\omega) & 0
 \end{array}
 \right],
\end{equation*}
and
\begin{eqnarray}
&&
L_{-}(\omega)=
\left(
\begin{array}{cc}
m-f(\tau)-\omega & \partial_r+\frac{2}{r}
\\[1ex]
-\partial_r & -m+f(\tau)-\omega
\end{array}
\right),
\nonumber
\\[2ex]
&&
L_{+}(\omega)=L_{-}(\omega)
-
2f'(\tau)\left(
\begin{array}{cc}
v^2&-v u
\\[1ex]
-v u&u^2
\end{array}
\right),
\nonumber
\end{eqnarray}
with $f(\tau)$ and $f'(\tau)$ evaluated at $\tau\equiv v^2-u^2$.

\section{Solitary waves: exact solutions and numerical methods}
\label{sec:cuevas-existence}

Solitary wave solutions of the form $\phi_\omega(x)e^{-{i}\omega t}$, $\omega\in(0,m)$, are known to exist in \eqref{eq:cuevas-Soler} and in other important systems based on the Dirac equation (see e.g. the review \cite{cuevas-ELS08}).
In the one-dimensional case, for pure power nonlinearity,
the solutions are available in a closed form; see Section~\ref{sect-solitary-waves-1d}.
However, for higher-dimensional cases, solitary wave and vortex solutions must be obtained by means of numerical methods. These methods can also be applied to 1D models with a general nonlinearity $f$
in (\ref{eq:cuevas-Soler1D}) when the solutions are not available in a closed form.

\subsection{One-dimensional Soler model: exact solutions}
\label{sect-solitary-waves-1d}

\begin{figure}[tb]
\begin{center}
\begin{tabular}{cc}
\includegraphics[width=.45\textwidth]{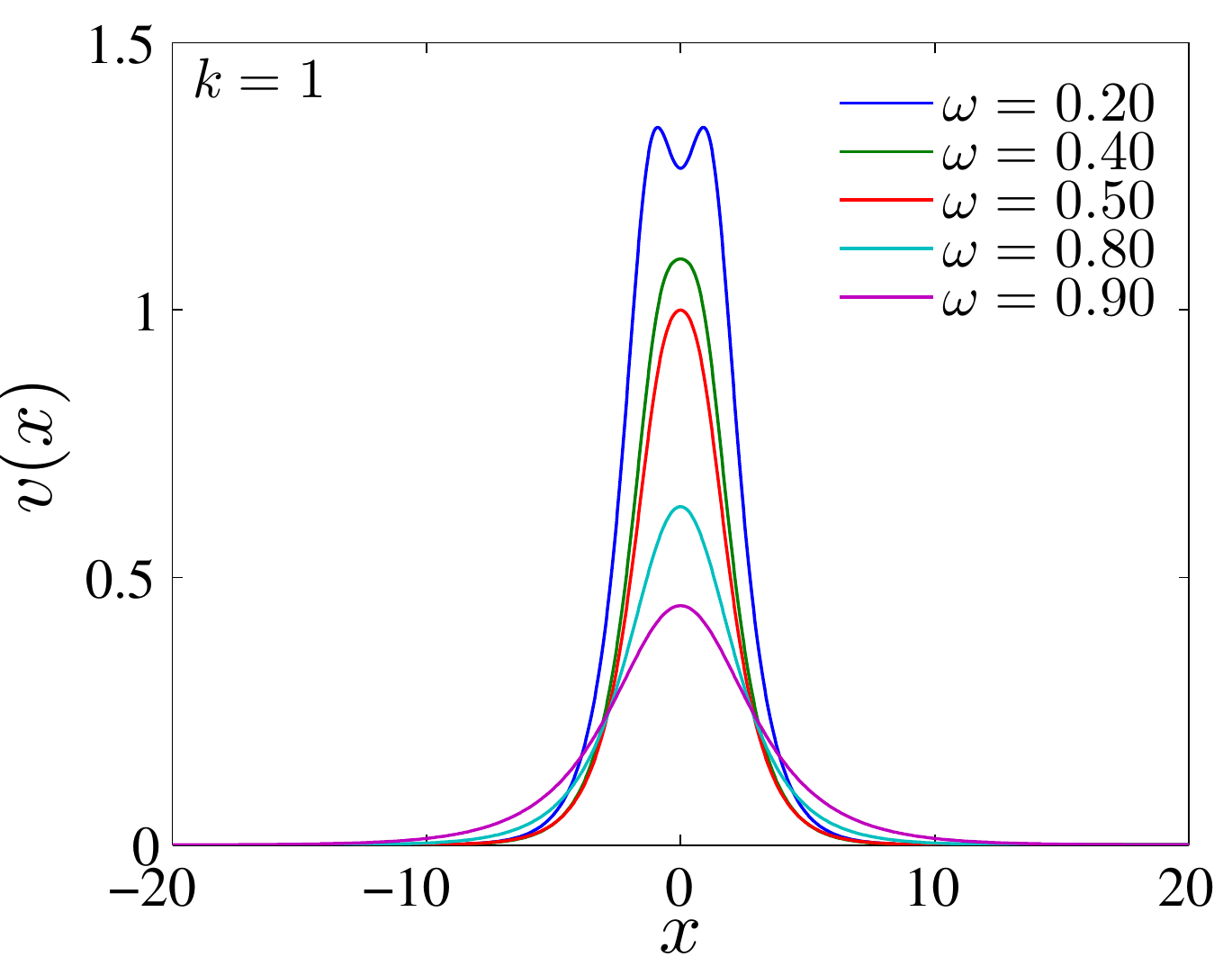} \hfill &
\includegraphics[width=.45\textwidth]{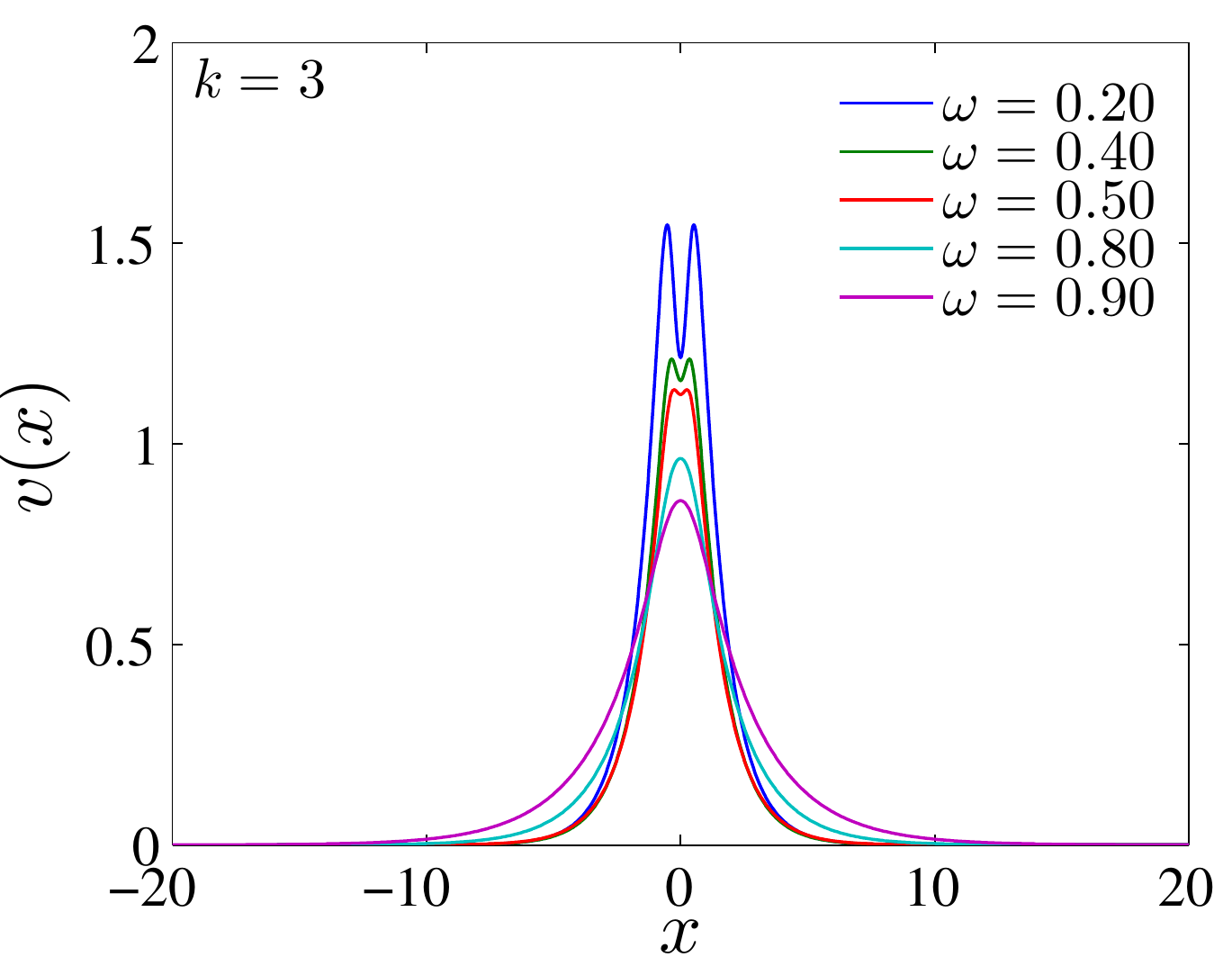} \\
\includegraphics[width=.45\textwidth]{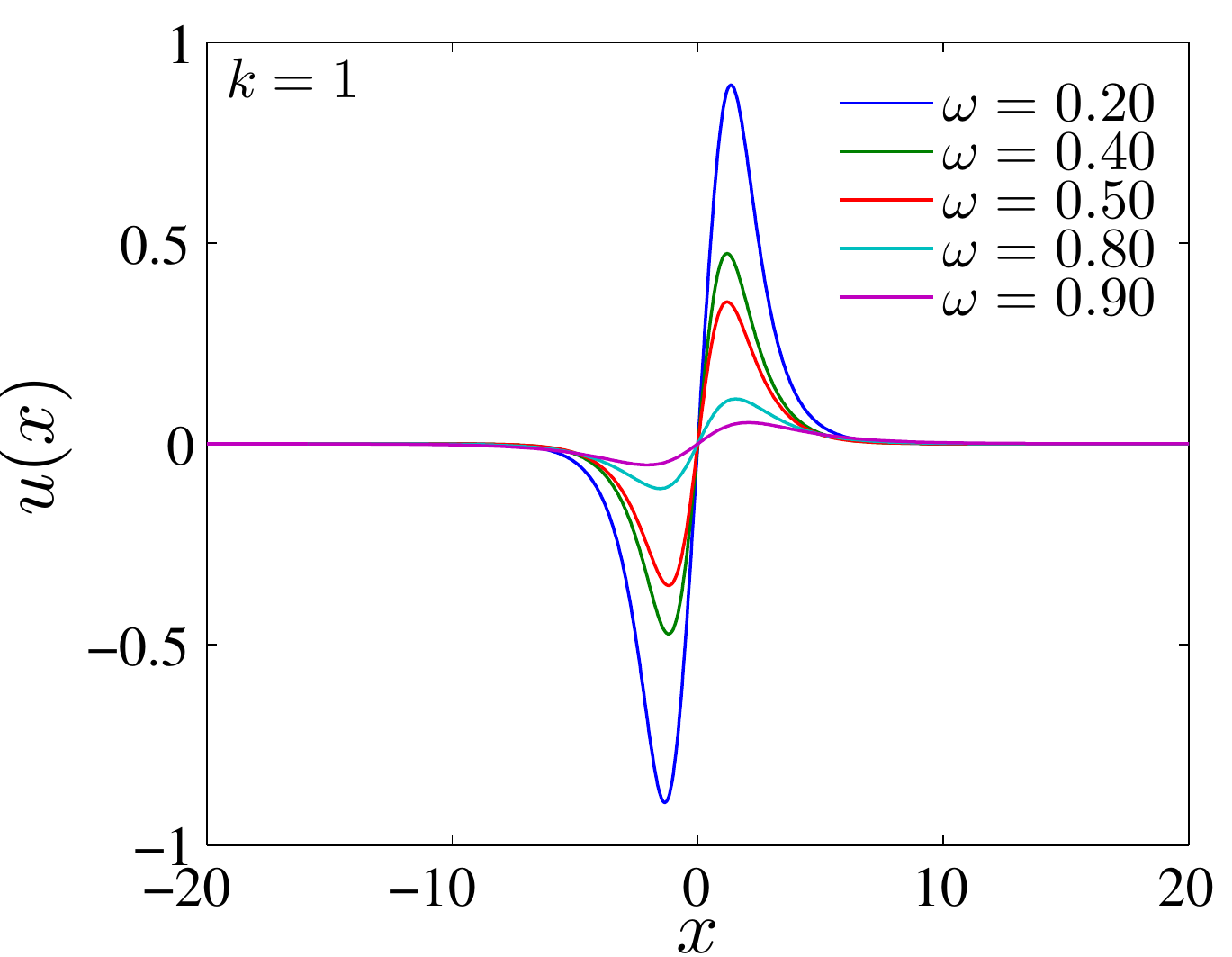} \hfill &
\includegraphics[width=.45\textwidth]{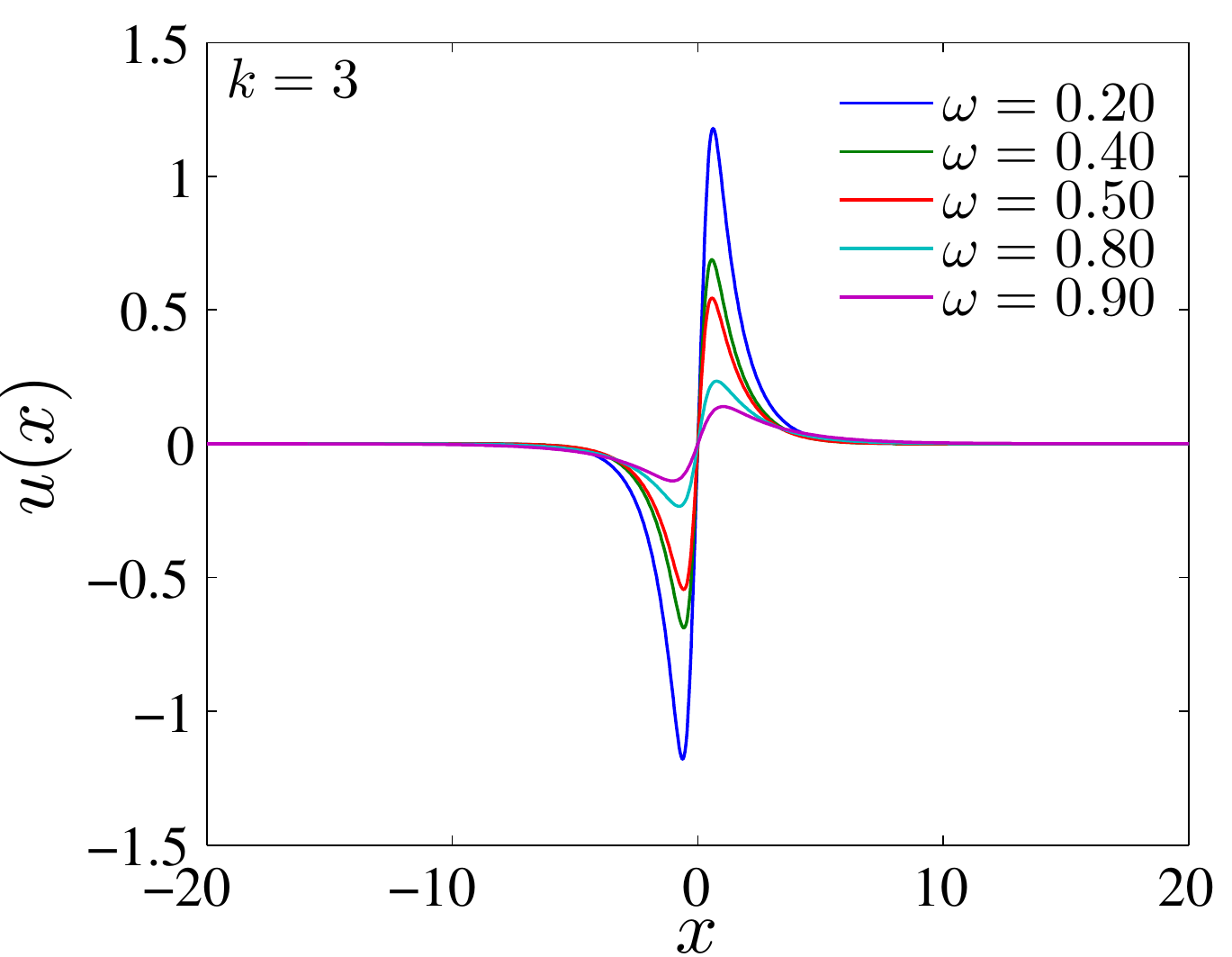} \\
\includegraphics[width=.45\textwidth]{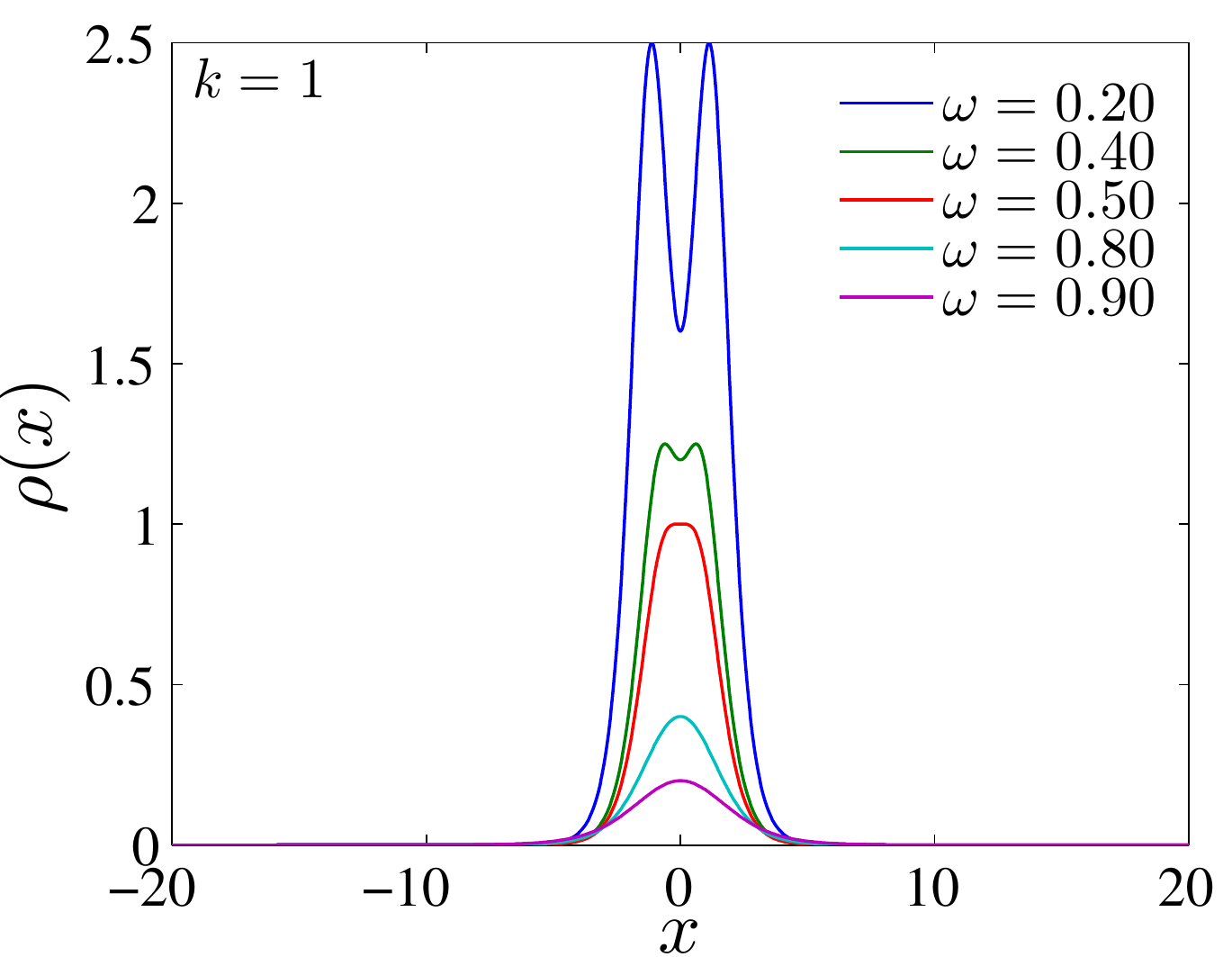} \hfill &
\includegraphics[width=.45\textwidth]{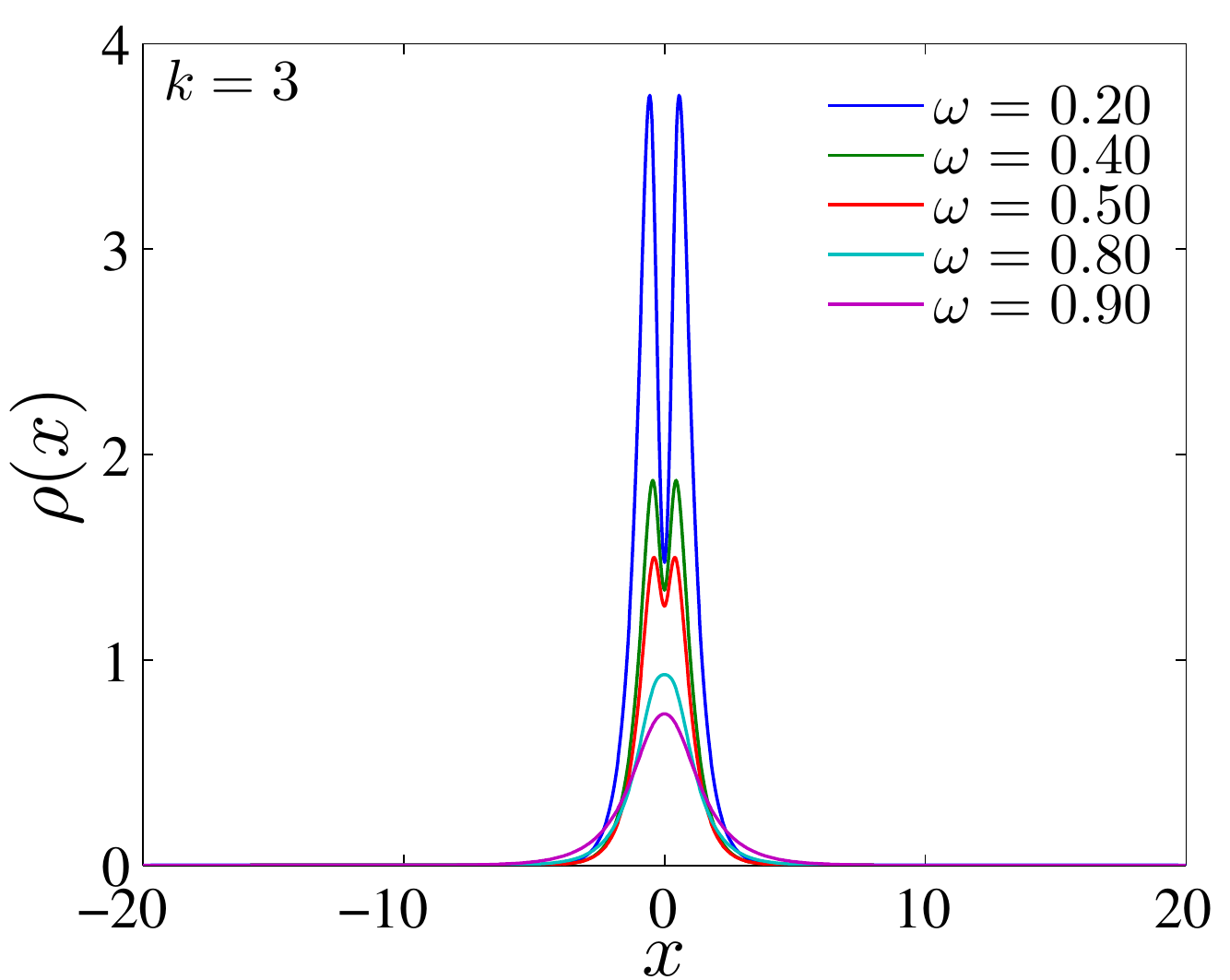} \\
\\
\end{tabular}
\end{center}
\caption{Profile of solitary waves in the 1D Soler model. Figures depict the first and second spinor components together with the solitary wave density. Left (right) panels correspond to $k=1$ ($k=3$)}
\label{cuevas-fig1}
\end{figure}

\begin{figure}[tb]
\begin{center}
\begin{tabular}{cc}
\includegraphics[width=.45\textwidth]{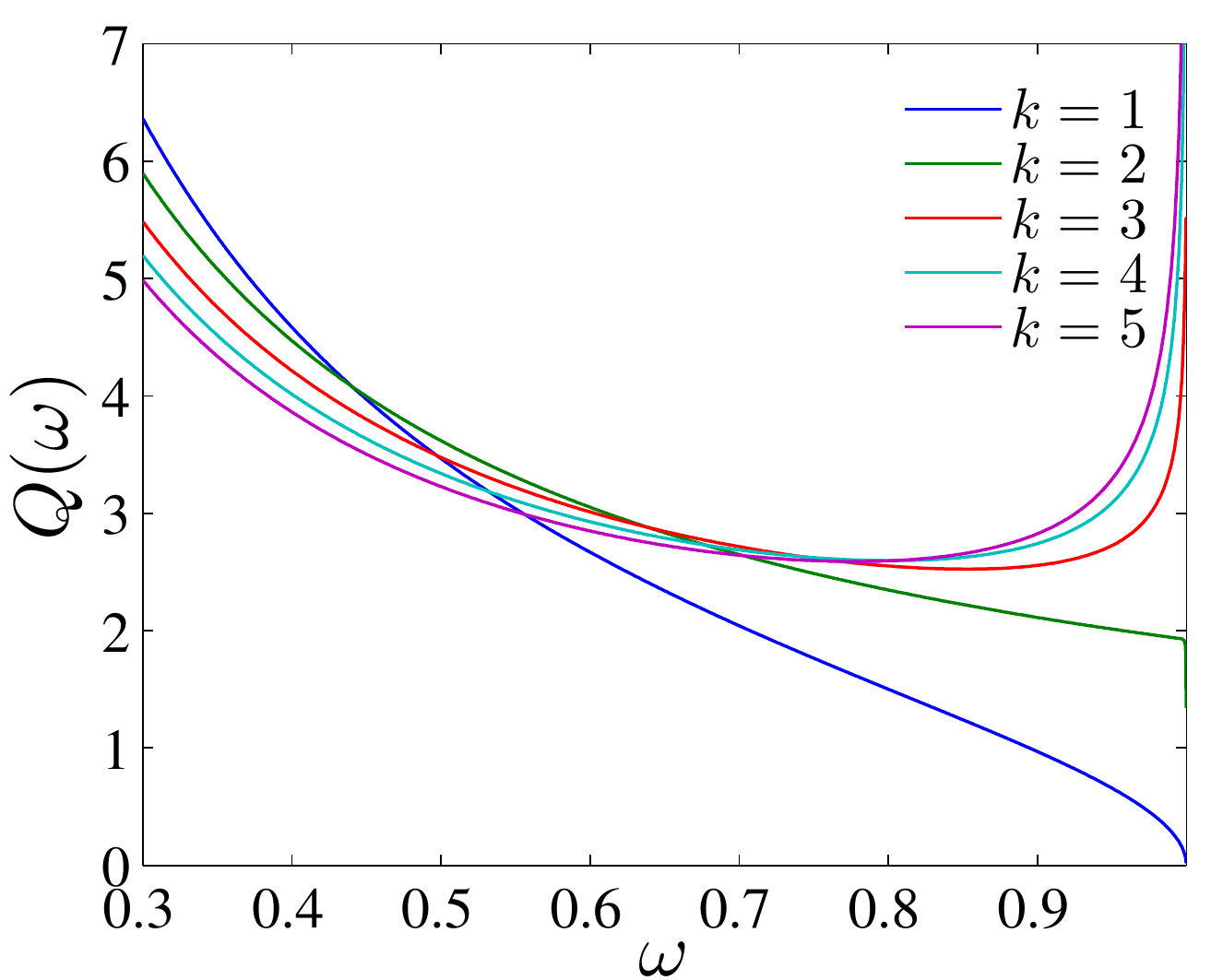} \hfill &
\includegraphics[width=.45\textwidth]{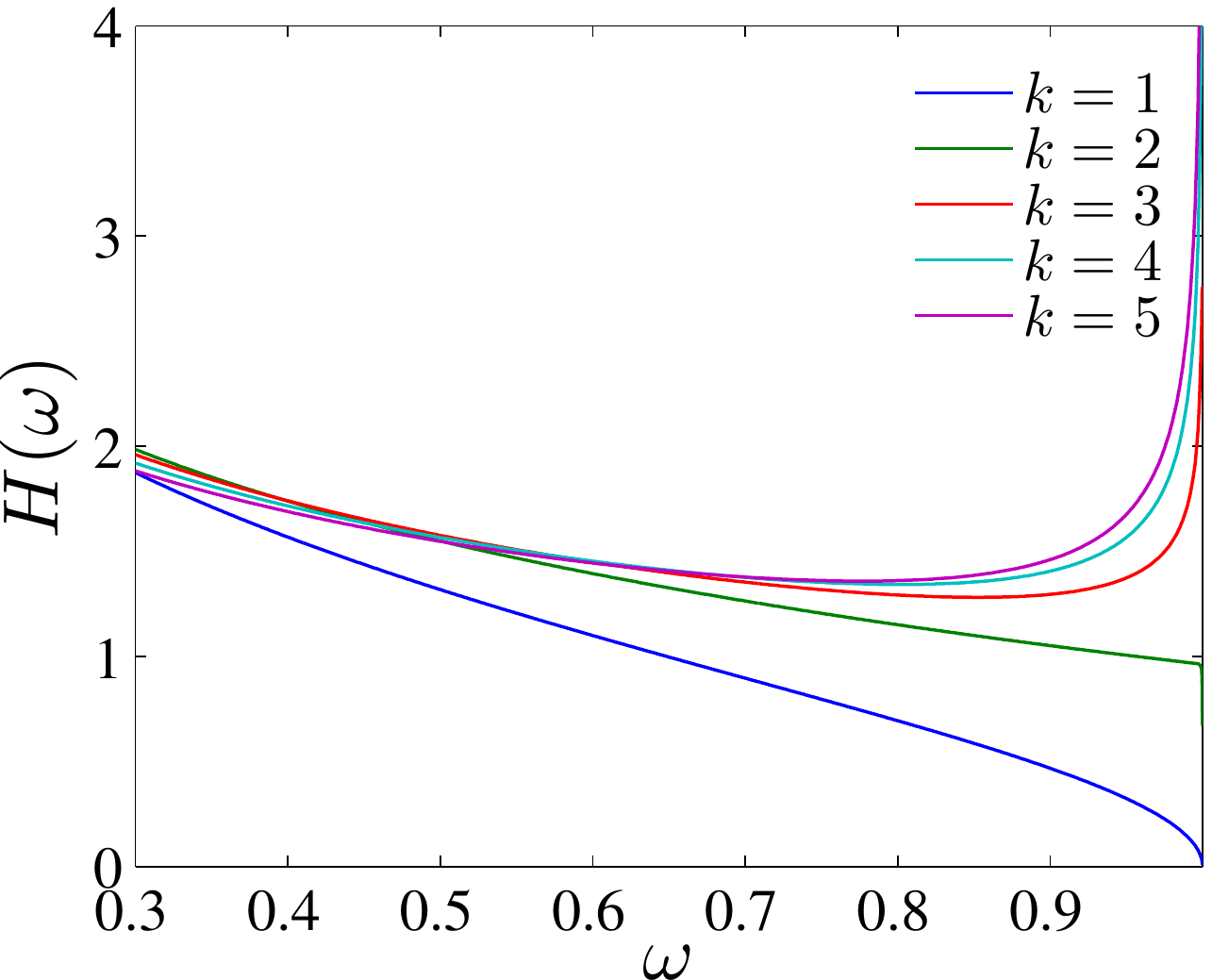} \\
\\
\end{tabular}
\end{center}
\caption{Charge and energy of solitary waves
(left and right panel, respectively) as functions of the frequency $\omega$
for $1\leq k\leq5$. Notice the existence of a minimum in the curve $Q(\omega)$ for $k>2$, which is related to the change in stability properties (see Section \ref{sec:cuevas-stability}).
}
\label{cuevas-fig2}
\end{figure}

In \cite{cuevas-LKG75} it was shown for the cubic nonlinearity, i.e. $k=1$ in (\ref{eq:cuevas-Soler1D}), and later in \cite{cuevas-CP06,cuevas-CKMS10,cuevas-MQC+12} for generic value $k>0$, that the solitary wave solutions can be found in a closed form for any $\omega\in(0,m)$:
\begin{eqnarray}\label{eq:cuevas-soliton1D}
 v(x)&=&\cosh(k\beta x)\sqrt{\frac{(m+\omega)}{m+\omega\cosh(2k\beta x)}}\left[\frac{(k+1)\beta^2}{m+\omega\cosh(2k\beta x)}\right]^{1/2k},\nonumber \\
 u(x)&=&\sinh(k\beta x)\sqrt{\frac{(m-\omega)}{m+\omega\cosh(2k\beta x)}}\left[\frac{(k+1)\beta^2}{m+\omega\cosh(2k\beta x)}\right]^{1/2k},
\end{eqnarray}
where $\beta=\sqrt{m^2-\omega^2}$. In the special case of $k=1$, waveforms in Eq.~(\ref{eq:cuevas-soliton1D}) reduce to
\begin{equation*}
 v(x)=\frac{\sqrt{2(m-\omega)}}{[1-\mu\tanh^2(\beta x)]\cosh(\beta x)},\quad
 u(x)=\frac{\sqrt{2\mu(m-\omega)}\tanh(\beta x)}{[1-\mu\tanh^2(\beta x)]\cosh(\beta x)} ,
\end{equation*}
with $\mu=(m-\omega)/(m+\omega)$. Fig. \ref{cuevas-fig1} shows the profiles of solitary waves given by the expression (\ref{eq:cuevas-soliton1D}) for $k=1$ and $k=3$. Notice that the first component of the spinor, $v(x)$, is spatially even, whereas the second component $u(x)$ is spatially odd. Moreover, $v^2(x)-u^2(x)>0$ for all $x\in\mathbb{R}$, so that the solitary waves satisfy the nonlinear Dirac equation \eqref{eq:cuevas-Soler} (with $n=1$) with both $f(s)=s$ and $f(s)=\vert{s}\vert$.
Evaluating $\rho''$ at $x=0$,
one can check that the charge density profiles $\rho(x)=\phi_\omega(x)\sp\ast\phi_\omega(x)$
(cf. \eqref{eq:cuevas-charge})
become double-humped for $\omega\leq\omega_h(k)$,
with $\omega_h(k)=m k/(k+1)$.
The dependence of the charge and energy with respect to $\omega$ for different values of $k\in\mathbb{N}$ are shown in Fig. \ref{cuevas-fig2}.

\subsection{Two-dimensional Soler model: numerical solutions} \label{subsec:cuevas-Soler2D}

No explicit solitary wave solutions
are known for the Soler model in 2D  \eqref{eq:cuevas-Soler2D}.
For this reason, one must rely on numerical results. We show in Section~\ref{subsubsec:cuevas-numerical} the numerical methods used for the numerical determination of stationary solutions in (\ref{eq:cuevas-stat2D}). These methods can easily be adapted for numerically solving the Soler 3D model (\ref{eq:cuevas-stat3D}) (in the particular case of zero vorticity) and for finding solitary wave solutions in 1D models where additional terms to the equation (\ref{eq:cuevas-stat1D}) have been added, such as external fields \cite{cuevas-MQC+12} or terms of preserving {$\mathcal{P}\mathcal{T}$} symmetry \cite{cuevas-CKS+16}.

\subsubsection{Brief summary of spectral methods}
\label{subsubsec:cuevas-numerical}

Prior to explaining the numerical methods used for calculating stationary solutions, we will proceed to present a summary of spectral methods needed for dealing with derivatives in continuum settings. For a detailed discussion on these methods, the reader is directed to \cite{cuevas-Boy01} and references therein.

Spectral methods arise due to the necessity of calculating spatial derivatives with higher accuracy than that given by finite difference methods. As shown in \cite{cuevas-CKS+15}, finite difference methods cannot be used for the stability and dynamics analysis of solitary waves in the Dirac equation.

In order to implement spectral derivatives, a differentiation matrix $\mathbf{D}\equiv\{D_{n,m}\}$ must be given together with $N$ collocation \footnote{Notice that this value of $N$ is not related to the dimension of the NLD, although the same symbol is used in both cases} (i.e. grid) points $\mathbf{x}\equiv\{x_n\}$,
$n=1,2,\ldots N$, which are not necessarily equi-spaced. Thus, if the {\em spectral} derivative of a function $\mathbf{f}(\mathbf{x})\equiv\{f_n(x_n)\}$ needs to be calculated, it can be cast as:

\begin{equation*}
 f'(x)=\partial_x f(x) \leftrightarrow f'_n=\sum_{m=1}^N D_{n,m}f_m ,
\end{equation*}
where $f_m\equiv f(x_m)$ and $f'_n\equiv f'(x_n)$. If $x\in[-L,L]$ and the boundary conditions are periodic, the Fourier collocation can be used. In this case,

\begin{equation}\label{eq:cuevas-colloc_Fourier}
 x_n=\frac{2L}{N}\left(n-\frac{N}{2}\right)\,,\quad n=1,2,\ldots N
\end{equation}
with $N$ even. The differentiation matrix is

\begin{equation*}
 D_{n,m}=\left\{\begin{array}{ll} 0 & \textrm{if } n=m , \\[2ex]
 \dfrac{\pi}{2L}\dfrac{(-1)^{n+m}}{\tan[(x_n-x_m)/2]} & \textrm{if } n\neq m . \end{array}\right.
\end{equation*}

Notice that doing the multiplication $\mathbf{D f}$ is equivalent to performing the following pair of Discrete Fourier Transform applications:

\begin{equation}\label{eq:cuevas-FFT}
 \mathbf{D f}=\mathcal{F}^{-1}\left(i\mathbf{k}\mathcal{F}(\mathbf{f})\right)\,,
\end{equation}
with $\mathcal{F}$ and $\mathcal{F}^{-1}$ denoting, respectively, the direct and inverse discrete Fourier transform \cite{cuevas-Tre00}. The vector wavenumber $\mathbf{k}=\{k_n\}$ is defined as:

\begin{equation*}
 k_n=\left\{\begin{array}{ll} \dfrac{n\pi}{L} & \textrm{if } n<N/2 , \\[2ex]
 0 & \textrm{if } n=N/2 . \end{array}\right.
\end{equation*}

The computation of the direct and inverse discrete Fourier transforms, which is useful in simulations, can be accomplished by the Fast Fourier Transform. In what follows, however, the differentiation matrix is used for finding the Jacobian and stability matrices. Notice that the grid for a finite difference discretization is the same as in the Fourier collocation; and, in addition, there is a differentiation matrix for the finite difference method, i.e.

\begin{equation}\label{eq:cuevas-difmat_fd}
 D_{n,m}=\frac{1}{2h}\left(\delta_{m,n+1}-\delta_{m,n-1}+\delta_{n,1}\delta_{m,N}-\delta_{n,N}\delta_{m,1}\right)\,, \qquad
 h=\frac{2L}{N},
\end{equation}
with $\delta$ being Kronecker's delta. It can be observed from the above discussion that in the Fourier spectral method, the banded differentiation matrix of the finite difference method is substituted by a dense matrix, or, in other words, a nearest-neighbor interaction is exchanged with a long-range one. The lack of sparsity of differentiation matrices is one of the drawbacks of spectral methods, especially when having to diagonalize large systems.
However, they have the advantage of needing (a considerably) smaller number of grid points $N$ for getting the same accuracy as with finite difference methods.

For fixed (Dirichlet) boundary conditions, the Chebyshev spectral methods are the most suitable ones. There are several collocation schemes, the Gauss--Lobatto being the most extensively used:
\begin{equation*}\label{eq:cuevas-gridCh}
 x_n=L\cos\left(\frac{n\pi}{N+1}\right)\,, \quad n=1,2,\ldots N\,,
\end{equation*}
with $N$ being even or odd. The differentiation matrix is
\begin{equation*}\label{eq:cuevas-diffCh}
 D_{n,m}=\left\{\begin{array}{ll} \dfrac{x_n}{2L(1-x_n^2)} & \textrm{if } n=m , \\[2ex]
 \dfrac{(-1)^{n+m}}{L\cos(x_n-x_m)} & \textrm{if } n\neq m . \end{array}\right.
\end{equation*}

The significant drawback of Chebyshev collocation is that the discretization matrix possesses a great number of spurious eigenvalues \cite{cuevas-Boy01}.
They are approximately equal to $N/2$. These spurious eigenvalues also have a significant non-zero real part, which increases when $N$ grows. This fact naturally reduces the efficiency of the method when performing numerical time-integration. However, it gives a higher accuracy than the Fourier collocation method when determining the spectrum of the stability matrix (see e.g. \cite{cuevas-CKS+15}).

Several modifications must be introduced when applying spectral methods to polar coordinates. They basically rely on overcoming the difficulty of not having Dirichlet boundary conditions at $r=0$ and the singularity of the equations at that point. In addition, in the case of the Dirac equation, the spinor components can be either symmetric or anti-symmetric in their radial dependence, so the method described in \cite{cuevas-Tre00,cuevas-HCC+08} must be modified accordingly. As shown in the previously mentioned references, the radial derivative of a general function $f(r,\theta)$ can be expressed as:
\begin{equation}\label{eq:cuevas-polar1}
 \partial_r f(r_n,\theta)=\sum_{m=1}^N D_{n,m}f(r_m,\theta)+D_{n,2N-m}f(r_m,\theta+\pi).
\end{equation}

Notice that in this case, the collocation points must be taken as
\begin{equation*}\label{eq:cuevas-gridpol}
 r_n=L\cos\left(\frac{n\pi}{2N+1}\right)\,, \quad n=1,2,\ldots 2N\,,
\end{equation*}
but only the first $N$ points are taken so that the domain of the radial coordinate does not include $r=0$. Analogously the differentiation matrix would
possess now $2N\times2N$ components, but only the upper half of the matrix, of size $N\times2N$ is used.

If the function that must be derived is symmetric or anti-symmetric, i.e. $f(r,\theta+\pi)=\pm f(r,\theta)$, with the upper (lower) sign corresponding to the (anti-)symmetric function, equation (\ref{eq:cuevas-polar1}) can be written as follows:
\begin{equation}\label{eq:cuevas-polar2}
 \partial_r f(r_n,\theta)=\sum_{m=1}^N \left[\left(D_{n,m}\pm D_{n,2N-m}\right)f(r_m,\theta)\right].
\end{equation}

Thus, the differentiation matrix has a different form depending on whether $f(r,\theta)$ is symmetric or anti-symmetric:

\begin{equation*}\label{eq:cuevas-polar3}
 \partial_r\mathbf{f}(\mathbf{r},\theta)=\mathbf{D^{(\pm)}f}\quad \textrm{if }\ f(r,\theta)=\pm f(r,\theta+\pi),
\end{equation*}
with $\mathbf{r}\equiv\{r_n\}$, $\mathbf{f}(\mathbf{r})\equiv\{f(r_n)\}$ and $\mathbf{D^{(\pm)}f}$ defined as in (\ref{eq:cuevas-polar2}).

\subsubsection{Fixed point methods}

Among the numerical methods available for solving nonlinear systems of equations we have chosen to use fixed point methods, such as the Newton--Raphson one \cite{cuevas-PFTV86}, which requires the transformation of the set of two coupled ordinary differential equations (\ref{eq:cuevas-stat2D}) into a set of $2N$ algebraic equations; this is performed by defining the set of collocation points $\mathbf{r}\equiv\{r_n\}$, and transforming the derivatives into multiplication of the differentiation matrices $\mathbf{D}^{(1)}$ and $\mathbf{D}^{(2)}$ (to be defined below) times the vectors $\mathbf{u}\equiv\{u_n\}$ and $\mathbf{v}\equiv\{v_n\}$, respectively, being $u_n\equiv u(r_n)$ and $v_n\equiv v(r_n)$ as explained in the previous section. Thus, the discrete version of (\ref{eq:cuevas-stat2D}) reads:
\begin{eqnarray*}
 F_n^{(1)} &\equiv& (m-\omega)v_n-g\tau_n^k v_n+\sum_m D_{n m}^{(2)}u_m+\frac{S+1}{r_n}u_n =0 , \nonumber \\
 F_n^{(2)} &\equiv& (m+\omega)u_n-g\tau_n^k u_n+\sum_m D_{n m}^{(1)}v_m+\frac{S}{r_n}v_n=0 ,
\end{eqnarray*}
with $\tau_n\equiv v_n^2-u_n^2$. It is important to notice that matrices $\mathbf{D}^{(1)}$ and $\mathbf{D}^{(2)}$ correspond to either $\mathbf{D}^{(+)}$ or $\mathbf{D}^{(-)}$, depending on the symmetry of $\mathbf{v}$ and $\mathbf{u}$, which, at the same time, depend on the value of the vorticity $S$. If $S$ is even,
then $\mathbf{v}$ and $\mathbf{u}$ are symmetric and antisymmetric, respectively,
being $\mathbf{D}^{(1)}=\mathbf{D}^{(+)}$ and $\mathbf{D}^{(2)}=\mathbf{D}^{(-)}$.
On the contrary, if $S$ is odd,
then $\mathbf{u}$ is symmetric
and
$\mathbf{v}$ is antisymmetric,
being $\mathbf{D}^{(1)}=\mathbf{D}^{(-)}$ and $\mathbf{D}^{(2)}=\mathbf{D}^{(+)}$.

In order to find the roots of the vector function $\mathbf{F}=(\{F^{(1)}_n\},\{F^{(2)}_n\})^T$, an analytical expression of the Jacobian matrix
\begin{equation*}
 \mathbf{J}=\left(\begin{array}{cc}
 \dfrac{\partial \mathbf{F^{(1)}}}{\partial \mathbf{u}} &
 \dfrac{\partial \mathbf{F^{(1)}}}{\partial \mathbf{v}} \\[3 ex]
 \dfrac{\partial \mathbf{F^{(2)}}}{\partial \mathbf{u}} &
 \dfrac{\partial \mathbf{F^{(2)}}}{\partial \mathbf{v}}
 \end{array}\right)=
 \left(\begin{array}{cc}
 (m-\omega)-g\tau^{k-1}[2k v^2+\tau] & 2kg\mathbf{v u}\tau^{k-1}+D^{(2)}+\dfrac{S+1}{r} \\[3 ex]
 -2kg\mathbf{v u}\tau^{k-1}+D^{(1)}-\dfrac{S}{r} & (m+\omega)-g\tau^{k-1}[\tau-2k v^2]
 \end{array}\right)
\end{equation*}
must be introduced, with the derivatives expressed by means of spectral methods and the matrix is evaluated at the corresponding grid points. The roots of $\mathbf{F}$, $\Phi=(\mathbf{v},\mathbf{u})^T$, are found by successive application of the iteration $\Phi\rightarrow{\Phi}-\mathbf{J}^{-1}\mathbf{F}$ until convergence is attained. In our case, we have chosen
as convergence condition that $\|\mathbf{F}\|_\infty<10^{-10}$.

Spectral stability is analyzed by
evaluating the functions appearing in matrix $\mathcal{A}\sb\omega$
of equation (\ref{eq:cuevas-stab2D}) at the collocation points and substituting the partial derivatives by the corresponding differentiation matrices. At this point, one must be very cautious because, as also occurred with the Jacobian, there will be two different differentiation matrices in our problem. Now $L_{-}(\omega)$ will be represented by the following matrix:
\begin{equation*}
L_{-}(\omega)=
\left(
\begin{array}{cc}
f(\tau)-\omega & D^{(2)}+\frac{S+1}{r}
\\[2ex]
-(D^{(1)}-\frac{S}{r}) & -f(\tau)-\omega
\end{array}
\right).
\end{equation*}

\subsubsection{Solitary waves and vortices}

\begin{figure}[tb]
\begin{center}
\begin{tabular}{cc}
\includegraphics[width=.45\textwidth]{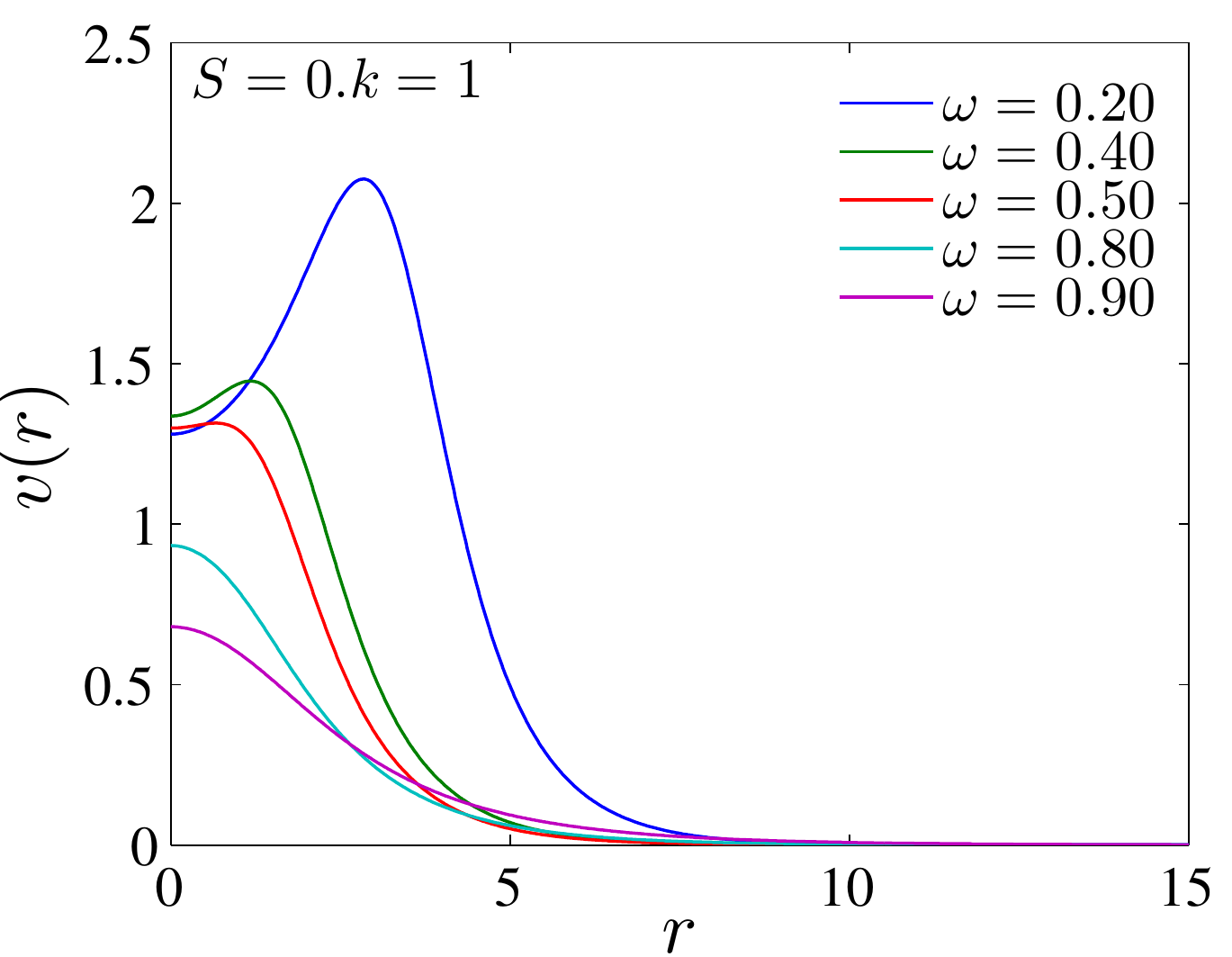} \hfill &
\includegraphics[width=.45\textwidth]{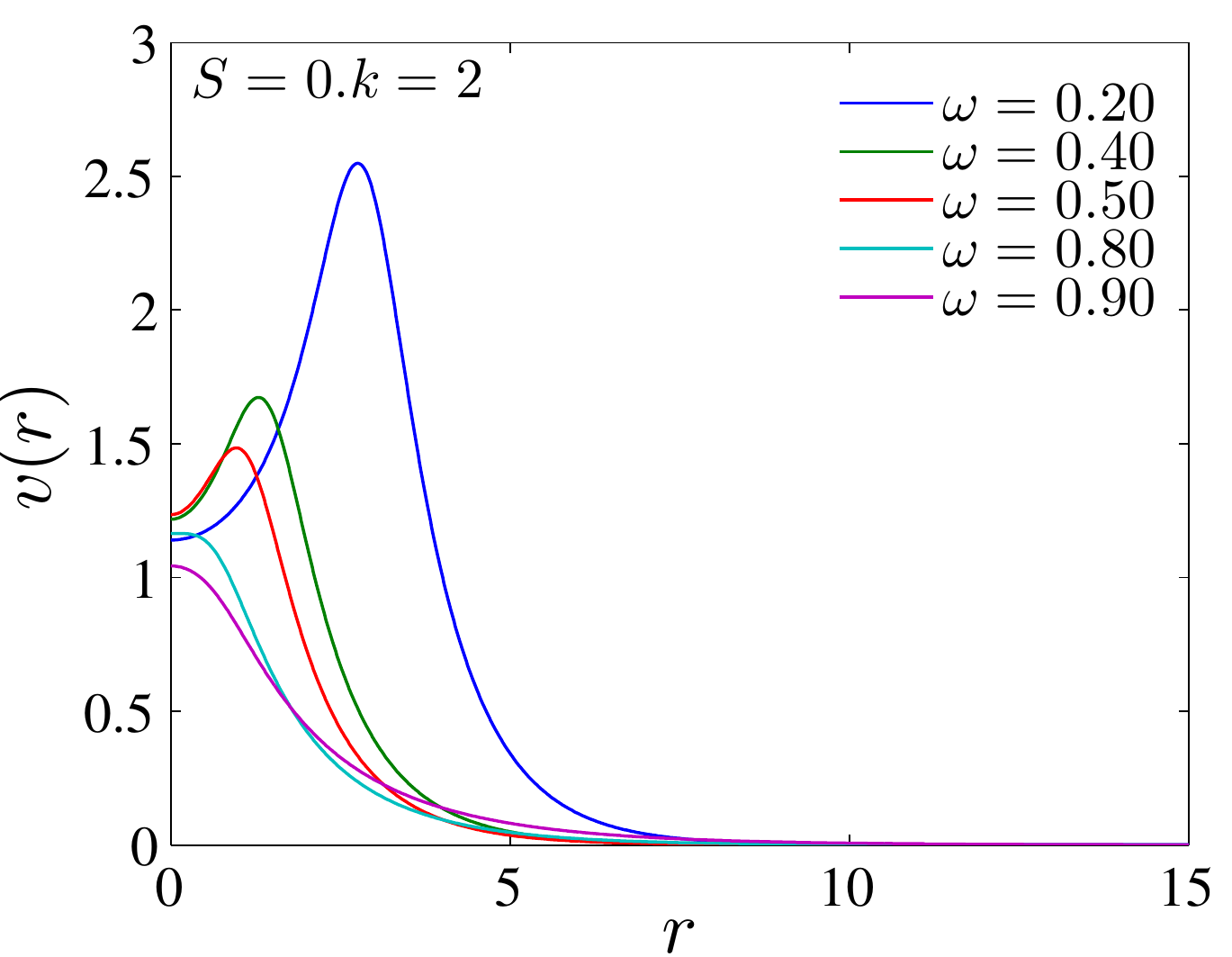} \\
\includegraphics[width=.45\textwidth]{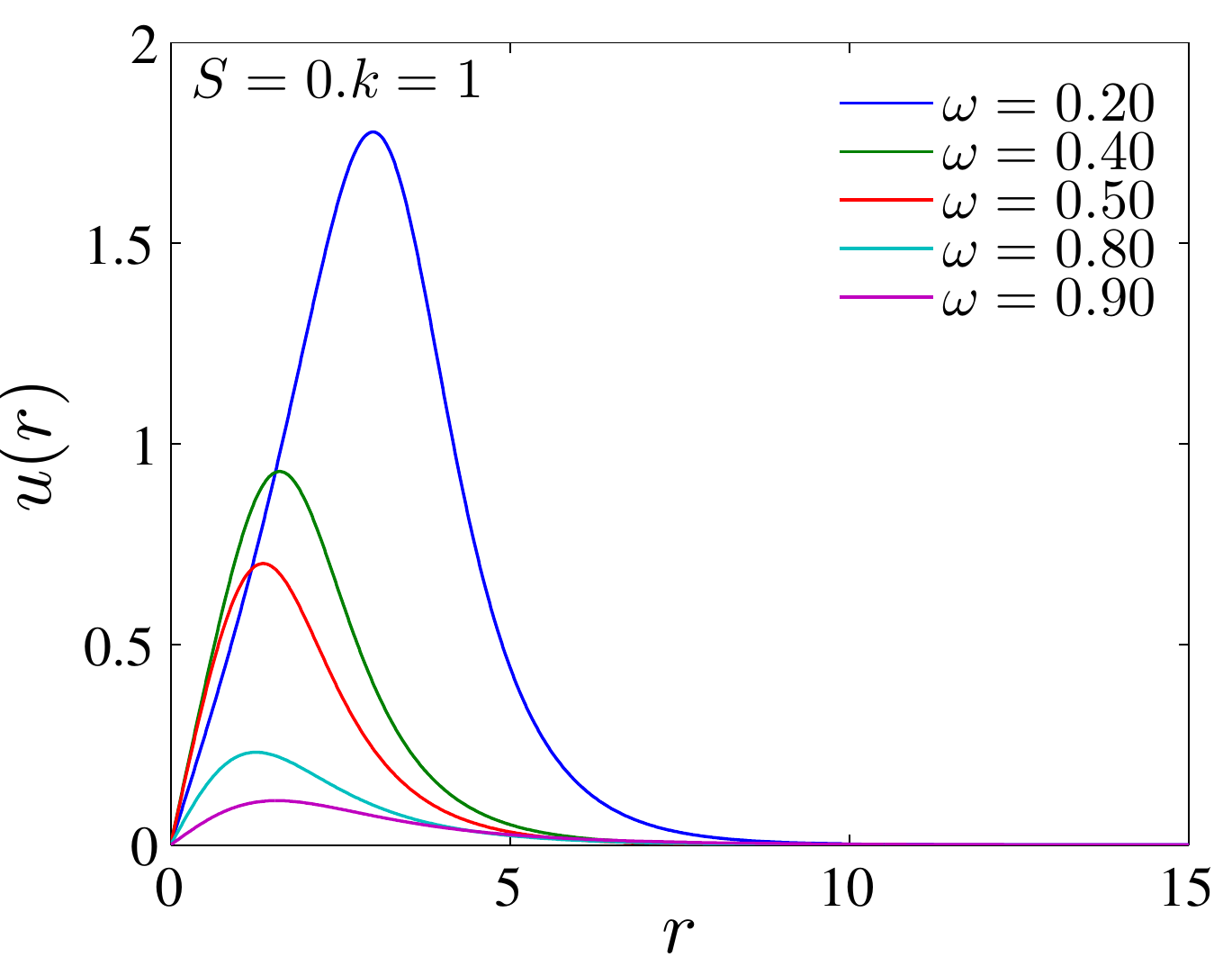} \hfill &
\includegraphics[width=.45\textwidth]{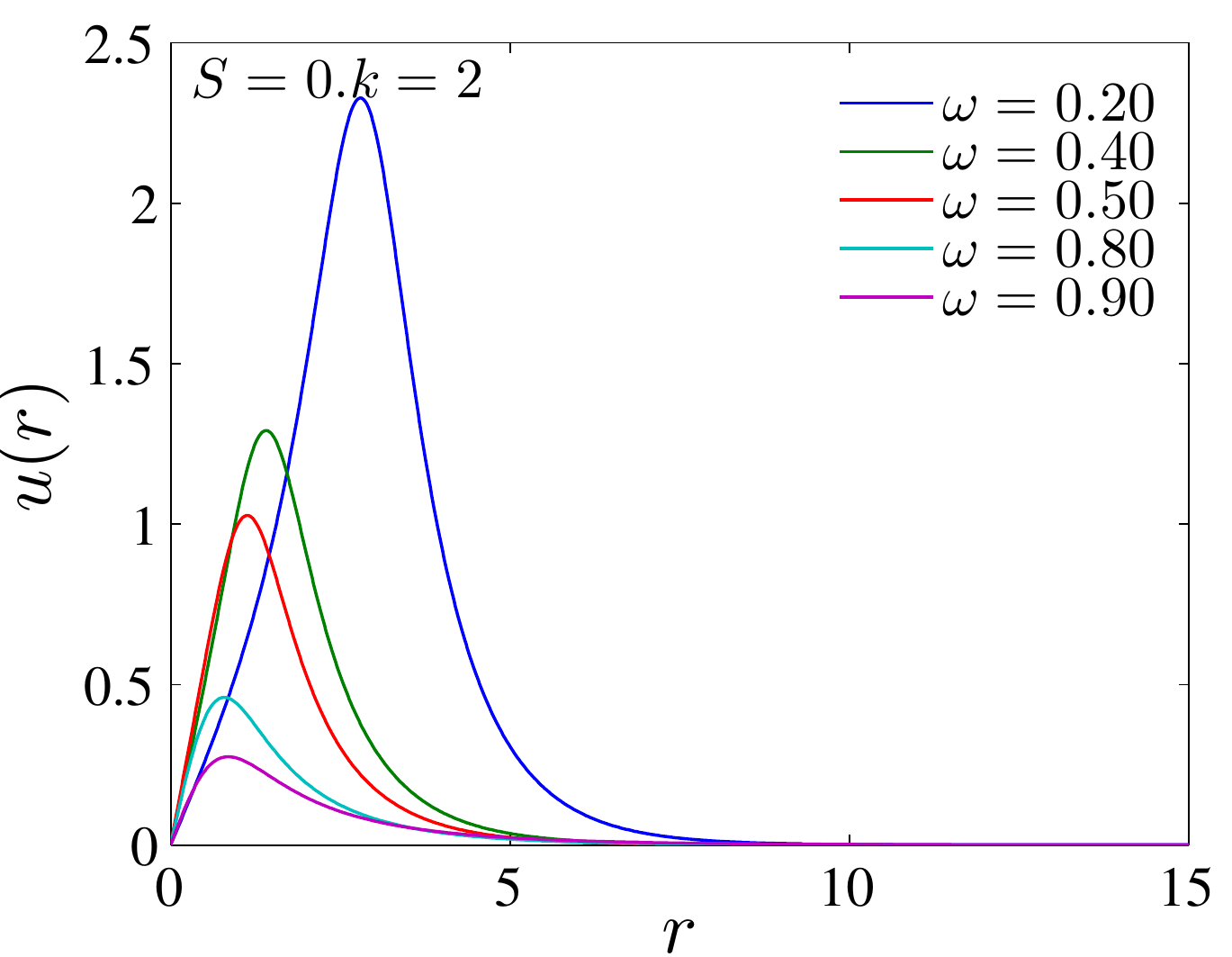} \\
\includegraphics[width=.45\textwidth]{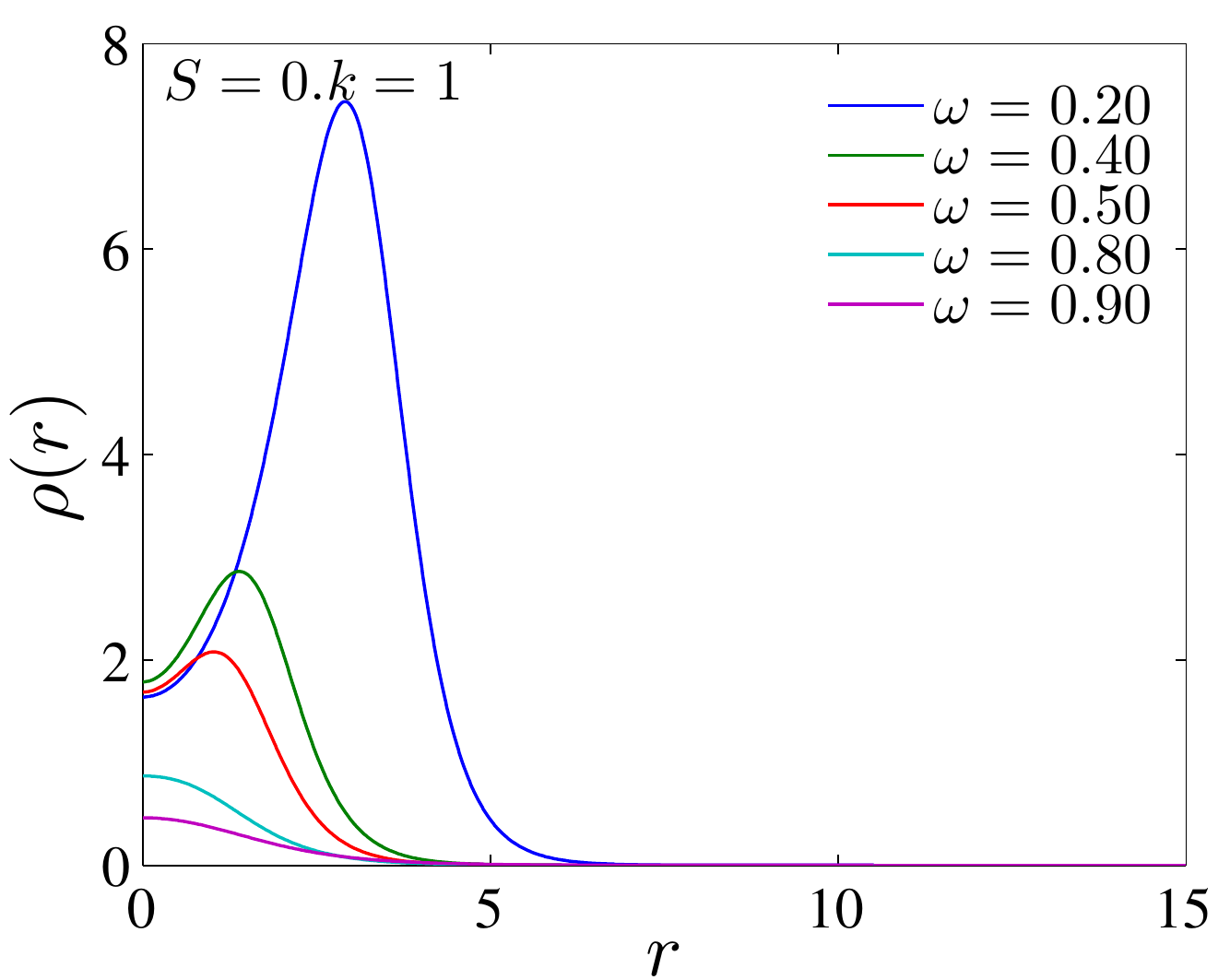} \hfill &
\includegraphics[width=.45\textwidth]{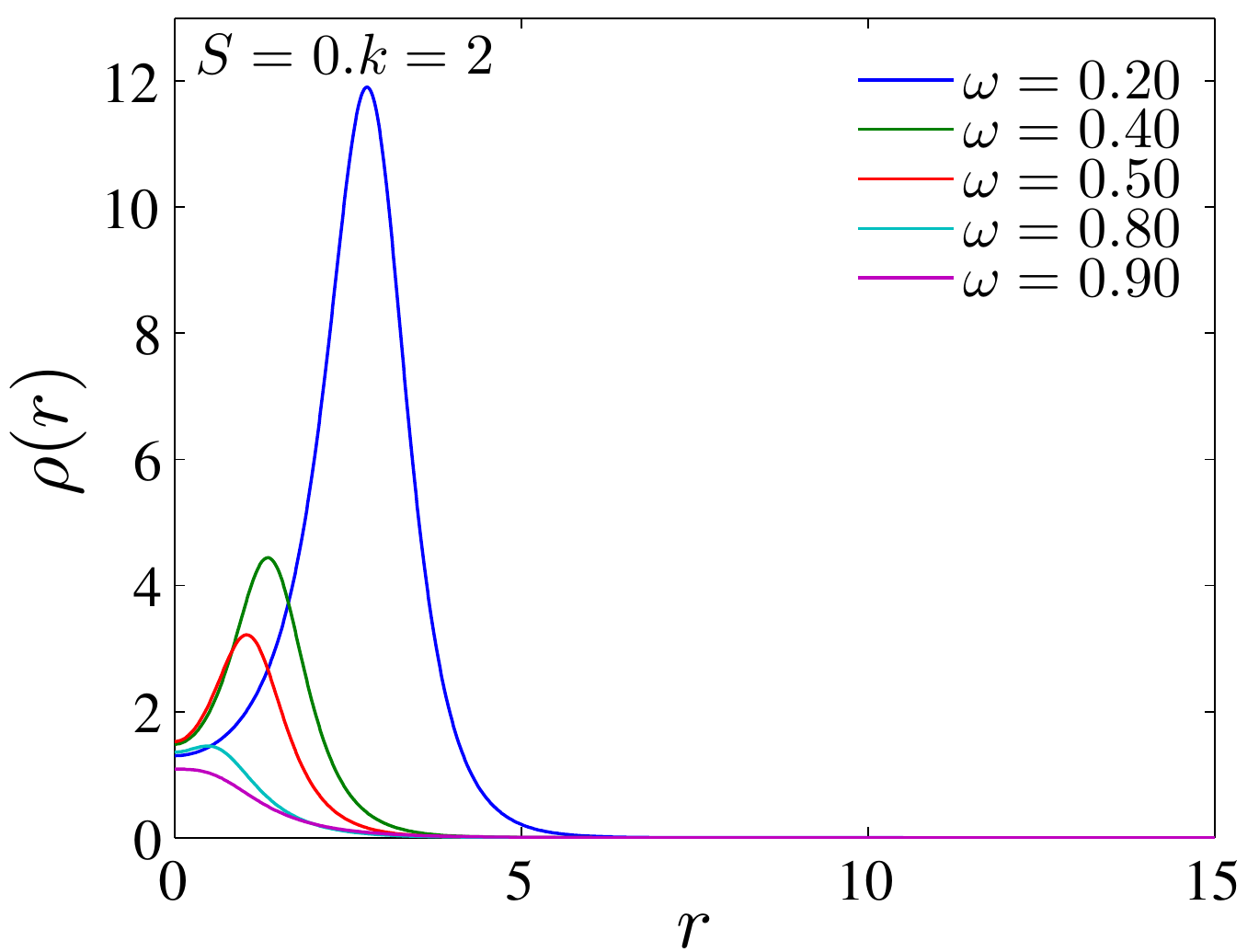} \\
\\
\end{tabular}
\end{center}
\caption{Radial profile of $S=0$ solitary waves in the 2D Soler model. Figures depict the first and second spinor components together with the solution density. Left (right) panels correspond to $k=1$ ($k=2$).}
\label{cuevas-fig3}
\end{figure}

\begin{figure}[tb]
\begin{center}
\begin{tabular}{cc}
\includegraphics[width=.45\textwidth]{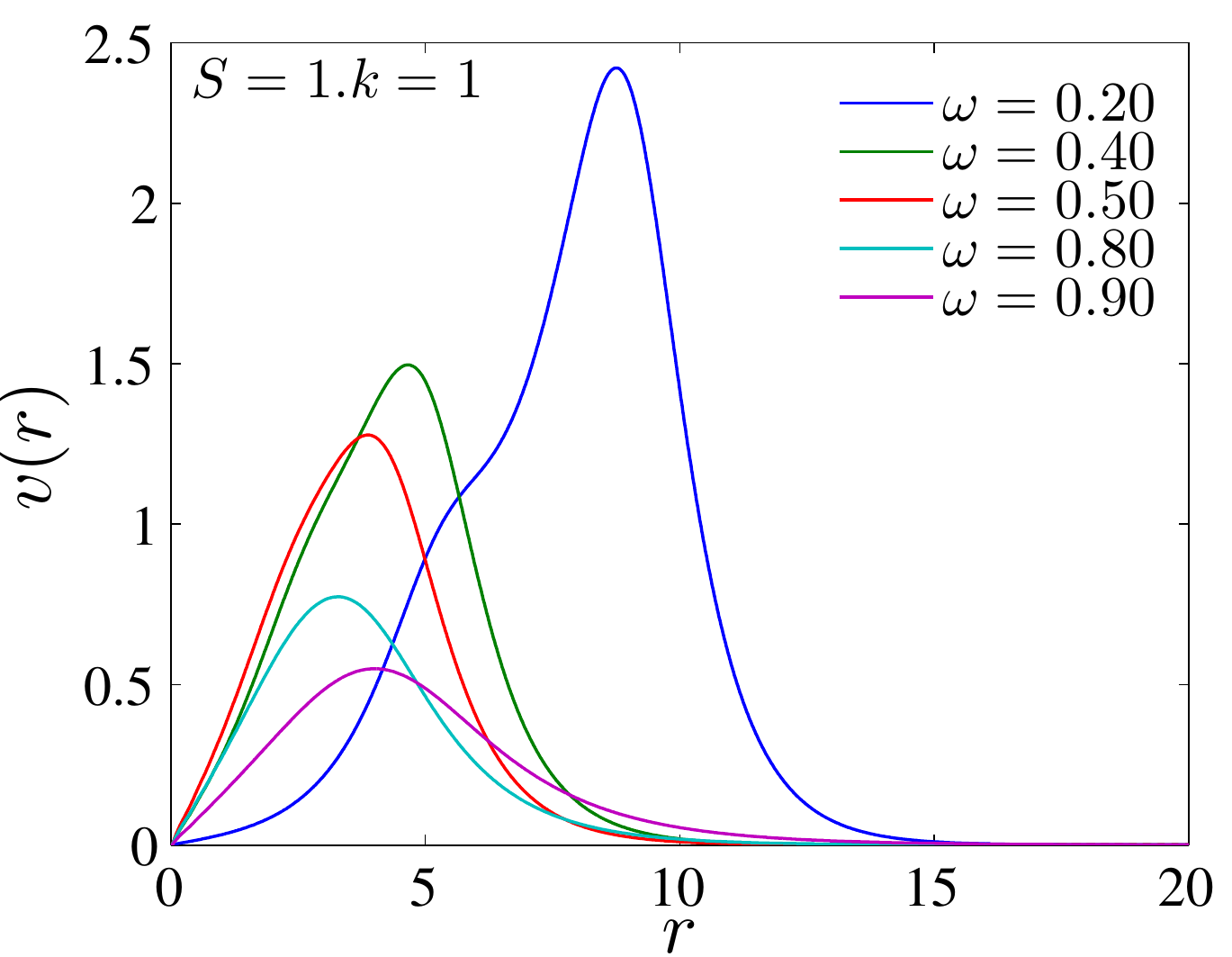} \hfill &
\includegraphics[width=.45\textwidth]{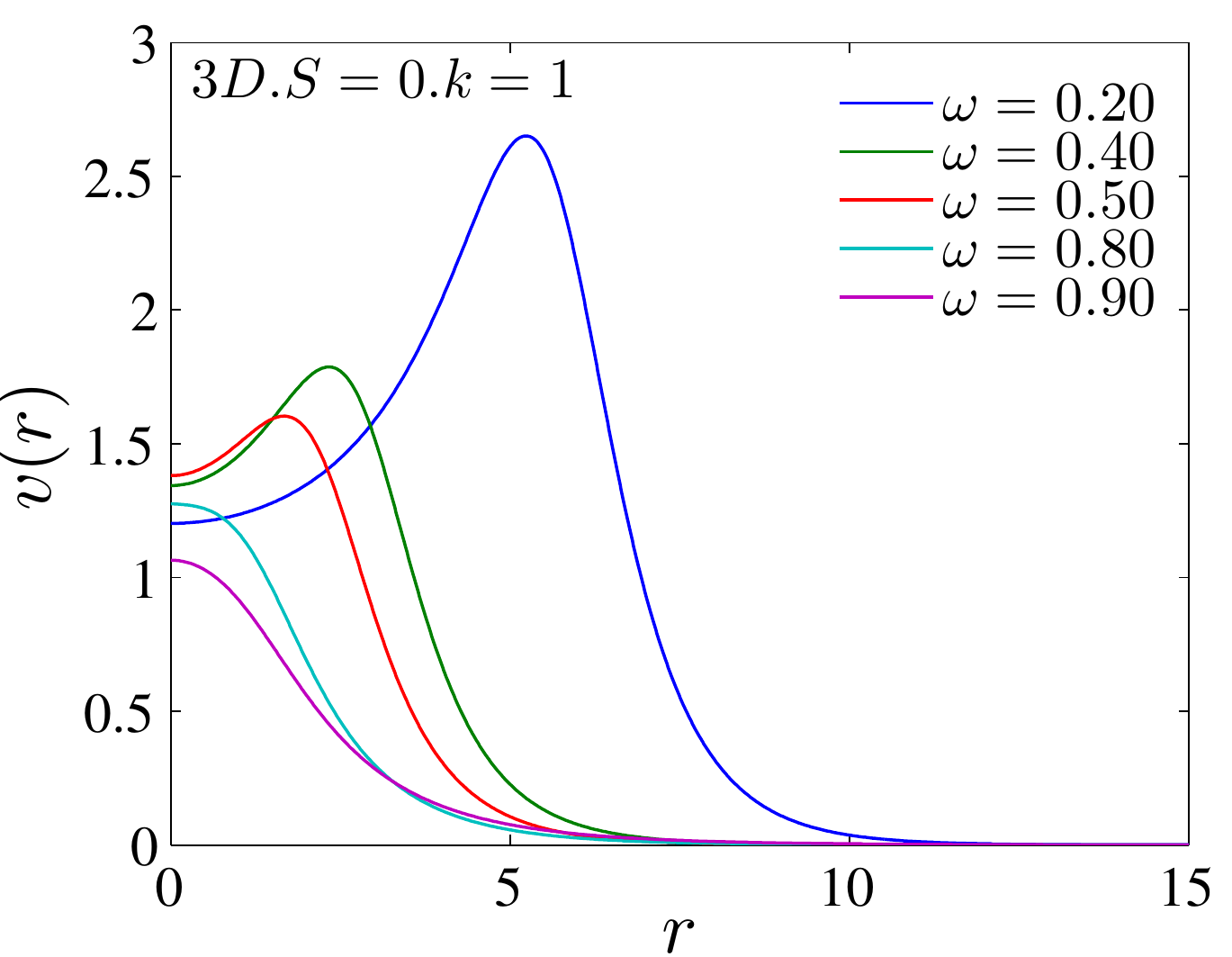} \\
\includegraphics[width=.45\textwidth]{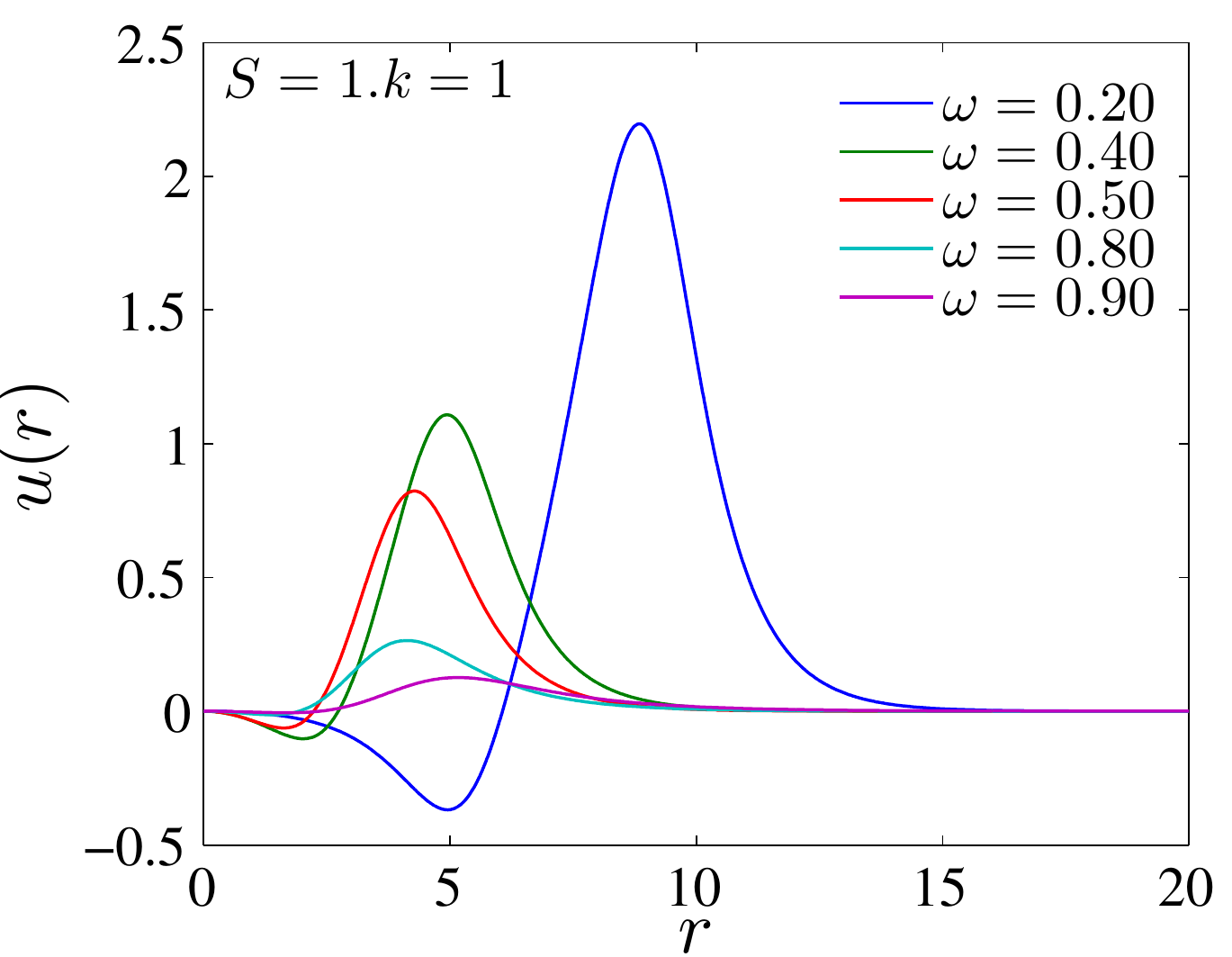} \hfill &
\includegraphics[width=.45\textwidth]{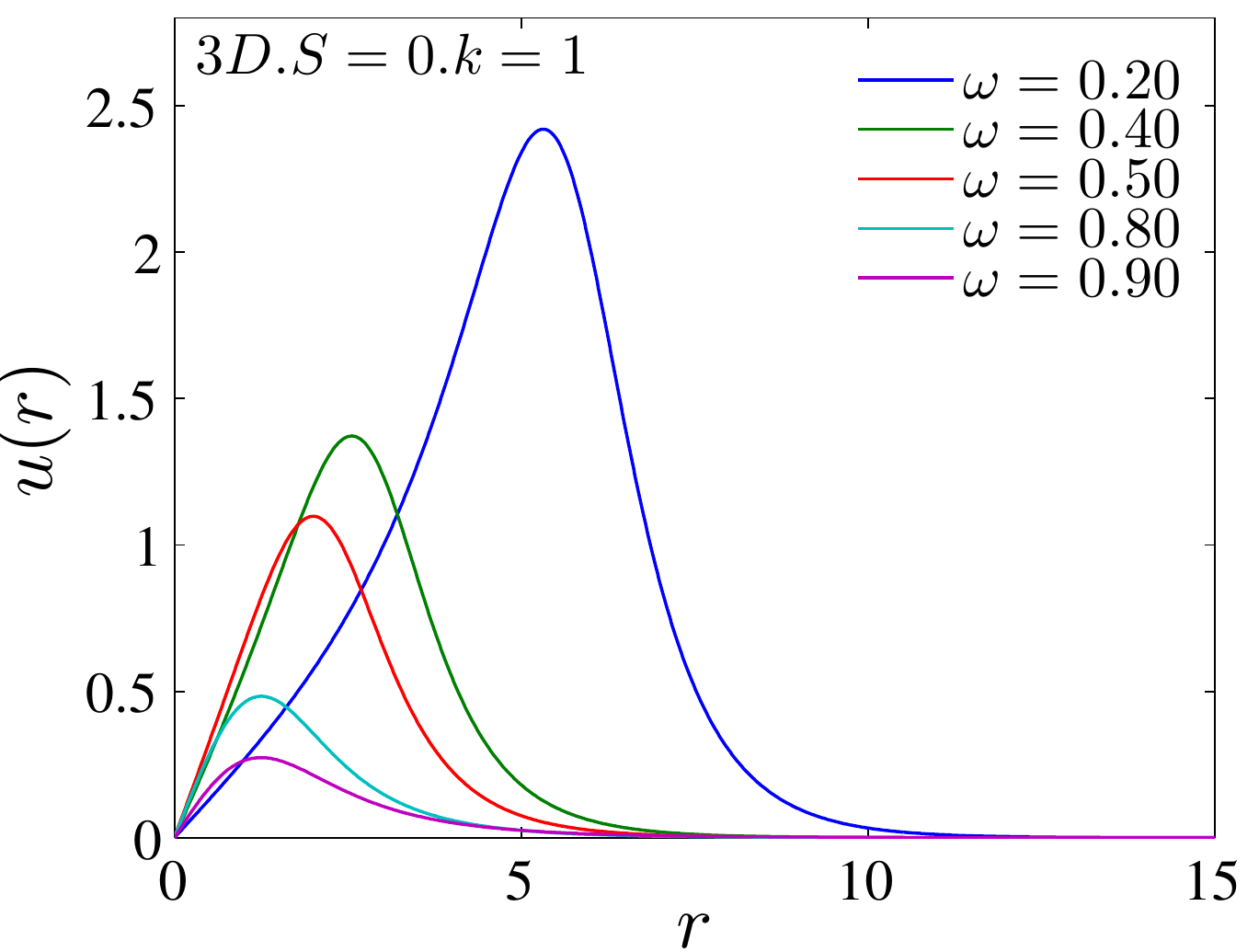} \\
\includegraphics[width=.45\textwidth]{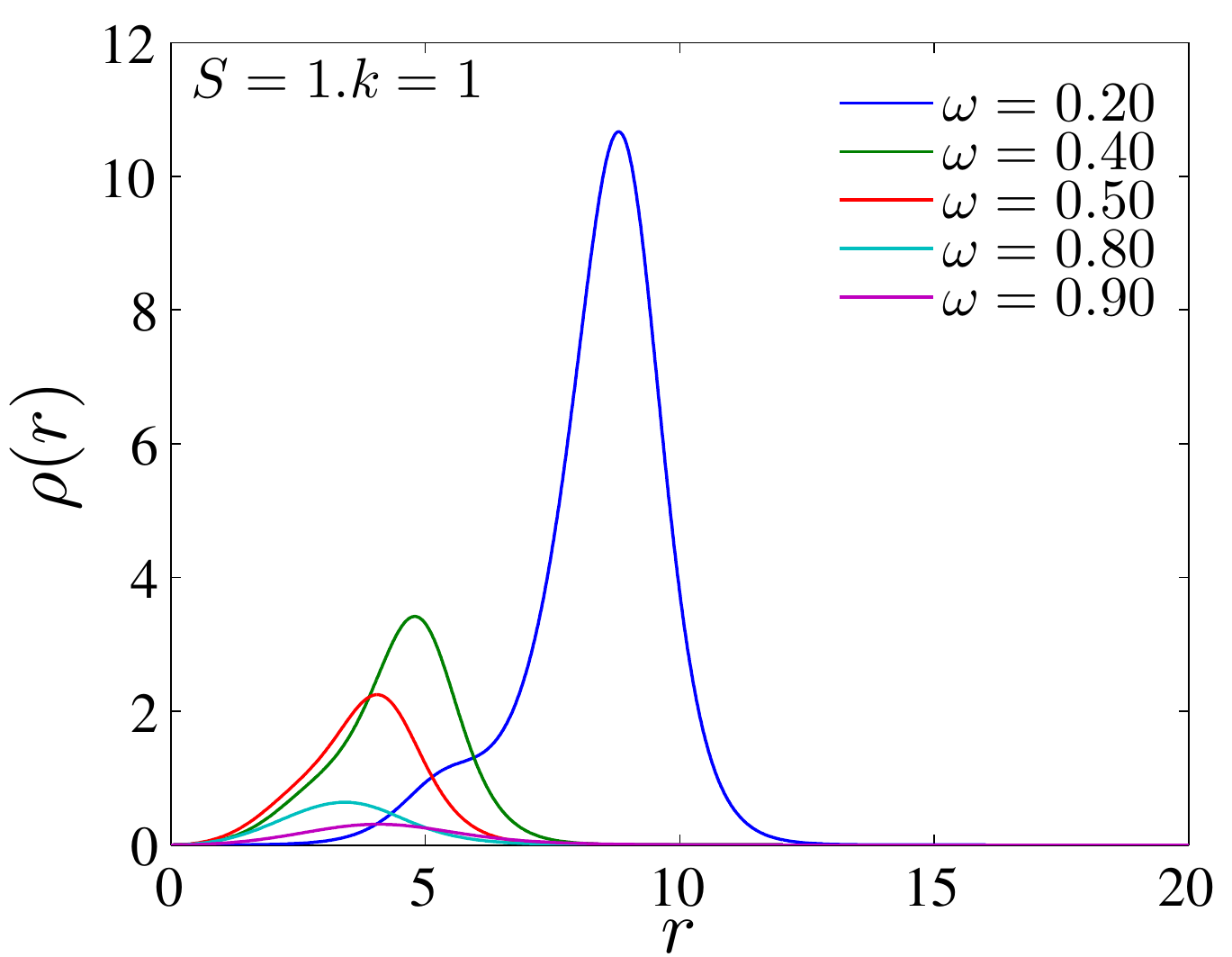} \hfill &
\includegraphics[width=.45\textwidth]{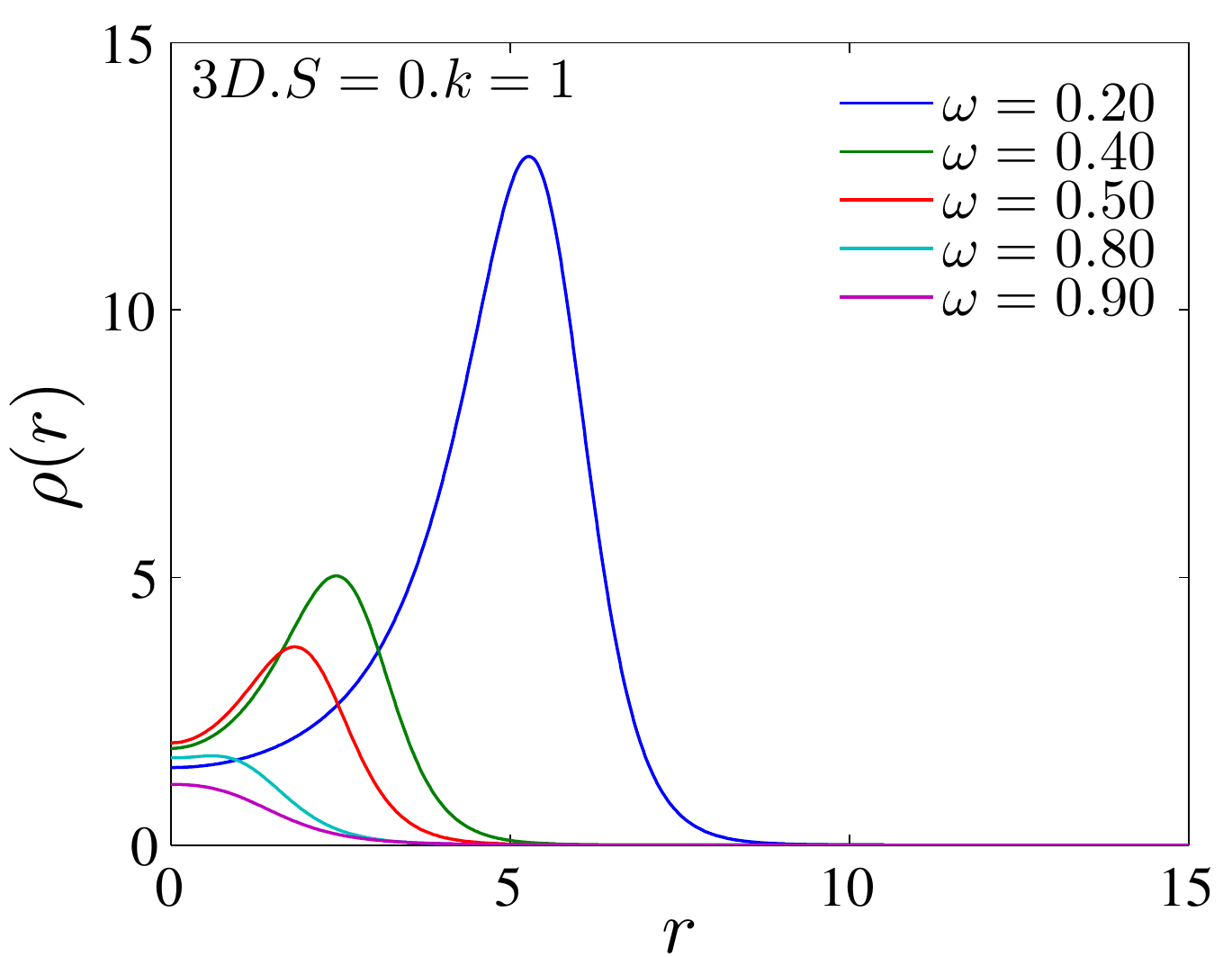} \\
\\
\end{tabular}
\end{center}
\caption{(Left panels) Radial profile of $S=1$ vortices in the cubic 2D Soler model. (Right panels) Radial profile of $S=0$ solitary waves in the cubic 3D Soler model. Figures depict the first and second spinor components together with the solution density.}
\label{cuevas-fig4}
\end{figure}

This section deals with the numerically found profiles for solitary waves ($S=0$) and vortices ($S=1$) in the two-dimensional Soler model. Fig. \ref{cuevas-fig3} shows, in radial coordinates, the profiles of each component of $S=0$ solitary waves with $k=1$ and $k=2$; the left panels of Fig. \ref{cuevas-fig4} depict those components for $S=1$ vortices. As explained in Section~\ref{subsubsec:cuevas-numerical}, the first spinor component is spatially symmetric whereas the second component is anti-symmetric as long as the vorticity of the first component, $S\in\mathbb{Z}$,
is even. The spatial symmetry is inverted if $S$ is odd. Notice also that in the $S=0$ case, the solution profile has a hump for $r>0$ whenever $\omega$ is below a critical value. It manifests as the transformation of the solitary wave density from a circle to a ring. The ring radius increases when $\omega$ decreases, becoming infinite when $\omega\rightarrow0$. For this reason, computations are progressively more demanding for smaller values of $\omega$.

\begin{figure}[tb]
\begin{center}
\begin{tabular}{cc}
\includegraphics[width=.45\textwidth]{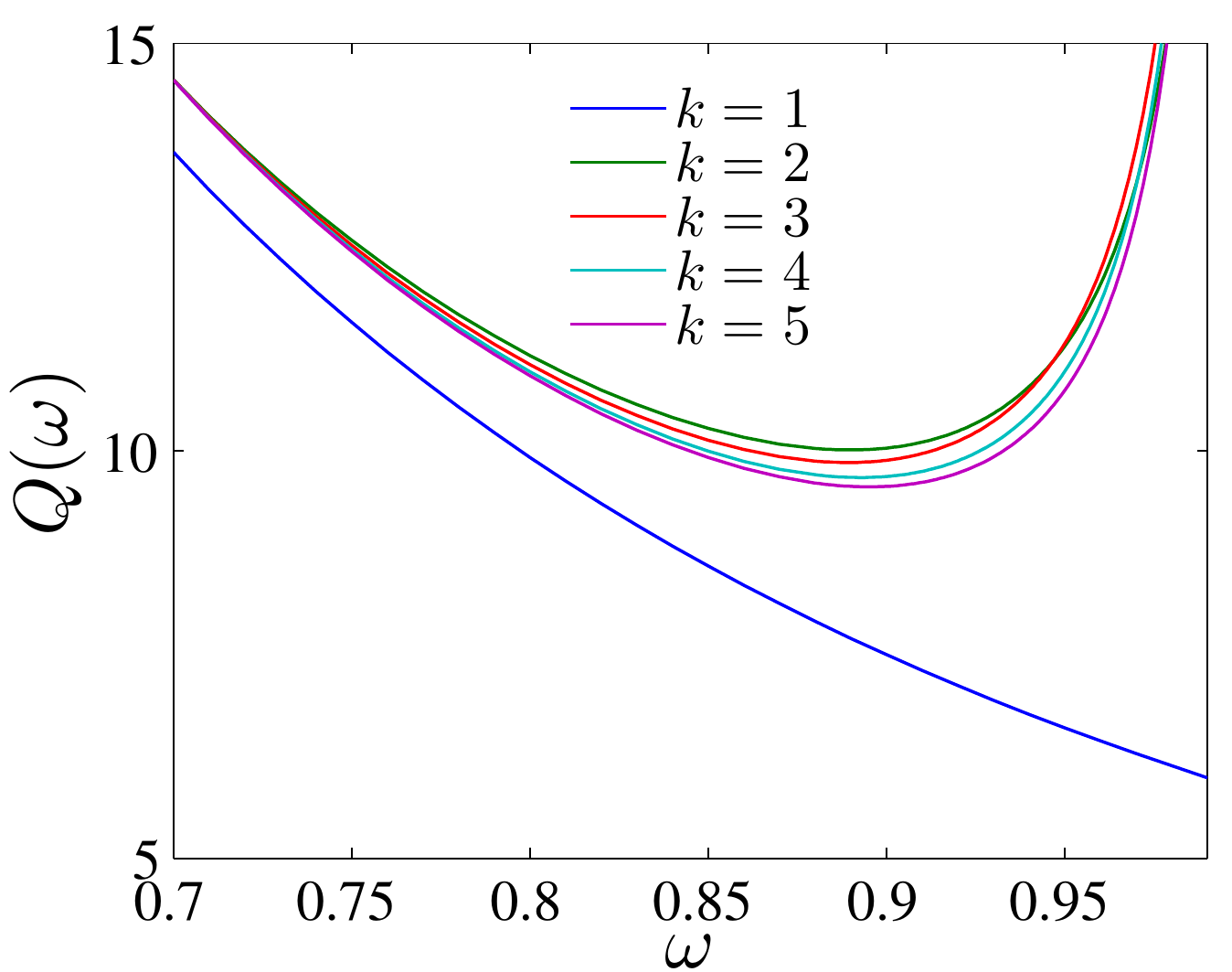} \hfill &
\includegraphics[width=.45\textwidth]{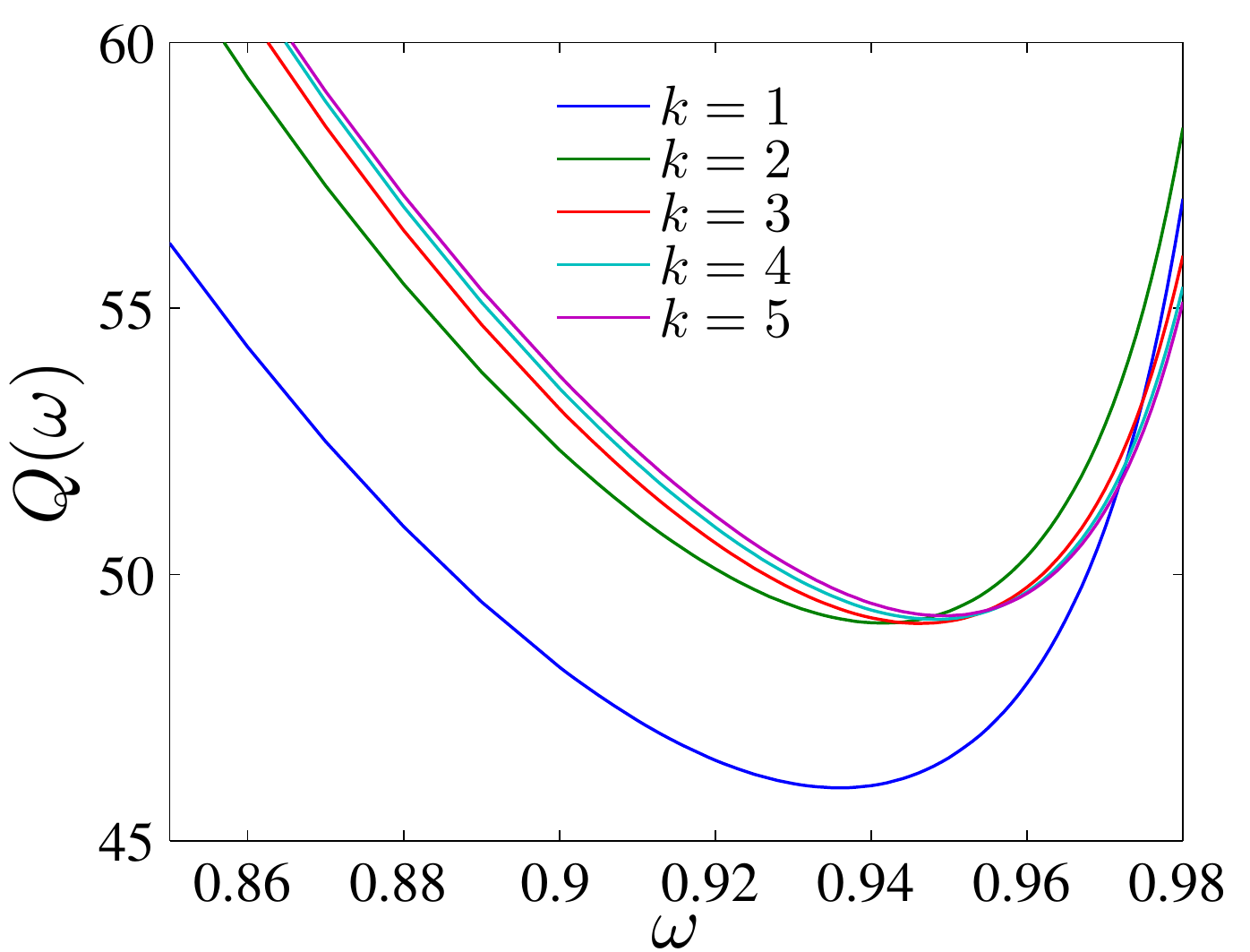} \\
\\
\end{tabular}
\end{center}
\caption{Dependence of the charge of $S=0$ solitary waves in the 2D and 3D Soler models (left and right panels, respectively) with respect to the frequency for $1\leq k\leq5$. Notice the existence of a minimum in the 2D (3D) curve for $k>1$ ($k\geq1$), which will be related to stability changes (see Section \ref{sec:cuevas-stability}).}
\label{cuevas-fig5}
\end{figure}

The right panels of Fig.~\ref{cuevas-fig4} show the radial profile of $S=0$ solitary waves in 3D. We have not included solitary waves with higher vorticity because, as explained in Subsection \ref{subsec:cuevas-Soler3D}, the Soler equation in radial coordinates can only be expressed in the $S=0$ case. Fig.~\ref{cuevas-fig5} shows the charge for the $S=0$ solitary waves in the 2D and 3D Soler models for different values of $k$.

It is worth mentioning that, despite the absence of an explicit
analytical form of 3D solitary waves,
their existence has been rigorously proven in
\cite{cuevas-CV86,cuevas-Mer88,cuevas-ES95}.

\section{Stability of solitary waves: theoretical results}\label{sec:cuevas-stability_theory}

In Section \ref{sec:cuevas-NLDmodels}, we presented the equation governing the linear stability analysis of stationary solutions. In the present section, we will show the theoretical background related to spectral and orbital stability. Many of the results proposed herein will be numerically checked in Section \ref{sec:cuevas-stability}.

\subsection{Spectral stability of solitary waves} \label{subsec:cuevas-spectral}

Prior to proceeding to the spectral stability analysis, we introduce some definitions.

The linearization of \eqref{eq:cuevas-Soler} at a solitary wave solution $\psi(t,x)=\phi_\omega(x)e^{-{i}\omega t}$ is represented by non-self-adjoint operators of the form
\begin{equation}\label{eq:cuevas-linearization}
J(D_m-\omega+V(x,\omega)),
\qquad
\mbox{
with $J$ skew-adjoint, $J^2=-1$,}
\end{equation}
where the matrix $J$ commutes with $D_m$ but not necessarily with the potential $V(x,\omega)$.

We say that the solitary wave is spectrally stable if the spectrum of its linearization operator has no points with positive real part. The spectral stability is the weakest type of stability; it does not necessarily lead to actual, dynamical one. The essential spectrum is easy to analyze: the application of Weyl's theorem
(see e.g. \cite[Theorem XIII.14, Corollary 2]{cuevas-RS78}
)
shows that the essential spectrum of the operator corresponding to the linearization at a solitary wave starts at $\pm(m-\vert{\omega}\vert){i}$ and extends to $\pm\infty{i}$. Thus, the spectral stability of the corresponding solitary wave would be a corollary of the absence of eigenvalues with positive real part in the spectrum of $J(D_m-\omega+V(\omega))$ in \eqref{eq:cuevas-linearization}. The major difficulties in identifying the point spectrum $\sigma\sb p\big(J(D_m-\omega+V(\omega))\big)$ are due to the spectrum of $D_m$ extending to both $\pm\infty$; this prevents us from using standard tools developed in the NLS context.

In the absence of linear stability (that is when the linearized system is not dynamically stable), one expects to be able to prove \emph{orbital instability}, in the sense of~\cite{cuevas-GSS87}; in \cite{cuevas-GO12}, such instability is proved in the context of the nonlinear Schr\"odinger equation; such results are still absent for the nonlinear Dirac equation.

Since the isolated eigenvalues depend continuously on the perturbation, it is convenient to trace the location of ``unstable'' eigenvalues (eigenvalues with positive real part) considering $\omega$ as a parameter. One wants to know how and when the ``unstable'' eigenvalues may emerge from the imaginary axis, particularly from the essential spectrum; that is, at which critical values of $\omega$ the solitary waves start developing an instability.
Below, we describe the possible scenarios.

\subsubsection*{Instability scenario 1: collision of eigenvalues}

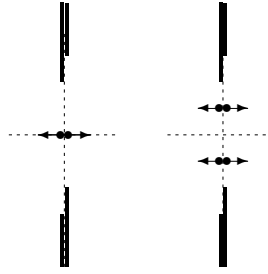
\begin{figure}
\sidecaption
\begin{picture}(180,100)(-35,-50)
\multiput(0,0)(0,0){1}{
\linethickness{0.2mm}
\put( 1.5, 0){\circle*{3}}
\put(-1.5, 0){\circle*{3}}
\put(0, 0){\vector( 1,0){10}}
\put(0, 0){\vector(-1,0){10}}
\linethickness{0.5mm}
\put(-0.9, 20){\line(0,1){30}}
\put(-0.9,-30){\line(0,-1){20}}
\put(0.9, 30){\line(0,1){20}}
\put(0.9,-20){\line(0,-1){30}}
\linethickness{0.1mm}
\multiput(0,-50)(0,3){30}{\line(0,1){1.2}}
\multiput(-21,0)(3,0){14}{\line(1,0){1.2}}
}
\multiput(60,0)(0,0){1}{
\linethickness{0.5mm}
\put(-0.9, 20){\line(0,1){30}}
\put(-0.9,-30){\line(0,-1){20}}
\put(0.9, 30){\line(0,1){20}}
\put(0.9,-20){\line(0,-1){30}}
\linethickness{0.1mm}
\multiput(0,-50)(0,3){30}{\line(0,1){1.2}}
\multiput(-21,0)(3,0){14}{\line(1,0){1.2}}
\put( 1.5,-10){\circle*{3}}
\put(-1.5,-10){\circle*{3}}
\put( 1.5, 10){\circle*{3}}
\put(-1.5, 10){\circle*{3}}
\put( 1.5,-10){\vector( 1,0){8}}
\put(-1.5,-10){\vector(-1,0){8}}
\put( 1.5, 10){\vector( 1,0){8}}
\put(-1.5, 10){\vector(-1,0){8}}
}
\end{picture}
\caption{Birth of ``unstable'' eigenvalues out of collisions of imaginary eigenvalues. When the frequency $\omega$ of the solitary wave
$\phi\sb\omega e^{-i\omega t}$
changes, the ``unstable'', positive-real-part eigenvalues
in the linearized equation
could be born from the collisions of discrete imaginary eigenvalues
}
\label{cuevas-fig6}
\end{figure}

The well-known Vakhitov--Kolokolov stability criterion \cite{cuevas-VK73} keeps
track of the collision of purely imaginary eigenvalues at the origin and
a subsequent birth of a positive and a negative eigenvalue. This
criterion was discovered in the context of nonlinear Schr\"odinger equations,
in relation to ground state solitary waves $\phi_\omega(x) e^{-{i}\omega t}$
(``ground state'' in the sense that $\phi_\omega(x)$ is strictly positive; for more
details, see \cite{cuevas-BL83}). When $\partial\sb\omega Q(\omega)<0$, with
$Q(\omega)=\|\phi_\omega\|\sb{L^2}^2$ being the charge of the solitary
wave (\ref{eq:cuevas-charge}), then the linearization at a solitary wave has purely imaginary
spectrum; when $\partial\sb\omega Q(\omega)>0$, there are two real (one positive,
one negative) eigenvalues of the linearization operator. The vanishing of the
quantity $\partial\sb\omega Q(\omega)$ at some value of $\omega$ indicates the
moment of the collision of eigenvalues, when the Jordan block corresponding
to the zero eigenvalue has a jump of two in its size. A nice feature of the
linearization at a ground state solitary wave in the nonlinear Schr\"odinger
equation is that its spectrum belongs to the imaginary axis, with some
eigenvalues possibly located on the real axis; thus, the collision of
eigenvalues at $\lambda=0$ is the only way the spectral instability could
develop. In the NLD context, such a collision does not necessarily occur at
$\lambda=0$; both situations as in Fig.~\ref{cuevas-fig6} are possible.

In \cite{cuevas-BCS15},
it was shown that
in NLD (and similar fermionic systems)
the collision of eigenvalues at the origin and
a subsequent transition to instability
is characterized not only by the Vakhitov--Kolokolov condition $d Q/d\omega=0$,
but also by the condition $E(\omega)=0$,
where $E$ is the value of the energy functional on the corresponding solitary wave.

\begin{theorem}
The algebraic multiplicity of the eigenvalue $\lambda=0$
of the linearization $\mathcal{A}\sb\omega$
at the solitary wave $\phi_\omega(x)e^{-{i}\omega t}$
has a jump of (at least) $2$ when
at a particular value of $\omega$
either $\partial\sb\omega Q(\phi_\omega)=0$ or $E(\phi_\omega)=0$,
with $Q(\phi_\omega)$ and $E(\phi_\omega)$ being
the charge and the energy of the solitary wave
$\phi_\omega(x)e^{-{i}\omega t}$.
\end{theorem}

The eigenvalues with positive real part
could also be born
from the collision of purely imaginary
eigenvalues at some point in the spectral gap
but away from the origin;
we have recently observed this scenario in the cubic Soler model in two spatial dimensions
\cite{cuevas-CKS+16a}.
Presently we do not have a criterion for such a collision of eigenvalues.

\subsubsection*{Instability scenario 2: bifurcations from the essential spectrum}

The most peculiar feature of the linearization
at a solitary wave in the NLD context is the possibility
of bifurcations of eigenvalues with nonzero real part
off the imaginary axis,
out of the bulk of the essential spectrum.

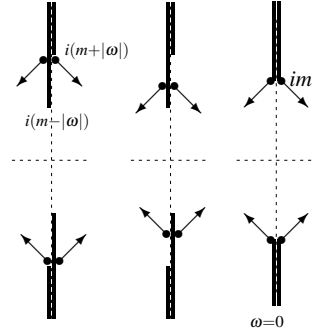
\begin{figure}
\sidecaption
\begin{picture}(140,120)(-0,-60)
\multiput(0,0)(0,0){1}{
\linethickness{0.5mm}
\put(-0.9, 20){\line(0,1){40}}
\put(-0.9,-40){\line(0,-1){20}}
\put(0.9, 40){\line(0,1){20}}
\put(0.9,-20){\line(0,-1){40}}
\linethickness{0.1mm}
\multiput(0,-60)(0,3){40}{\line(0,1){1.2}}
\multiput(-15,0)(3,0){10}{\line(1,0){1.2}}
\put( 3,-38){\circle*{3}}
\put(-1.5,-38){\circle*{3}}
\put( 1.5, 38){\circle*{3}}
\put( -3, 38){\circle*{3}}
\put( 3,-38){\vector(1,1){10}}
\put(-1.5,-38){\vector(-1,1){10}}
\put( 1.5, 38){\vector(1,-1){10}}
\put( -3,38){\vector(-1,-1){10}}
\put( 5,40){$\scriptstyle i(m+|\omega|)$}
\put( -10,13){$\scriptstyle i(m-|\omega|)$}
}
\multiput(45,0)(0,0){1}{
\linethickness{0.5mm}
\put(-0.9, 20){\line(0,1){40}}
\put(-0.9,-40){\line(0,-1){20}}
\put(0.9, 40){\line(0,1){20}}
\put(0.9,-20){\line(0,-1){40}}
\linethickness{0.1mm}
\multiput(0,-60)(0,3){40}{\line(0,1){1.2}}
\multiput(-15,0)(3,0){10}{\line(1,0){1.2}}
\put(1.5, 28){\circle*{3}}
\put(3, -28){\circle*{3}}
\put(-3, 28){\circle*{3}}
\put(-1.5,-28){\circle*{3}}
\put( 3,-28){\vector(1,1){10}}
\put(-1.5,-28){\vector(-1,1){10}}
\put( 1.5, 28){\vector(1,-1){10}}
\put( -3,28){\vector(-1,-1){10}}
}
\multiput(85,0)(0,0){1}{
\linethickness{0.5mm}
\put(-0.9, 30){\line(0,1){30}}
\put(0.9, 30){\line(0,1){30}}
\put(-0.9,-30){\line(0,-1){25}}
\put(0.9,-30){\line(0,-1){25}}
\linethickness{0.1mm}
\multiput(0,-50)(0,3){35}{\line(0,1){1.2}}
\multiput(-15,0)(3,0){10}{\line(1,0){1.2}}
\put( 5,28){$im$}
\put( 2,-30){\circle*{3}}
\put(-2,-30){\circle*{3}}
\put( 2, 30){\circle*{3}}
\put(-2, 30){\circle*{3}}
\put( 2,-30){\vector(1,1){10}}
\put(-2,-30){\vector(-1,1){10}}
\put( 2, 30){\vector(1,-1){10}}
\put(-2, 30){\vector(-1,-1){10}}
\put(-10, -63){$\scriptstyle\omega=0$}
}
\end{picture}
\caption{
Possible bifurcations from the essential spectrum.
Theoretically,
when $|\omega|<m$,
the nonzero-real-part eigenvalues
could be born from the embedded thresholds
at $\pm i(m+|\omega|)$,
from the embedded eigenvalue in the bulk of the essential spectrum
between the threshold and the embedded threshold,
and from the collision of the thresholds
at $\pm i m$
when $\omega=0$.
}
\label{cuevas-fig7}
\end{figure}

The article \cite{cuevas-BC16a} gives a thorough analytical study
of eigenvalues of the Dirac operators,
focusing on whether and how such eigenvalues
can bifurcate from the essential spectrum.
Generalizing the Jensen--Kato approach \cite{cuevas-JK79}
to the context of the Dirac operators,
it was shown
in \cite[Theorem 2.15]{cuevas-BC16a}
that for $|\omega|<m$
the bifurcations from the essential spectrum
are only possible
from embedded eigenvalues (Fig.~\ref{cuevas-fig7}, center), with the following exceptions:
the bifurcation could start at the embedded thresholds
located at $\pm {i}(m+\vert{\omega}\vert)$
(Fig.~\ref{cuevas-fig7}, left),
or they could start at $\lambda=\pm i m$ when $\omega=0$
(Fig.~\ref{cuevas-fig7}, right;
this situation correspond to the collision of thresholds).
Indeed, bifurcations from the embedded thresholds
have been observed in a one-dimensional
NLD-type model of coupled-mode equations
\cite{cuevas-BPZ98,cuevas-CP06}.
The bifurcations from
the collision of thresholds
at $\pm{i} m$ (when $\omega=0$)
were demonstrated in~\cite{cuevas-KS02}
in the context of the perturbed massive Thirring model.

One can use the Carleman--Berthier--Georgescu estimates \cite{cuevas-BG87}
to prove that there are no embedded eigenvalues
(hence no bifurcations)
in the portion of the essential spectrum
outside of the embedded thresholds \cite{cuevas-BC16a}.

As to the bifurcations from the embedded eigenvalues
before the embedded thresholds,
as in Fig.~\ref{cuevas-fig7} (center),
we do not have any such examples in the NLD context,
although such examples could be produced for
Dirac operators of the form \eqref{eq:cuevas-linearization}
(with $V$ kept self-adjoint).

\subsubsection*{Instability scenario 3: bifurcations from the nonrelativistic limit}

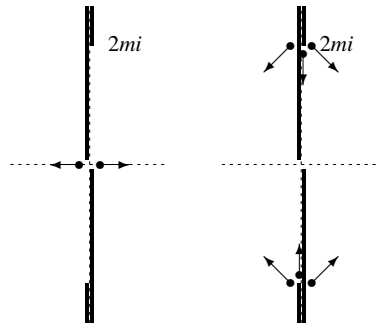
\begin{figure}
\sidecaption
\begin{picture}(140,120)(0,-70)
\multiput(-20,-10)(0,0){1}{
\linethickness{0.5mm}
\put(7,43){$2mi$}
\put(-0.9, 2){\line(0,1){58}}
\put(-0.9,-45){\line(0,-1){15}}
\put(0.9, 45){\line(0,1){15}}
\put(0.9,-2){\line(0,-1){58}}
\linethickness{0.1mm}
\multiput(0,-60)(0,3){40}{\line(0,1){1.2}}
\multiput(-30,0)(3,0){20}{\line(1,0){1.2}}
\put( 4, 0){\circle*{3}}
\put(-4, 0){\circle*{3}}
\put( 5,0){\vector( 1,0){10}}
\put(-5,0){\vector(-1,0){10}}
}
\multiput(60,-10)(0,0){1}{
\linethickness{0.5mm}
\put(7,43){$2mi$}
\put(-0.9, 2){\line(0,1){58}}
\put(-0.9,-45){\line(0,-1){15}}
\put(0.9, 45){\line(0,1){15}}
\put(0.9,-2){\line(0,-1){58}}
\linethickness{0.1mm}
\multiput(0,-60)(0,3){40}{\line(0,1){1.2}}
\multiput(-30,0)(3,0){20}{\line(1,0){1.2}}
\put( 4,-45){\circle*{3}}
\put(-4,-45){\circle*{3}}
\put( 4, 45){\circle*{3}}
\put(-4, 45){\circle*{3}}
\put( 4,-45){\vector( 1,1){10}}
\put(-4,-45){\vector(-1,1){10}}
\put( 4, 45){\vector( 1,-1){10}}
\put(-4, 45){\vector(-1,-1){10}}
\put( 0.7, 42){\circle*{3}}
\put( 0.7, 42){\vector( 0,-1){12}}
\put(-0.7,-42){\circle*{3}}
\put(-0.7,-42){\vector( 0,1){12}}
}
\end{picture}
\caption{Bifurcations from $\lambda=0$ and hypothetical bifurcations from $\lambda=\pm 2 m{i}$ in the nonrelativistic limit, $\omega\!\lesssim\! m$.
The nonzero-real-part eigenvalues could be present in the spectrum
of the linearization at a solitary wave $\phi\sb\omega e^{-i\omega t}$
for $\omega$ arbitrarily close to $m$;
these eigenvalues would have to be located near $\lambda=0$
or near the embedded threshold at $\lambda=\pm 2 m i$.
}
\label{cuevas-fig8}
\end{figure}

The nonzero-real-part eigenvalues
could be present in the spectrum
of the linearization operators at
small amplitude solitary waves
for all $\omega\lesssim m$,
being born ``from the nonrelativistic limit''.
It was proved in \cite[ Theorem 2.19]{cuevas-BC16a},
under very mild assumptions,
that the bifurcations of eigenvalues
for $\omega$ departing from $\pm m$
are only possible from the thresholds
$\lambda=0$ and $\lambda=\pm 2 m {i}$;
see Fig.~\ref{cuevas-fig8}.

We now undertake a detailed study of these bifurcations;
let us concentrate on the case $\lambda=0$.
It is of no surprise that the behaviour of eigenvalues
of the linearized operator
near $\lambda=0$,
in the nonrelativistic limit $\omega\lesssim m$,
follows closely the pattern which one observes in the
nonlinear Schr\"odinger equation with the same nonlinearity.
In other words,
if the linearizations of the nonlinear Dirac equation
at solitary waves with $\omega\lesssim m$
admit a family of eigenvalues $\Lambda\sb\omega$
which continuously depends on $\omega$,
such that
$\Lambda\sb\omega\to 0$ as $\omega\to m$,
then this family
is merely a deformation
of an eigenvalue family $\Lambda\sb\omega\sp{\mathrm{NLS}}$
of the linearization of the nonlinear Schr\"odinger equation
with the same nonlinearity
(linearized at corresponding solitary waves).
To make this rigorous,
one considers the spectral problem for the linearization
at a solitary wave with $\omega\lesssim m$,
applies the rescaling
with respect to $m-\omega\ll 1$,
and uses the reduction based on the Schur complement method,
recovering in the nonrelativistic limit
$\omega\to m$
the linearization of the nonlinear Schr\"odinger equation, and
then applying the Rayleigh--Schr\"o\-din\-ger perturbation theory;
in \cite{cuevas-CGG14}, this approach was developed
to prove the linear instability of small amplitude solitary waves
$\phi_\omega(x)e^{-{i}\omega t}$
in the ``charge-supercritical'' NLD,
in the nonrelativistic limit
$\omega\lesssim m$.

\begin{theorem}\label{theorem:nd}
Assume that $f(s)=|s|^k$, where $k\in\mathbb{N}$
satisfies
$k>2/n$ (and $k<2$ for $n=3$).
Then there is $\omega_1<m$
such that the solitary wave solutions
$\phi_\omega(x)e^{-{i}\omega t}$
(in the form of the Wakano Ansatz \eqref{eq:cuevas-spinor3D})
to NLD
are linearly unstable for
$\omega\in(\omega_1,m)$.
More precisely, let $\mathcal{A}\sb\omega$
be the linearization of the nonlinear Dirac equation at a solitary wave
$\phi_\omega(x)e^{-{i}\omega t}$. Then for $\omega\in(\omega_1,m)$
there are eigenvalues
\[
\pm\lambda\sb\omega\in\sigma\sb{p}(\mathcal{A}\sb\omega),
\qquad
\lambda\sb\omega>0,
\qquad
\lambda\sb\omega=O(m-\omega).
\]
\end{theorem}

Let us remark here that the restriction
in the above theorem
that $k$ is a natural number which
was needed to make sure that the solitary wave family
of the form of the Wakano Ansatz indeed exists.
Theorem~\ref{theorem:nd} extends to
$f(s)=a|s|^k+O(|s|^{K})$, $a>0$,
with $k\in(2/n,\,2/(n-2))$
($k>2/n$ when $n\le 3$) and
$K>k$.
The existence of the corresponding
families of solitary waves
was later proved in \cite{cuevas-BC16}.
In that article,
a general construction was given
for small amplitude solitary waves in the nonlinear Dirac equation,
deriving the asymptotics which
we will need in the forthcoming
stability analysis of such solitary waves.
This is a general result proved for nonlinearities
which are not necessarily smooth,
thus applicable to e.g. critical and subcritical nonlinearities.

We point out that the instability stated in Theorem~\ref{theorem:nd}
is in a formal agreement
with the Vakhitov--Kolokolov stability criterion \cite{cuevas-VK73};
one has
$dQ(\omega)/d\omega>0$ for $\omega\lesssim m$.
Conversely, we expect that
the presence of eigenvalues with nonzero real part
in the vicinity of $\lambda=0$
for $\omega\lesssim m$,
is prohibited by the
Vakhitov--Kolokolov stability criterion
$\frac{d Q(\omega)}{d\omega}<0,
$
$\omega\lesssim m.
$

Similarly to how the NLS corresponds to the nonrelativistic
limit of NLD,
in the nonrelativistic limit
of the Dirac--Maxwell system
one arrives at the
Choquard equation \cite{cuevas-Lieb77};
see \cite{cuevas-CS12} and the references therein.
The Choquard equation
is known to be spectrally (in fact, even orbitally) stable \cite{cuevas-CL82};
we expect that this implies absence of unstable eigenvalues
bifurcating from the origin
in the Dirac--Maxwell system.

As we pointed out above, in the nonrelativistic limit $\omega\lesssim m$,
there could be eigenvalue families of the
linearization of the nonlinear Dirac operator
bifurcating not only from the origin,
but also from the embedded threshold
(that is, such that
$\lim\sb{\omega\to m}\Lambda\sb i(\omega)=\pm 2 m{i}$).
Rescaling and using the Schur complement approach
shows that there could be at most $N/2$ such families
bifurcating from each of $\pm 2m{i}$,
with $N$ the number of components
of a spinor field (in 3D Dirac, one takes $N=4$).
Could these eigenvalues go off the imaginary axis into the complex plane?
While for the nonlinear Dirac equations with a general nonlinearity
the answer to this question is unknown,
in the Soler model
we can exclude this scenario.
One can show that
there are exact eigenvalues
$\lambda\sb\pm(\omega)=\pm 2\omega {i}$,
each being of multiplicity $N/2$;
thus, we know exactly what happens to the eigenvalues
which bifurcate from $\pm 2m{i}$, and
expect no bifurcations of eigenvalues
off the imaginary axis.
The details are given in \cite{cuevas-BC17}.

\bigskip

Let us finish with a very important result: the existence of eigenvalues $\pm 2\omega{i}$ of the linearization at a solitary wave in the Soler model \eqref{eq:cuevas-Soler} is a consequence of having bi-frequency solitary wave solutions in the Soler model, in any dimension and for any nonlinearity. For more details, see \cite{cuevas-BC17}.

\subsection{Orbital and asymptotic stability of solitary waves}

The spectral analysis is one aspect of global analysis of the dynamical
stability. In principle any spectral instability around a stationary solution
should lead to a dynamical instability, namely the stationary solution is
orbitally unstable. The contrapuntal statement that a stable stationary state
has a spectrally stable linearized operator needs to be analyzed carefully.

If the Dirac operator $D_m$ is perturbed by some zero-order external
potential, the perturbation theory provides tools which allow one to analyze
the linear stability of linearized operators of the form
\eqref{eq:cuevas-linearization}. Still some important restrictions on the potential
appear (decay, regularity, and absence of resonances). Even if the
perturbation analysis needs some work, it is much less involved compared to
the complete spectral characterization of the linearized operator. This opens
the gates to the analysis of the nonlinear stability.

Prior to a
bibliographical review of the available works in this direction, we make a
remark. While in many models the orbital stability is obtained by using the
energy as some kind of a Lyapunov functional, this is no longer possible for
models of Dirac type since the energy is sign-indefinite. Even if there are
some conserved quantities which allow one to control certain negative
directions of the Hessian of the energy, the latter are in infinite number
(``infinite Morse index'') and in most cases the conservation laws are not
enough. The route ``use linear stability to prove the asymptotic stability''
seems to be the only one available for the sign-indefinite systems such as
nonlinear Dirac, Dirac--Hartree--Fock, and others. As a result, due to the
strong indefiniteness of the Dirac operator (the energy conservation does not
lead to any bounds on the $H^{1/2}$-norm), we do not know how to prove the
\emph{orbital stability} \cite{cuevas-GSS87} but via proving the asymptotic
stability first. The only exceptional case in nonlinear Dirac-type systems
seems to be the completely integrable massive Thirring model in one spatial
dimension~\cite{cuevas-Thi58}, where additional conserved quantities arising from
the complete integrability allow one to prove orbital stability of solitary
waves \cite{cuevas-PS14,cuevas-CPS16}. Note that these conserved quantities are used not to
control the negative directions but rather to construct a new Lyapunov
functional. More precisely, by \cite{cuevas-PS14}, there is a functional $R$ defined
on $H^1(\mathbb{R},\mathbb{C}^2)$ (which contains terms dependent on powers of components of
$\psi\in\mathbb{C}^2$ of order up to six) which is (formally) conserved for solutions
to the massive Thirring model, and it is further shown that there is
$\omega_0\in(0,m]$ such that for $\omega\in(-\omega_0,\omega_0)$ the solitary
wave amplitude is a local minimizer of $R$ in $H^1$ under the charge and
momentum conservation, and hence the corresponding solitary wave is orbitally
stable in $H^1(\mathbb{R},\mathbb{C}^2)$. Moreover, in \cite{cuevas-CPS16}, using the global
existence of $L^2$-solutions for the (cubic) massive Thirring model
\cite{cuevas-Can11}, the orbital stability of solitary waves in $L^2(\mathbb{R})$ has been
shown, with the proof based on the auto-B\"acklund transformation. Now we
turn to the asymptotic stability. In \cite{cuevas-CPS17}, the asymptotic stability
was proved for the small energy perturbations to solitary waves in the
Gross--Neveu model. The model is taken with particular pure-power
nonlinearities when all the assumptions on the spectral and linear stability
of solitary waves have been verified directly. This is, referring to the
previous discussion, also the ``proof of concept'': it is shown that there
are translation-invariant systems based on the Dirac operator which are
asymptotically stable; this is in spite of the energy functional being
unbounded from below.

First results on asymptotic stability were obtained in
\cite{cuevas-Bou06,cuevas-Bou08} in the case $n=3$, in the external potential. There, the
spectrum of the linear part of the equation $D_m+V$ is supposed to be, beside
the essential spectrum $\mathbb{R}\setminus(-m,m)$, formed by two simple eigenvalues; let us denote them by
$\lambda_0$ and $\lambda_1$, with $\lambda_0<\lambda_1$.
From the associated eigenspaces, there is a bifurcation of
small solitary waves for the nonlinear equation. The corresponding linearized
operators are exponentially localized small perturbations of $D_m+V$, so that
the perturbation theory allows a precise knowledge of the resulting spectral
stability. Depending on the distance from $\lambda_0$ to $\lambda_1$ compared to the distance from $\lambda_0$ to
the essential spectrum, the resulting point spectrum for the
linearized operator may be discrete and purely imaginary and hence spectrally
stable, or instead it may have nonzero-real-part eigenvalues if a ``nonlinear Fermi
Golden Rule'' assumption is satisfied (similarly to the Schr\"odinger case, see~\cite{cuevas-Sig93,cuevas-BP95,cuevas-SW99}); in the latter case, linear and dynamical
instabilities occur. In the former case, the linear stability follows from
the spectral one via the perturbation theory. In any case, using the
dispersive properties for perturbations of $D_m$, there is a stable manifold
of real codimension $2$. Due to the presence of nonzero discrete modes, even
in the linearly stable case, the dynamical stability is not guaranteed.
Before considering the results on the dynamics outside this manifold, for
perturbations along the remaining two real directions, one could ask what
might happen if $D_m+V$ has only one eigenvalue. The answer follows quite
immediately with the ideas from \cite{cuevas-Bou06,cuevas-Bou08}. In this case, there is
only one family of solitary waves and it is asymptotically stable. Notice
that the asymptotic profile is possibly another solitary wave but close to
the perturbed one. In the one-dimensional case, this was studied properly in
\cite{cuevas-PS14}. Note that the one-dimensional framework suffers from relatively
weak dispersion which makes the analysis of the stabilization process more
delicate. As for the dynamics outside the above-mentioned stable manifold,
the techniques rely on the analysis of nonlinear resonances between discrete
isolated modes and the essential spectrum where the dispersion takes place.
This requires the normal form analysis in order to isolate the leading
resonant interactions. The former is possible only if the `` nonlinear Fermi
Golden Rule'' is imposed.
Such an analysis was done in \cite{cuevas-BC12} but in a
slightly different framework: instead of considering the perturbative case
the authors chose the translation-invariant case, imposing a series
of assumptions that lead to the spectral stability of solitary waves. These
assumptions are verified in some perturbative context with $V\neq 0$. This
case is analyzed in \cite{cuevas-CT16}. The asymptotic stability approach from
\cite{cuevas-PS14,cuevas-BC12,cuevas-CPS17} is developed under important restrictions on the types
of admissible perturbations. These restrictions are needed to avoid the
translation invariance and, most importantly, to prohibit the perturbations
in the direction of exceptional eigenvalues $\pm 2\omega{i}$ of the
linearization operator at a solitary wave $\phi_\omega(x)e^{-{i}\omega t}$. These
eigenvalues are a feature of the Soler model (see
\cite{cuevas-Gal77,cuevas-DR79}); they are present in the spectrum for any
nonlinearity $f$ in the Soler model \eqref{eq:cuevas-Soler}, see \cite{cuevas-BC12a,cuevas-Gal77,cuevas-DR79}.
These eigenvalues are embedded into the essential spectrum when
$\vert{\omega}\vert>m/3$ and violate the ``nonlinear Fermi Golden Rule'': they do
not ``interact'' (that is, do not resonate) with the essential spectrum; the
energy from the corresponding modes does not disperse to infinity. This
does not allow the standard approach to proving the asymptotic stability.

\section{Stability of solitary waves: numerical results}\label{sec:cuevas-stability}

Once the theoretical background on linear stability has been presented, we review in this section some very recent numerical results on this topic. To this aim, we first include a brief introduction to the Evans function formalism \cite{cuevas-Eva72}, and then, detailed results based on numerical analysis of BdG-like spectral stability are shown for both 1D and 2D Soler models.

Let us recall some notation regarding the spectral stability, as we will make an extensive use of them in what follows. The {\em essential spectrum} corresponds to $\lambda\in {i}(-\infty,|\omega|-m] \cup {i}[-|\omega|+m,\infty)$. Embedded eigenvalues can be in the region $\lambda\in \pm{i}[-|\omega|+m,|\omega|+m]$ of the essential spectrum; for abbreviation, we denote this region as the {\em embedded spectrum} and the remaining part of the essential spectrum as {\em non-embedded spectrum}.

In what follows, without lack of generality we will take $g=m=1$ unless stated otherwise.

\subsection{Evans function approach to the analysis of spectral stability}
\label{subsec:cuevas-Evans}

The study of the spectral stability of the cubic 1D Soler model was performed in
\cite{cuevas-BC12a}, with the aid of the Evans function technique. This was the
first definitive linear stability result (as well as the first
\emph{definite} stability result) in the context of the nonlinear Dirac
equation.

Let us give more details. In order to compute
$\sigma\left(\mathcal{A}\sb\omega\right)$
we can employ the Evans function which
provides an efficient tool to locate the point spectrum. The Evans function
was first introduced by J.W. Evans \cite{cuevas-Eva72,cuevas-Eva72a,cuevas-Eva72b,cuevas-Eva75} in his
study of the stability of nerve impulses. In his work, Evans defined
$D(\lambda)$ to represent the determinant of eigenvalue problems
associated with traveling waves of a class of nerve impulse models.
$D(\lambda)$ was constructed to detect the intersections of the subspace of
solutions decaying exponentially to the right and the subspace of solutions
decaying exponentially to the left. Jones \cite{cuevas-Jon84} used Evans' idea to
study the stability of a singularly perturbed FitzHugh--Nagumo system. Jones
called it the Evans function, and the notation $E(\lambda)$ is now common.
The first general definition of the Evans function was given by Alexander et
al. \cite{cuevas-AGJ90} in their study of the stability for traveling waves of a
semi-linear parabolic system. Pego and Weinstein \cite{cuevas-PW92} expanded on
Jones' construction of Evans function to study the linear instability of
solitary waves in the Korteweg--de Vries equation (KdV), the
Benjamin--Bona--Mahoney equation (BBM), and the Boussinesq equation.
Generally, the Evans function for a differential operator $\mathcal{D}$ is an
analytic function such that $E(\lambda) = 0$ if and only if $\lambda$ is an
eigenvalue of $\mathcal{D}$, and the order of zero is equal to the algebraic
multiplicity of the eigenvalue.

Let us give a simple example which illustrates the nature of the Evans function.
Consider the stationary Schr\"odinger equation
\begin{equation} \label{eq:cuevas-simple_example_evans}
-\lambda^2 u(x)=H u(x),
\end{equation}
where $H=-\partial\sb x^2+V$ with $V\in C(\mathbb{R})$, $\mathop{\mathrm{supp}}(V)\subset (-1,1)$.
For $\lambda\in\mathbb{C}\setminus\{0\}$,
$\mathrm{Re}(\lambda)>0$,
it has the solutions $J_{+}(\lambda, x)$ and $J_{-}(\lambda, x)$,
$x\in\mathbb{R}$,
defined by their behaviour at $\pm\infty$:
\[
J_{+}(\lambda, x)=e^{-\lambda x}, \quad x\ge 1;
\qquad \quad J_{-}(\lambda, x)=e^{+\lambda x}, \quad x\le -1.
\]
We should note that $J_+$ and $J_-$ decay exponentially
as $x\to\pm\infty$, respectively, and they have the same asymptotics at $\pm\infty$ as the solutions to the equation
\begin{equation*} \label{eq:cuevas-simple_example_jost}
-\lambda^2 u(x)=H_0 u(x),
\end{equation*}
where $H_0=-\partial\sb x^2$,
which agrees with $H$ on $\mathbb{R}\setminus[-1,1]$.
We call $J_+$ and $J_-$ the Jost solution to (\ref{eq:cuevas-simple_example_evans}) and define the Evans function to be the Wronskian
of $J_{+}$ and $J_{-}$:
\[
E(\lambda)=W(J_+, J_-)(x,\lambda)
=J_+(x,\lambda) \partial\sb x J_-(x,\lambda)
- J_-(x,\lambda)\partial\sb x J_+(x,\lambda),
\]
where the right-hand side depends only on $\lambda$.
Vanishing of $E$ at some particular $\lambda\in\mathbb{C}$,
$\mathrm{Re}(\lambda)<0$
shows that the Jost solutions $J_{+}$ and $J_{-}$ are linearly dependent, and
there is $c\in\mathbb{C}\setminus\{0\}$
such that
\[
\phi(x)=
\begin{cases}
J_{+}(x,\lambda),\qquad x\ge 0
\\
c J_{-}(x,\lambda),\qquad x< 0
\end{cases}
\]
is $C^1$ and thus is an eigenfunction
corresponding to an eigenvalue $\lambda^2$
of $H$.

The construction for the one-dimensional Soler model is done by
decomposing $L^2(\mathbb{R},\mathbb{C}^4)$
into two invariant subspaces
for the operator $\mathcal{A}\sb\omega$ introduced in (\ref{eq:cuevas-JL}):
the ``even'' subspace, with even first and third components and
with odd second and fourth components, and the ``odd'' subspace, with odd
first and third components and with even second and fourth components; the
direct sum of the ``even'' and ``odd'' subspaces coincides with
$L^2(\mathbb{R}, \mathbb{C}^4)$.
The Evans function corresponding to the ``even'' subspace is defined by
\begin{equation} \label{eq:cuevas-1dEvans}
E_{even}(\lambda)=\mathop{\mathrm{det}}\left(R_1, R_3, J_1, J_2\right),
\end{equation}
where $R_j(x)$,
$1\le j\le 4$, are the solutions to the equation $\lambda R=\mathcal{A}\sb\omega R$
with the initial data
\[
R_j|_{x=0}=\mathbf{e}_j,
\qquad 1\le j\le 4,
\]
where $\mathbf{e}_j$, $1\le j\le 4$,
is the standard basis in $\mathbb{C}^4$.
$J_1$ and $J_2$ are the Jost solution of $\mathcal{A}\sb\omega$,
which are defined as the solutions to $\lambda \Psi=\mathcal{A}\sb\omega\Psi$
with the same asymptotics at $+\infty$ as the solutions to $\lambda \Psi=\mathbf(\mathbf{D}_m-\omega)\Psi$
which decay as $x\to+\infty$,
where
\[
\mathbf{D}_m=\begin{bmatrix}D_m & 0 \\ 0 & D_m\end{bmatrix}, \quad D_m=\begin{bmatrix} m & \partial_x \\ -\partial_x & -m \end{bmatrix}
=-{i}(-\sigma_2)\partial\sb x+m\sigma_3.
\]
The Evans function corresponding to the ``odd'' subspace is constructed by
using in \eqref{eq:cuevas-1dEvans} functions $R_2$ and $R_4$
instead of $R_1$ and $R_3$.
We note that, by Liouville's formula,
the right-hand side in \eqref{eq:cuevas-1dEvans} does not depend on $x$.

\begin{figure}
\begin{center}
\includegraphics[width=.45\textwidth]{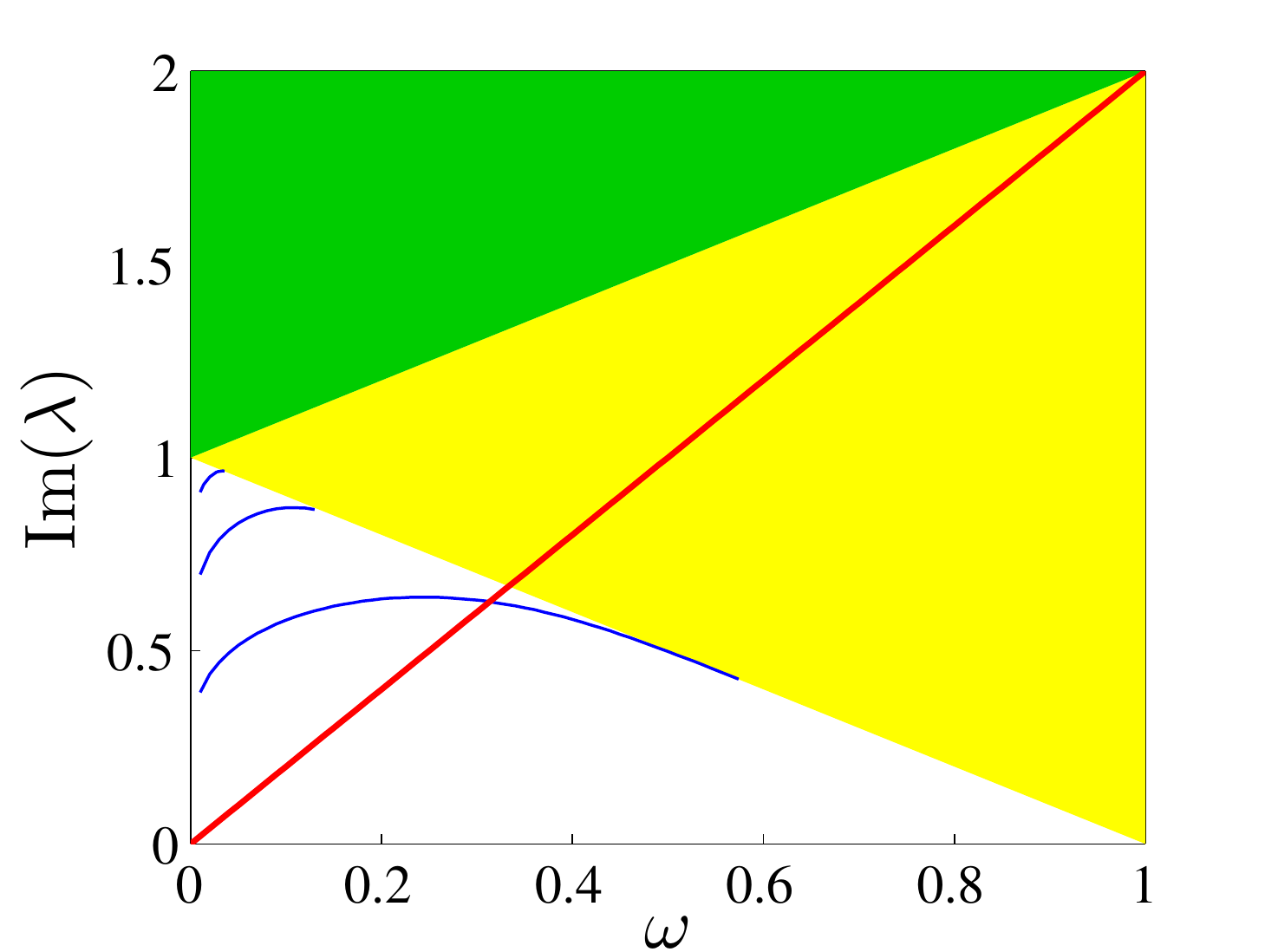}
\end{center}
\caption{Eigenvalues corresponding to zeros of the Evans function in the upper half of the spectral gap as a function of $\omega$.
Yellow area represents the part of the continuous spectrum
that corresponds to ${i} L_{\pm}(\omega)$,
while the green area represents the (doubly-covered) part of the continuous spectrum
corresponding to both ${i} L_{\pm}(\omega)$ and $-{i} L_{\pm}(\omega)$.
The eigenvalues $\lambda=2\omega i$ (red straight line) are embedded into the essential spectrum for $\omega>m/3$.}
\label{cuevas-fig9}
\end{figure}

Fig. \ref{cuevas-fig9} shows the zeros of the Evans function
which are plotted alongside with the essential spectrum for the linearization at the solitary waves in the 1D Soler model.

Later, in \cite{cuevas-CKS+16a}, it was observed that the linearized operator admits
invariant subspaces which correspond to spinorial spherical harmonics. This
allows one to factorize the operator, essentially reducing the consideration
to a one-dimensional setting, and to perform a complete numerical analysis of
the linearized stability in the nonlinear Dirac equation in two spatial
dimensions and give partial results in three dimensions, basing our approach on
both the Evans function technique and the linear stability analysis using
spectral methods.

For the two-dimensional Soler model, we can use the same process as the
one-dimensional case to construct the Evans function. Recall (see
\eqref{eq:cuevas-stab2D_partial}) that $\mathcal{A}\sb\omega$ acts invariantly on
$\mathcal{X}_q$ for each $q\in \mathbb{Z}$ and
$\mathcal{A}_{\omega,q}
=\mathcal{A}\sb\omega|_{\mathcal{X}_q}$. We consider the case $S=0$. The
Evans function for each $\mathcal{A}_{\omega,q}$ is defined by
\begin{equation*} \label{eq:cuevas-2dEvans}
E_q(\lambda)=\mathop{\mathrm{det}}(R^{+}_q, R^{-}_q, Y_1, Y_2).
\end{equation*}
Here $R^{+}_q$ and $R^{-}_q$ are linearly independent solutions to the equation $\lambda R=\mathcal{A}_{\omega,q} R$ with the following linearly independent initial data at $r=0$
\begin{equation*}
\begin{bmatrix}
-\lambda-(\omega+f_0)\frac{{i} q}{|q|}
\\
0
\\
{i}|q|
\\
q
\end{bmatrix}
\quad \text{and} \quad
\begin{bmatrix}
0
\\
-{i}\lambda\frac{q}{|q|}+\omega+f_0
\\
|q|
\\
{i} q
\end{bmatrix},
\end{equation*}
where $f_0=m-g\left(u^2(0)-v^2(0)\right)$. The Jost solutions $Y_1$ and $Y_2$ of $\mathcal{A}_{\omega,q}$ are defined as the solution to $\lambda Y=\mathcal{A}_{\omega,q} Y$ with the same asymptotics at $+\infty$ as the solutions to $\lambda Y=\mathbf{D}_q Y$ where
\[
\mathbf{D}_q=\left[
\begin{array}{cc}
-\sigma_1\frac{{i} q}{r} & {D}_m-\omega I_2
\\[2ex]
- {D}_m+\omega I_2 & -\sigma_1\frac{{i} q}{r}
\end{array}
\right],
\qquad
q\in\mathbb{Z}.
\]

\subsection{Bogoliubov--de Gennes analysis: The one-dimensional case}\label{subsec:cuevas-stability1D}

\begin{figure}[tb]
\begin{center}
\begin{tabular}{cc}
\includegraphics[width=.45\textwidth]{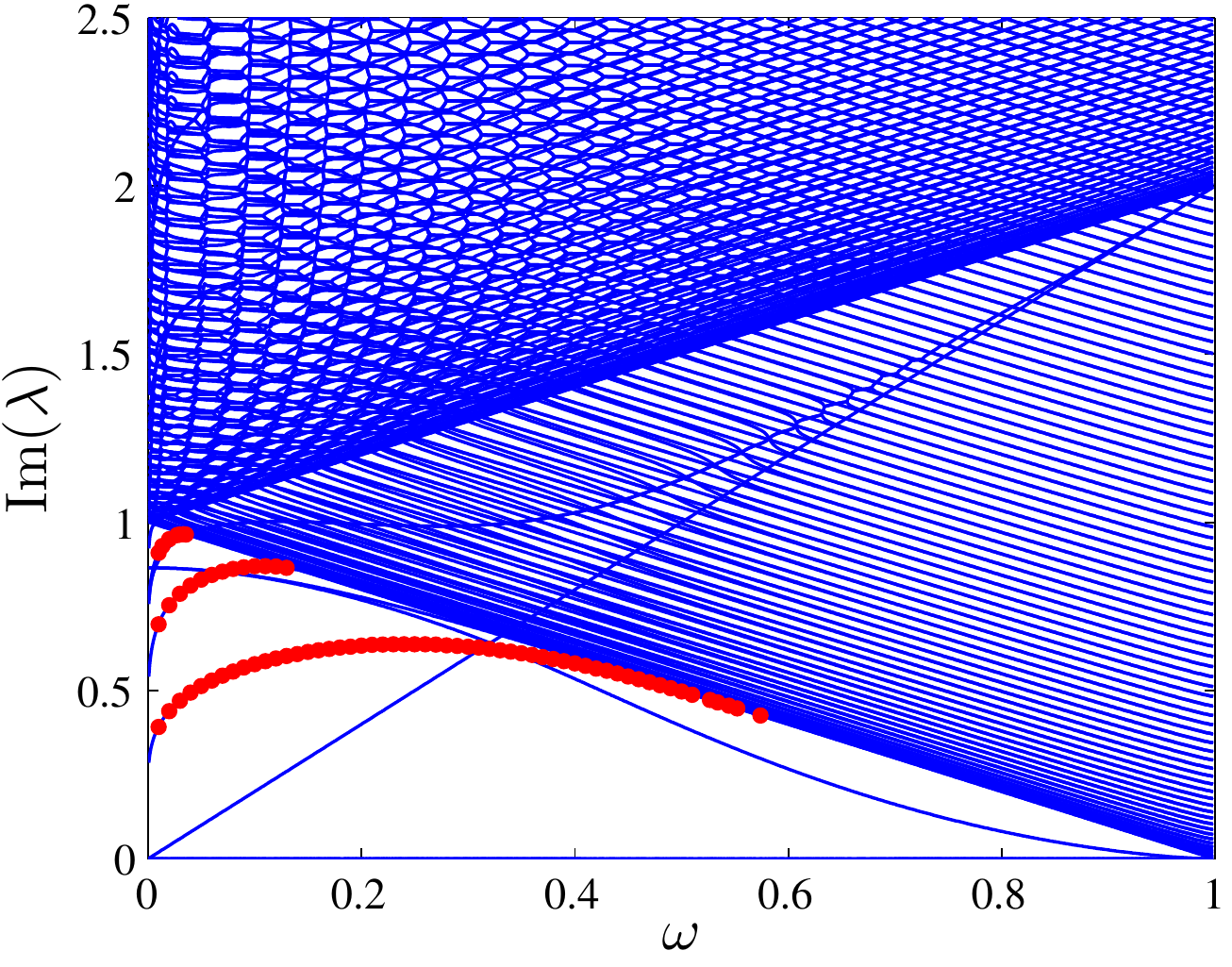} \hfill &
\includegraphics[width=.45\textwidth]{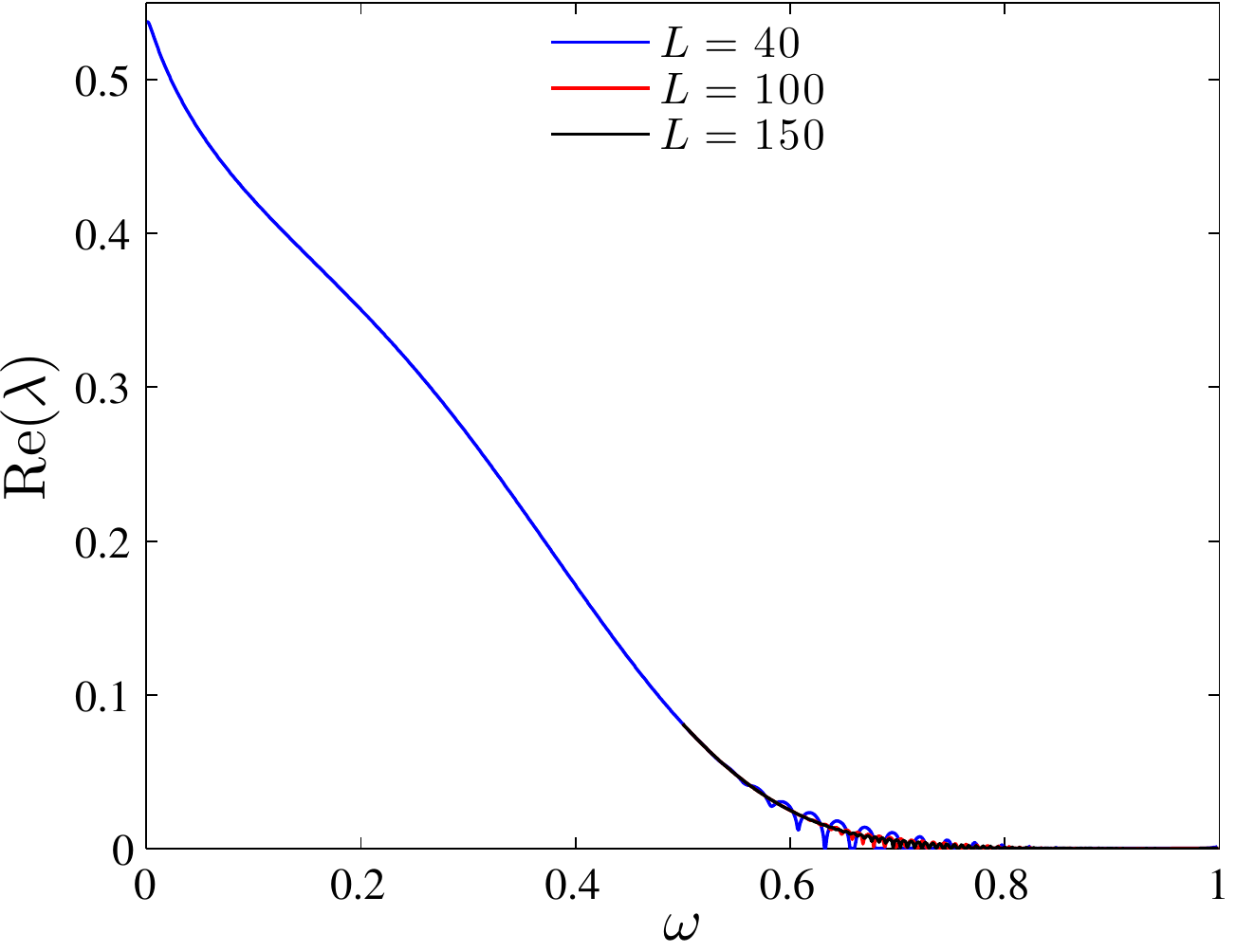} \\
\\
\end{tabular}
\end{center}
\caption{Spectrum of the stability matrix (\ref{eq:cuevas-stab1D}) for solitary waves in domain $[-L,L]$ with $L=40$ obtained using finite differences with $N=800$ grid points in the cubic ($k=1$) case. For the sake of simplicity, only the positive real and imaginary parts of the eigenvalues are shown. Dots correspond to Evans function predictions. Right panel displays only the maximum values of $\mathrm{Re}(\lambda)$ (i.e. the growth rates) and includes the values for $L=40$, $L=100$, and $L=150$.}
\label{cuevas-fig10}
\end{figure}

Let us recall from the analysis shown in Section \ref{sec:cuevas-stability_theory} that near the non-relativistic limit ($\omega\lesssim m$),
the stability of solitary waves
formally agrees with the Vakhitov--Kolokolov stability criterion
$\partial\sb\omega Q(\phi\sb\omega)<0$ \cite{cuevas-VK73}.
In particular, there is no positive eigenvalue emerging from $\lambda=0$
for $\omega\lesssim m$
as long as $k\leq2$ (and, consequently, the solitary waves
are spectrally stable),
while in the case $k>2$
there is a pair of (a positive and a negative) eigenvalues
which result in linear instability.
As it turns out, in the one-dimensional case,
the Vakhitov--Kolokolov stability criterion
agrees with the observed stability of solitary waves
not only in the nonrelativistic limit,
but for all frequencies $\omega\in(0,m)$,
as our numerical calculations show below. Evans function analysis presented above also shows that solitary waves do not present oscillatory instabilities (i.e. there are no complex $\lambda$'s with nonzero real part) in the 1D case; the instability could only develop when eigenvalues collide and bifurcate from the origin. Additionally, for any $k$, the existence of an eigenvalue $\lambda=\pm2\omega{i}$ is a consequence of the $\mathbf{SU}(1,1)$-invariance of the Soler model \cite{cuevas-Gal77,cuevas-DR79}. This mode, which does not give rise to any instability, is embedded into the essential spectrum for
$\omega\in(m/3,m)$ (see Fig.\ref{cuevas-fig10}).

Let us mention that it was shown in~\cite{cuevas-MQC+12,cuevas-SQM+14} that attempts to apply Derrick's argument~\cite{cuevas-Der64} to stability of solitary waves
in the context of the nonlinear Dirac equation \cite{cuevas-Bog79,cuevas-SV86} -- in particular, the so-called Bogolubsky criterion -- do not seem to work.
This is not particularly surprising, given that Derrick's empirical argument, based on singling out one family of perturbations of a solitary wave
and checking whether the solitary wave corresponds to the energy minimum on this curve, was introduced in the context of the second order systems, appealing to our Newtonian-world intuition. Apparently, this approach does not necessarily work in the context of the first order systems, such as the Dirac equation.

Another surprising result was explored by some
of the present authors in \cite{cuevas-CKS15}. It corresponds to the BdG analysis using finite difference discretization of the 1D Soler model; that is, the spatial derivatives $\partial_x f(x)$ in (\ref{eq:cuevas-Soler1D}) are substituted by the central difference $(f_{n+1}-f_{n-1})/(2h)$. This method is tantamount to using the collocation points of (\ref{eq:cuevas-colloc_Fourier}) and (\ref{eq:cuevas-difmat_fd}) with $N$ collocation points, a domain $x\in[-L,L]$ and $h=2L/N$. Fig. \ref{cuevas-fig10} shows the stability eigenvalues for $k=1$ and $L=80$ with a discretization step $h=0.1$. Although there are instabilities caused by eigenvalue collisions in the non-embedded spectrum, we neglected them, as they disappear in the limit of $h\rightarrow0$ and $L\rightarrow\infty$. The solitary waves were found to be unstable for small $\omega$, with a growth rate that decreases when $\omega$ is increased. The source of instabilities is a localized mode (with non-zero real part of its eigenvalue even when $\omega\rightarrow0$) that enters the essential spectrum at $\omega\approx0.037$ i.e. it embeds into the essential spectrum. Once inside the linear modes band, this localized mode causes multiple ``bubbles", but at $\omega\approx0.632$, it returns to the imaginary eigenvalue axis and the solitary wave becomes stable. Nevertheless, this stability is ephemeral, as the solitary waves become unstable again at $\omega\approx0.634$. From this point, there is a succession of instability bubbles, whose amplitude (i.e., the maximal growth rate associated with them) decreases with $\omega$. In order to observe the behavior of bubbles when the domain is enlarged, the same figure compares the growth rates for $L=40$, $100$ and $150$. It is observed that the number of bubbles increases with $L$, but their width decreases. In any case, the envelope of the bubbles tends to zero asymptotically when $\omega$ approaches 1, in a similar way as it was observed for dark solitons in the discrete nonlinear Schr\"odinger equation (DNLS) setting~\cite{cuevas-JK99}. The convex nature of the relevant (apparent) envelope curve is inconclusive in connection with the stability aspect; it is unclear, based on those computations, whether the curve, as $h \rightarrow 0$, still intersects the axis and no longer features an unstable mode past a critical value of $\omega$. The alternative scenario is that the approach to the stable NLS limit of $\omega \rightarrow 1$ is merely asymptotic.

In order to find out a strategy which assures a spectral accuracy of BdG stability analysis which is also correlated to the Evans' function analysis, we used spectral collocation methods in \cite{cuevas-CKS+15}. We utilized two case examples of such methods therein: the Fourier Spectral Collocation Method, which implicitly enforces periodic boundary conditions, and the Chebyshev Spectral Collocation Method, which enforces (homogeneous) Dirichlet boundary conditions (see Subsection \ref{subsec:cuevas-Soler2D}). The advantage of the Finite Difference Method with respect to the other ones concerns the fact that the resulting stability matrix is sparse. In the computations performed in that work and that will be presented below, $N=800$ collocation points were taken in a domain $[-L,L]$, with a discretization parameter $h=1/(2L)$; this value coincides with the distance between grid points in the Fourier collocation and finite difference methods, but not in the Chebyshev collocation as the grid points are not equidistant. Increasing the node numbers to $N=1200$ does not seem to qualitatively improve the findings.

\begin{figure}[tb]
\begin{center}
\begin{tabular}{cc}
\includegraphics[width=.45\textwidth]{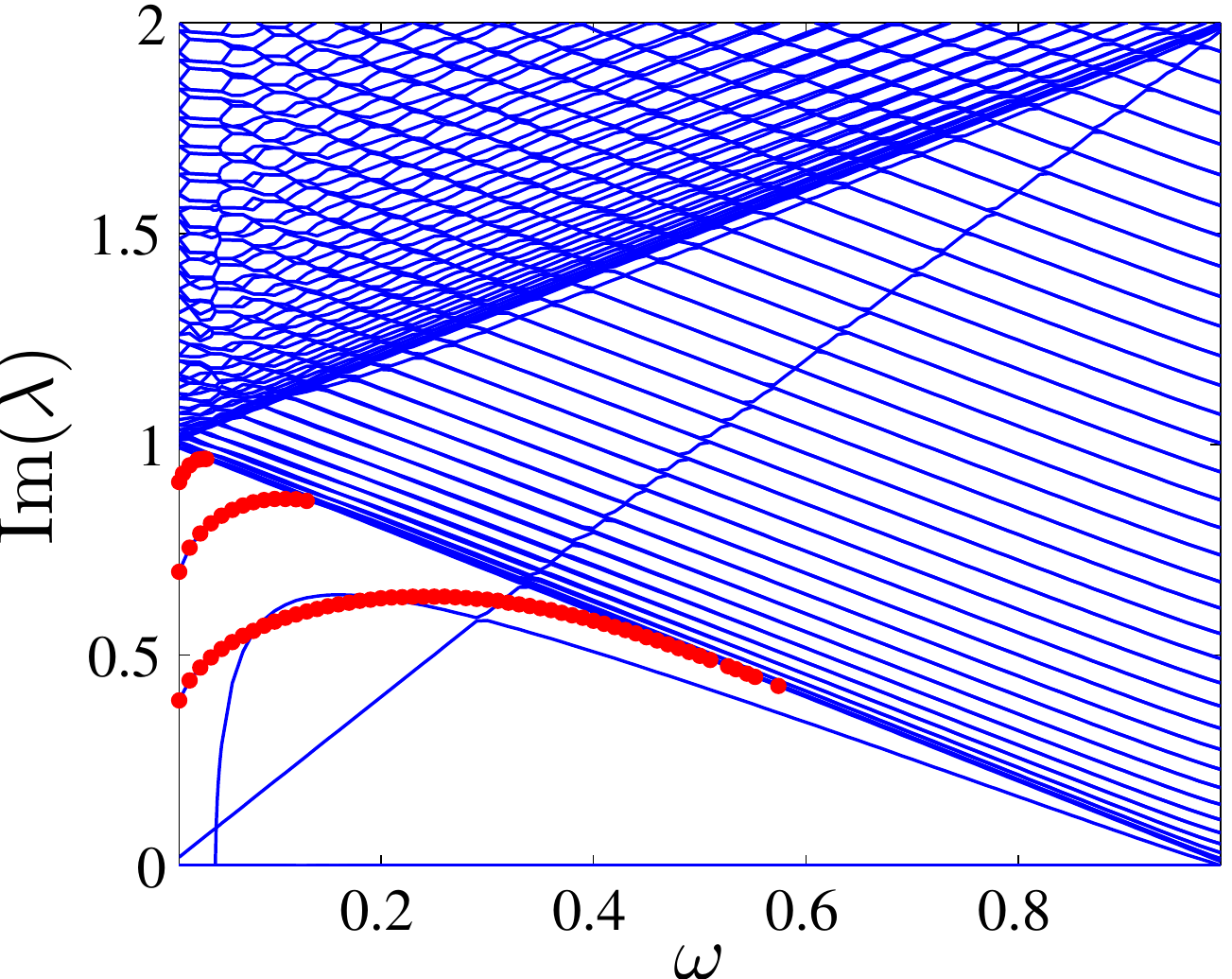} \hfill &
\includegraphics[width=.45\textwidth]{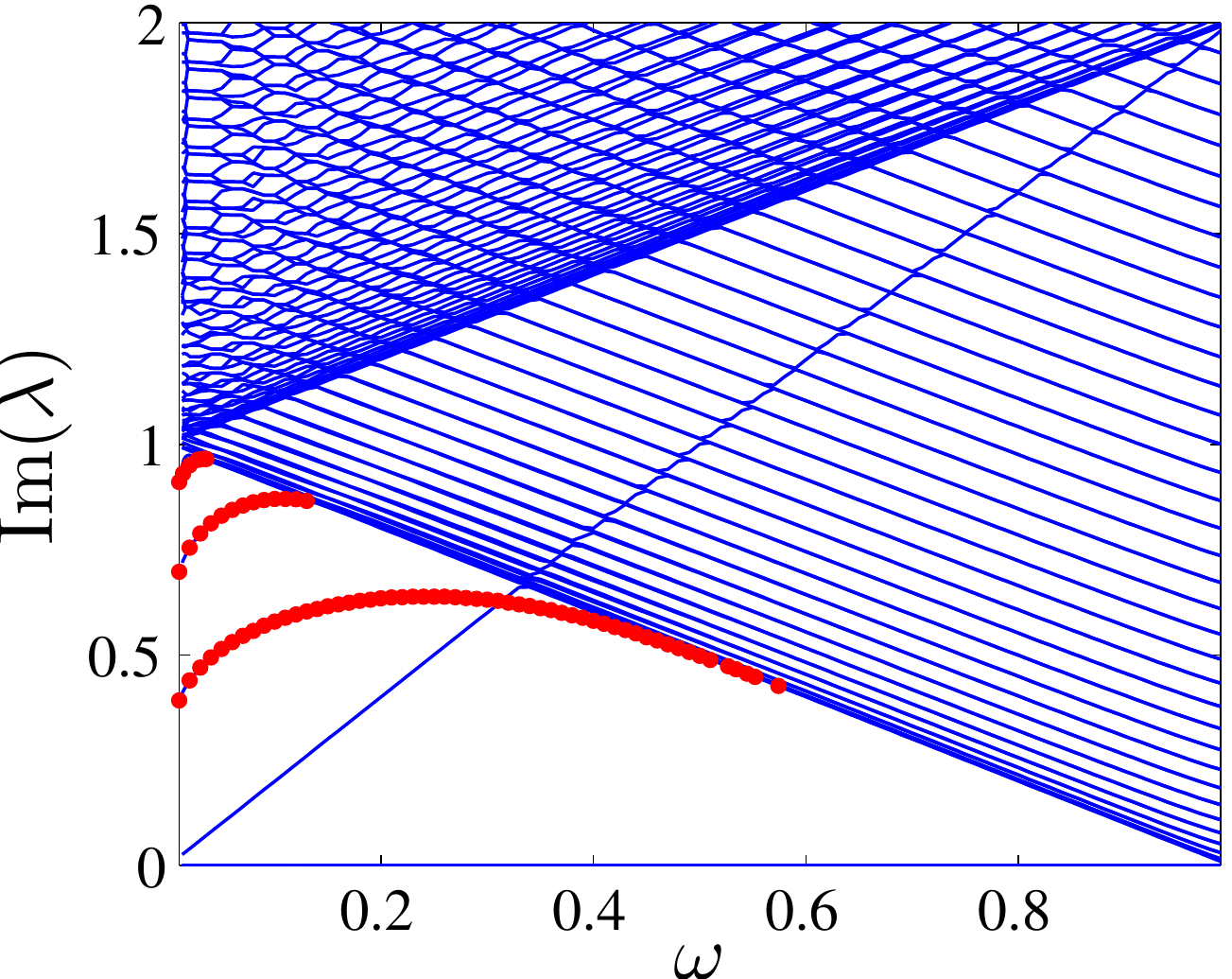} \\
\end{tabular}
\end{center}
\caption{Imaginary part of the spectrum of the stability matrix (\ref{eq:cuevas-stab1D}) for solitary waves in domain $[-L,L]$ with $L=40$ obtained using Fourier (left) and Chebyshev (right) spectral collocation method with $N=800$ grid points in the cubic ($k=1$) case. For the sake of simplicity, only the positive imaginary parts of the eigenvalues are shown. Dots correspond to Evans function predictions.}
\label{cuevas-fig11}
\end{figure}

\begin{figure}[tb]
\begin{center}
\begin{tabular}{cc}
\includegraphics[width=.45\textwidth]{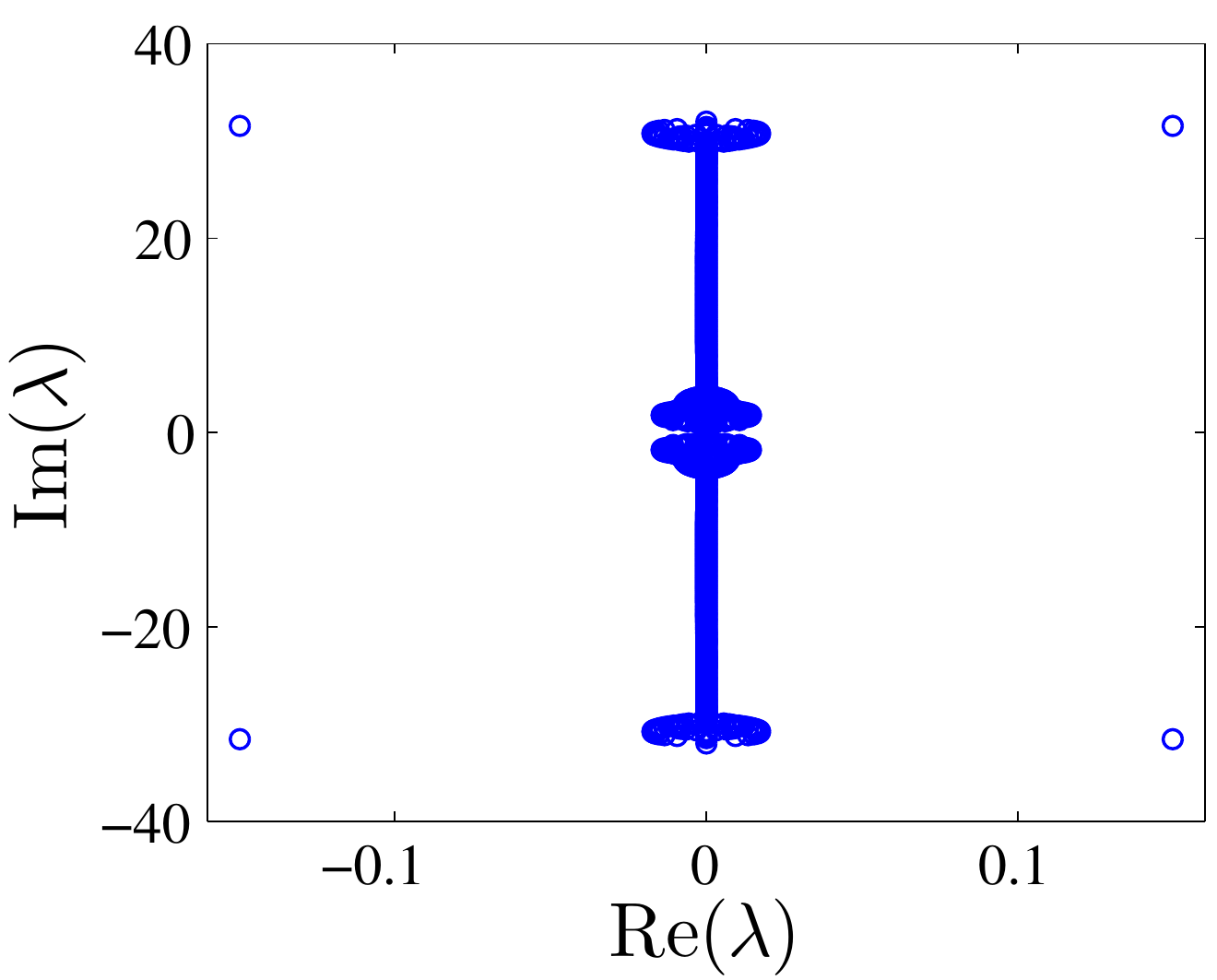} \hfill &
\includegraphics[width=.45\textwidth]{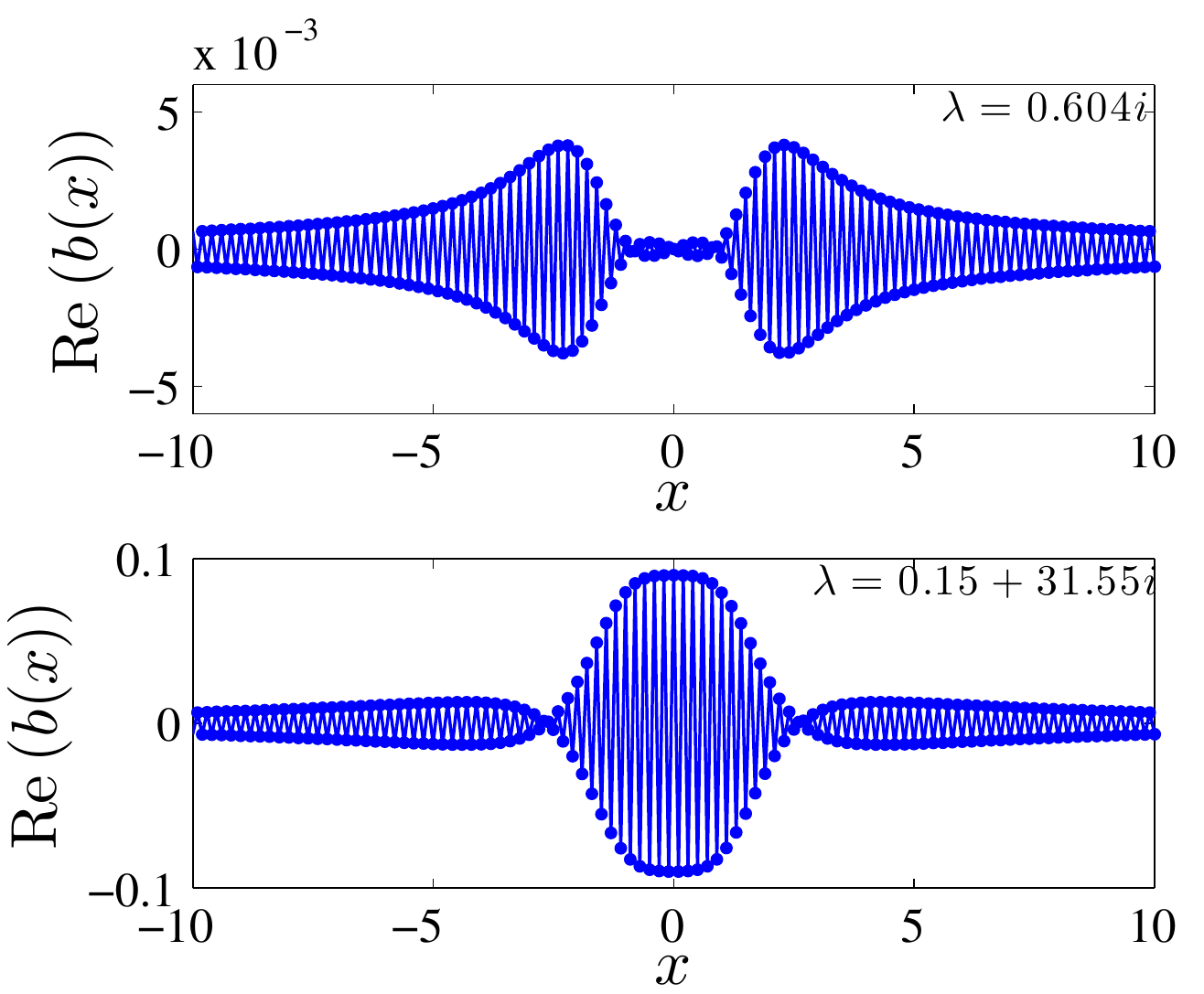} \\
\end{tabular}
\end{center}
\caption{Spectral plane of a solitary wave with $\omega=0.1$
(cubic case, $L=40$, and $N=800$)
obtained using the Fourier spectral collocation method (left panel).
The typical profile of two modes corresponding to spurious eigenvalues is depicted in the right panel. In particular, we have included the mode with $\mathrm{Re}(\lambda)=0$ which does not arise in the Evans' function analysis of Subsection \ref{subsec:cuevas-Evans} together with the largest real part eigenvalue, which is also spurious.}
\label{cuevas-fig12}
\end{figure}

\begin{figure}[tb]
\begin{center}
\includegraphics[width=.45\textwidth]{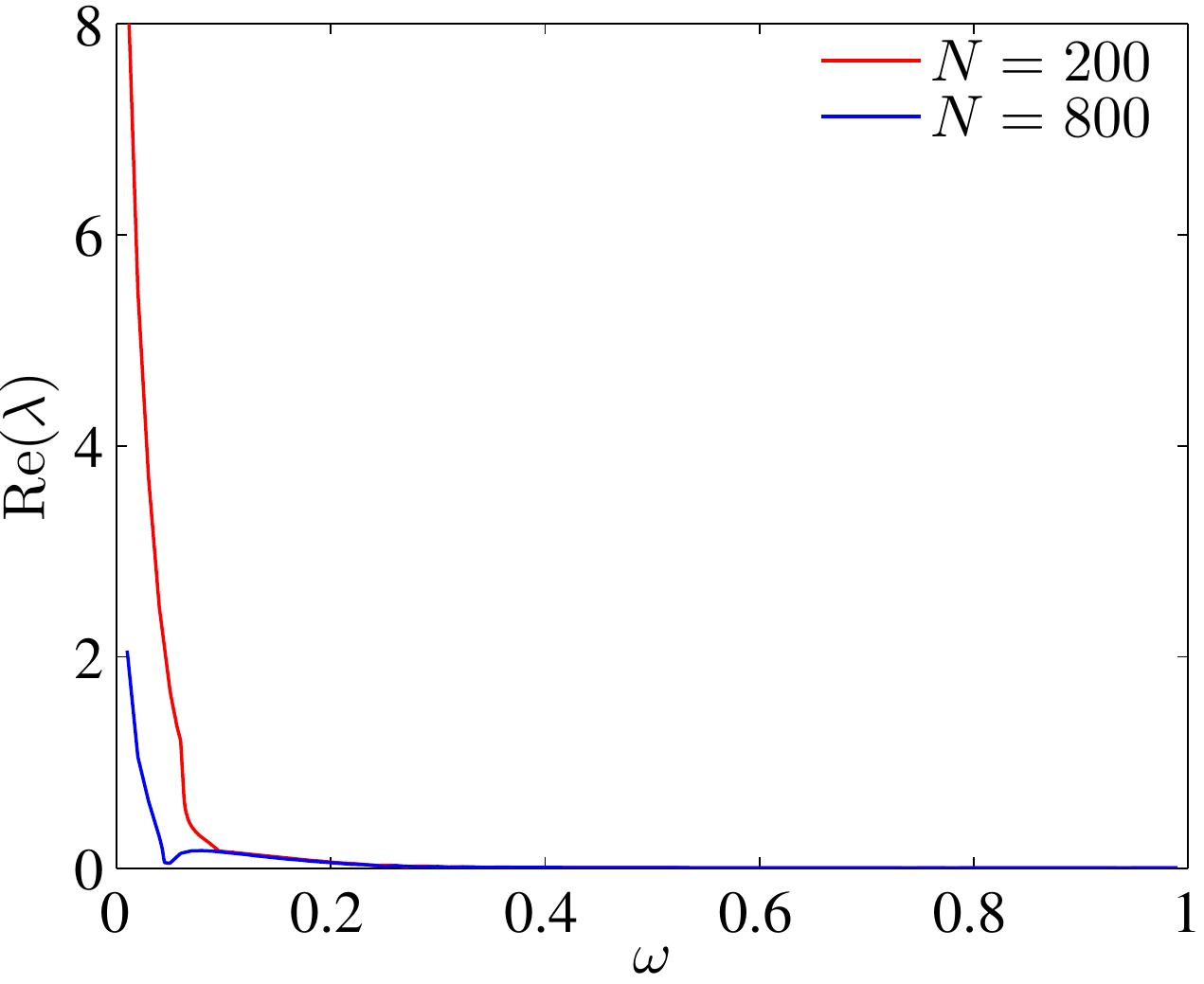}
\end{center}
\caption{Growth rates (i.e. maximum of the real part of the eigenvalues) for a solitary wave with $L=40$ in the cubic case using the Fourier spectral collocation method. The number of grid points is either $N=800$ or $N=200$.}
\label{cuevas-fig13}
\end{figure}

In Fig.~\ref{cuevas-fig11} we examine the dependence of the imaginary part of the eigenvalues $\lambda$ with respect to the frequency $\omega$ of the solution for both spectral methods in the cubic case of $k=1$. In addition to the $\lambda=\pm 2\omega{i}$ mode, the different methods have additional modes which can be compared also with the Evans function analysis outcome of Fig.~\ref{cuevas-fig9}. We thus find that the comparison of the Fourier spectral collocation method with the Evans function analysis (Fig.~\ref{cuevas-fig9}) seems qualitatively (and even quantitatively) to yield very good agreement with the exception of a mode that seems to initially grow steeply (for small $\omega$) and subsequently to slowly asymptote to the band edge (as $\omega$ increases). This mode is shown in the right panel of Fig.~\ref{cuevas-fig12}, while the left panel of the figure illustrates a prototypical example of the Fourier spectral collocation method spectrum for $\omega=0.1$. From the above panel, we can immediately infer that this mode is, in fact, spurious and an outcome of the discretization as it carries a staggered profile that cannot be supported in the continuum limit. In the left panel of the same figure, we can see the existence of additional spurious modes forming bubbles of complex eigenvalues. However, the fact that these bubbles are occurring at the eigenvalues of the continuous spectrum assures us that these are spurious instabilities due to the finite size of the domain and ones which disappear in the $L \rightarrow \infty$, $h \rightarrow 0$ limit. This is confirmed by Fig.~\ref{cuevas-fig13} which shows that as we decrease $h$ (and increase the number of lattice sites, approaching the continuum limit for a given domain size) the growth rate of such spuriously unstable eigenmodes accordingly decreases.

\begin{figure}[tb]
\begin{center}
\begin{tabular}{cc}
\includegraphics[width=.45\textwidth]{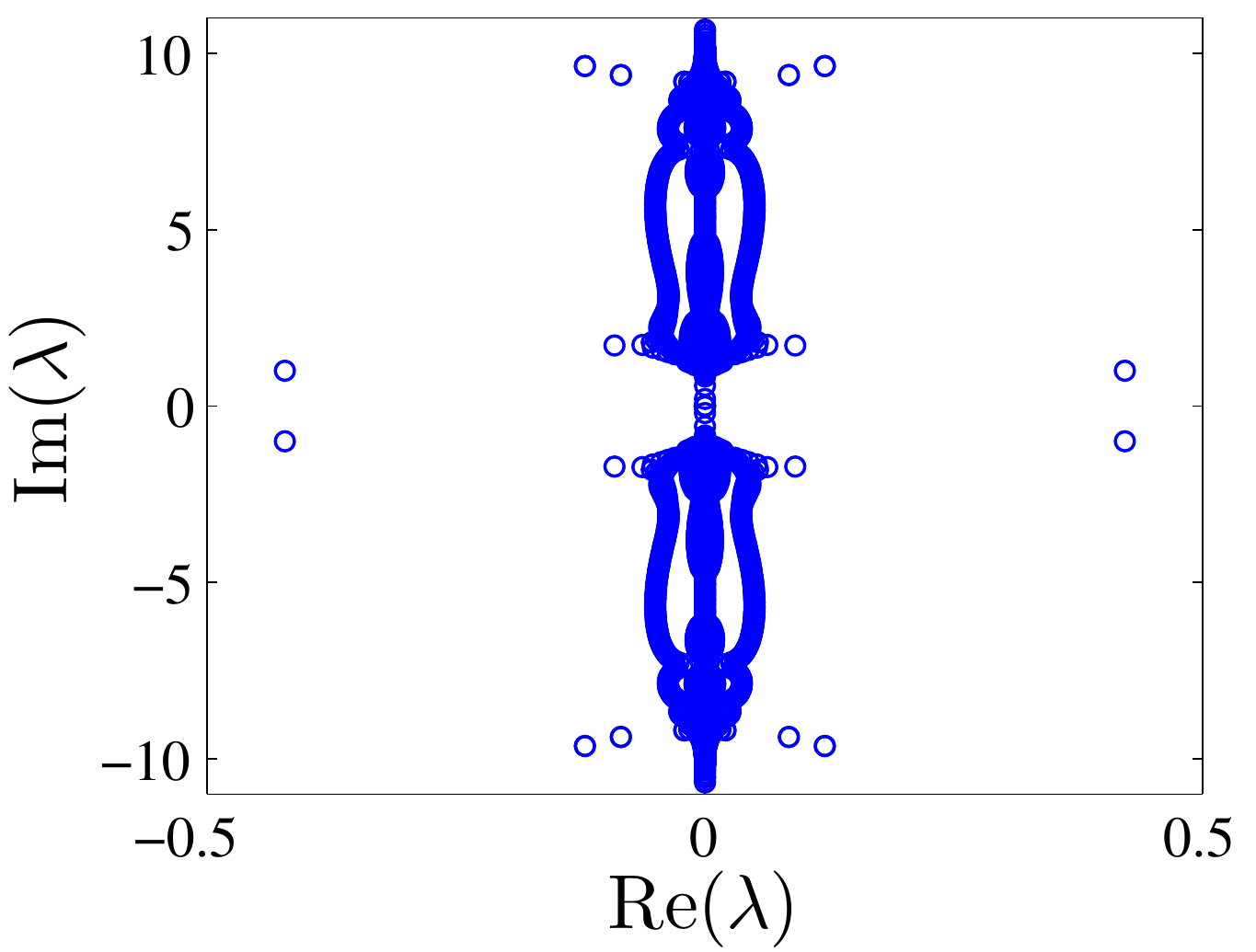} \hfill &
\includegraphics[width=.45\textwidth]{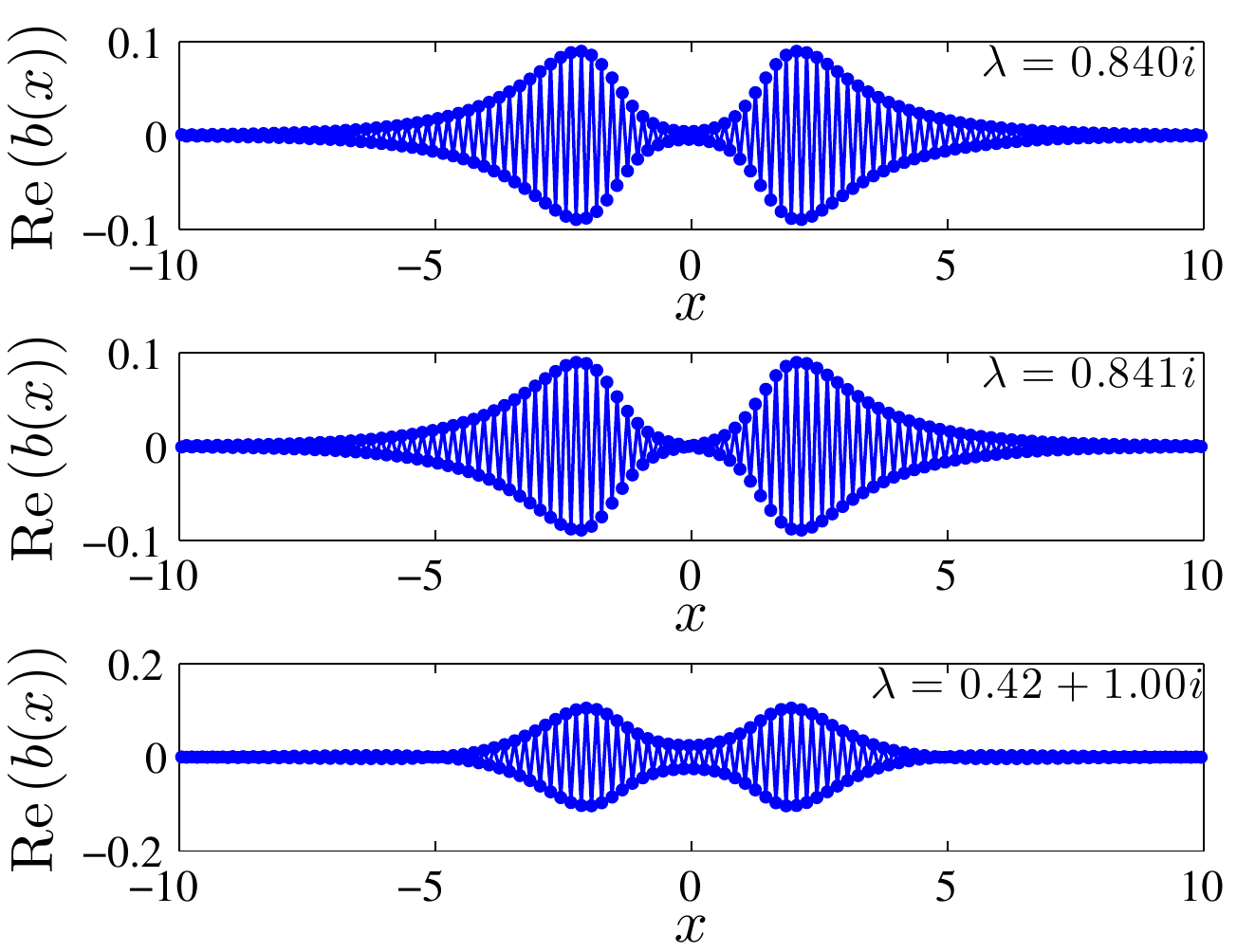} \\
\end{tabular}
\end{center}
\caption{Spectral plane of a solitary wave with $\omega=0.1$, $L=40$ and $N=800$ in the cubic case, using finite difference discretization (left panel). The typical profile of three modes corresponding to spurious eigenvalues is depicted in the right panel. In particular, we have included the two modes with $\mathrm{Re}(\lambda)=0$ which do not arise in the Evans function analysis together with the embedded spurious mode.}
\label{cuevas-fig14}
\end{figure}

Remarkably, the finite difference spectrum of Fig.~\ref{cuevas-fig10} is the one that seems most ``distant'' from the findings of the Evans function method. While all four of the internal modes of the latter spectrum seem to be captured by the finite difference method, three additional modes create a nontrivial disparity. Two of them are in fact ``benign'' and maintain an eigenvalue below the band edge of the continuous spectrum for all values of
$\omega\in (0,m)$. However, as explained in~\cite{cuevas-CKS15}, we also observe the existence of an eigenmode embedded in the essential spectrum. Unfortunately, this mode is accompanied by a real part in the corresponding eigenvalue and hence gives rise to a spurious instability. While the origin of this mode starting from the so-called anti-continuum limit will be thoroughly explained in Section \ref{sec:cuevas-DNLD}, the persistence and especially the instability inducing nature of such a mode remains an open problem as the continuum limit is approached. Fig.~\ref{cuevas-fig14} presents a graph analogous to Fig.~\ref{cuevas-fig12} but for the finite difference method. The undesirable unstable mode, as well as additional spurious modes are explicitly indicated through the eigenvector profiles of the right panel.

\begin{figure}[tb]
\begin{center}
\begin{tabular}{cc}
\includegraphics[width=.45\textwidth]{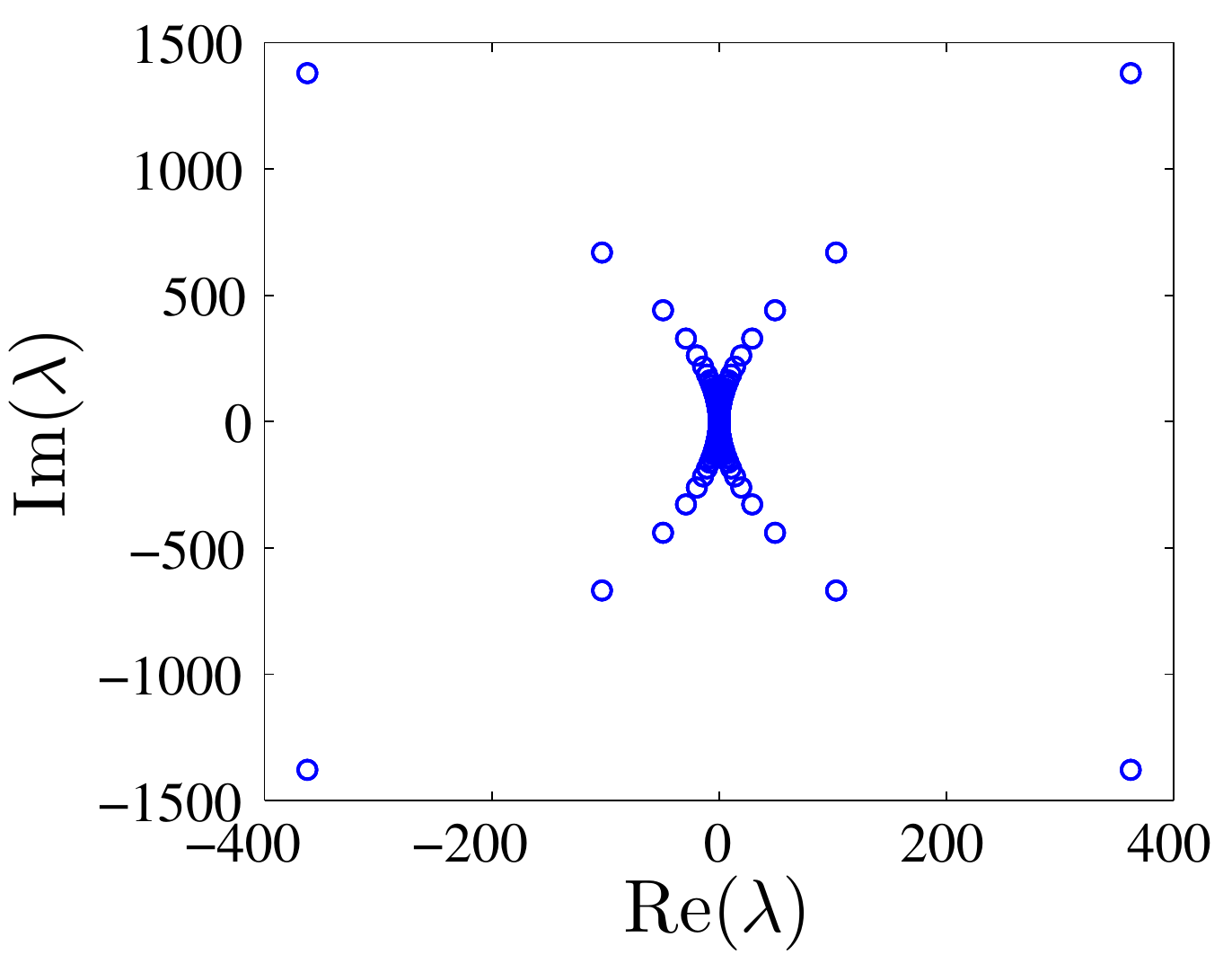} \hfill &
\includegraphics[width=.45\textwidth]{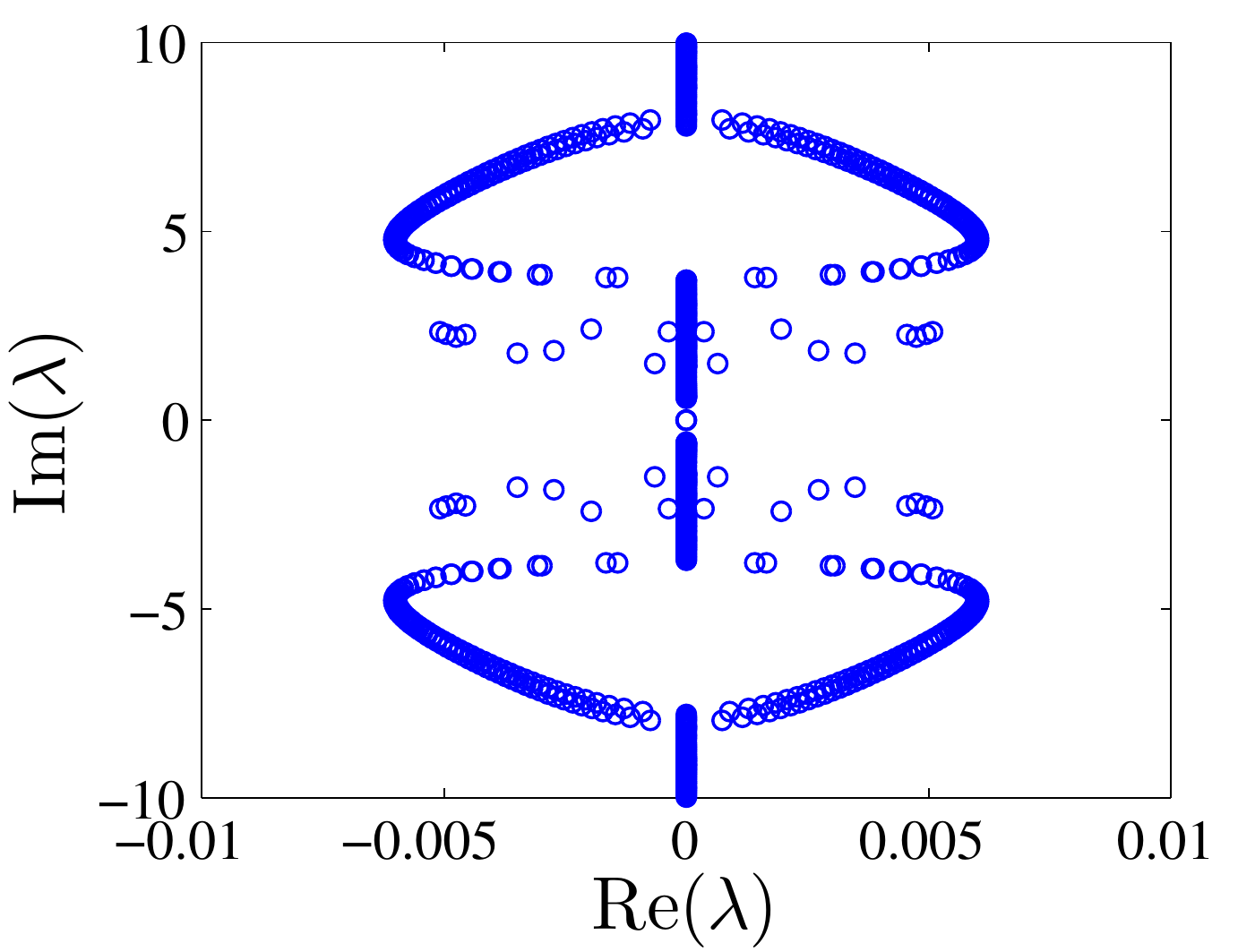} \\
\end{tabular}
\end{center}
\caption{Spectral plane of a solitary wave with $\omega=0.4$, $L=40$, and $N=800$ in the cubic case, using the Chebyshev spectral collocation method. The right panel is a zoom of that on the left, illustrating the weak, spurious instabilities (which disappear as the continuum limit is approached).}
\label{cuevas-fig15}
\end{figure}

The scenario of the Chebyshev spectral collocation method bears advantages and disadvantages in its own right. Although it gives an accurate result for the imaginary part of the eigenvalues, their real part grows for large $\mathrm{Im}(\lambda)$, as is also shown in Fig.~\ref{cuevas-fig15}. Additionally, as indicated in \cite{cuevas-Boy01}, approximately half of the values of the spectrum are spurious within the Chebyshev collocation methods, so they should be excluded from consideration. Furthermore, one can observe that in this case as well, spurious instability bubbles arise (see the right panel of Fig.~\ref{cuevas-fig15}), yet we have checked that these disappear in the continuum limit of $h \rightarrow 0$.

\begin{figure}[tb]
\begin{center}
\includegraphics[width=.45\textwidth]{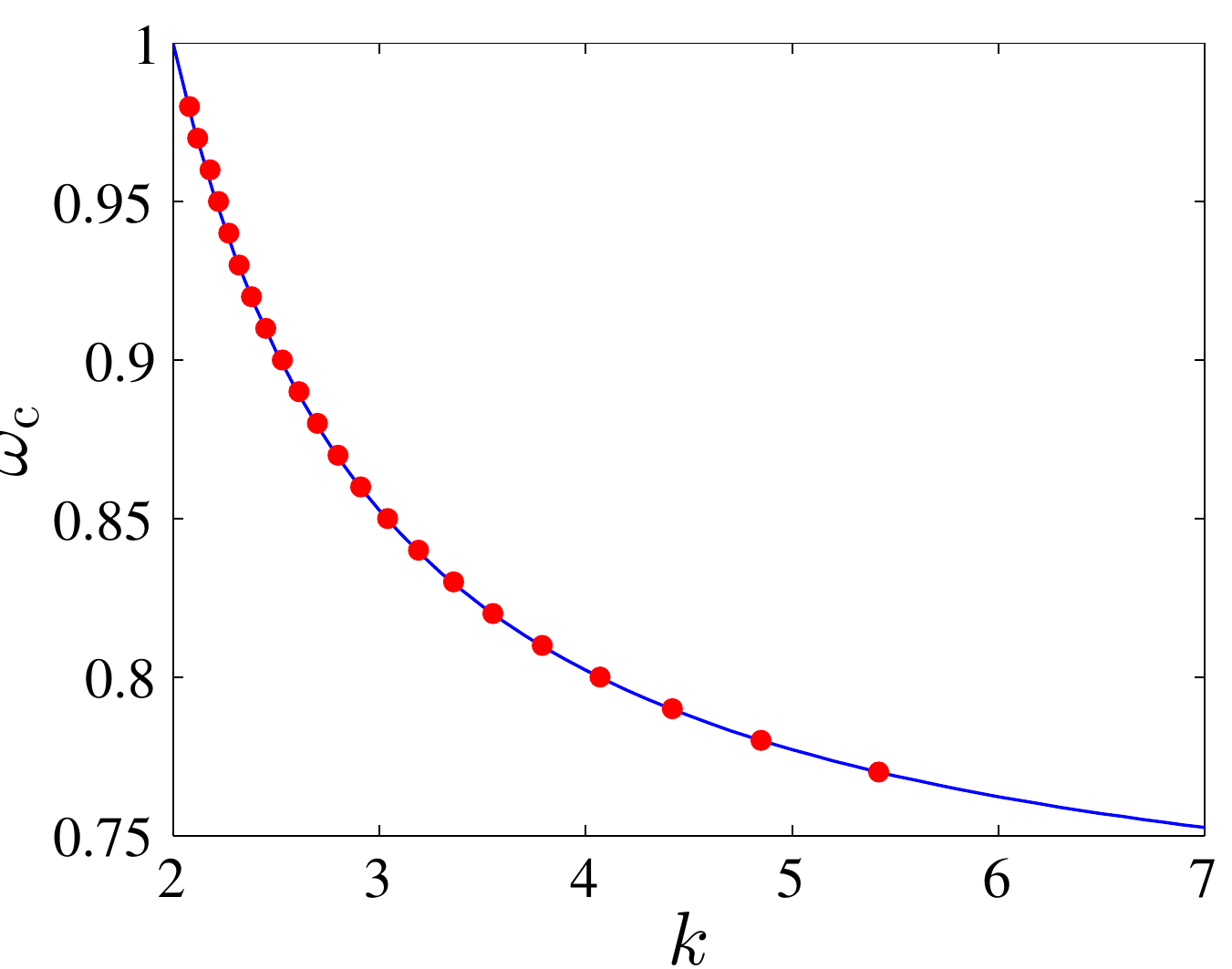}
\end{center}
\caption{Exponential bifurcation loci in the $\omega_c$-$k$ plane for the 1D Soler model.
The solitary waves
under the curve are linearly (spectrally) stable,
while the ones above the curve are linearly unstable. Full line corresponds to the application of the Vakhitov--Kolokolov criterion (i.e. points for which $\partial_\omega Q(\omega)=0$), whereas the dots correspond to the stability calculations.}
\label{cuevas-fig16}
\end{figure}

As a final aspect of the spectral considerations that we provide herein, we have examined the instability that arises e.g. from the Chebyshev spectral collocation method for larger values of $k$. Recall that the Chebyshev spectral collocation method predicts (at least as regards the point spectrum out of the non-embedded spectrum) that there is no instability for any $\omega$ in the case of $k=1$, in agreement with the Evans function analysis and \cite{cuevas-PS16}. The method identifies an instability for such point spectrum eigenvalues {\it only} for $k>2$. The relevant instability predicted numerically in the $k$-$\omega$ plane is illustrated in Fig.~\ref{cuevas-fig16}. We note that this instability is precisely captured by the Vakhitov--Kolokolov criterion, i.e. it precisely corresponds to the condition
$\partial_\omega Q(\omega)=0$,
in agreement with \cite{cuevas-BCS15}. Hence, by analogy with the nonrelativistic limit $\omega \rightarrow m=1$, we expect this to be an instability associated with the collapse of the latter model (however, we will observe a key dynamical difference, in comparison to the NLS, in Section \ref{sec:cuevas-dynamics}). Nevertheless, it is relevant to point out here that the NLD, contrary to the NLS, does not exhibit an instability for all $\omega$ when $k>2$. The instability is instead limited to $\omega > \omega_c(k)$, as characterized by the curve of Fig.~\ref{cuevas-fig16}. Hence, it can be inferred that the
instability is mitigated by the relativistic limit of the NLD and
only occurs in an interval of frequency values including
the non-relativistic limit $\omega \rightarrow m=1$, yet
not encompassing the full range of available frequencies in the
relativistic case.

\subsection{Bogoliubov--de Gennes analysis: The two- and three- dimensional cases}

From the experience acquired with the study of the stability of solitary waves in one spatial dimension, it is clear that a Chebyshev spectral collocation method must be followed in order to analyze the stability in higher-dimensional solitary waves. This is the approach followed in the present section, which summarizes the results of \cite{cuevas-CKS+16a}.

Let us remember that the spectrum of $\mathcal{A}\sb\omega$ is the union of spectra of the one-dimensional spectral problems (\ref{eq:cuevas-stab2D_partial}): $\sigma\left(\mathcal{A}\sb\omega\right)=\mathop{\cup}_{q\in\mathbb{Z}}\sigma\left(\mathcal{A}_{\omega,q}\right)$. In our numerics we have analyzed values of $q\in[-6,6]$, although the main phenomenology is captured by $|q|\leq4$ and those are the values shown in the next figures for the sake of better visualization.

\begin{figure}[tb]
\begin{center}
\begin{tabular}{cc}
\includegraphics[width=.45\textwidth]{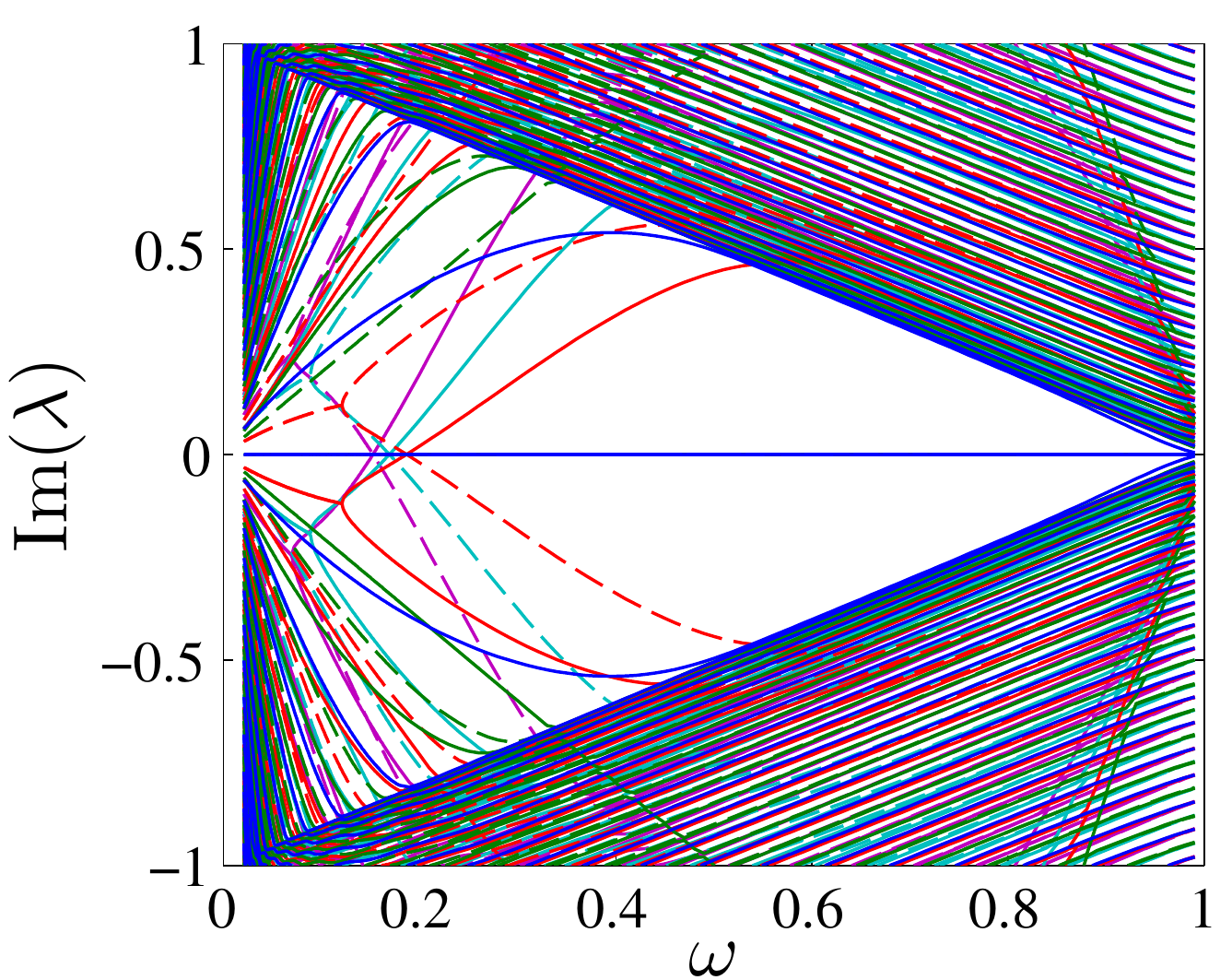} \hfill &
\includegraphics[width=.45\textwidth]{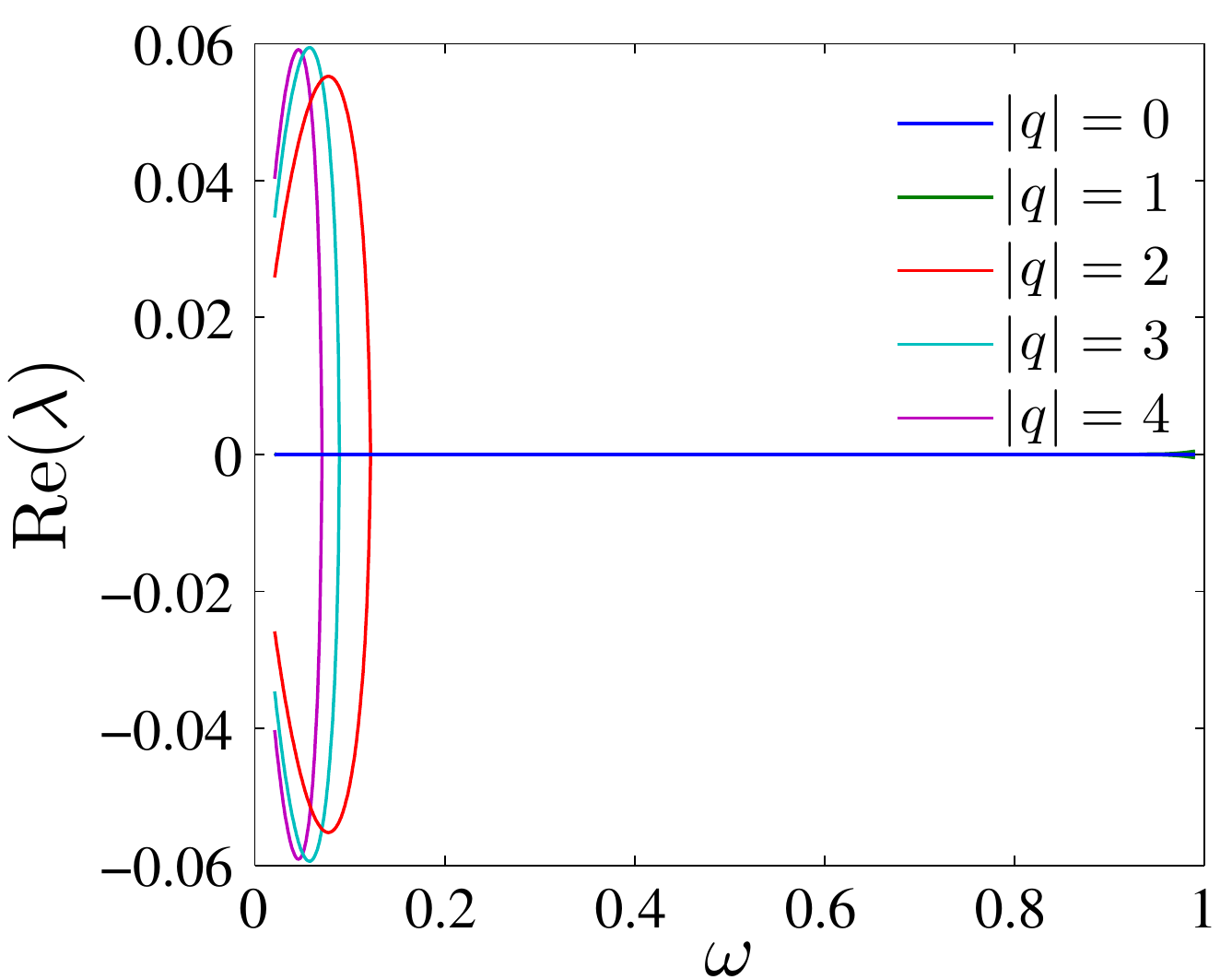} \\
\includegraphics[width=.45\textwidth]{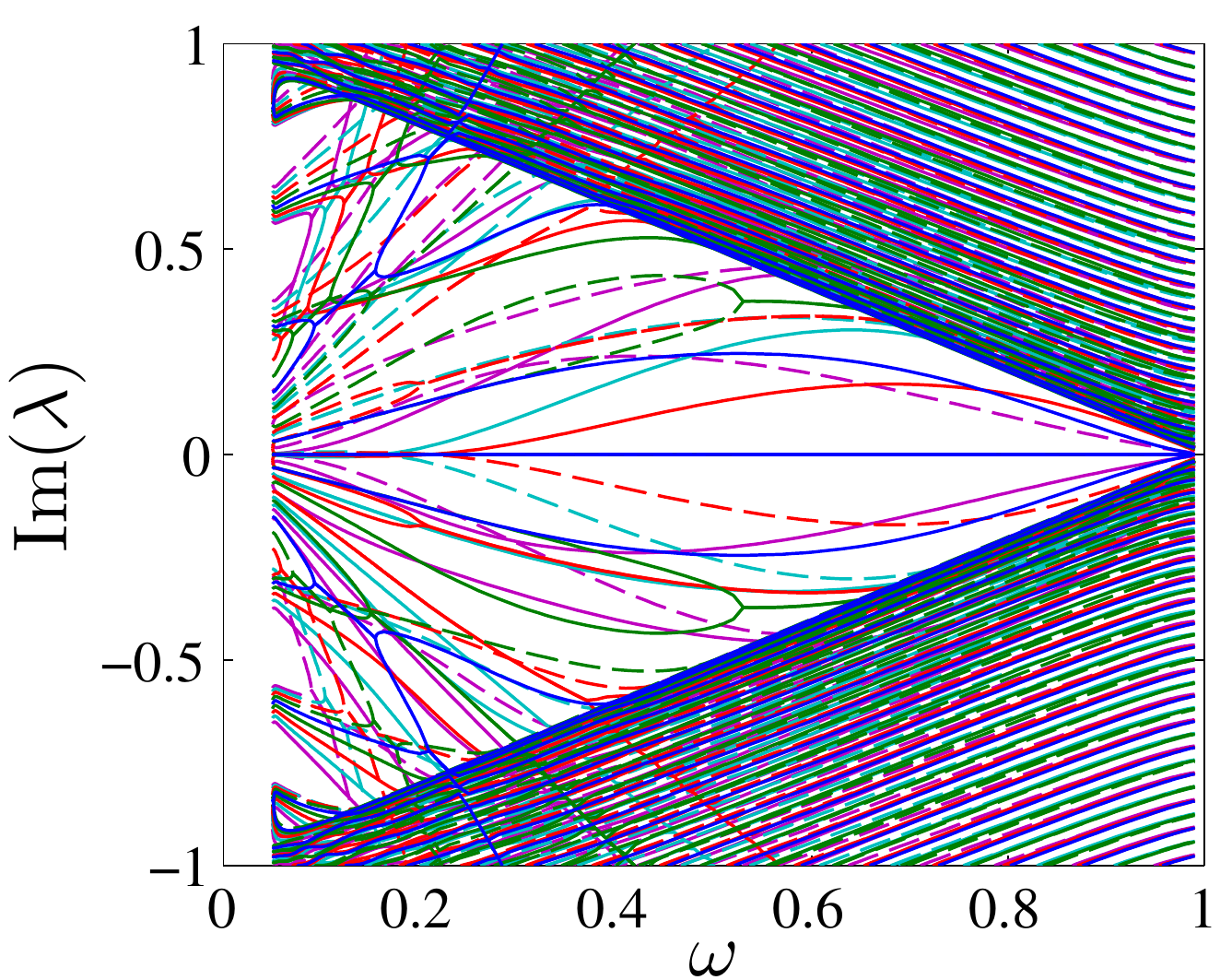} \hfill &
\includegraphics[width=.45\textwidth]{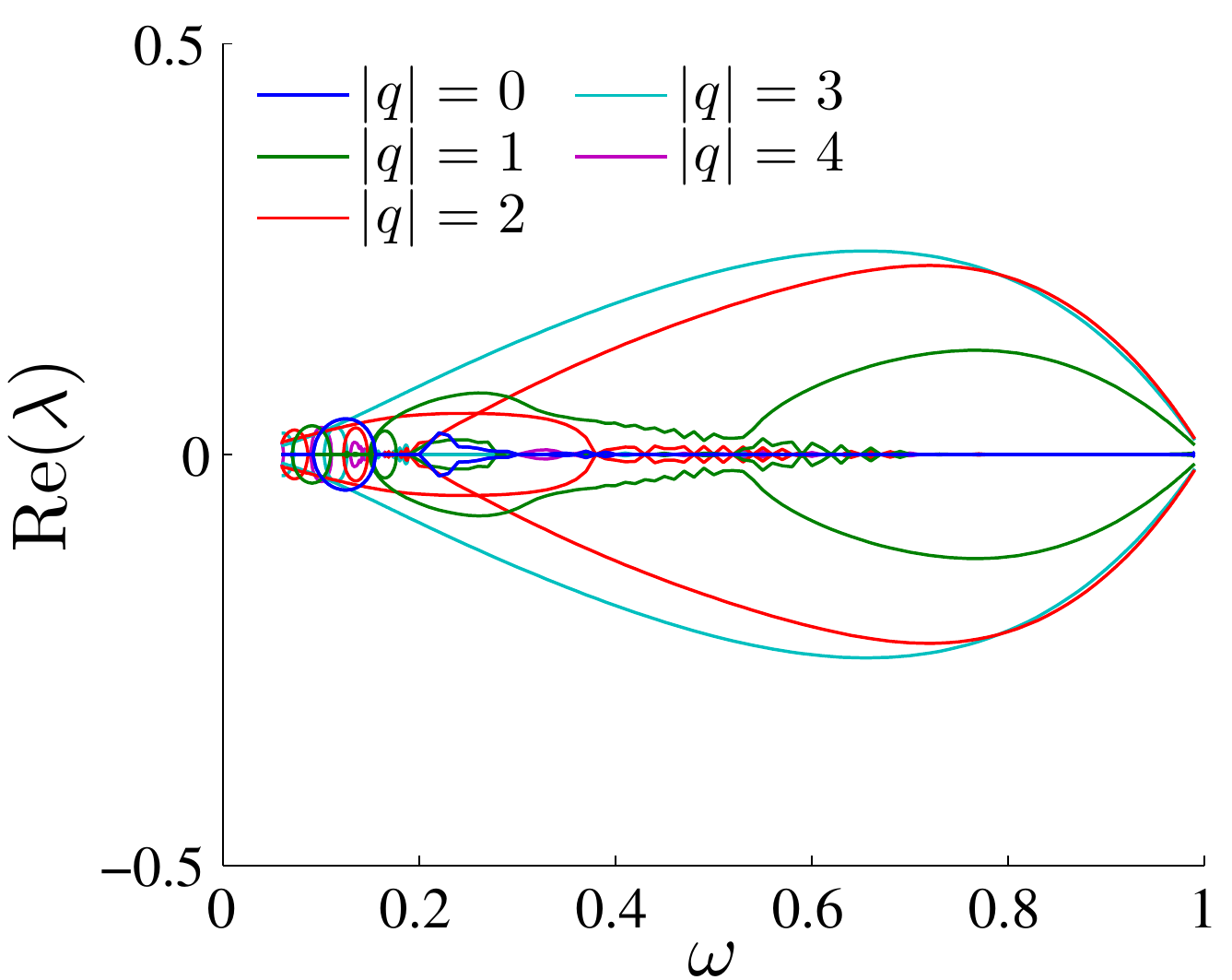} \\
\end{tabular}
\end{center}
\caption{Dependence of the (left) imaginary and (right) real part of the eigenvalues with respect to the frequency $\omega$ of solitary waves in the 2D Soler model with cubic ($k=1$) nonlinearity. Top (respectively, bottom) panels correspond to $S=0$ solitary waves ($S=1$ vortices). For the sake of clarity, we only included the values $|q|\le 4$. Full (dashed) lines in left panels represent the eigenvalues for $q\geq0$ ($q<0$). The correspondence between colors and $|q|$ is indicated in the legend of right panels.}
\label{cuevas-fig17}
\end{figure}

We start by considering the stability of $S=0$ solitary waves in the cubic ($k=1$) case. Top panels of Fig.~\ref{cuevas-fig17} show the dependence of the real and imaginary parts of the eigenvalues with respect to the stationary solution frequency $\omega$. From the spectral dependencies we can deduce several features of the 2D Soler model. First of all, it is known that the 2D NLS is charge-critical, and the zero eigenvalues are degenerate~\cite{cuevas-SS99}: they have higher algebraic multiplicity. In the NLD case, however, this degeneracy is resolved: in the $S=0$ case, as $\omega$ starts decreasing, two eigenvalues (corresponding to $q=0$) start at the origin when $\omega=1$ and move out of the origin for $\omega\lesssim 1$. The absence of the algebraic degeneracy of the zero eigenvalue prevents solitary waves from NLS-like self-similar blow-up which is possible in the charge-critical NLS \cite{cuevas-Mer90}. Secondly, the $\mathbf{U}(1)$ symmetry and the translation symmetry of the model result in zero eigenvalues with $q=0$ and $|q|=1$, respectively (in both $S=0$ and $S=1$ cases). Thirdly, as in the 1D Soler model, the eigenvalues $\lambda=\pm 2\omega{i} $, which are associated with the $\mathbf{SU}(1,1)$ symmetry of the model, are also present. This eigenvalue pair corresponds to $q=\mp1$, i.e., to an excited linearization eigenstate. Finally, contrary to the 1D case, where the solitary waves corresponding to any $\omega<1$ are spectrally stable, the $S=0$ solitary wave is linearly unstable for $\omega<0.121$ because of the emergence of nonzero-real-part eigenvalues via a Hamiltonian Hopf bifurcation in the $|q|=2$ spectrum at $\omega=0.121$. Another Hopf bifurcation occurs corresponding to $|q|=3$ (at $\omega=0.0885$), then yet another one corresponding to $|q|=4$ for lower $\omega$.

Vortices with $S>0$ are unstable for every $\omega$, because of the presence in the spectrum of quadruplets of complex eigenvalues. These quadruplets emerge (and disappear) for different values of $q$ via direct (inverse) Hopf bifurcations (see bottom panels of Fig.~\ref{cuevas-fig17}). The spectrum for $S\geq2$ vortex is quite similar to that of $S=1$; for this reason, we do not analyze it further. Notice that the eigenvalues $\lambda=\pm2\omega i$ generally correspond to the
particular mode with $q=\mp(2S+1)$.

\begin{figure}[tb]
\begin{center}
\begin{tabular}{cc}
\includegraphics[width=.45\textwidth]{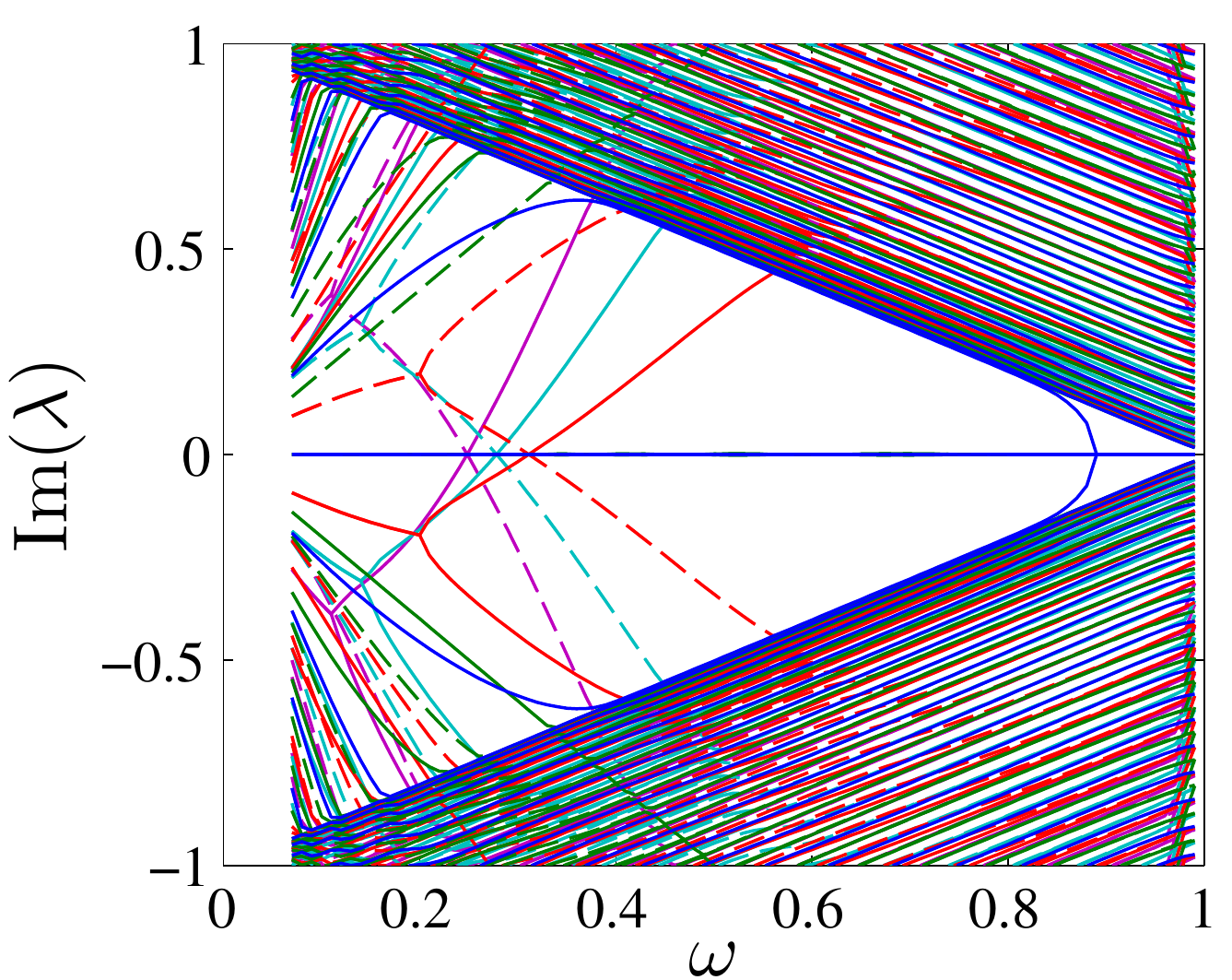} \hfill &
\includegraphics[width=.45\textwidth]{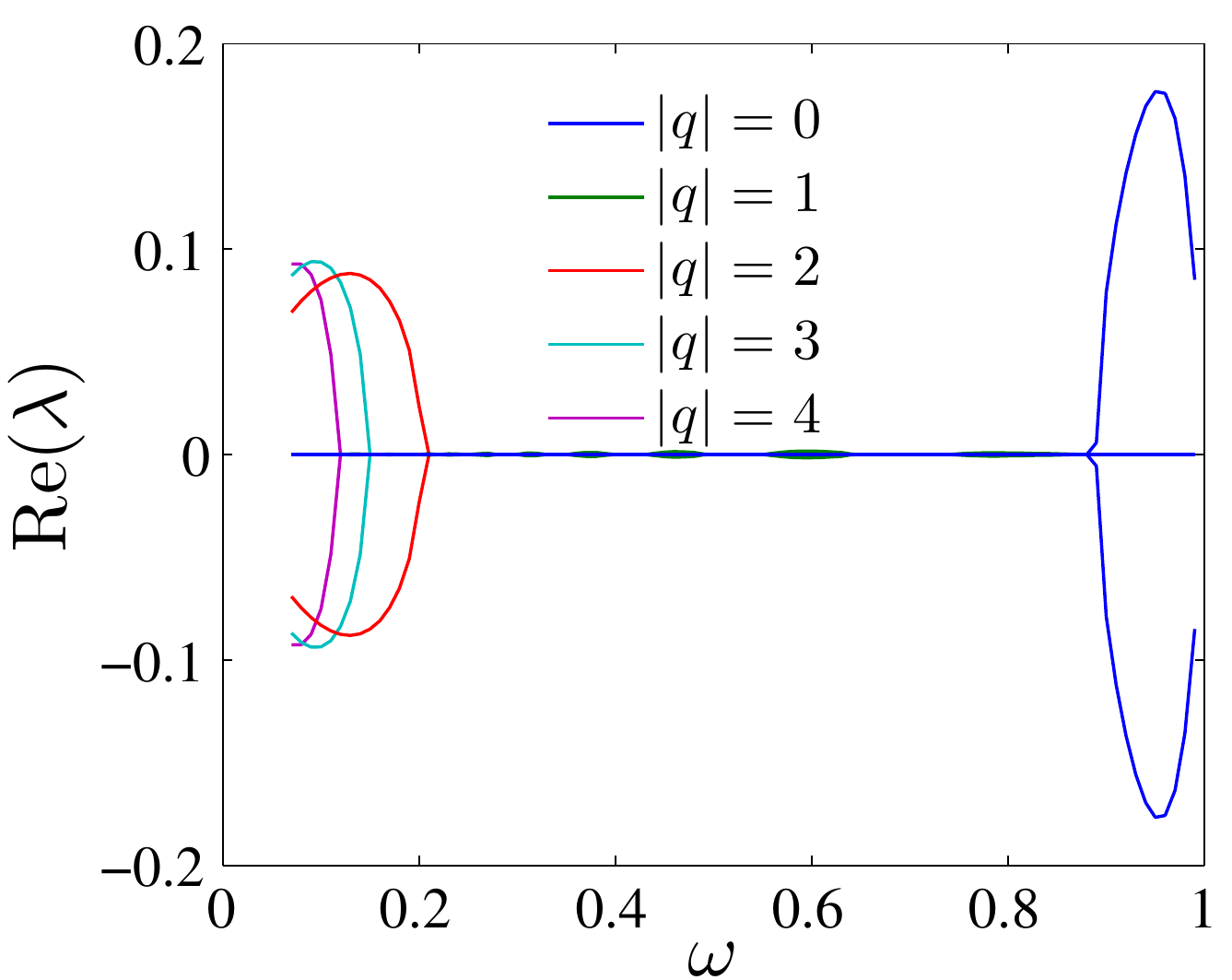} \\
\end{tabular}
\end{center}
\caption{Dependence of the (left) imaginary and (right) real part of the eigenvalues with respect to the frequency $\omega$ of $S=0$ solitary waves in the 2D Soler model with quintic ($k=2$) nonlinearity. For the sake of clarity, we only included the values $|q|\le 4$. Full (dashed) lines in the left panel represent the eigenvalues for $q\geq0$ ($q<0$). The correspondence between colors and $|q|$ is indicated in the legend of the right panel.}
\label{cuevas-fig18}
\end{figure}

\begin{figure}[tb]
\begin{center}
\begin{tabular}{cc}
\includegraphics[width=.45\textwidth]{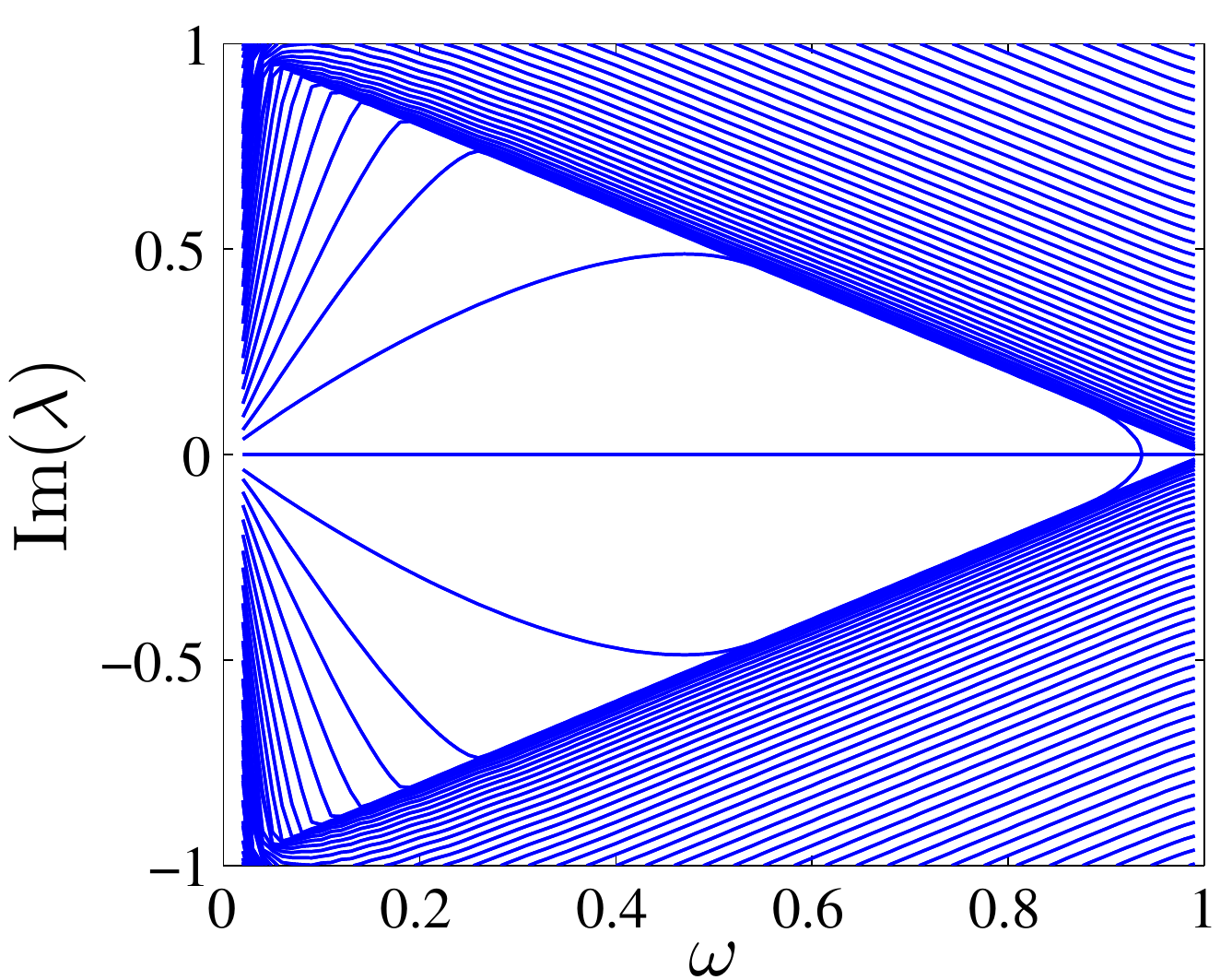} \hfill &
\includegraphics[width=.45\textwidth]{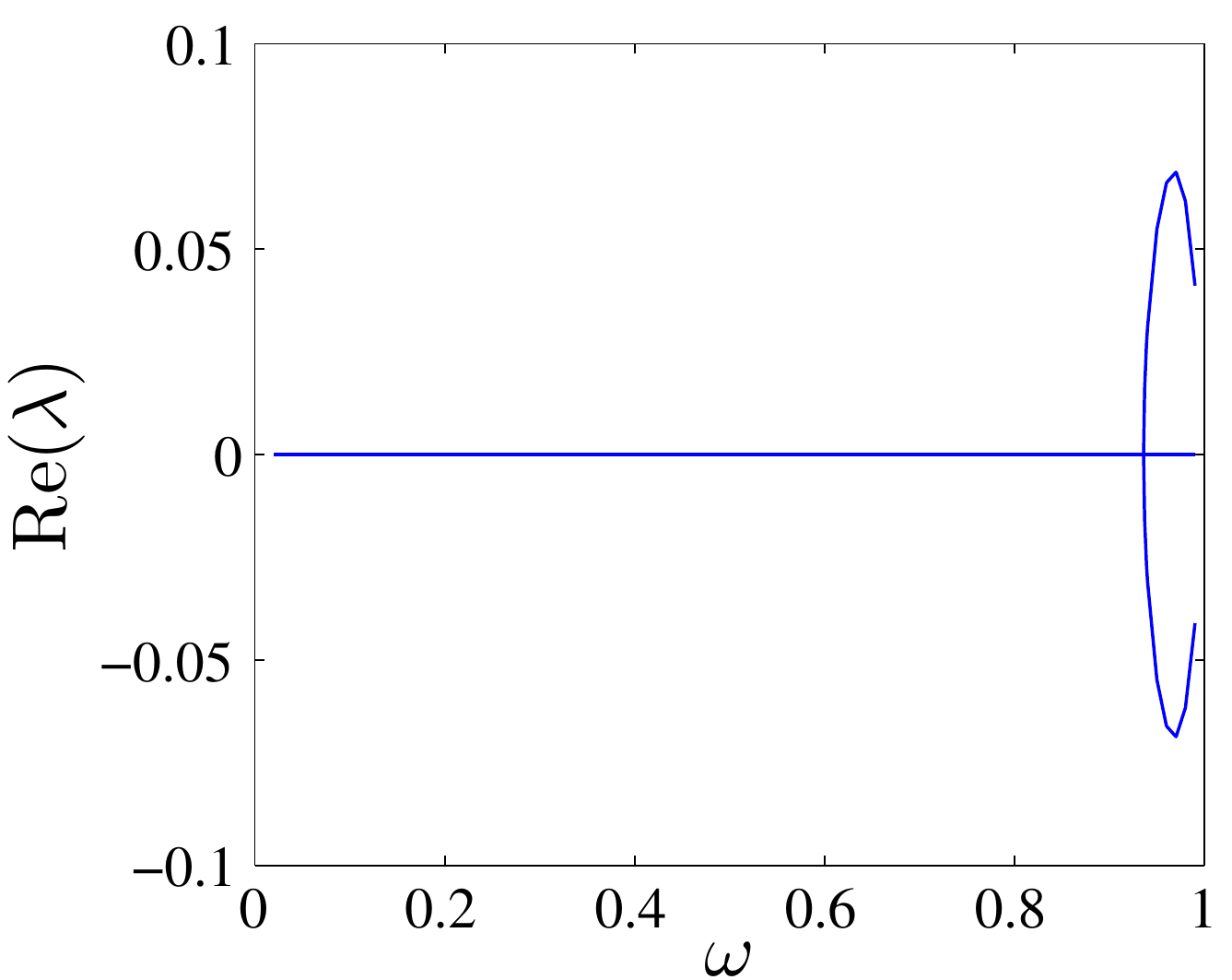} \\
\end{tabular}
\end{center}
\caption{Dependence with respect to $\omega$ of the (left) imaginary and (right) real part of the eigenvalues of the one-dimensional invariant ($q=0$) subspace of $S=0$ solitary waves in the 3D Soler model with cubic ($k=1$) nonlinearity.}
\label{cuevas-fig19}
\end{figure}

\begin{figure}
\begin{center}
\includegraphics[width=.45\textwidth]{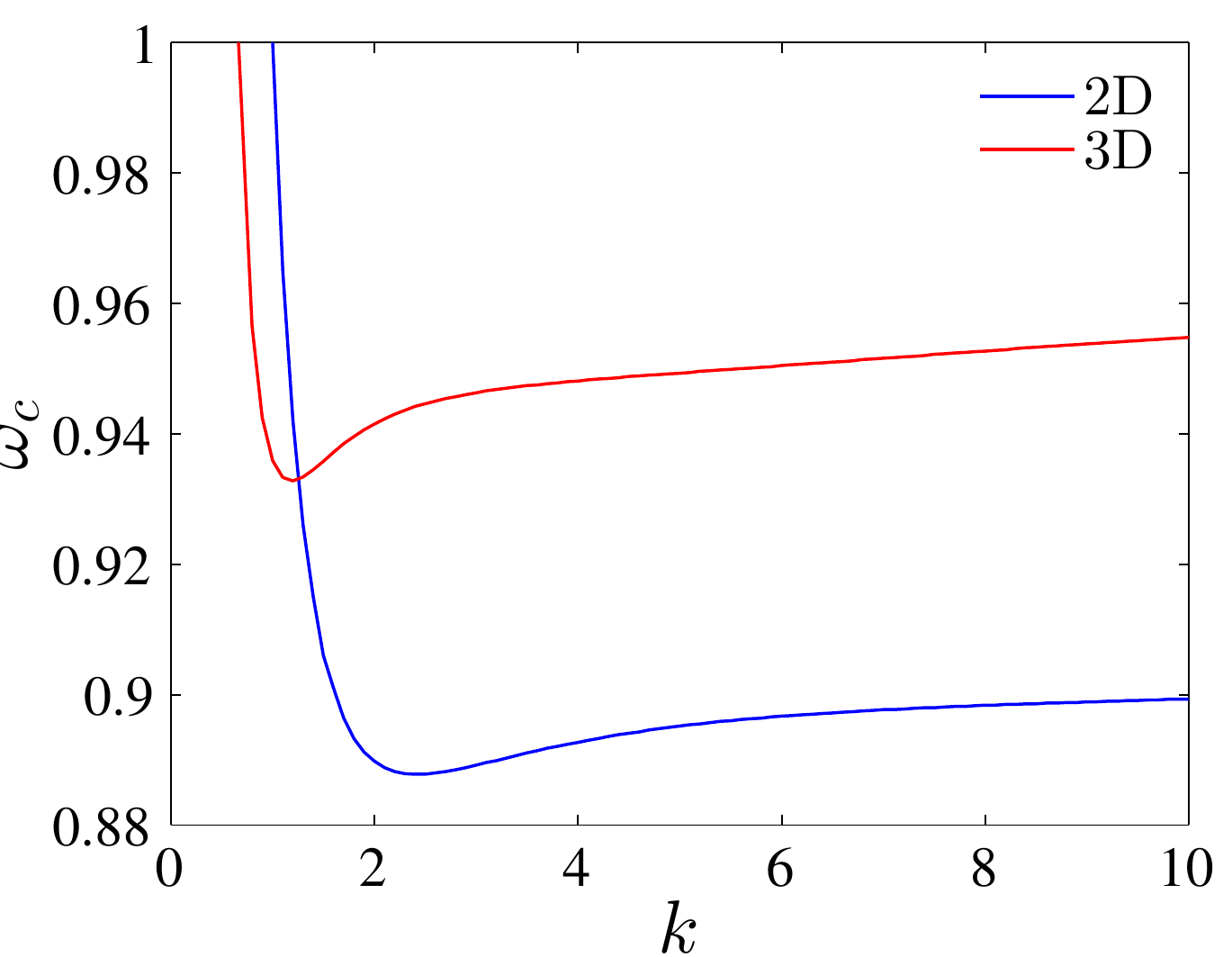} \hfill
\end{center}
\caption{Critical frequency for radially-symmetric ($q=0$) exponential bifurcations in the 2D and 3D Soler model, as a function of the exponent $k$. For $(\omega_c,1)$, the $S=0$ solitary waves are linearly unstable. For $k\leq2/n$, with $n$ being the system dimension, there is no linear instability for $\omega\lesssim1$, according to the Vakhitov--Kolokolov criterion (see Section \ref{sec:cuevas-stability_theory}).}
\label{cuevas-fig20}
\end{figure}

It is especially interesting that a wide parametric (over frequencies) interval of stability of solitary waves with $S=0$ can also be observed
in the quintic ($k=2$) NLD case (see Fig.~\ref{cuevas-fig18}); while the quintic NLS solitary waves blow up (even in one dimension), the quintic NLD solitary waves are stable even in two dimensions, except for the interval $\omega<0.312$ where the coherent structures experience the same Hopf bifurcation as in the cubic case, and for $\omega>0.890$ where an exponential instability created by radial $q=0$ perturbations emerges. This exponential instability is predicted by the Vakhitov--Kolokolov criterion as it coincides with the point at which $\partial_\omega Q(\omega)=0$ (see Fig.~\ref{cuevas-fig5}). Perhaps even more remarkably, Fig.~\ref{cuevas-fig19} illustrates that this stability of NLD solitary waves against radial perturbations can be found in suitable frequency intervals even in 3D \cite{cuevas-CGG14}. Both of the above cases (quintic 2D and cubic 3D Soler models) are charge-supercritical i.e., the charge goes to infinity in the nonrelativistic limit $\omega\to 1$. Contrary to the pure-power supercritical NLS whose solitary waves remain linearly unstable for all frequencies,
solitary waves in the Soler model become spectrally stable when $\omega$ drops below some dimension-dependent critical value $\omega_c=\omega_c(n,k)$, with $n$ being the number of spatial dimensions. Fig.~\ref{cuevas-fig20} shows those critical frequencies (associated with $q=0$ and radially-symmetric collapse) as a function of the nonlinearity parameter $k$ for $n=2$ and $n=3$. For $\omega\in(\omega_c,1)$, the NLD solitary waves are linearly unstable. Below $\omega_c$ the linear instability disappears. For $k\le 2/n$, there is no linear instability for $\omega\lesssim 1$. In the particular case of cubic ($k=1$) 3D Soler model,
we have that $\omega_c\approx 0.936$. This value was identified by Soler in his original paper \cite{cuevas-Sol70} as the value at which both the energy and charge of solitary waves have a minimum. Hence, we indeed find that the radially-symmetric collapse-related instability ceases to be present below this critical point.

\section{Dynamics}\label{sec:cuevas-dynamics}

Once the stability properties of solitary waves and vortices of the Soler model have been elucidated, it is now natural to turn our attention towards
the observation of their dynamical properties. In the one-dimensional case, we will analyze some integration schemes in order to observe their suitability for simulation of solitary waves in nonlinear Dirac equations. In addition, the dynamics of unstable solutions in equations with high-order instabilities (i.e. $k>1$) will be shown. Finally, the
dynamics of unstable solitary waves and vortices for the 2D Soler model
will be considered.

\subsection{One-dimensional solutions}

This subsection is divided into two parts. In the first one, we will show the evolution of stable solitary waves within several numerical integrators in the cubic ($k=1$) Soler model. The second part deals with the evolution of unstable solitary waves with $k>1$. Most of the results presented herein are taken from \cite{cuevas-CKS+15}.

\subsubsection{Stable solutions}

We turn here our attention to the implications of spectral collocation methods to the nonlinear dynamical evolution problem. We focus on the case of $k=1$. Given the large (yet spurious) growth rate of the modes emerging from the Chebyshev spectral collocation method and the spurious point spectrum instability of the finite difference method, for our dynamical considerations, we will focus our attention to the Fourier spectral collocation method results. As discussed in Subsection \ref{subsec:cuevas-stability1D}, in that method too, there exist spurious modes which, as expected, are found to affect the corresponding dynamics. As a dynamical outcome of these modes, the solitary waves are found to be destroyed after a suitably long evolution time, although the time for this feature is controllably longer in comparison to the one observed in \cite{cuevas-SQM+14}. This, in turn, suggests the expected stability of the solitary wave solutions, in accordance with what was proposed in Section~\ref{sec:cuevas-stability}.

As a prototypical diagnostic of the dynamical stability of solitary waves in a finite domain $[-L,L]$,
we have monitored the $L^2$-error in a similar fashion as in~\cite{cuevas-SQM+14}:
\begin{equation*}
\varepsilon_2(t)=\Big(\int\ \left| \rho(t,x)-\rho(0,x) \right|^2 \,dx\Big)^{1/2},
\end{equation*}
with $\rho=\psi\sp\ast\psi$ being the charge density.

\begin{figure}[ht]
\begin{center}
\begin{tabular}{cc}
\includegraphics[width=.45\textwidth]{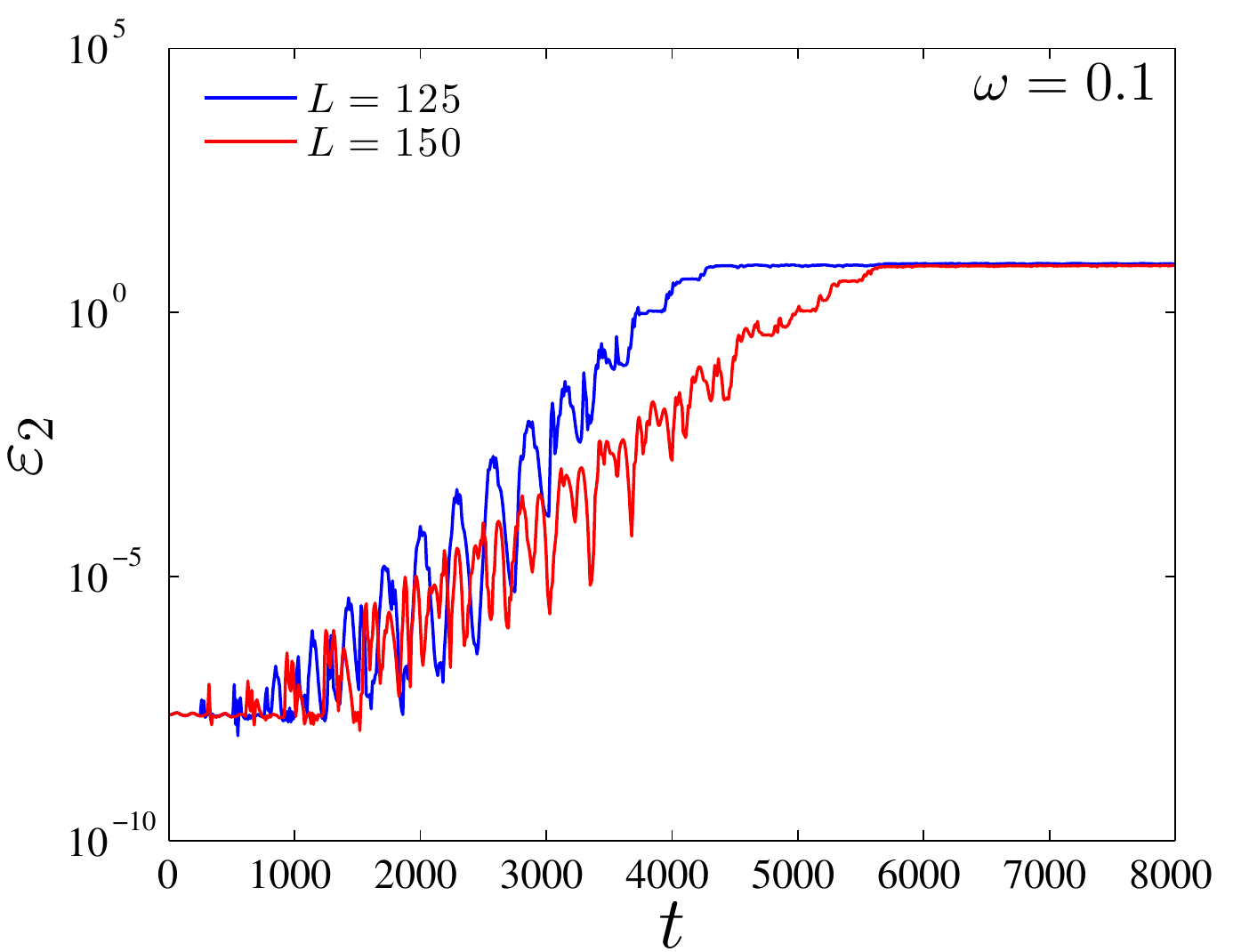} \hfill &
\includegraphics[width=.45\textwidth]{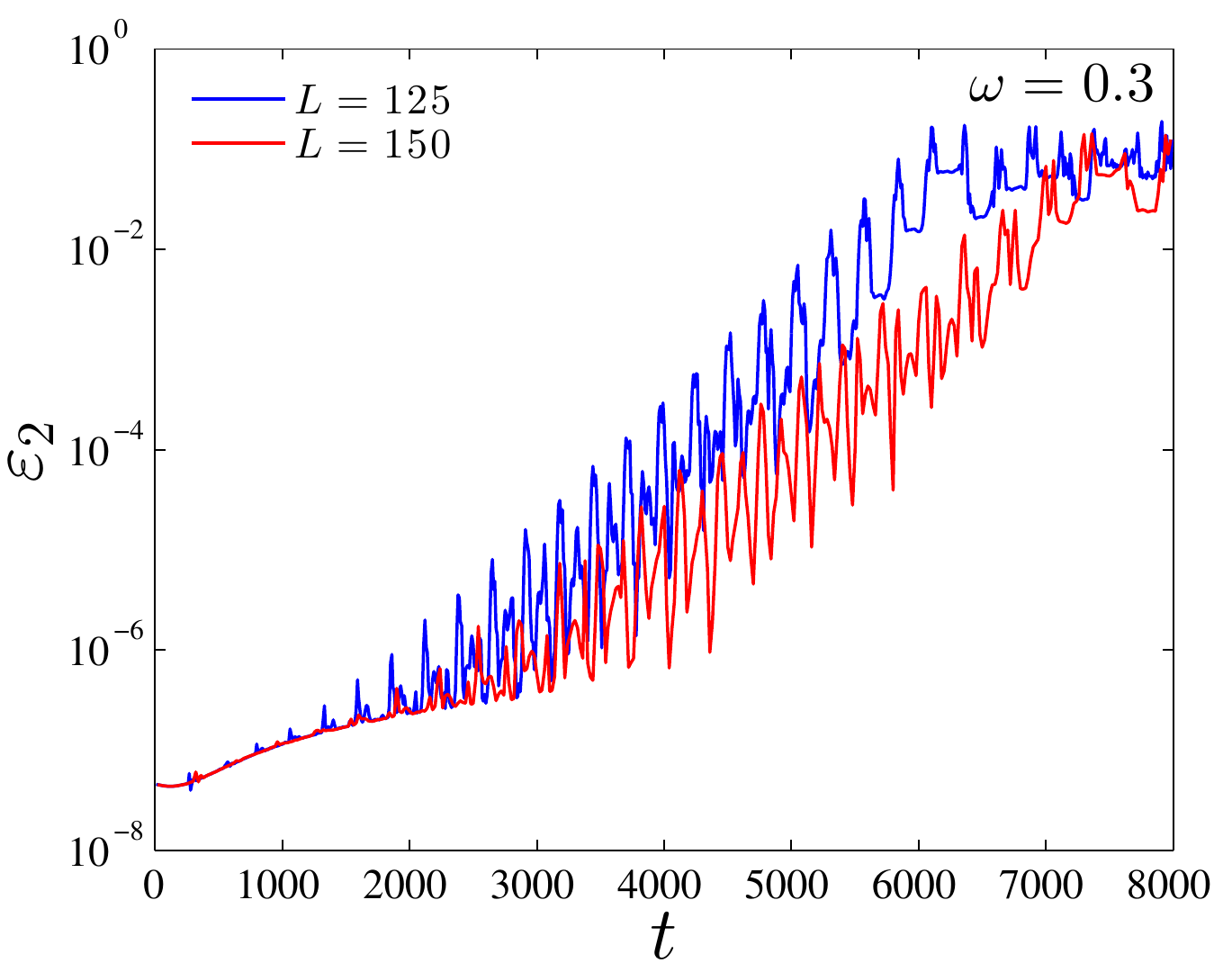} \\
\includegraphics[width=.45\textwidth]{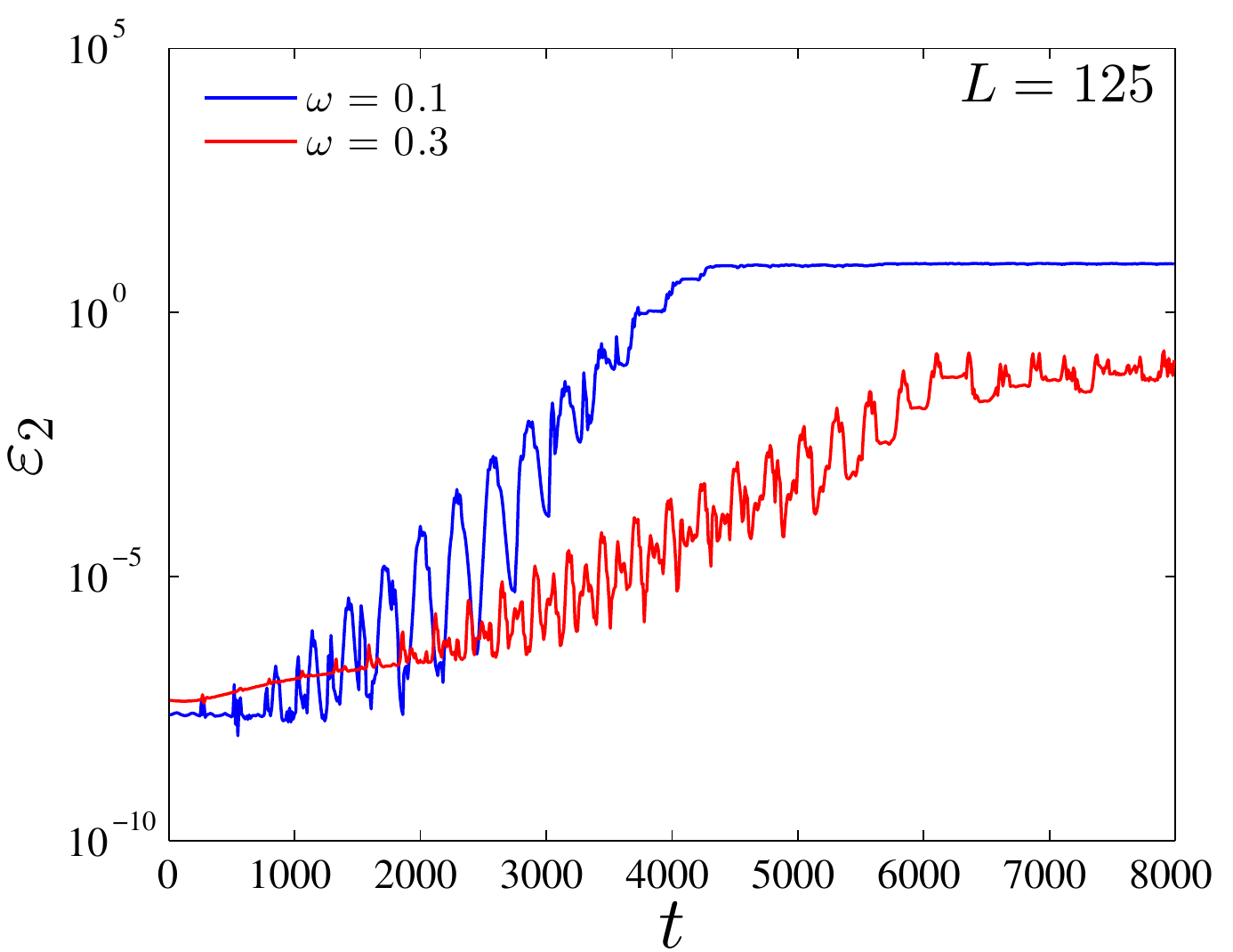} \hfill &
\includegraphics[width=.45\textwidth]{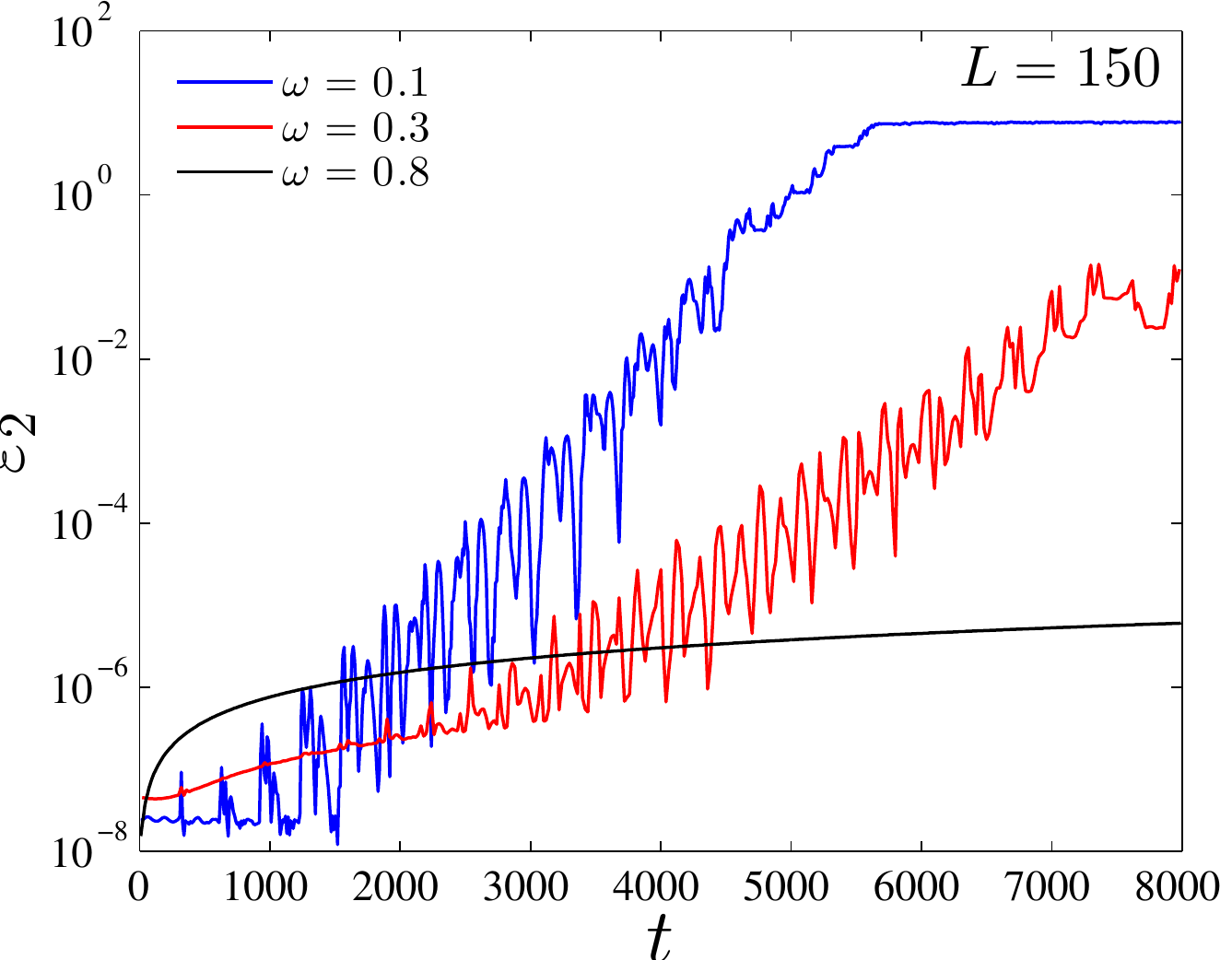} \\
\end{tabular}
\end{center}
\caption{Stable solitary waves simulations in cubic 1D Soler model using a 4th-order Runge--Kutta integrator with a Fourier spectral collocation method. The norm error is compared for different domain sizes and frequencies. In every case, the time step of the integrator is $\Delta t=0.05$.}
\label{cuevas-fig21}
\end{figure}

A first approach to the dynamics problem is accomplished by choosing a fixed-step 4th order Runge--Kutta method. We observe that the lifetime is longer when the frequency $\omega$ is fixed and the domain length $L$ is increased. This is associated with the decrease of the size of spurious instability bubbles, as we approach the infinite domain limit. A similar decrease of the growth rate is observed for a given $L$, when the discretization spacing $h$ is decreased (i.e., as the continuum limit is approached), in accordance with the spectral picture of Fig.~\ref{cuevas-fig13}. In addition, if $L$ is fixed, the lifetime is longer when $\omega$ is increased. This is summarized in Fig.~\ref{cuevas-fig21}. This is, of course, in consonance with earlier observations such as those of~\cite{cuevas-SQM+14}, however, our ability to expand upon the lifetimes as the domain and discretization parameters are suitably tuned suggests that in the infinite domain, continuum limit such instabilities could be made to disappear upon suitable selection
of the numerical scheme. As a final comment, we note that the growth rates observed in Fig.~\ref{cuevas-fig21} are consonant with the maximal (yet spurious) instability growth identified in Fig.~\ref{cuevas-fig13}. This is yet another indication that this growth featured in the time dynamics is a spurious by-product of the discretization scheme, rather than a true feature of the corresponding continuum problem.

\begin{table}[tb]
\caption{Comparison between the critical times for which $\varepsilon_2>10^{-3}$ using the Fourier Spectral Collocation method with a 4th-order Runge--Kutta integrator ($t_1$) and the Operator Splitting Method of \cite{cuevas-SQM+14} ($t_2$).}
\label{tab1}
\begin{center}
\begin{tabular}{p{2cm} p{2cm} p{2cm} p{2cm} p{2cm}}
\noalign{\smallskip}\svhline\noalign{\smallskip}
& \multicolumn{2}{c}{$\omega=0.1$} & \multicolumn{2}{c}{$\omega=0.5$} \\
\noalign{\smallskip}\svhline\noalign{\smallskip}
$L$ & $t_1$ & $t_2$ & $t_1$ & $t_2$ \\
\noalign{\smallskip}\svhline\noalign{\smallskip}
50 & 1220 & 121 & 5620 & 6614 \\
75 & 1320 & 122 & 8480 & 8724 \\
100 & 1990 & 122 & 14660 & 9937 \\
125 & 2540 & 120 & 14660 & 11670 \\
150 & 3120 & 122 & 14660 & 13560 \\
\hline\noalign{\smallskip}
\end{tabular}
\end{center}
\end{table}

In Table~\ref{tab1} we compare the critical time for which $\varepsilon_2>10^{-3}$ within the Fourier spectral collocation method and the corresponding time for the 4th order operator splitting algorithm used in~\cite{cuevas-SQM+14} for which we have the frequencies $\omega=0.1$ and $\omega=0.5$ and different domain lengths $L$. As can be seen from the comparison, although in some cases (e.g. for $\omega=0.5$ and $L=50$) the observed destabilization may happen later for the scheme of~\cite{cuevas-SQM+14}, generally the Fourier spectral collocation method code explored herein allows to enhance the wave lifetime, in some cases by an order of magnitude. This can be further improved by tweaking parameters such as $h$ and the time spacing of the integrator $\Delta t$, as discussed above. Hence, our conclusion is that despite the artificial instabilities existing in the spectral picture and their dynamical manifestation, it is anticipated that the continuum, real line variant of the problem is spectrally stable for all $\omega\in (0,m)$ in the case of $k=1$.

A tweak to the problem could be, on the one hand, to use adaptive step-size integrators \cite{cuevas-HNW93}. The case of 4th-5th order Dormand--Prince integrator \cite{cuevas-DP80} does not improve  significantly the solitary wave lifetime. On the other hand, when using a 2nd-3rd order Runge--Kutta integrator supplemented by a TR-BDF2 scheme (i.e. a trapezoidal rule step as a first stage and a backward differentiation formula as a second stage) \cite{cuevas-SH96}, many of the spurious eigenvalues can be damped out and the lifetimes are strongly enhanced.

\subsubsection{Unstable solutions for high-order nonlinearity}

\begin{figure}
\begin{center}
\begin{tabular}{cc}
\includegraphics[width=.45\textwidth]{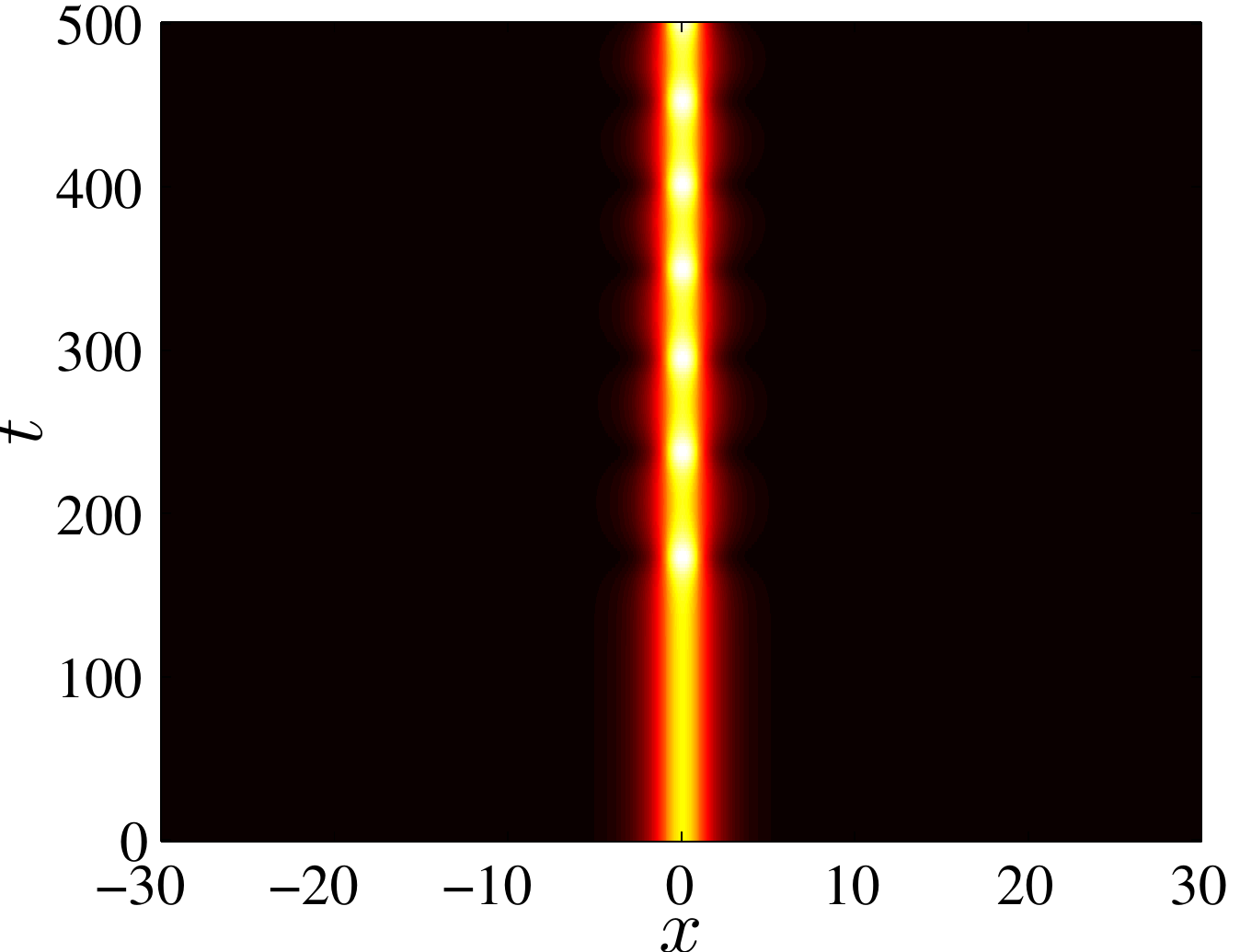} &
\includegraphics[width=.45\textwidth]{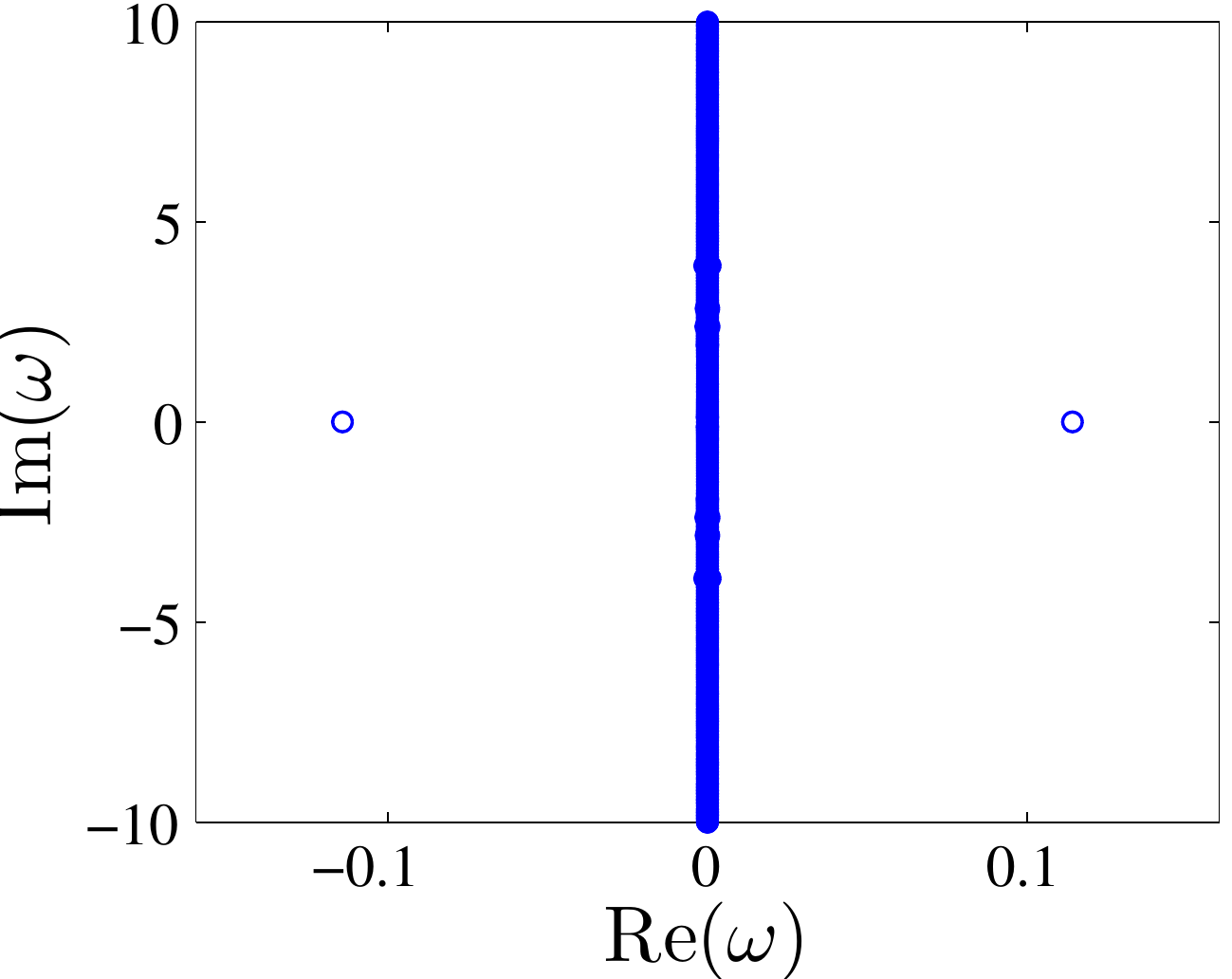} \\
\end{tabular}
\end{center}
\caption{(Left) Time evolution of a 1D solitary wave with nonlinearity exponent $k=3$ and frequency $\omega=3$. (Right) Spectral plane of the solitary wave whose evolution in traced in the left panel.}
\label{cuevas-fig22}
\end{figure}

Having observed that the solitary wave solutions of the problem with $k=1$ (and, in fact, with any $k<2$) are
dynamically stable, we now turn our attention to the dynamics associated with the instability
in the case $k>2$,
for $\omega > \omega_c(k)$, as per Fig.~\ref{cuevas-fig16}. Figure~\ref{cuevas-fig22} shows the evolution of an exponentially unstable solitary wave with $k=3$ and $\omega=0.9$. We can observe the existence of oscillations around a stable fixed point. This fixed point approximately corresponds to the solitary wave with frequency $\omega\approx0.82$, for which the solution is spectrally stable. This is in stark contrast with the supercritical dynamics of the Nonlinear Schr\"odinger equation. There, the instability directly leads to collapse and an indefinite growth of the amplitude of the solution. On the contrary, in the case of the Soler model, for any value of $k$ for which the solution may become unstable, there exists (for the same $k$) an interval of spectrally stable states of the same type.
Hence, the solution to the Soler model
does not escape towards collapse but rather departs from the vicinity of the unstable fixed point solution and finds itself orbiting around a center, i.e., a stable solitary wave structure.

\subsection{Two-dimensional solutions}

This subsection reviews the results on the dynamics of 2D solitary waves and vortices shown in \cite{cuevas-CKS+16a}. In order to simulate their dynamics, Chebyshev spectral methods and finite difference methods are not the most suitable ones, because of the presence of many spurious eigenvalues, and the dimensionality of the problem makes the TR-BDF2 schemes difficult to implement because of the high memory requirements.
Thus, it seems that the optimal way to proceed is to use a Fourier spectral collocation method,
which, as shown for the 1D problem, works fairly well as long as the frequency $\omega$
is not close to zero.

Consequently, periodic boundary conditions must be supplied to our problem. This is less straightforward when working in polar coordinates in the domain $(0,L)\times[0,2\pi)$. For this reason, we opt to work with a purely 2D problem in rectangular coordinates in the domain $(-L,L]\times(-L,L]$. The simulations we show below have been performed with a Dormand--Prince numerical integrator using such a spectral collocation scheme with the aid of Fast Fourier Transforms (\ref{eq:cuevas-FFT}).

\begin{figure}
\begin{center}
\begin{tabular}{c}
\includegraphics[width=.9\textwidth]{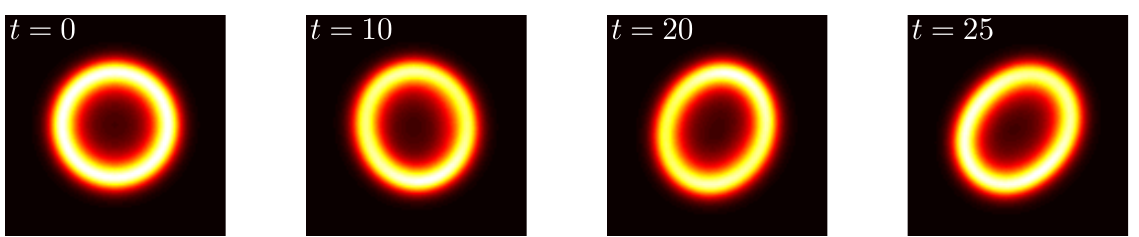} \\
\includegraphics[width=.9\textwidth]{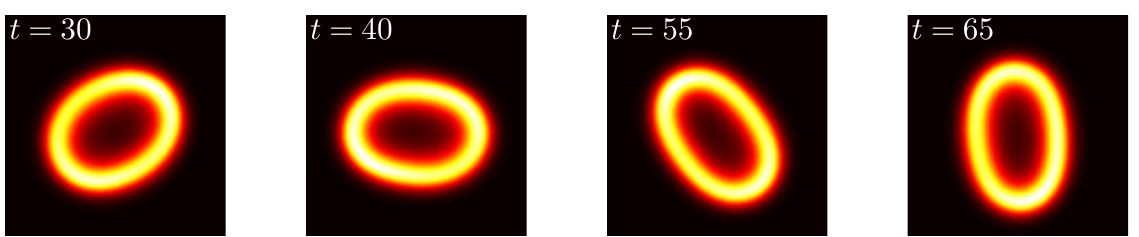} \\
\end{tabular}
\end{center}
\caption{Snapshots showing the evolution of the density of an unstable $S=0$ solitary wave with $\omega=0.12$ in the cubic 2D Soler model. The solitary wave which initially had a circular shape becomes elliptical and rotates around the center of the original solitary wave.}
\label{cuevas-fig23}
\end{figure}

\begin{figure}[tb]
\begin{center}
\begin{tabular}{cc}
\includegraphics[width=.45\textwidth]{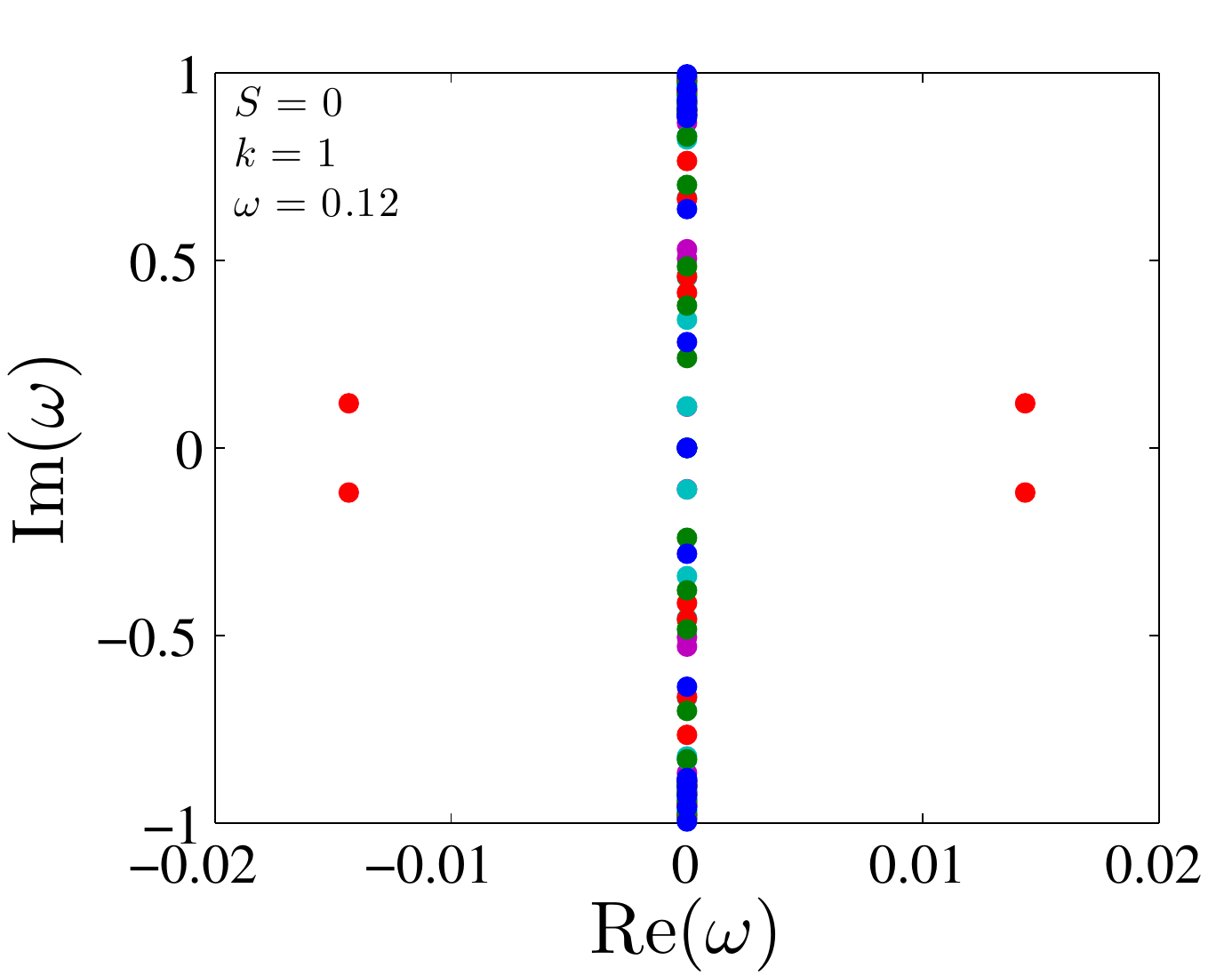} \hfill &
\includegraphics[width=.45\textwidth]{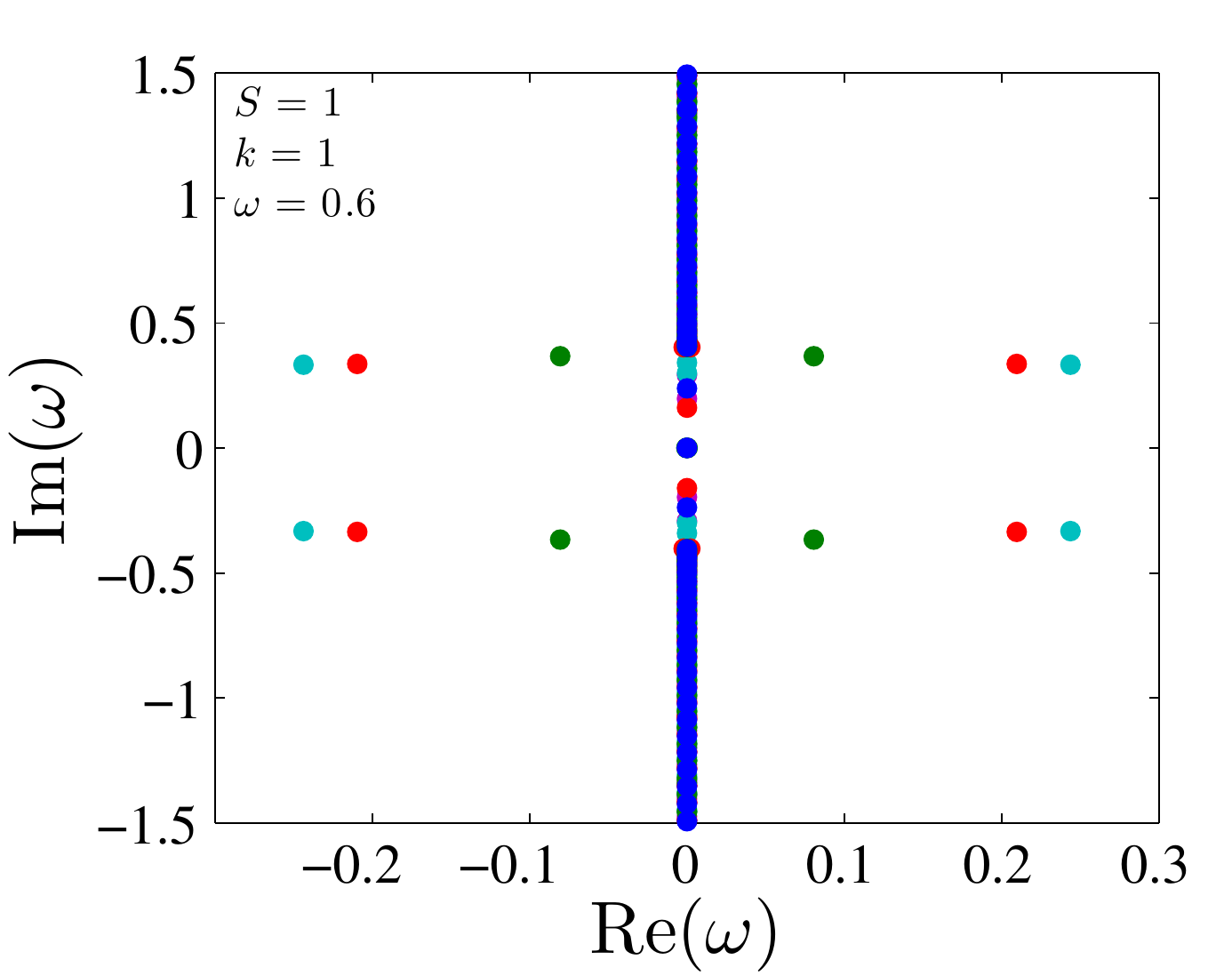} \\
\end{tabular}
\end{center}
\caption{Spectral planes of the unstable solitary waves whose dynamics are depicted in Figs.~\ref{cuevas-fig23} and \ref{cuevas-fig25} (left and right panels, respectively). Each color represents a different value of $q$ as in Fig.~\ref{cuevas-fig17}. }
\label{cuevas-fig24}
\end{figure}

\begin{figure}
\begin{center}
\begin{tabular}{c}
\includegraphics[width=.9\textwidth]{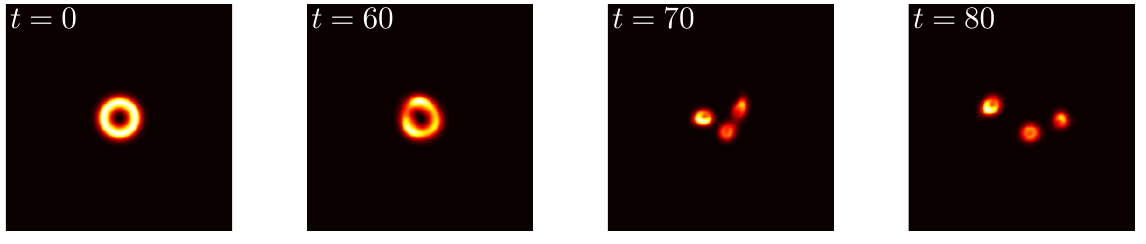} \\
\includegraphics[width=.9\textwidth]{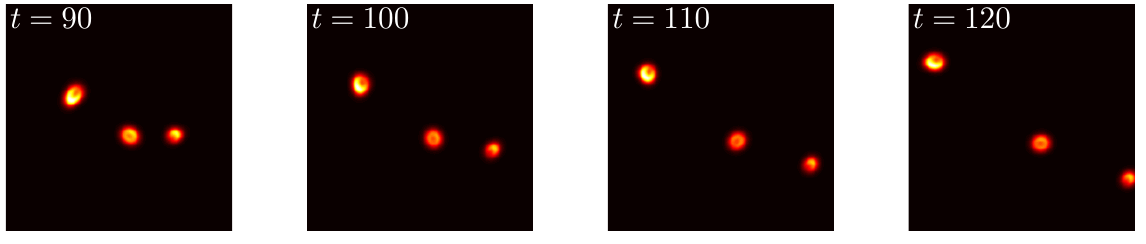} \\
\end{tabular}
\end{center}
\caption{Snapshots showing the evolution of the density of an unstable $S=1$ vortex with $\omega=0.6$ in the cubic 2D Soler model.}
\label{cuevas-fig25}
\end{figure}

\begin{figure}
\begin{center}
\begin{tabular}{c}
\includegraphics[width=.9\textwidth]{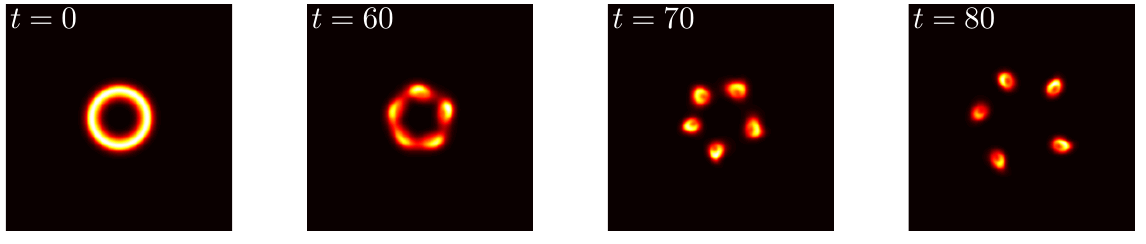} \\
\includegraphics[width=.9\textwidth]{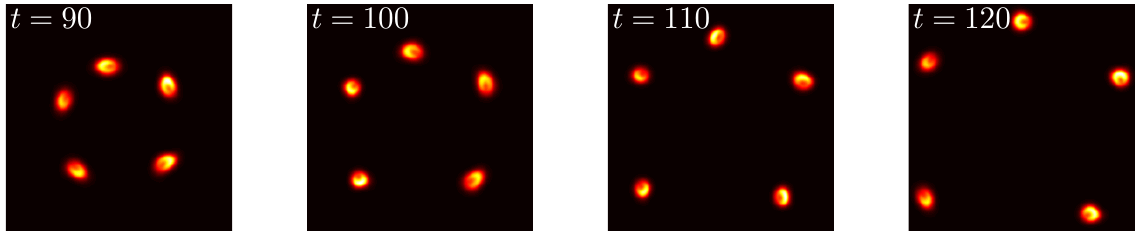} \\
\end{tabular}
\end{center}
\caption{Snapshots showing the evolution of the density of an unstable $S=2$ vortex with $\omega=0.6$ in the cubic 2D Soler model.}
\label{cuevas-fig26}
\end{figure}

A prototypical example of the evolution of unstable $S=0$ solitary waves for $k=1$ is shown in Fig.~\ref{cuevas-fig23}. As can be observed, the radial symmetry in the density of $S=0$ solitary waves is spontaneously broken and, as a result,
the coherent structures become elliptical and rotate around the center of the circular density of the original solitary wave in line with the expected amplification of the $q=2$ unstable eigenmode (see spectrum at the left panel of Fig.~\ref{cuevas-fig24}). The dynamical outcome of $S=1$ vortices for $k=1$ is shown in Fig.~\ref{cuevas-fig25}, whose instability (see spectral plane at right panel of Fig.~\ref{cuevas-fig24}) leads to the splitting into three smaller ones; in particular, the first spinor component splits into structures without angular dependence, whereas the second component splits into corresponding ones with angular dependence $\propto e^{i \theta}$,
in accordance with the Ansatz of Eq.~(\ref{eq:cuevas-spinor2D}). This preserves the total vorticity across the two components. The instability of an $S=2$ vortex eventually leads to the emergence of five similar structures to the previous case, again preserving the total vorticity (see Fig.~\ref{cuevas-fig26}).

\begin{figure}
\sidecaption
\includegraphics[width=.5\textwidth]{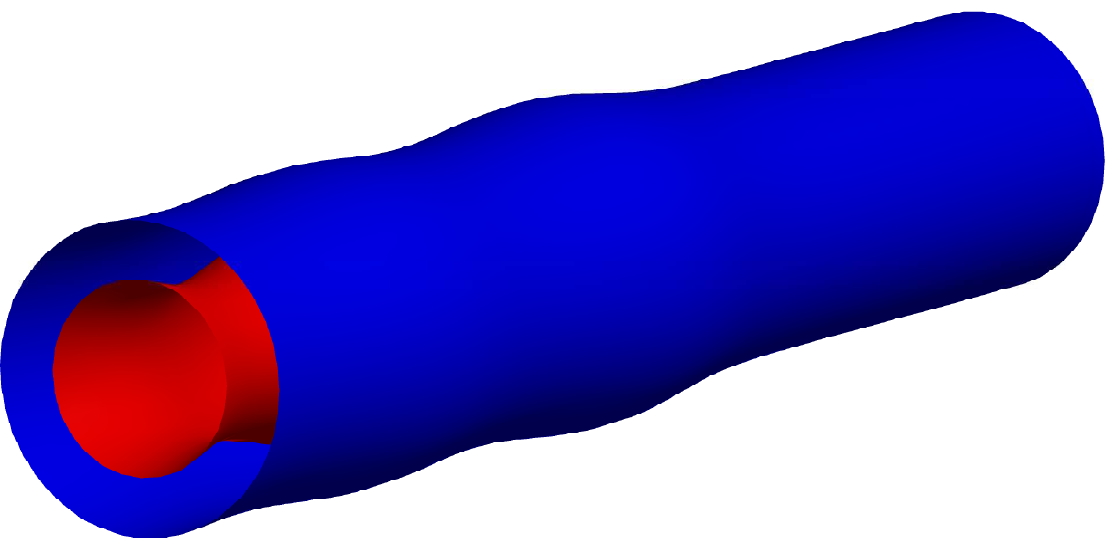}
\caption{Isosurface for the density of the $S=0$ solitary wave in the quintic 2D Soler model with $\omega=0.94$.}
\label{cuevas-fig27}
\end{figure}

Finally, we have analyzed the outcome of the instabilities caused by radially-symmetric perturbations in the $k=2$ case for $\omega>\omega_c$ (see Fig.~\ref{cuevas-fig27}). We can observe the typical behavior of such solutions, i.e. the density width (and amplitude) oscillate leading to a ``breathing'' structure, but there is no collapse. This phenomenology is reminiscent of the 1D case of the previous subsection (as in Fig.~\ref{cuevas-fig22}).

\section{The (one-dimensional) Discrete Nonlinear Dirac equation}\label{sec:cuevas-DNLD}

In the present section, we will take a somewhat different path from the previous ones. In particular, we will consider the existence, stability, and dynamics of solitary waves in a discrete version of the 1D Soler model. This equation is simply obtained, as explained in Subsection \ref{subsec:cuevas-stability1D}, by substituting the spatial derivatives $\partial_x \psi(t,x)$ in (\ref{eq:cuevas-Soler1D}) by the central difference $\epsilon(\psi_{n+1}(t)-\psi_{n-1}(t))$, with $\psi_n(t)$ being a discretized spinor and $\epsilon$ a coupling constant so that the continuum limit is attained in the $\epsilon\rightarrow\infty$ limit.

A significant part of our motivation for considering a discrete nonlinear Dirac equation is due, on the one hand, to the possibility to deploy the technology of the so-called anti-continuum (AC) limit of MacKay--Aubry~\cite{cuevas-MA94}, in order to appreciate the stability properties near the limit of uncoupled adjacent sites and, on the other hand, to the feature that in the continuum limit of, in principle, infinite coupling, our conclusions are expected to connect to what is known for the corresponding PDE (Soler) models that have been explored previously in the chapter. Admittedly, the discretization that is selected herein is, arguably, not the most natural possible one (in that we utilize next-nearest neighbors in order to discretize the first derivative terms by centered differences). Nevertheless, it is identified that it is the most suitable one for the present setting and discrete solitary waves are systematically obtained from the AC limit. Moreover, a very recent development worth noting is that spin-orbit Bose--Einstein condensates have recently been considered in the realm of an optical lattice~\cite{cuevas-HZK+14}, which is often thought (in the so-called superfluid regime) as being tantamount to a discretization of the original continuum problem, through a suitable Wannier function reduction~\cite{cuevas-AKKS02}. This suggests that considering discrete variants of Dirac models may be a natural step for near future considerations.

In what follows, we consider the existence and stability of discrete solitary waves from the anti-continuum ($\epsilon=0$) to the continuum ($\epsilon\rightarrow\infty$) limit of the Discrete nonlinear Dirac Equation in the form of a discretized 1D Soler model. Applying the discretization explained at the beginning of the section, stationary states satisfy the following equation, where the cubic nonlinearity $k=1$ has been chosen
(and, without loss of generality, we took $g=m=1$):
\begin{equation}\label{eq:cuevas-stat1Ddisc}
 \begin{split}
 \omega v_n &= \,\,\epsilon(u_{n+1}-u_{n-1})-\left[\left(v_n^2-u_n^2\right)-1\right]v_n, \\
 \omega u_n &= -\epsilon(v_{n+1}-v_{n-1})+\left[\left(v_n^2-u_n^2\right)-1\right]u_n,
 \end{split}
\end{equation}
with $n\in\{-(N-1)/2\ldots(N-1)/2\}$, and $N$ being the number of lattice points \footnote{We use the same notation $N$ for the number of lattice points and
for the number of spinor components although we believe this should not lead to any confusion}.  Stability of stationary discrete solitary wave solutions is found by diagonalizing the matrix obtained by discretizing the problem (\ref{eq:cuevas-stab1D}). The dispersion relation of discrete linear modes is given by

\begin{equation*}
 \lambda(q)=\pm i\left[\omega\pm\sqrt{1+4\epsilon^2\sin^2q}\right] .
\end{equation*}

Then, the essential spectrum extends over the interval $\lambda\in i[-\sqrt{1+4\epsilon^2},\omega-1] \cup i[-\omega+1,\sqrt{1+4\epsilon^2}]$, with its embedded part being $\lambda\in i[-\omega-1,\omega-1] \cup i[-\omega+1,\omega+1]$.

\begin{figure}
\begin{center}
\begin{tabular}{cc}
\includegraphics[width=.45\textwidth]{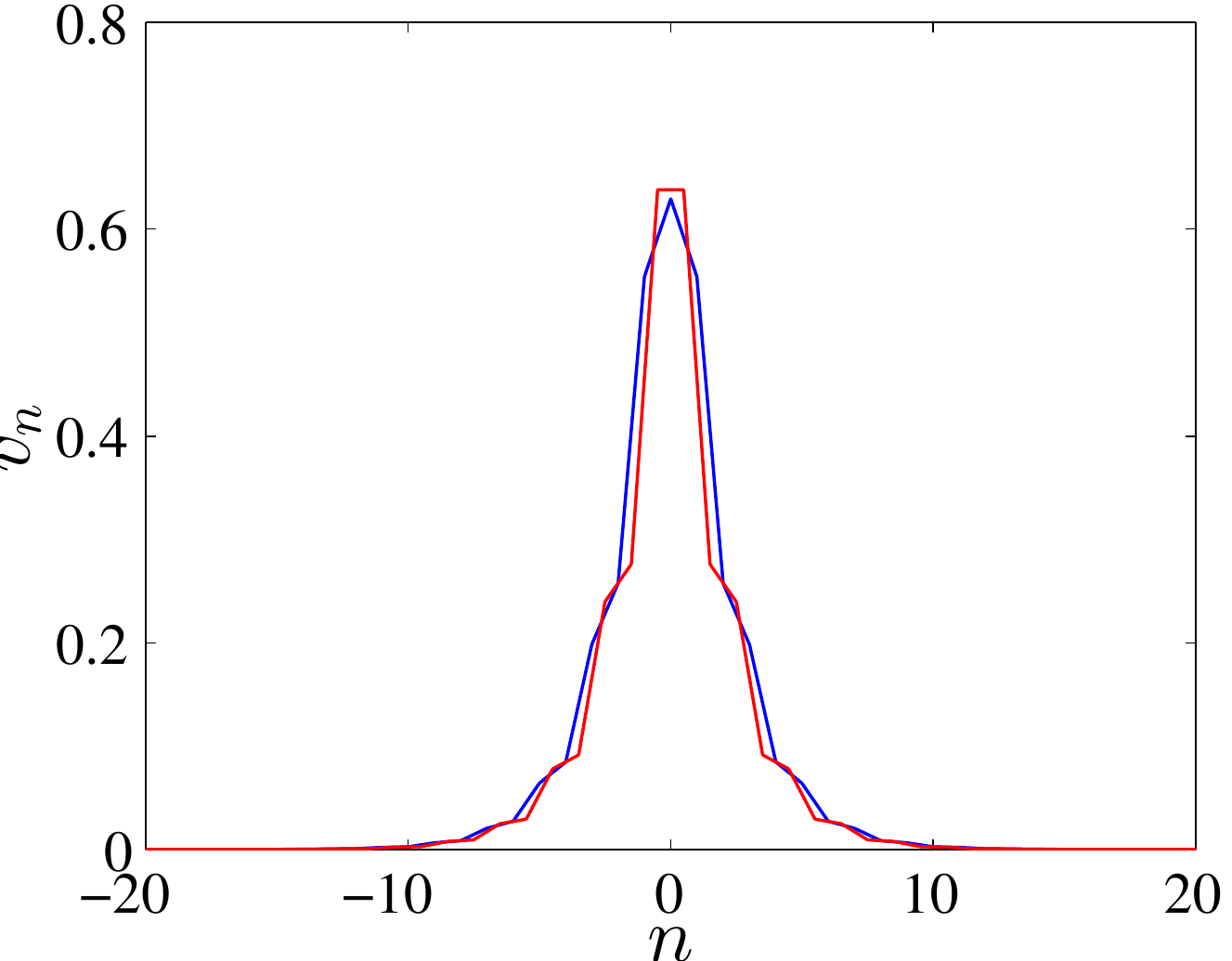} & \hfill
\includegraphics[width=.45\textwidth]{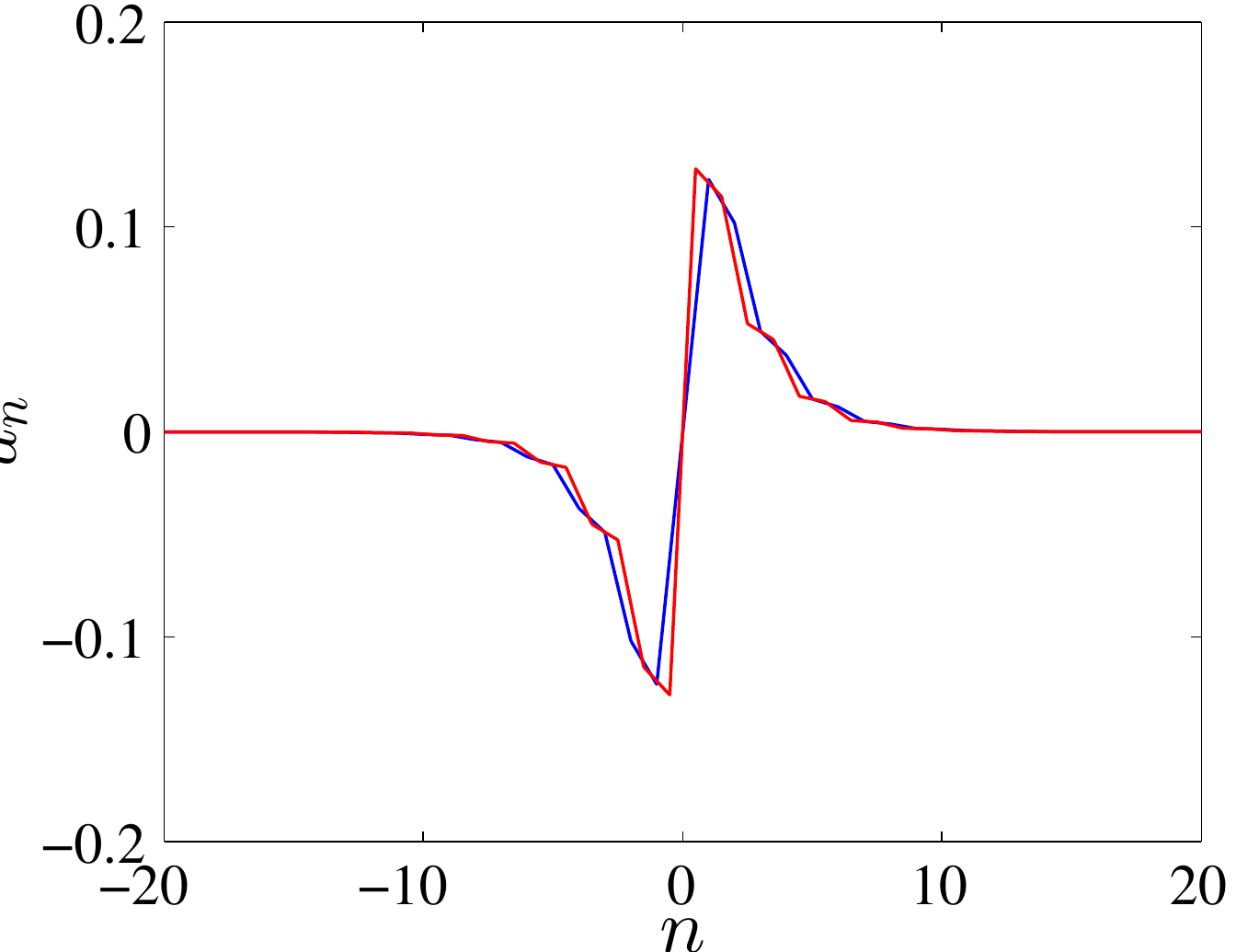} \\
\includegraphics[width=.45\textwidth]{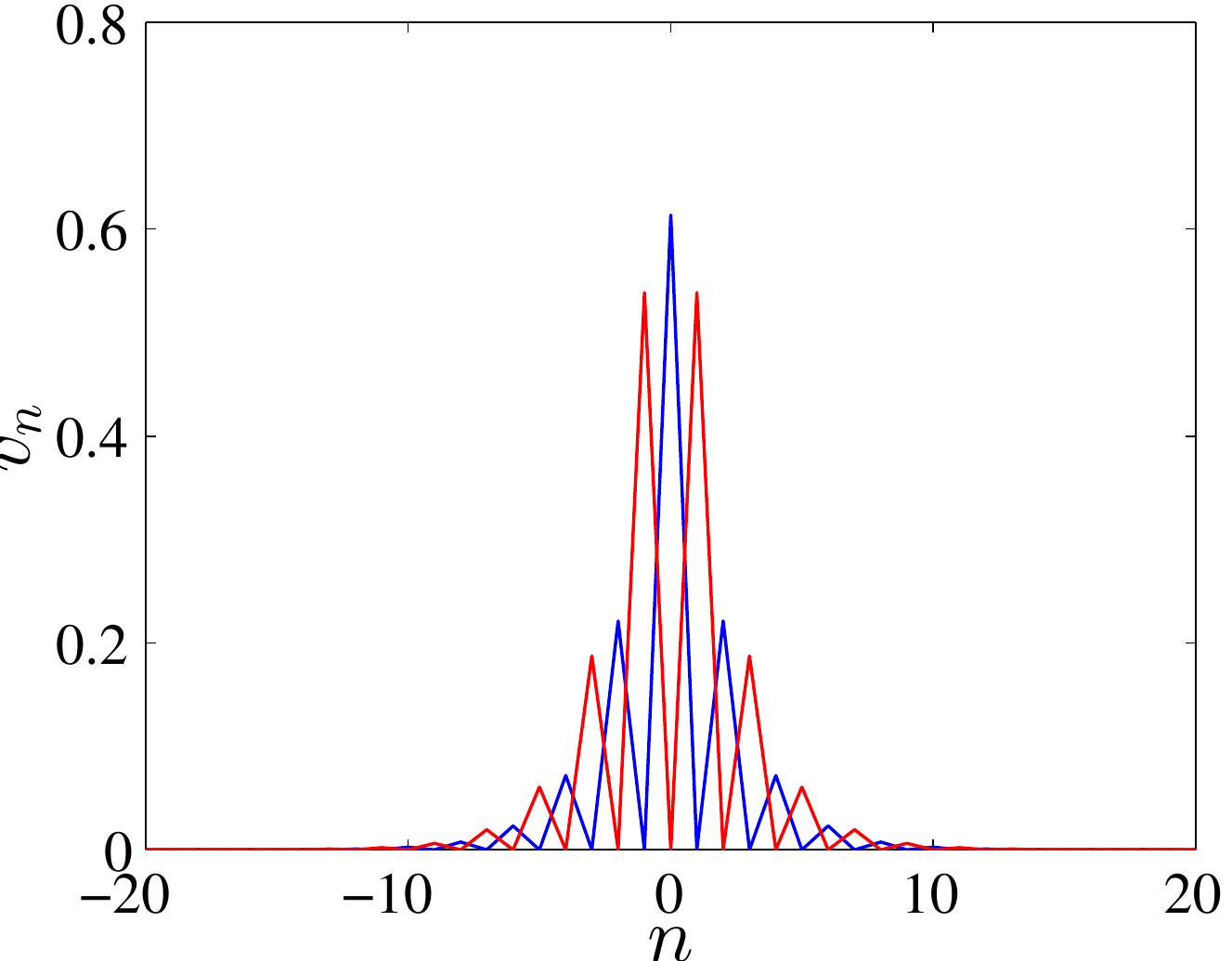} & \hfill
\includegraphics[width=.45\textwidth]{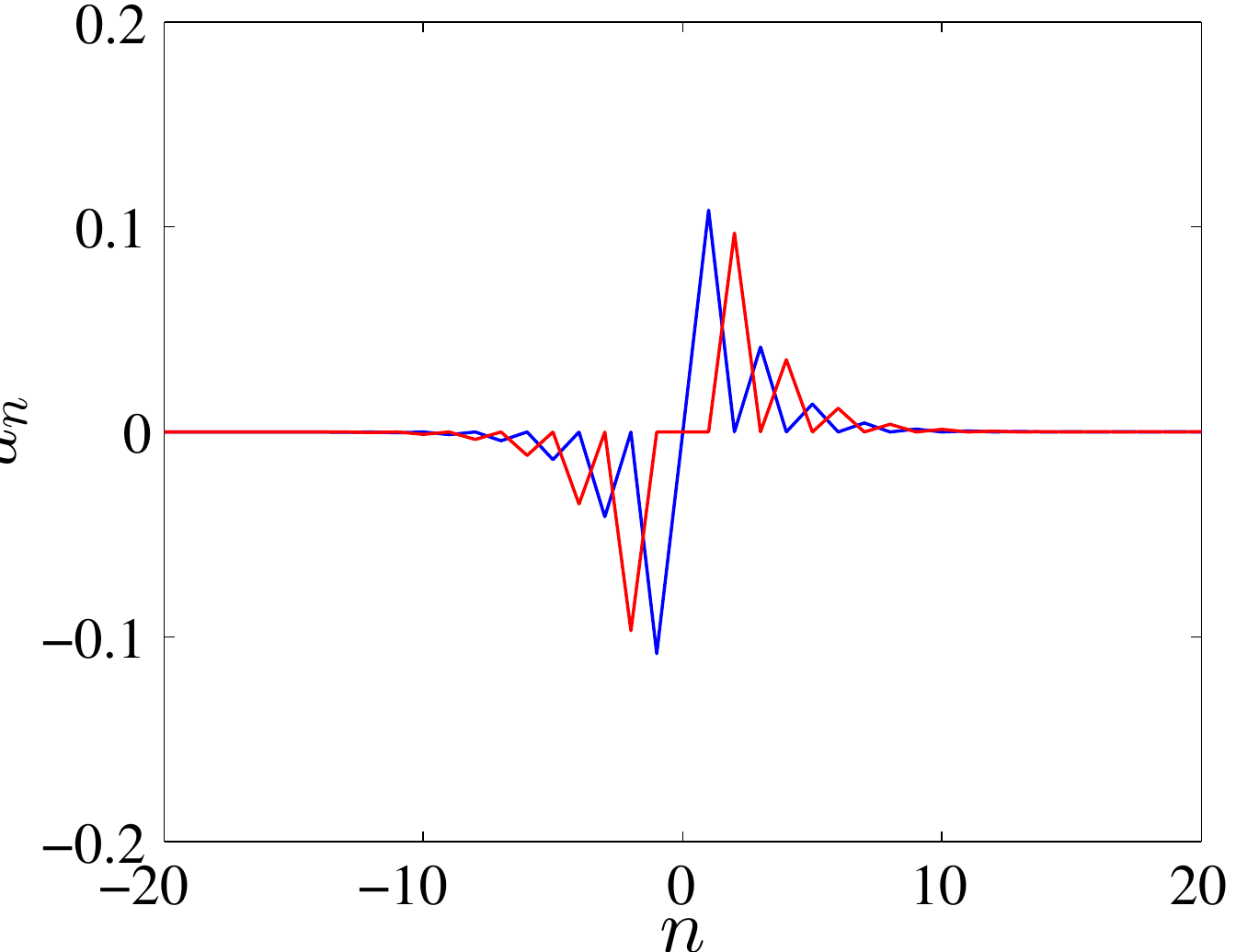} \\
\end{tabular}
\end{center}
\caption{Profiles of both spinor components for the discrete 3-site (top) and 1-site (bottom) soliton (blue line) and its complimentary solution (red line). In all cases, $\omega=0.8$ and $\epsilon=0.5$ (i.e. close to the anti-continuum limit).}
\label{cuevas-fig28}
\end{figure}

We start by considering the three-site soliton, which in the anti-continuum (AC) limit is given by $u_n=0\ \forall$ $n$ and $v_{-1}=v_0=v_1=\sqrt{1-\omega}$, $v_n=0$ if $|n|\geq2$ (see Fig.~\ref{cuevas-fig28}).
The motivation for starting with this particular solution, as opposed to the more ``canonical'' single site one,
is that this is the state that is found by extending the continuum solitary wave all the way to the AC limit. We will, however, discuss the continuation of the single-site (and the 2-site) solution in detail in what follows.

Let us explain below the general behavior for $\omega>1/3$. Outside this range, the solitary waves are generically unstable and hence we do not consider them further here.

\begin{figure}
\begin{center}
\begin{tabular}{cc}
\includegraphics[width=.45\textwidth]{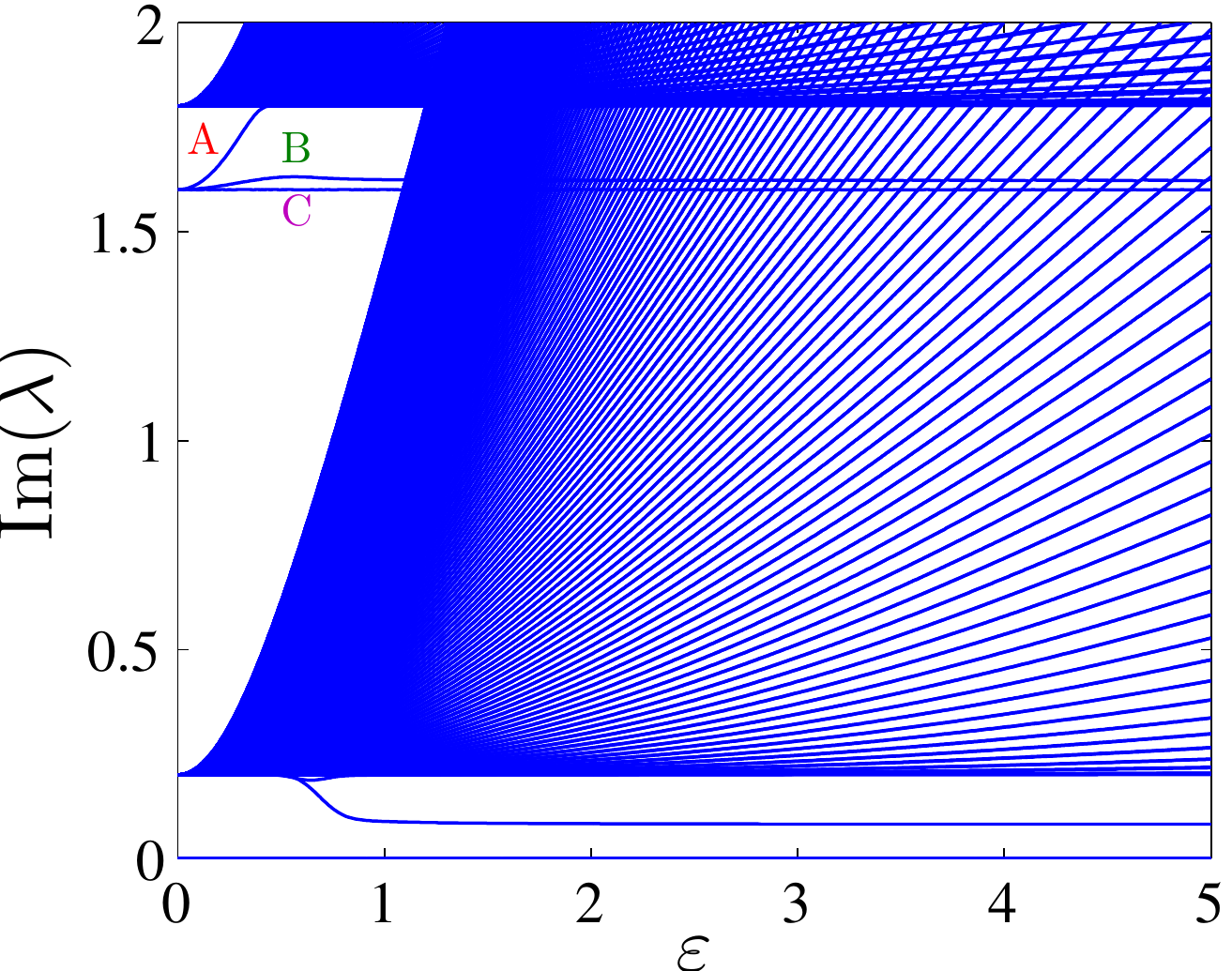} & \hfill
\includegraphics[width=.45\textwidth]{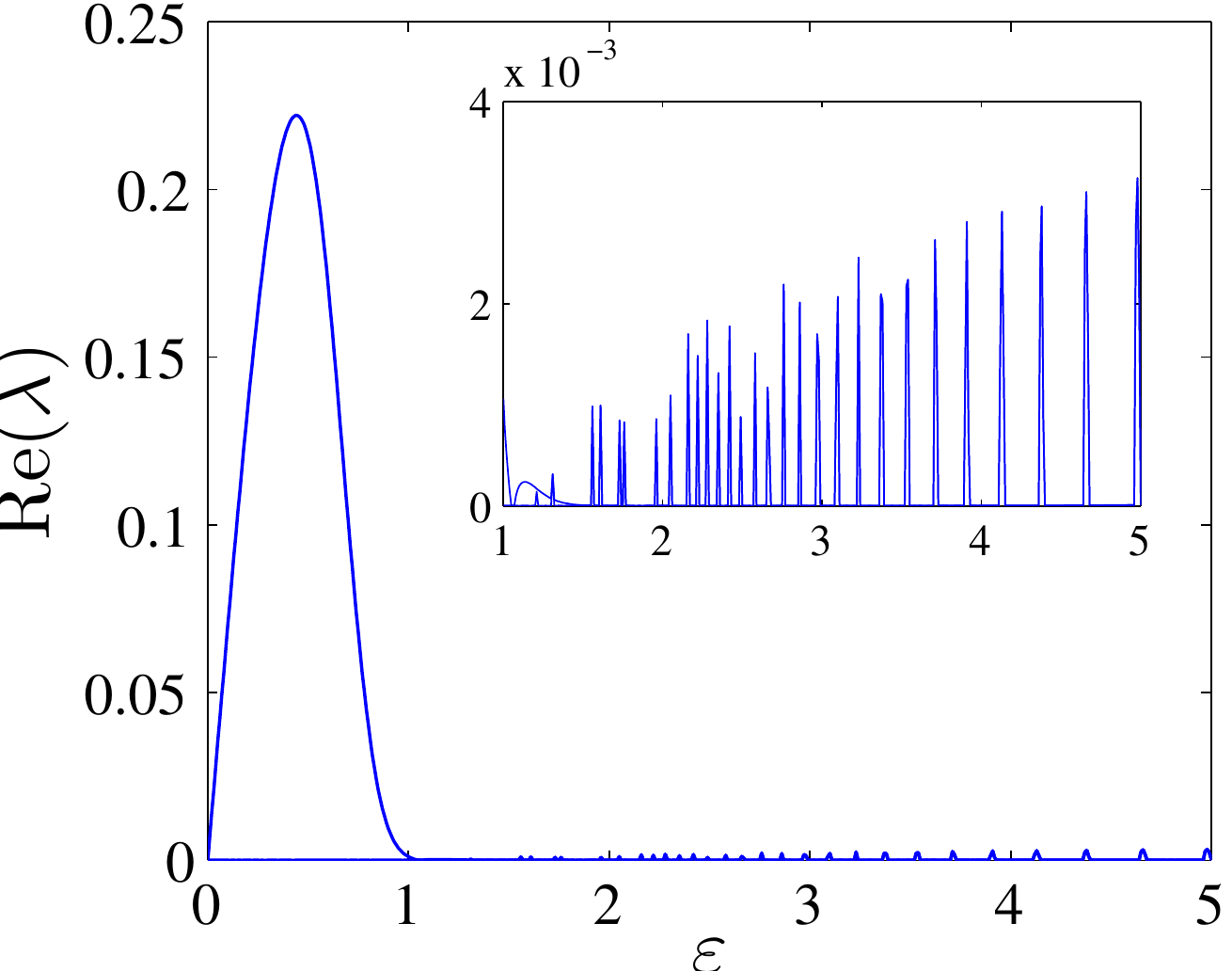} \\
\includegraphics[width=.45\textwidth]{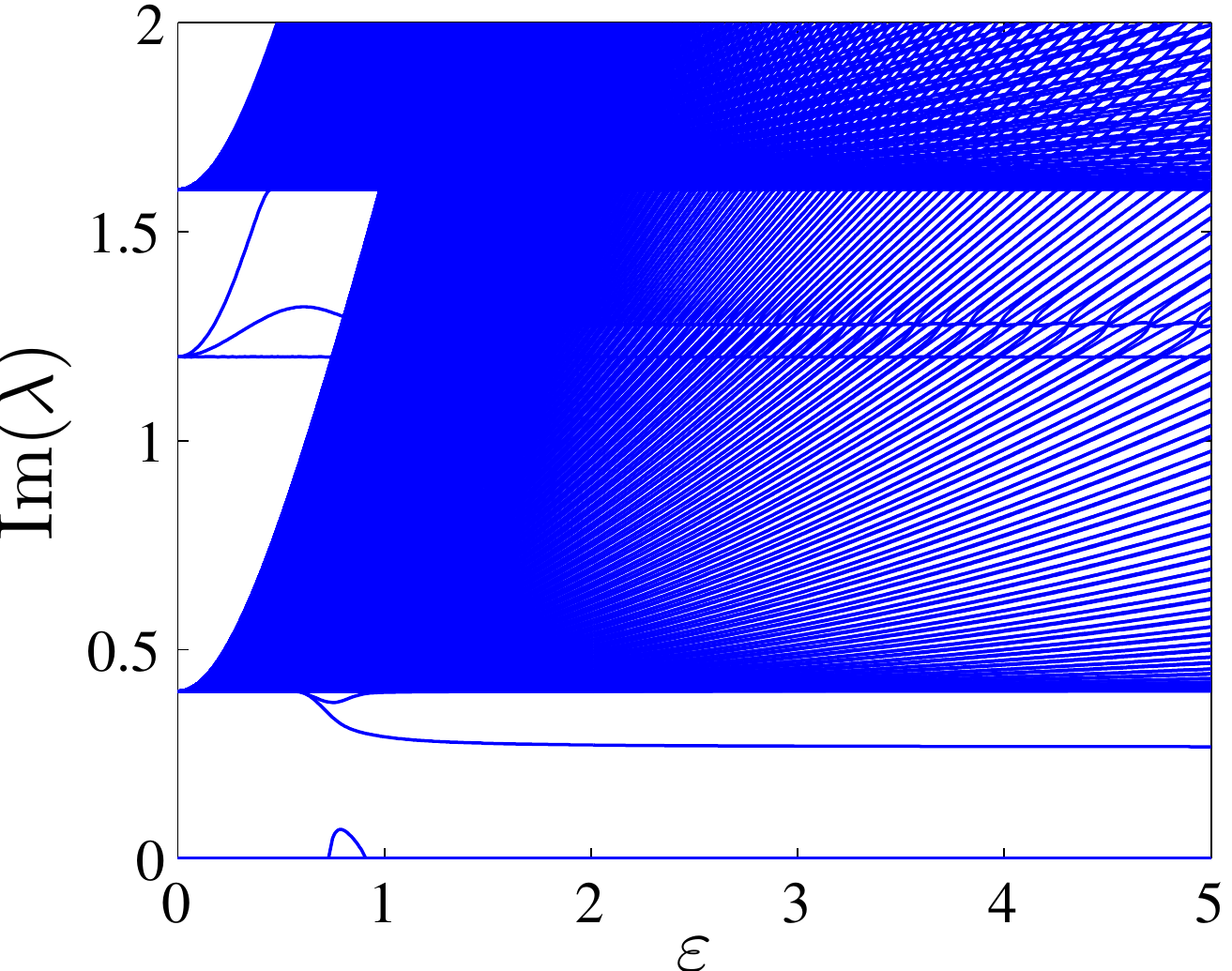} & \hfill
\includegraphics[width=.45\textwidth]{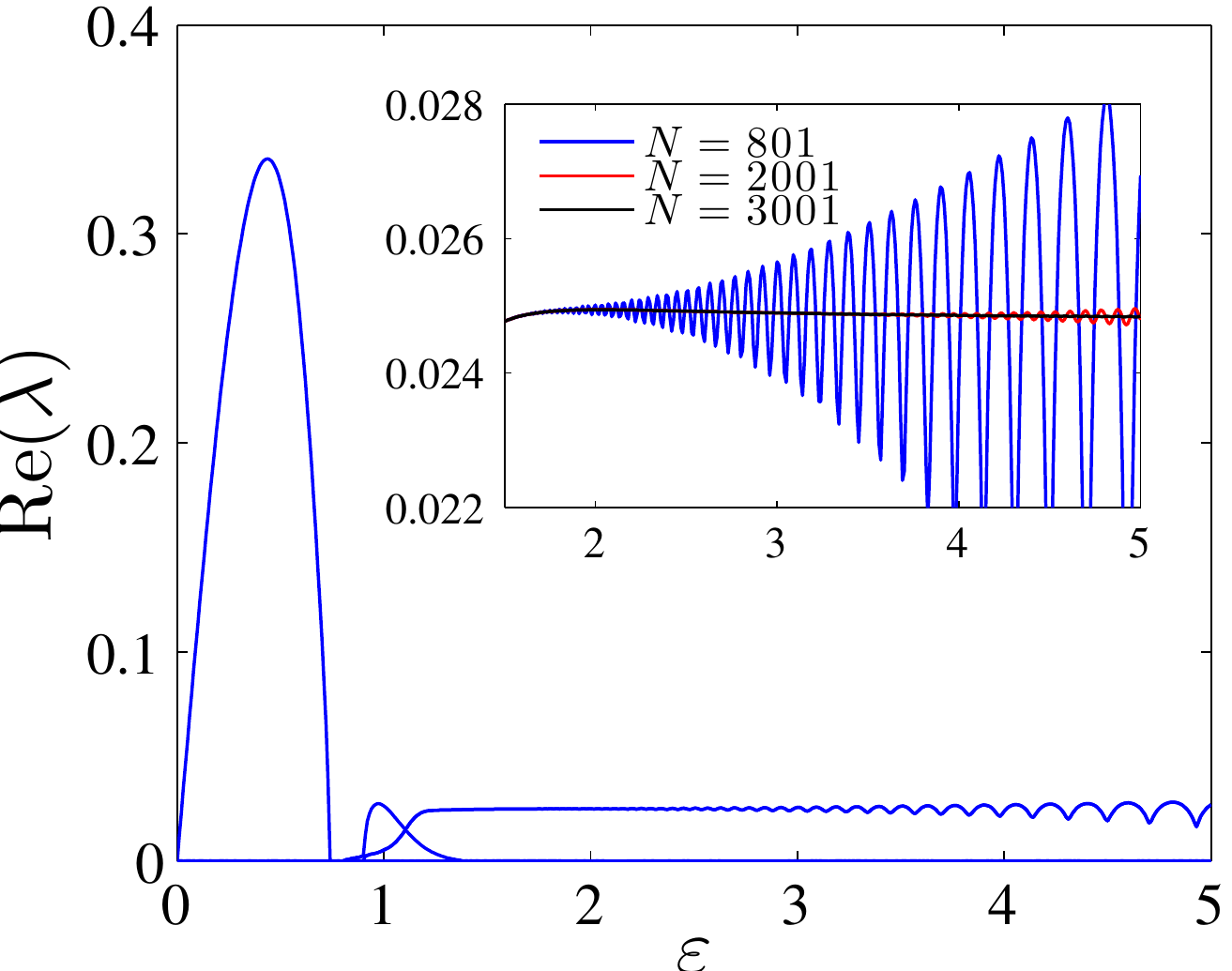} \\
\end{tabular}
\end{center}
\caption{Dependence of the stability eigenvalues with respect to the coupling $\epsilon$ for discrete 3-site solitons with $\omega=0.8$ (top) and $\omega=0.6$ (bottom). Only the positive real and imaginary parts of the eigenvalues are shown. The size of the system is $N=801$. In the top left panel, letters (A, B, C) indicate the three relevant modes whose properties are indicated in the text. In the top right, the inset is a magnification of the relevant $\mathrm{Re}(\lambda)$ shown in the figure but at a different scale. The inset in the bottom right panel shows the oscillations of the growth rate for different system sizes when $\omega=0.6$. Notice that oscillation amplitude decreases rapidly as the number of lattice nodes increases.}
\label{cuevas-fig29}
\end{figure}

In the AC limit the individual sites are decoupled, and the eigenvalues of their respective $4 \times 4$ matrices can be explicitly computed for both the cases of excited and non-excited sites. Thus it is straightforward to see from the stability matrix that in that limit, this three-site solution possesses 3 pairs of modes at $\lambda=0$, 3 pairs at $\lambda=\pm2\omega{i}$, $(N-3)$ pairs at $\lambda=\pm i(1+\omega)$ and $(N-3)$ pairs at $\lambda=\pm i(1-\omega)$. When the coupling is switched on (see Fig.~\ref{cuevas-fig29}), the wave becomes exponentially unstable because of one among the 3 pairs at $\omega=0$ that detaches from the origin yielding a real eigenvalue pair in a similar way as occurs e.g. for the two-site structure in the DNLS equation~\cite{cuevas-Kev09}. The other two vanishing eigenvalue pairs remain at the origin. In addition, the eigenmodes at $\lambda=\pm2\omega{i}$ detach into three pairs that will subsequently collide with the essential spectrum; let us denote those modes as A, B, C (from upper to lower imaginary part of the eigenvalue). Mode C remains exactly at $\lambda=\pm2\omega{i}$ for every coupling. Notice that mode C is always below the essential spectrum for $\omega<1/3$. The imaginary part of the eigenvalue of mode A rapidly increases entering the non-embedded spectrum at the point where the real part of the eigenvalue responsible for the exponential instability reaches its maximum. The exponential instability mentioned previously disappears close to (but not at) the point where mode C enters the embedded spectrum. However, when the coupling increases, the exponential instability appears again with a similar (non-monotonic) behavior as the previous one, except for the presence of smaller growth rates and of a slower decrease in the growth rate (past the point of the maximal growth rate). The most complex parametric dependence is the one experienced by mode B. The latter enters the embedded spectrum for a value of $\epsilon$ higher than that for which mode C enters therein. Then, the system becomes oscillatorily unstable and undergoes a Hopf bifurcation (in the case of finite systems, due to the quantization of the continuous spectrum, this translates into a series of instability bubbles; for a similar scenario in the DNLS see e.g.~\cite{cuevas-JK99}). As a consequence, there are many oscillations in the real part of mode B when the coupling is high; the amplitude of those oscillations decreases when the system size increases, as shown in the inset of bottom right panel of Fig.~\ref{cuevas-fig29}. When the frequency increases (say $\omega\gtrsim0.67$) the real part of mode B does not asymptote to a nearly constant value as the coupling strength increases, but, on the contrary, a series of oscillation bubbles around 0 appears (see top panels of Fig.~\ref{cuevas-fig29}).

\begin{figure}
\begin{center}
\begin{tabular}{cc}
\includegraphics[width=.45\textwidth]{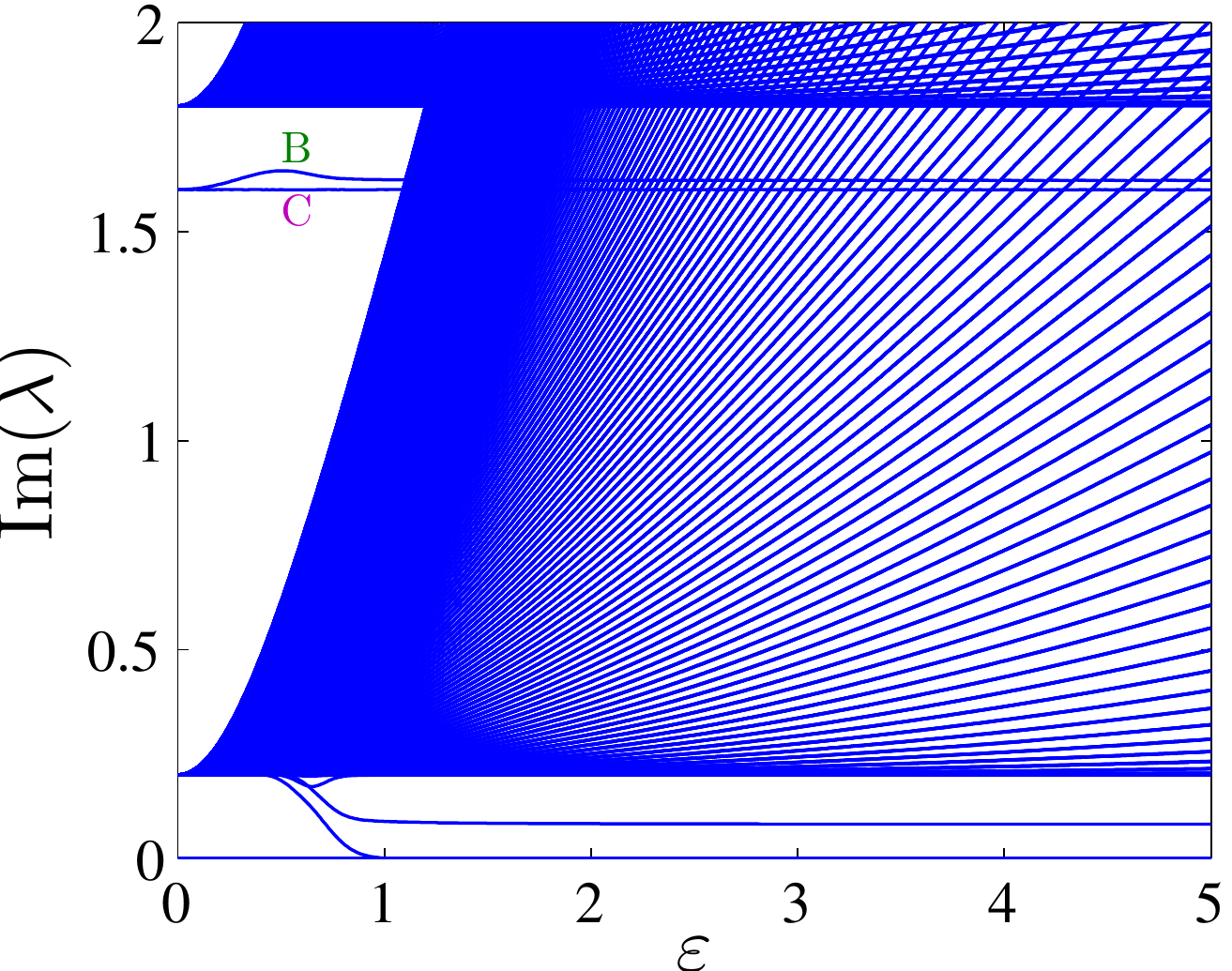} & \hfill
\includegraphics[width=.45\textwidth]{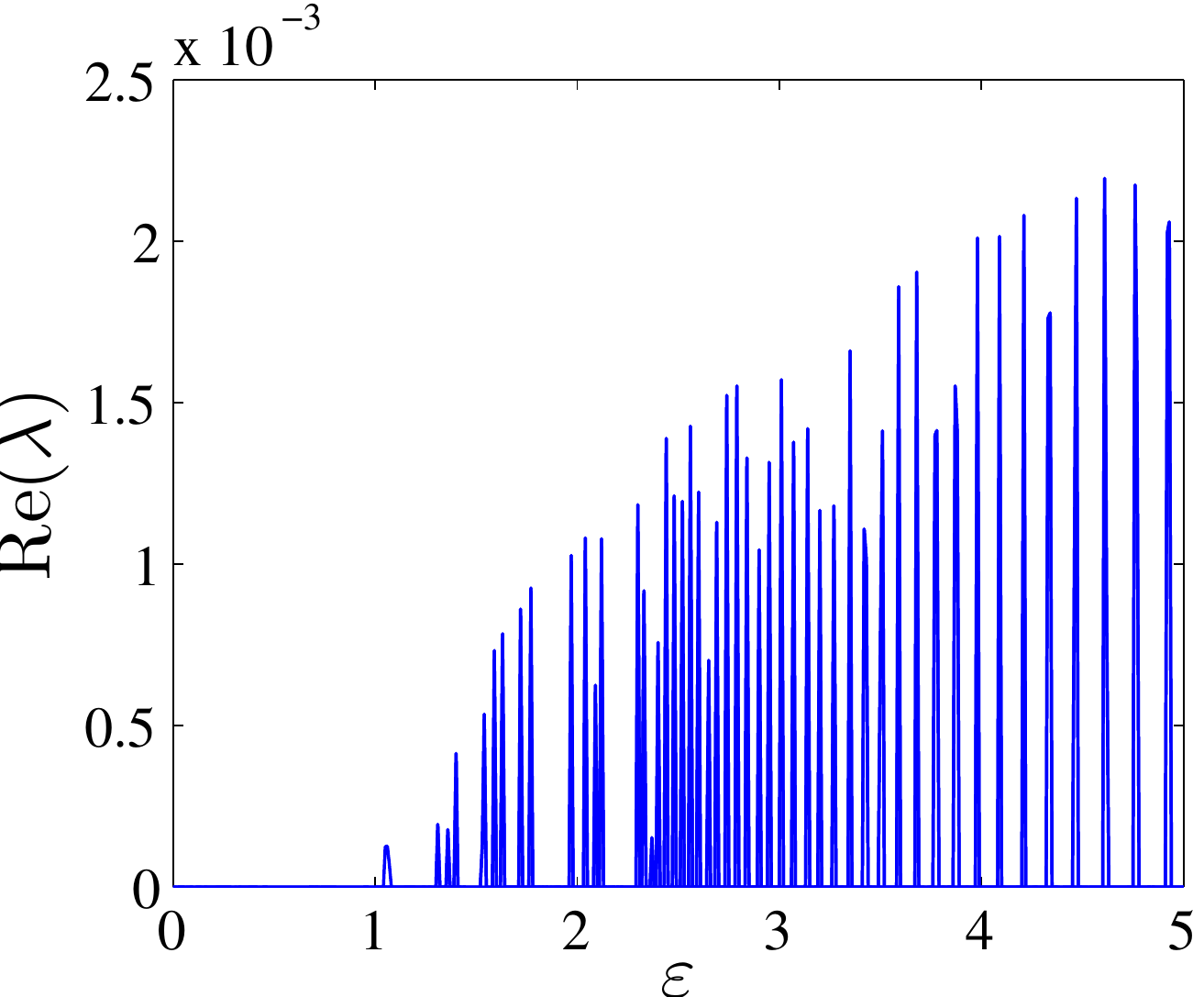} \\
\includegraphics[width=.45\textwidth]{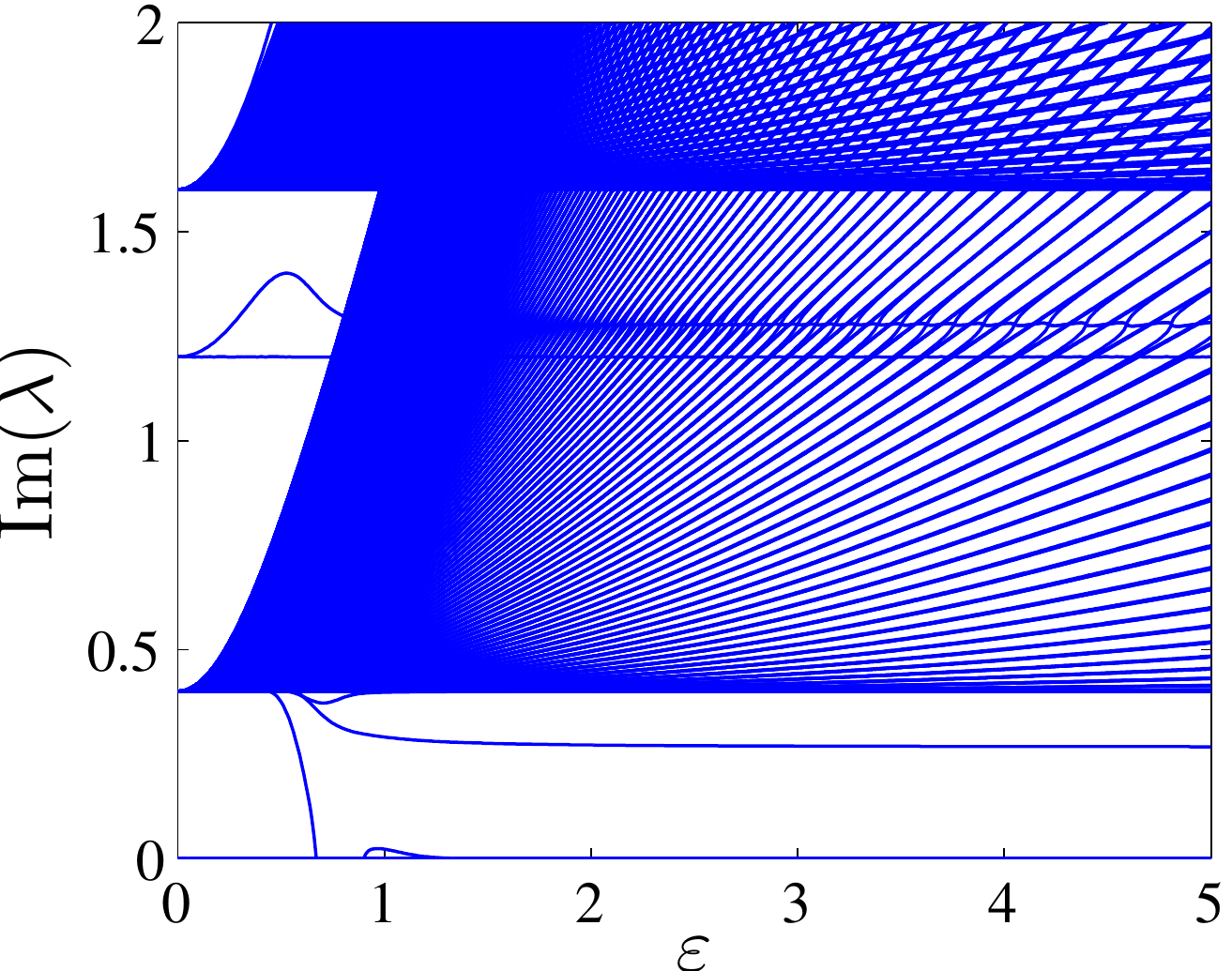} & \hfill
\includegraphics[width=.45\textwidth]{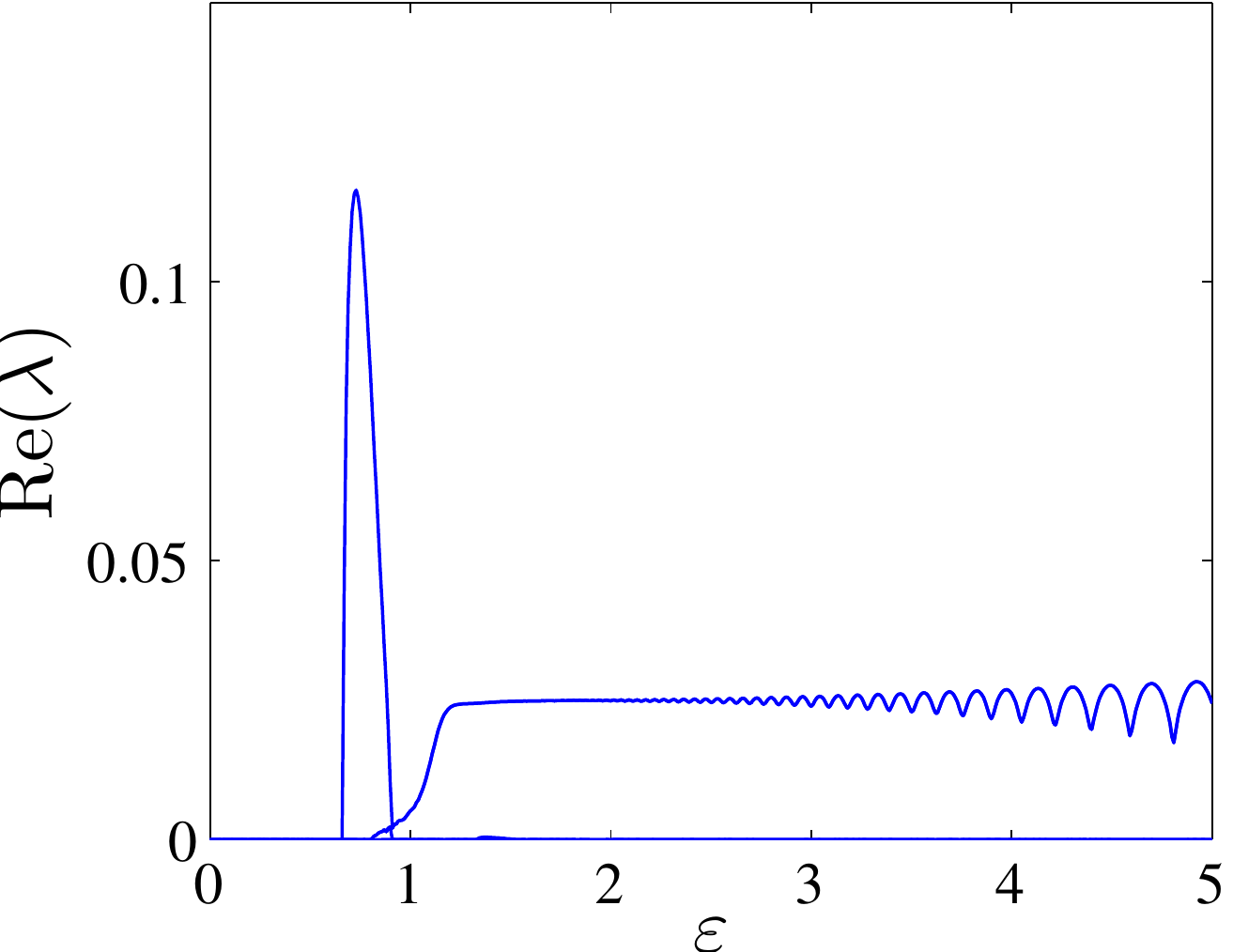} \\
\end{tabular}
\end{center}
\caption{Same as Fig.~\ref{cuevas-fig29} but for the case of discrete 2-site solitons.}
\label{cuevas-fig31}
\end{figure}

\begin{figure}
\begin{center}
\begin{tabular}{cc}
\includegraphics[width=.45\textwidth]{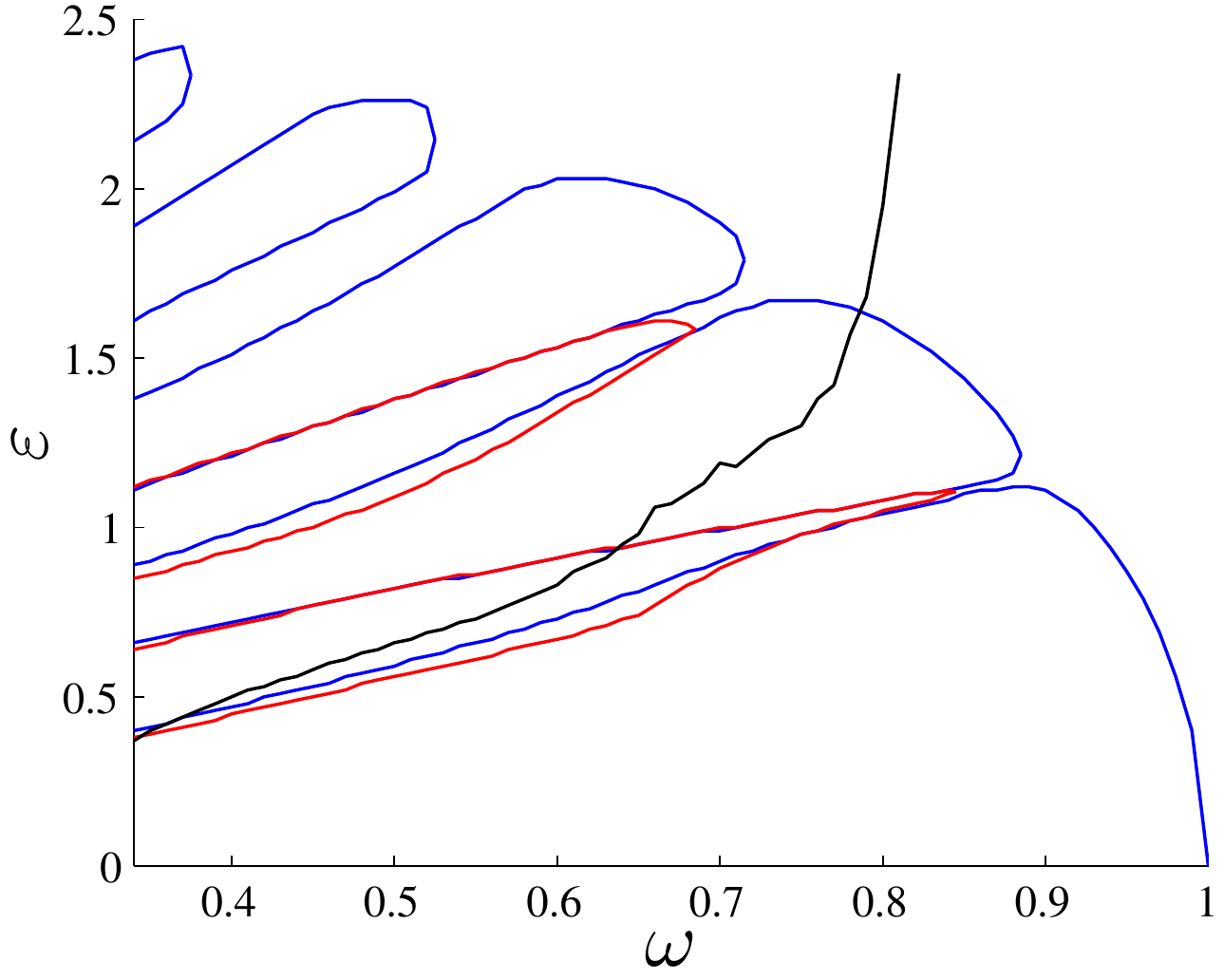} & \hfill
\includegraphics[width=.45\textwidth]{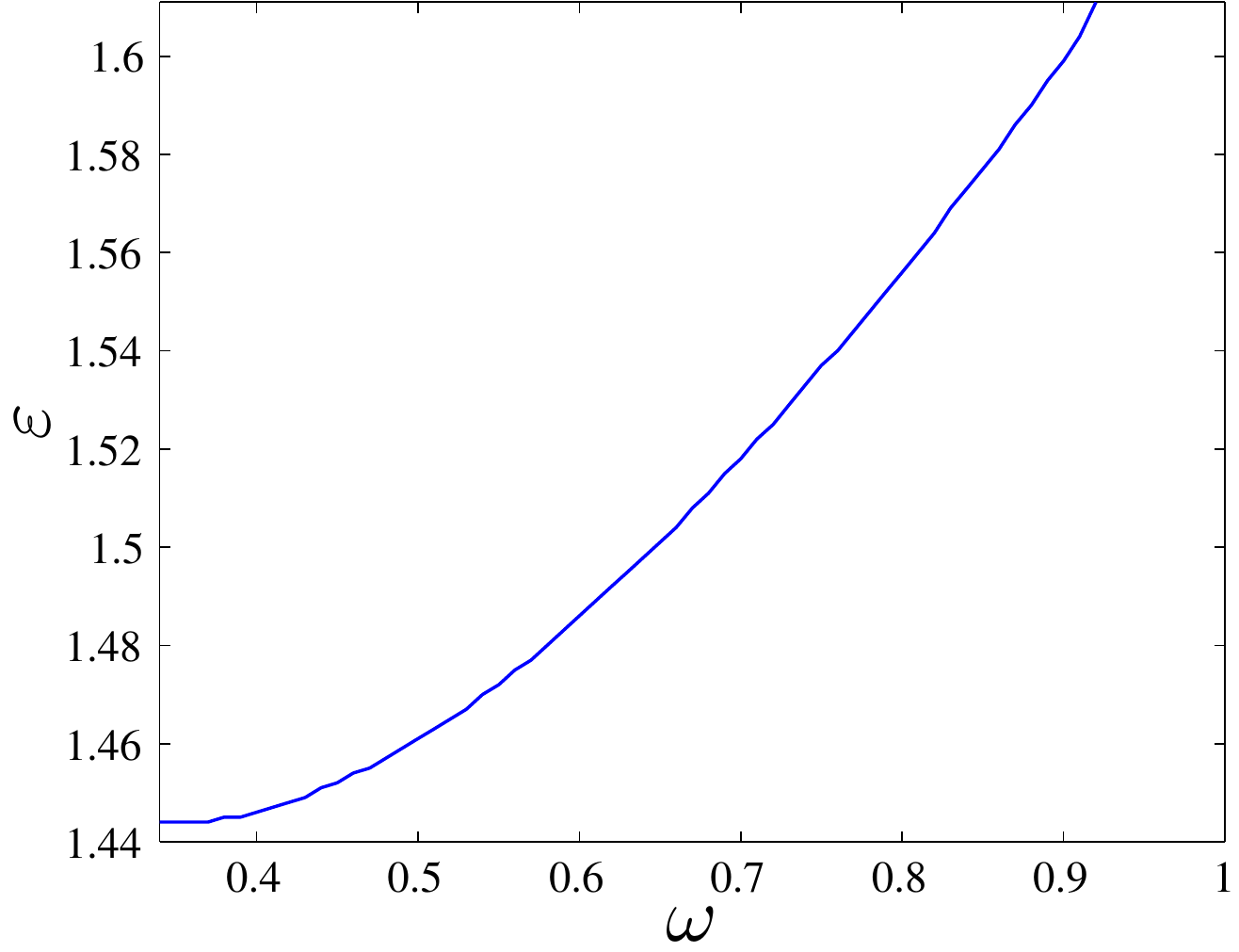} \\
\end{tabular}
\end{center}
\caption{$\epsilon$ vs $\omega$ plane where different unstable regimes for discrete 3-site and 2-site (left) and 1-site solitons (right) are displayed. 3-site solitons are unstable inside the regions with blue edges of the left panel, whereas 2-site solitons are unstable inside the red lines. Above the black lines, both 3-site and 2-site solutions are oscillatorily unstable. 1-site solitons are exponentially unstable above the blue line of right panel, and stable below this line.}
\label{cuevas-fig32}
\end{figure}

A complementary scenario is experienced by the two-site soliton, given in the AC limit by $u_n=0\ \forall$ $n$ and $v_0=v_1=\sqrt{1-\omega}$, $v_n=0$ elsewhere (see the red line on the top panels of Fig.~\ref{cuevas-fig28}). At this limit, the 2-site structure possesses 2 pairs of modes at $\lambda=0$, 2 pairs at $\lambda=\pm2\omega{i}$, $(N-2)$ pairs at $\lambda=\pm i(1+\omega)$ and $(N-2)$ pairs at $\lambda=\pm i(1-\omega)$. When the coupling is switched on (see Fig.~\ref{cuevas-fig31}), the structure remains stable because of the persistence of both pairs at $\lambda=0$. Mode A does not exist for this case; on the other hand, the oscillatory instabilities caused by mode B also exist for the 2-site case. When increasing the coupling, the solitary wave experiences a bifurcation leading to an exponential instability and becomes unstable, contrary to the 3-site soliton (notice that in typical Klein--Gordon and -- e.g. saturable -- DNLS settings, such stability exchanges take place between 2-site and 1-site breathers or solitary waves \cite{cuevas-Aub06,cuevas-HMSK04,cuevas-MCKC06}). Here, there are exponential stability exchanges between 2-site and 3-site solitons, although the bifurcations of the two families of solutions do not perfectly coincide (nevertheless, in a number of such exchanges, the corresponding stabilization/destabilization thresholds are fairly proximal). This scenario is summarized in the left panel of Fig.~\ref{cuevas-fig32}. We should note in passing that these near-exchanges of stability suggest a scenario similar to the ones occurring e.g. in the saturable or cubic-quintic DNLS model where the near-exchange of stability of the 1- and 2-site solitons (in that case) is mediated through a series of pitchfork and reverse pitchfork bifurcations of asymmetric solution branches~\cite{cuevas-CTCM06,cuevas-VJ06}.

\begin{figure}
\begin{center}
\begin{tabular}{cc}
\includegraphics[width=.45\textwidth]{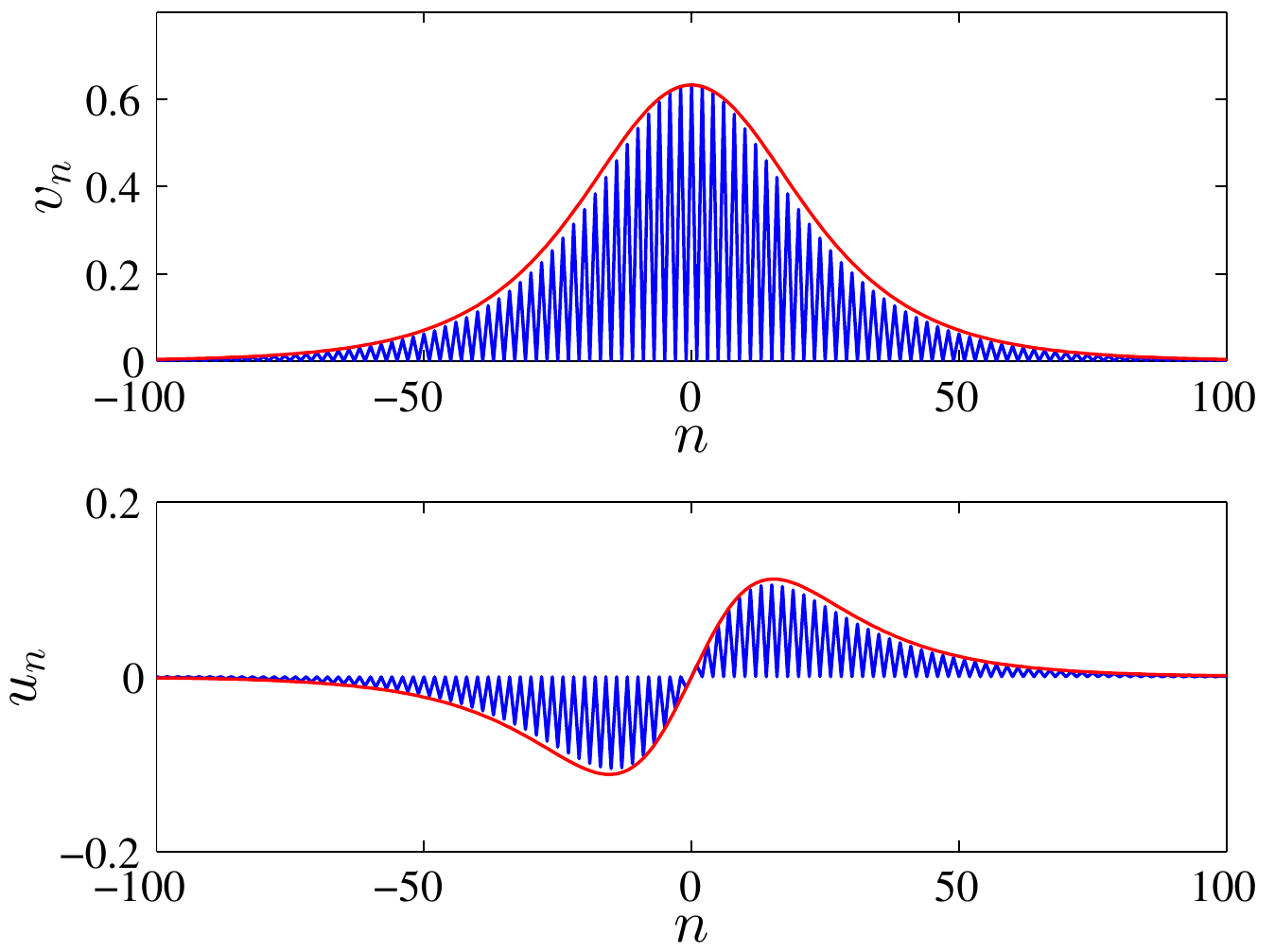} & \hfill
\includegraphics[width=.45\textwidth]{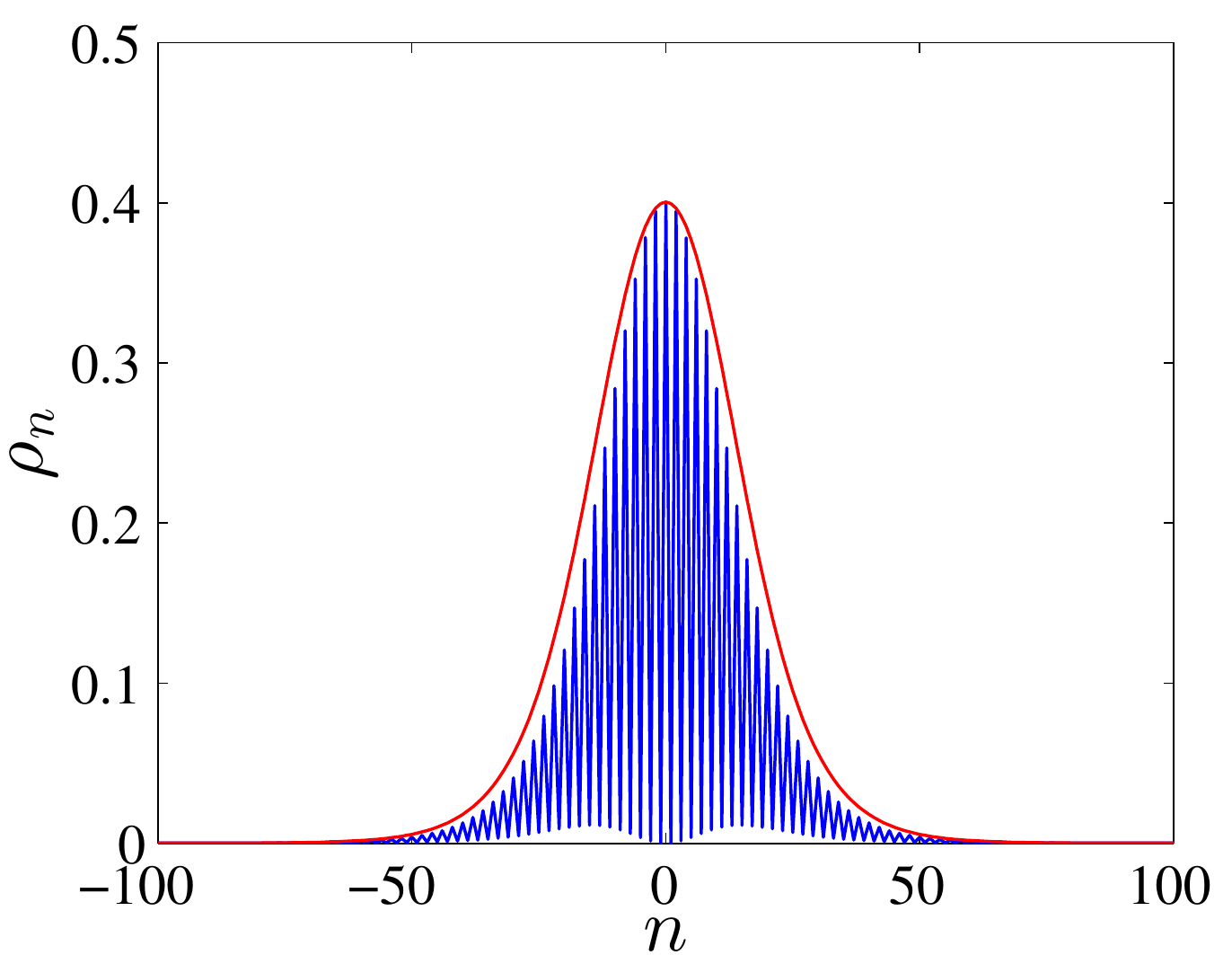} \\
\end{tabular}
\end{center}
\caption{(Left) The spinor components of a discrete 1-site (blue) and 3-site solitons (red) with $\omega=0.8$ and $\epsilon=5$ (i.e. close to the continuum limit). (Right) Charge density $\rho_n=|u_n|^2+|v_n|^2$ for the solitary waves at the left. It is clear that the 1-site soliton does {\it not} asymptote to the 3-site soliton, but rather to a different envelope which is an homoclinic solution of (\ref{eq:cuevas-cont1s}).}
\label{cuevas-fig33}
\end{figure}

There is an interesting kind of solution that also exists from the AC limit and can be extended all the way to the continuum limit, namely the one-site soliton (see bottom panels of Fig.~\ref{cuevas-fig28}). This has the following property which is, in fact, preserved upon continuation for any value of the coupling (see bottom panels of Fig.~\ref{cuevas-fig33}): $v_n=0$ for odd $n$ and $u_n=0$ for even $n$; however, the charge density of the soliton is qualitatively different from that of the three-site solitons. In the AC limit $\epsilon=0$, $v_0=\sqrt{1-\omega}$, and $v_n=0$ for the rest of sites (with $u_n=0\ \forall$ $n$). The form of this solution can be identified as we approach the continuum limit, by transforming the discrete NLD equation (\ref{eq:cuevas-stat1Ddisc}) into the new set of equations:

\begin{equation}\label{eq:cuevas-disc1s}
 \begin{split}
 \epsilon(u_{n+1}-u_{n-1}) - v_n^3 + (1-\omega) v_n =& 0,\  \\
 \epsilon(v_{n+2}-v_n) + u_{n+1}^3 + (1+\omega) u_{n+1} =& 0,
 \end{split}
\end{equation}
for even $n$. There, by neglecting the irrelevant (in this setting) inactive odd sites for one of the fields, and the even ones for the other, the envelope of the solitary waves can be seen to approach the homoclinic orbits of the following system of ODEs that is found by obtaining the continuum limit of (\ref{eq:cuevas-disc1s}):

\begin{equation}\label{eq:cuevas-cont1s}
 \begin{split}
 \partial_x v =& -u^3 - (1+\omega) u , \\
 \partial_x u =& v^3 - (1-\omega) v .
 \end{split}
\end{equation}

Equation (\ref{eq:cuevas-cont1s}), which possesses homoclinic solutions for a wide range of frequencies $\omega$, is tantamount to the Dirac equation with Lorentz-symmetry-breaking nonlinearity appearing in models of binary waveguide arrays \cite{cuevas-TLB14} or graphene nanoribbons \cite{cuevas-HWC15}.

\begin{figure}
\begin{center}
\begin{tabular}{cc}
\includegraphics[width=.45\textwidth]{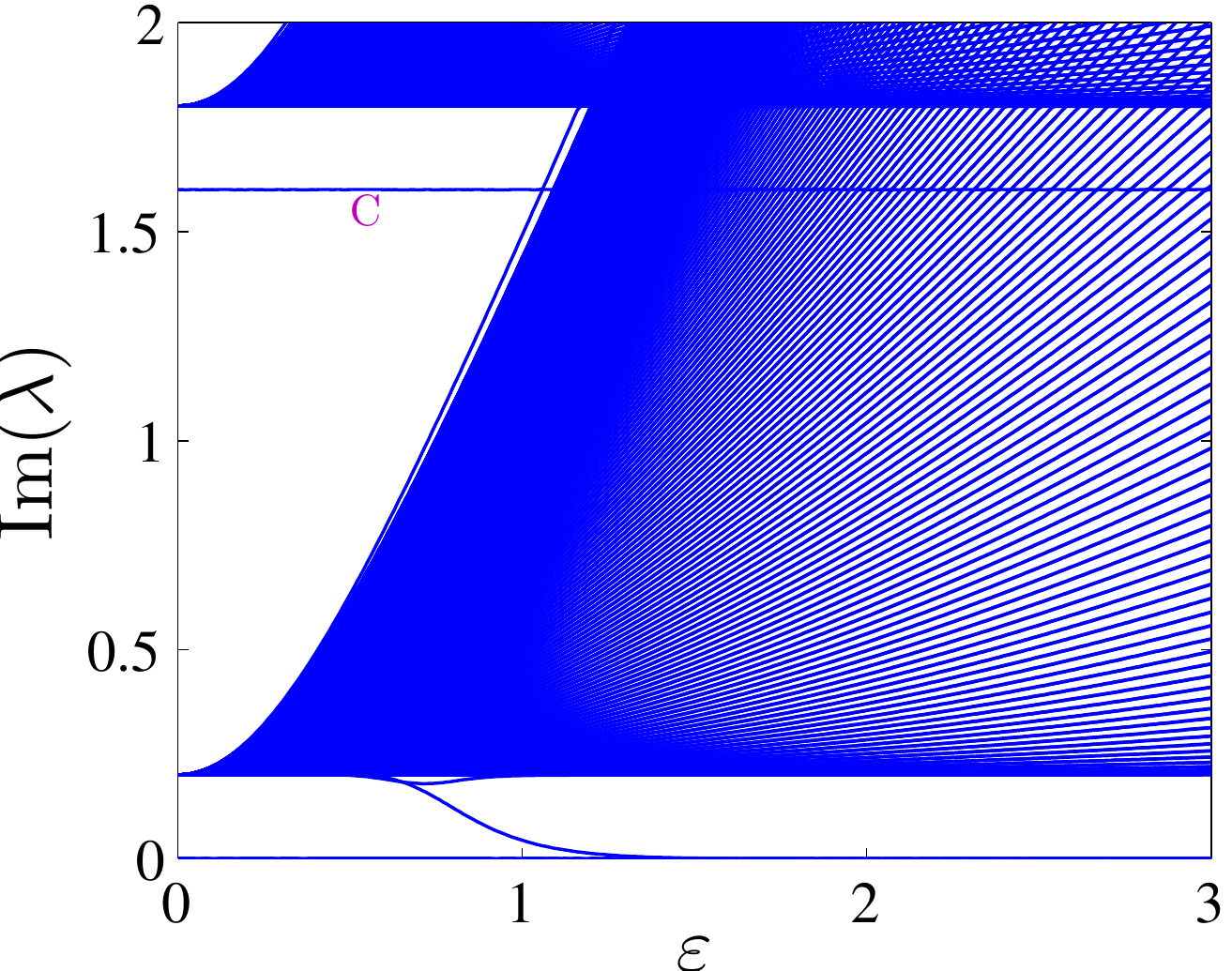} & \hfill
\includegraphics[width=.45\textwidth]{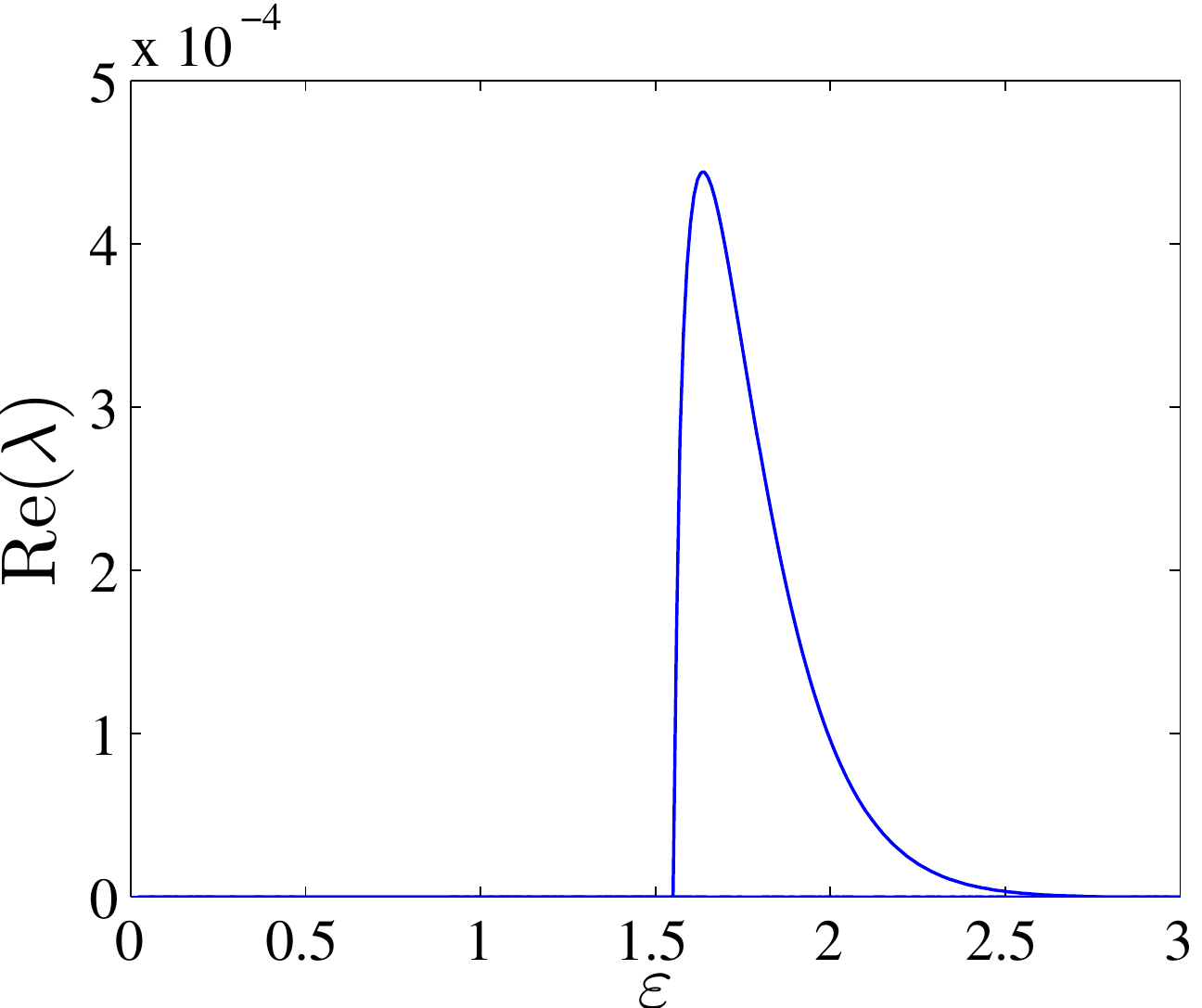} \\
\includegraphics[width=.45\textwidth]{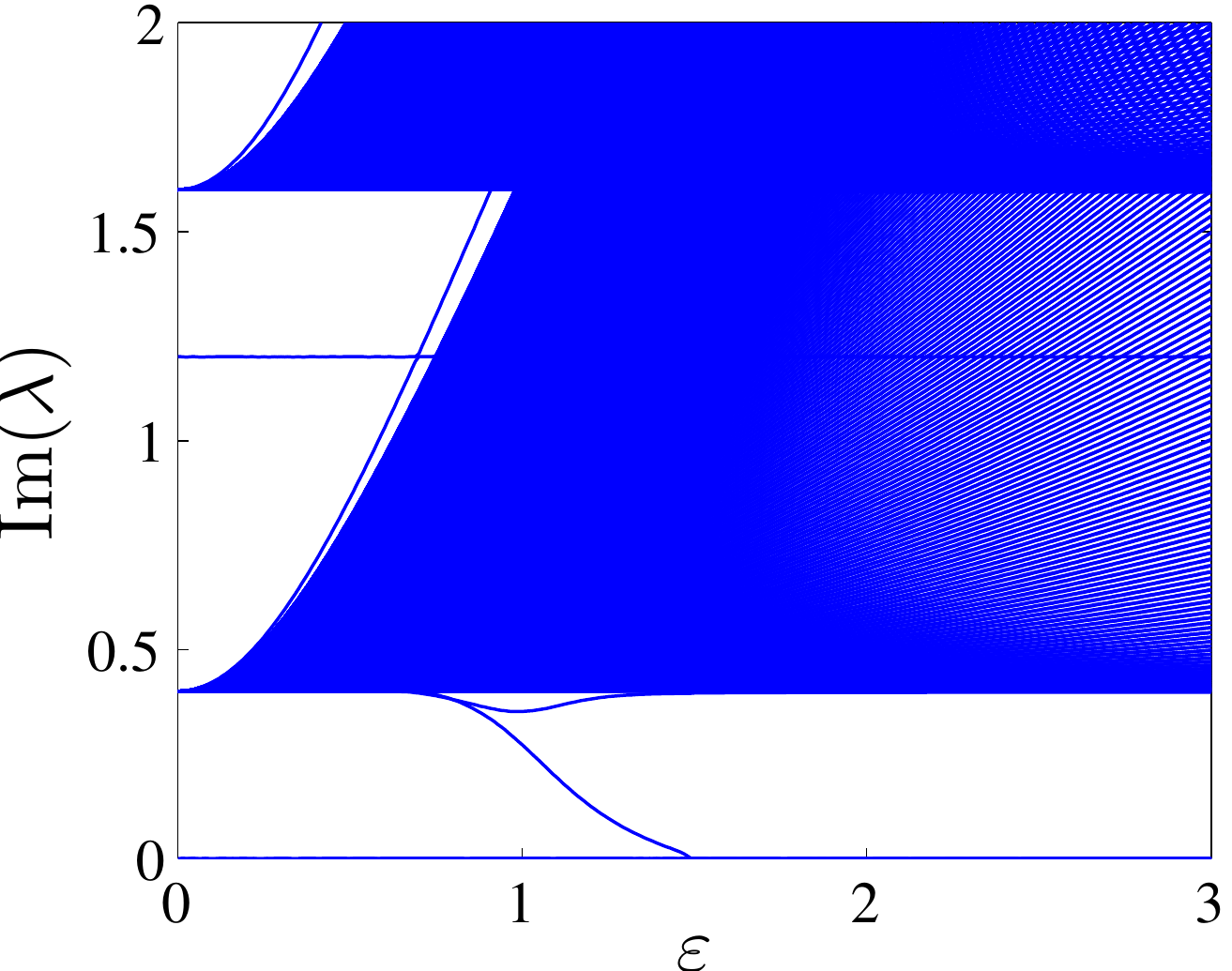} & \hfill
\includegraphics[width=.45\textwidth]{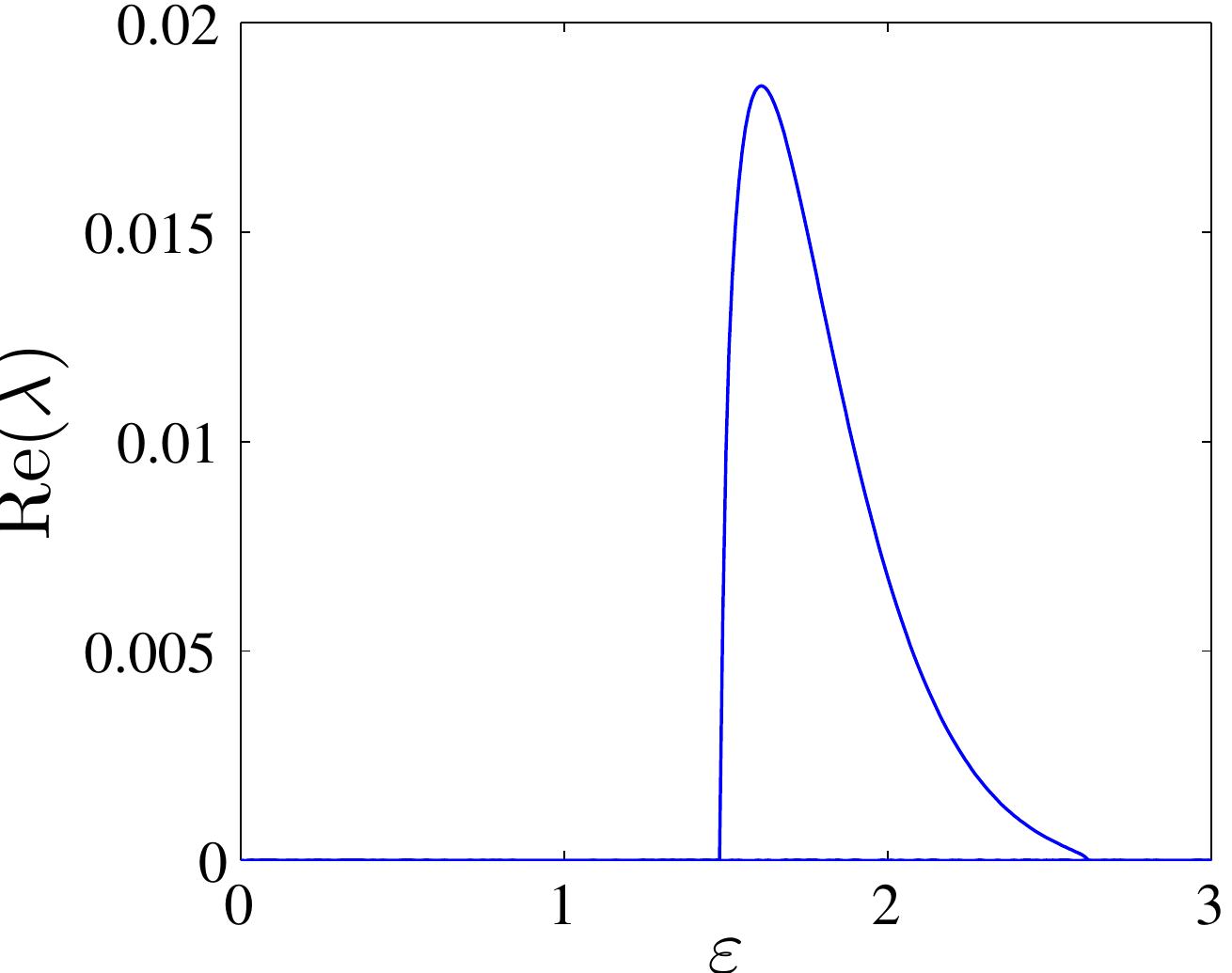} \\
\end{tabular}
\end{center}
\caption{Same as Fig.~\ref{cuevas-fig29} but for the case of discrete 1-site solitons.}
\label{cuevas-fig34}
\end{figure}

The spectrum of the one-site solitons at $\epsilon=0$ consists of a single pair of eigenvalues at $\lambda=0$ and another single pair at $\lambda=\pm2\omega{i}$; apart from these, there are $N-1$ pairs at $\lambda=\pm i(1+\omega)$ and $\lambda=\pm i(1-\omega)$. When the coupling is switched on, as there is only a single pair of eigenmodes at $\lambda=0$, the solitary wave does not experience exponential instabilities; in addition, the non-existence of mode B prevents the existence of destabilizing Hopf bifurcations arising in 3-site and 2-site solitons (see Fig.~\ref{cuevas-fig34}). The only observed instability is an exponential one arising for a finite value of coupling and caused by a mode that bifurcates from the essential spectrum as the coupling strength increases; the growth rate of this bifurcation depends non-monotonically on $\epsilon$ and tends asymptotically to zero when reaching the continuum limit. Similar to the 3-site structures, there is a complementary family of solitary waves consisting of 2-site structures with a hole in between, characterized by $v_0=v_2$ (see red line in the bottom panels of Fig.~\ref{cuevas-fig28}).

We close this section by mentioning that a brief study of the dynamics of unstable discrete solitary waves was performed in \cite{cuevas-CKS15}. As stated therein, a systematic classification of the dynamical implications of the different identified instabilities and of the various possible configurations is a
fairly extensive undertaking and
deserves a separate numerical investigation in its own right.

\section{{$\mathcal{P}\mathcal{T}$}-symmetric Nonlinear Dirac equation}\label{sec:cuevas-PT}

This section is devoted to the existence, stability, and dynamics of solitary waves in a modified 1D Soler model where an additional parity-time
symmetry preserving perturbation has been introduced.

\subsection{$\mathcal{P}\mathcal{T}$ symmetry and Nonlinear Dirac equation}

The study of open systems bearing gain and loss (especially so in a balanced form) is a topic that has emerged over the past two decades as a significant
theme of study~\cite{cuevas-Ben07,cuevas-BFGJ12,cuevas-FJZ08,cuevas-MECM11}. While the realm of {$\mathcal{P}\mathcal{T}$}-symmetry introduced by Bender and collaborators was originally intended as an alternative to the standard Hermitian quantum mechanics, its most canonical realizations (beyond the considerable mathematical analysis of the theme in its own right at the level of operators and spectral theory in mathematical physics) emerged elsewhere in physics. More specifically, in optical systems~\cite{cuevas-RDM05,cuevas-KGM08,cuevas-MECM08} the analogy between the paraxial approximation of Maxwell's equations and the Schr{\"o}dinger equation formed the basis on which the possibility of {$\mathcal{P}\mathcal{T}$}-symmetric realizations initially in optical waveguide experiments was proposed and then experimentally implemented~\cite{cuevas-GSD+09,cuevas-RME+10,cuevas-RBM+12}. The success of this program motivated further additional initiatives in other directions of experimental interest, including, but not limited to, {$\mathcal{P}\mathcal{T}$}-symmetric electronic circuits~\cite{cuevas-SLZ+11,cuevas-SLL+12}, mechanical systems~\cite{cuevas-BBPS13} and whispering-gallery microcavities~\cite{cuevas-POL+14}.

Our aim in this Section is to present a connection between the budding area of research of {$\mathcal{P}\mathcal{T}$}-symmetric systems and the nonlinear Dirac equation. To this aim, we choose a generalized {$\mathcal{P}\mathcal{T}$}-symmetric 1D Soler model, which in covariant form reads as follows:

\begin{equation}\label{eq:cuevas-SolerPT2}
 \left(i\gamma\sp\mu\partial\sb\mu-m+g\left(\bar\psi\psi\right)^{k}+\gamma\gamma^5\right)\psi=0.
\end{equation}

Alternatively, written in the standard form as a function of the bispinor components $\psi=[\psi_1(x),\psi_2(x)]^T$ \cite{cuevas-CKS+16}, the equations
assume the following form:

\begin{equation}\label{eq:cuevas-SolerPT}
 \begin{split}
 i\partial_t\psi_1 &= \,\,\partial_x\psi_2-g(|\psi_1|^2-|\psi_2|^2)^k\psi_1+m\psi_1+i\gamma\psi_2, \\
 i\partial_t\psi_2 &= -\partial_x\psi_1+g(|\psi_1|^2-|\psi_2|^2)^k\psi_2-m\psi_2+i\gamma\psi_1.
 \end{split}
\end{equation}

Equations (\ref{eq:cuevas-SolerPT}) are {$\mathcal{P}\mathcal{T}$}-symmetric because they are invariant under the transformation
\[
{\mathcal{P}}:\;
x \rightarrow -x,
\qquad
\psi_1 \rightarrow \psi_1,
\qquad
\psi_2 \rightarrow -\psi_2
\]
and
\[
{\mathcal{T}}:\;
t \rightarrow -t,
\qquad
i \rightarrow -i,
\qquad
\psi_1 \rightarrow \psi_1,
\qquad
\psi_2 \rightarrow \psi_2.
\]
This transformation assumes that $\psi_1(t,x)$ is spatially even and that $\psi_2(t,x)$ is spatially odd.
The key addition to this model in comparison to the original Soler model (\ref{eq:cuevas-Soler1D}) is the inclusion of the gain-loss term proportional to $\gamma$ in the (implicit) form of the Dirac matrix $\gamma^5$ (cf.~\cite{cuevas-Ben07}) multiplying the spinor $\psi$ in (\ref{eq:cuevas-SolerPT2}). In our case of two-component spinors, the role of $\gamma^5$ is played by the Pauli matrix $\sigma_1$ (\ref{eq:cuevas-Pauli}). We note in passing that in its linear form, the model can be converted under a suitable transformation (associated with the so-called $\mathcal{C}$-operator) to a Hamiltonian one with a reduced mass of $\tilde{m}=\sqrt{m^2-\gamma^2}$~\cite{cuevas-Ben07,cuevas-BJR05}.

It is straightforward to see that in the linear case (of $g=0$), plane waves $\psi_1(t,x)= A e^{i (\kappa x - \omega t)}$ and $\psi_2(t,x)=i B e^{i (\kappa x - \omega t)}$ are solutions provided the dispersion relation $\omega= \pm \sqrt{m^2 + \kappa^2 - \gamma^2}$ is satisfied. Not only does the above formula have the characteristic Dirac form, but it also is consistent with the equivalence of the linear {$\mathcal{P}\mathcal{T}$}-Dirac equation with effective mass $\tilde{m}=\sqrt{m^2 - \gamma^2}$, as per the above discussion.

In the present {$\mathcal{P}\mathcal{T}$}-NLD we have been unable to identify explicit stationary solitary waves in the massive $m\neq0$ case, so we must rely on numerical analysis. In the same vein, it does not appear to be straightforward to generalize the transformation of~\cite{cuevas-Ben07,cuevas-BJR05} to the present nonlinear setting.

To determine the stability of stationary solitary wave solutions,
we consider infinitesimal perturbations of the form:
\begin{equation*}\label{eq:cuevas-linstab1D}
 \begin{split}
 \psi_1(t,x) &= e^{-i \omega t} \left[v(x) + \delta (a_1(x) e^{\lambda t} + b_1^{*}(x) e^{\lambda^{*} t}) \right],\\
 \psi_2(t,x) &= e^{-i \omega t} \left[u(x) + \delta (a_2(x) e^{\lambda t} + b_2^{*}(x) e^{\lambda^{*} t}) \right],
 \end{split}
\end{equation*}
where $\delta$ denotes a formal small parameter. The relevant linearization equations are derived to order ${\rm O}(\delta)$
[by substitution of the above Ansatz into Eqs.~(\ref{eq:cuevas-SolerPT})] and are subsequently solved as a matrix eigenvalue problem

\begin{equation*}
 \lambda[a_1(x),a_2(x),b_1(x),b_2(x)]^T=\mathcal{M}[a_1(x),a_2(x),b_1(x),b_2(x)]^T,
\end{equation*}
with $\mathcal{M}$ being
\begin{equation*}\label{eq:cuevas-stabmat1D}
 \mathcal{M}=\left(\begin{array}{cc} L_1 & L_2 \\ \\ -L_2^* & -L_1^* \end{array}\right)-i\gamma
 \left(
 \begin{array}{cc}
 \gamma^5 & 0
 \\[2ex]
 0 & \gamma^5
 \end{array}
 \right)
\end{equation*}
and
\begin{equation*}
\begin{split}
 L_1 &=
 \left(\begin{array}{cc}
 f(|v|^2-|u|^2)-m+\Lambda & -\partial_x \\ \\
 \partial_x & m-f(|v|^2-|u|^2)+\Lambda
 \end{array}\right) \\
 &+
 f'(|v|^2-|u|^2)
 \left(\begin{array}{cc}
 |v|^2 & -v^*u \\ \\
 -v^*u & |u|^2
 \end{array}\right),
\\
 L_2 &= f'(|v|^2-|u|^2)
 \left(\begin{array}{cc}
 v^2 & -v u \\ \\
 -v u & u^2
 \end{array}\right).
\end{split}
\end{equation*}

   From the dynamical equations (\ref{eq:cuevas-SolerPT}) it is straightforward to show that the charge is not preserved. Instead, the following ``moment equation'' is satisfied:
\begin{equation}\label{eq:cuevas-chargederiv}
 \frac{dQ}{dt}=4\gamma\int\mathrm{Re}(V^*U)\,dx\,.
\end{equation}
Note that in the case of a standing wave state, $dQ/dt=0$ and charge is conserved.

Although the charge is not generally conserved, remarkably there is a conserved quantity in the form of the energy:
\begin{equation*}\label{eq:cuevas-energyPT}
 E=\frac{1}{2}\int \left[ \psi_1^*\partial_x \psi_2-\psi_2^*\partial_x \psi_1+m(|\psi_1|^2-|\psi_2|^2)-
 \frac{g}{k+1}[|\psi_1|^2-|\psi_2|^2]^{k+1}\right]\,dx\,.
\end{equation*}

Notice that there is no $\gamma$ dependence in this formula. As a matter of fact, this is the same definition for the energy as in the $\gamma=0$ limit (cf. Eq.(\ref{eq:cuevas-hamiltonian})). But, intriguingly, $dE/dt=0$ even for $\gamma\neq0$. This is rather unusual in our experience in {$\mathcal{P}\mathcal{T}$}-symmetric models and is effectively related to the special form of introducing {$\mathcal{P}\mathcal{T}$}-symmetry through the  matrix $\gamma^5$. We note that in this form, it is not transparent
(unlike e.g. in the Schr{\"o}dinger {$\mathcal{P}\mathcal{T}$}-symmetric models~\cite{cuevas-Ben07})
which component corresponds to the gain and which one to the loss. Effectively, isolating the time-dependence and the $\gamma$-dependent term in the equations (i.e., $i \partial_t \psi_1 = i \gamma \psi_2$ and $i \partial_t \psi_2= i \gamma \psi_1$ and momentarily ignoring the rest of the terms), it appears as if both components bear both gain and loss.

\subsection{Numerical results}

In the numerical computations presented herein, we have utilized spectral collocation methods in order to approximate the spatial derivatives of Eq.~(\ref{eq:cuevas-SolerPT}). As discussed in Subsection \ref{subsec:cuevas-Soler2D}, the Chebyshev collocation is, arguably, the most suitable method
for approximating the relevant derivatives as it gives a better spectral accuracy. However, because of the particular ({$\mathcal{P}\mathcal{T}$}-symmetric) structure
of the system, the implementation of fixed-point methods, requires a high amount of computer memory which poses implementation challenges. Furthermore, the method has the drawback that the double humped solitary waves, cannot be well resolved (i.e. the humps cannot be observed) because of the Chebyshev collocation including more points at the edge of the system in comparison to the center. Consequently, in what follows, a Fourier collocation scheme has been implemented. In suitable limit cases, we have checked that the results are similar for the different implementations and that no extra spurious eigenmodes arise in comparison to the standard case. We note that hereafter we will focus on the case of $k=1$ for our numerical implementation and $g=m=1$ has been fixed.

\begin{figure}
\begin{center}
\begin{tabular}{cc}
\includegraphics[width=.45\textwidth]{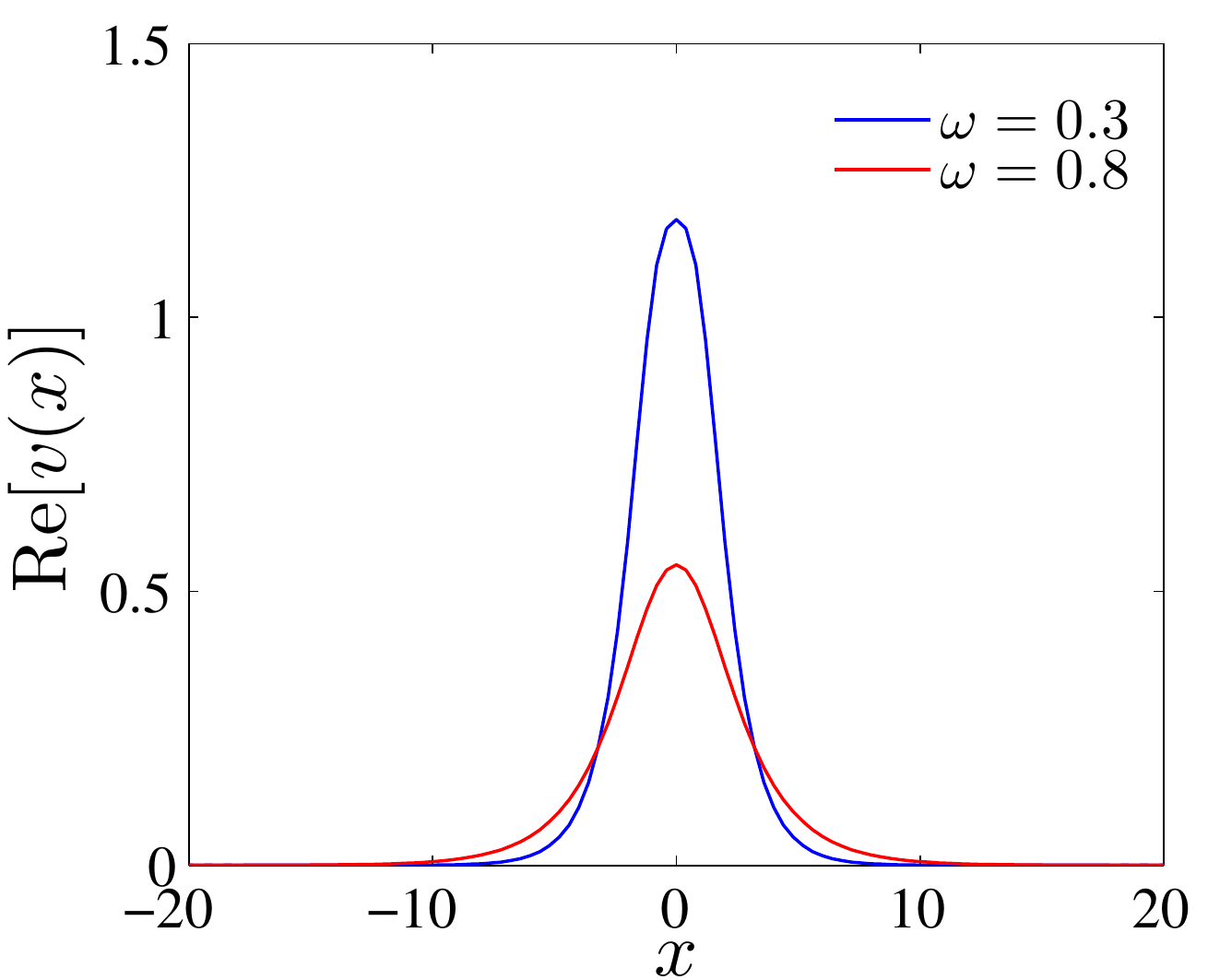} & \hfill
\includegraphics[width=.45\textwidth]{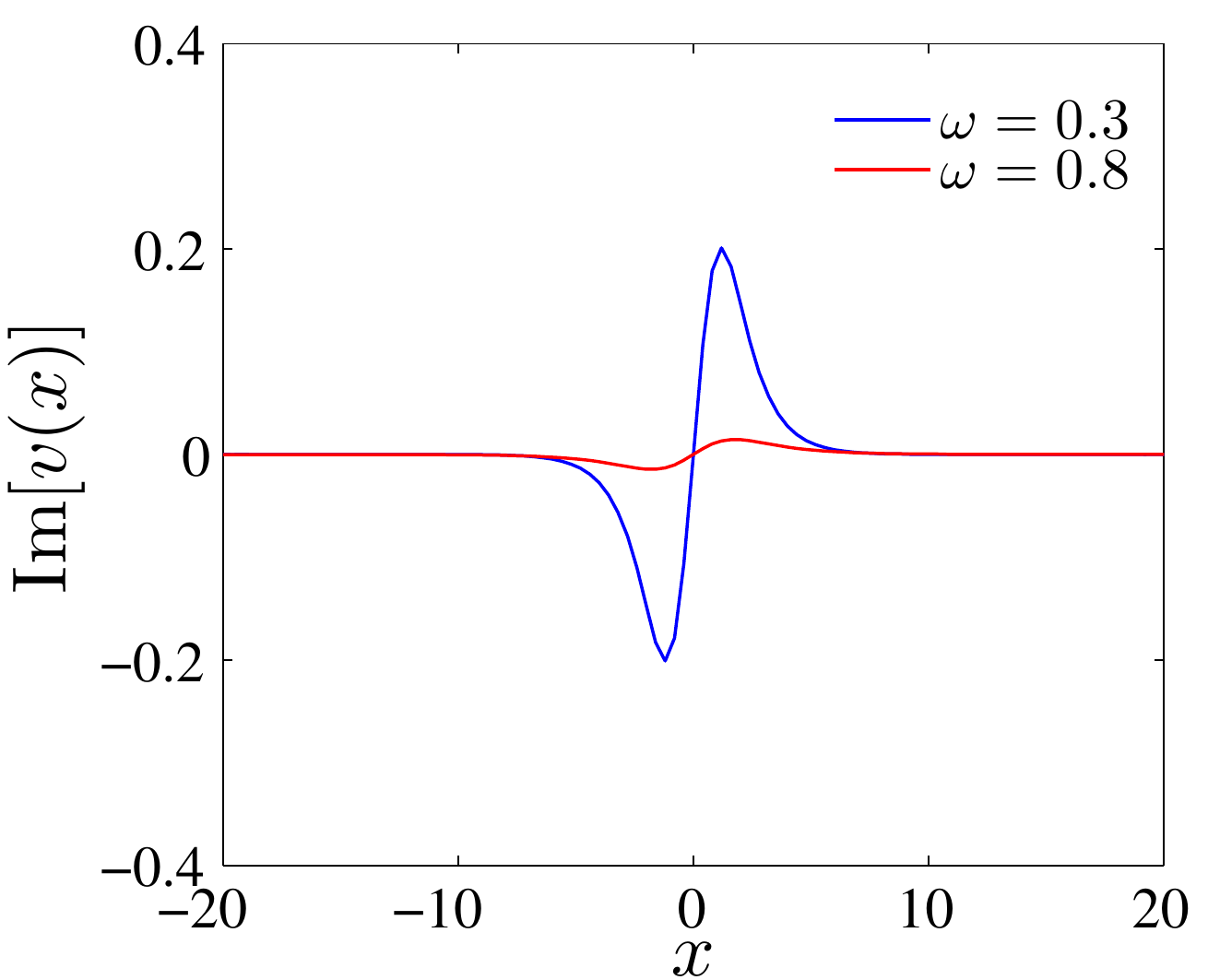} \\
\includegraphics[width=.45\textwidth]{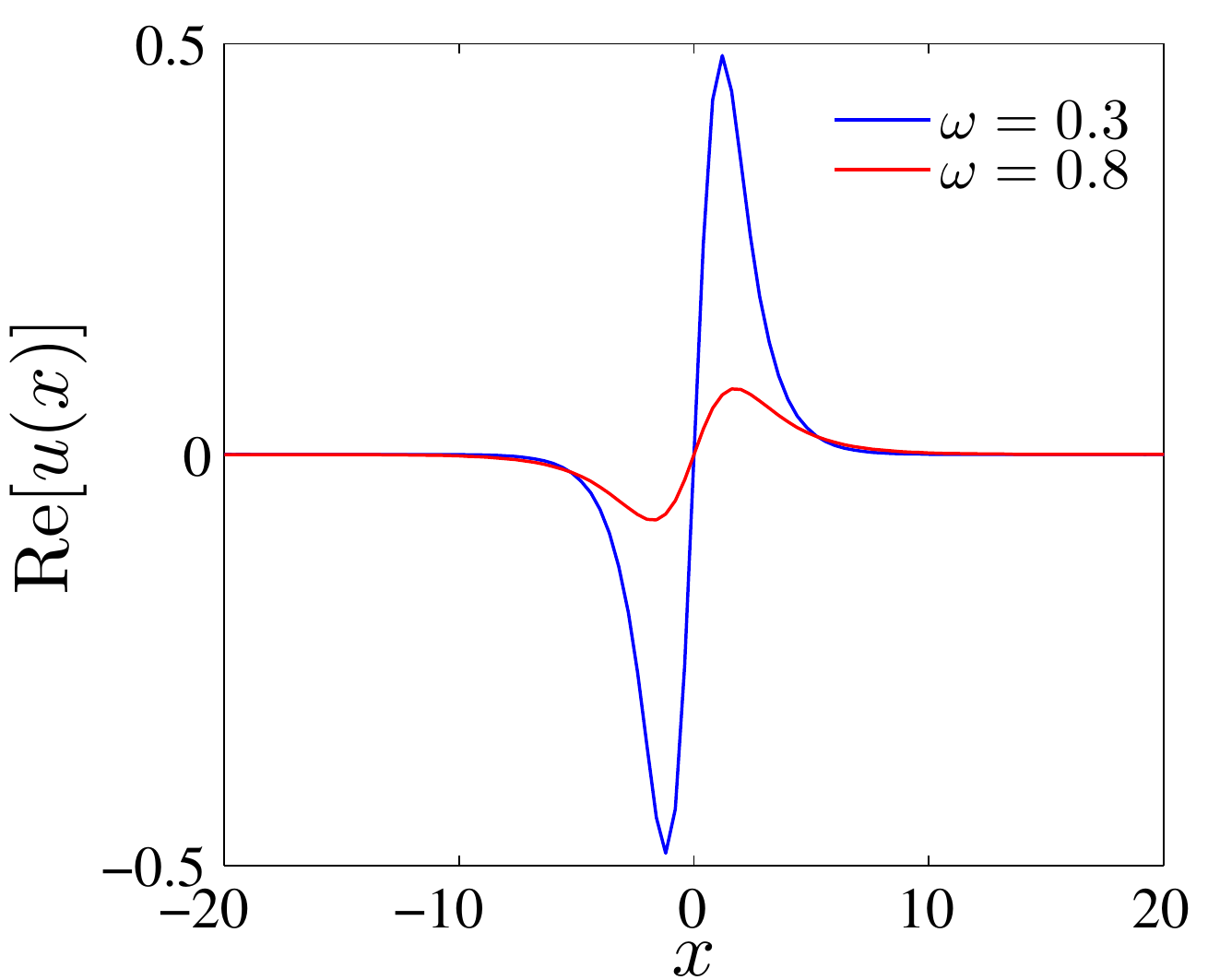} & \hfill
\includegraphics[width=.45\textwidth]{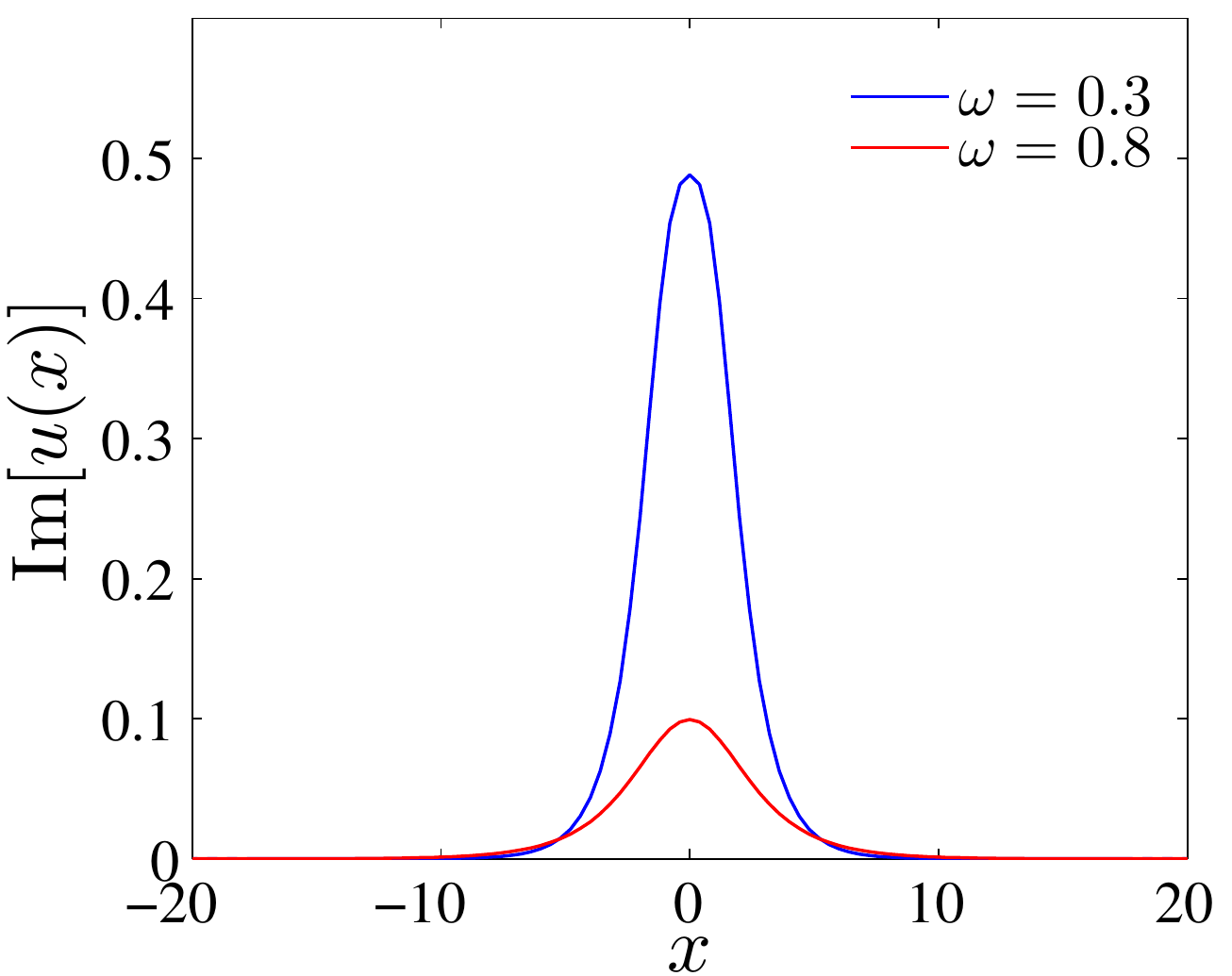} \\
\multicolumn{2}{c}{\includegraphics[width=.45\textwidth]{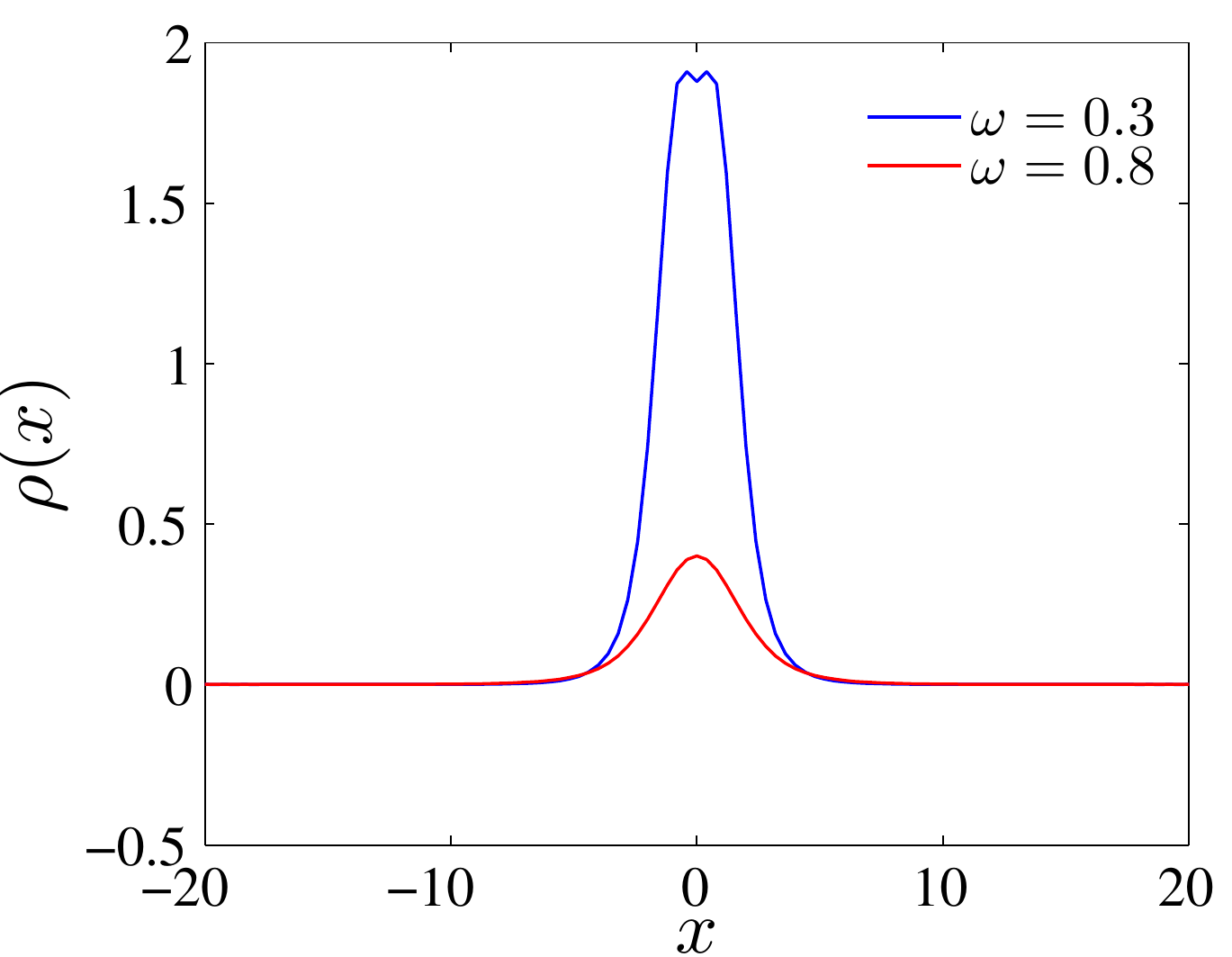}} \\
\end{tabular}
\end{center}
\caption{Real and imaginary part of each spinor component for {$\mathcal{P}\mathcal{T}$}-symmetric solitary waves with $\omega=0.8$ and $\omega=0.3$. Here $\gamma=0.3$ in every case. The charge density of the solitary wave is depicted in the bottom panel.}
\label{cuevas-fig35}
\end{figure}

\begin{figure}
\begin{center}
\begin{tabular}{cc}
\includegraphics[width=.45\textwidth]{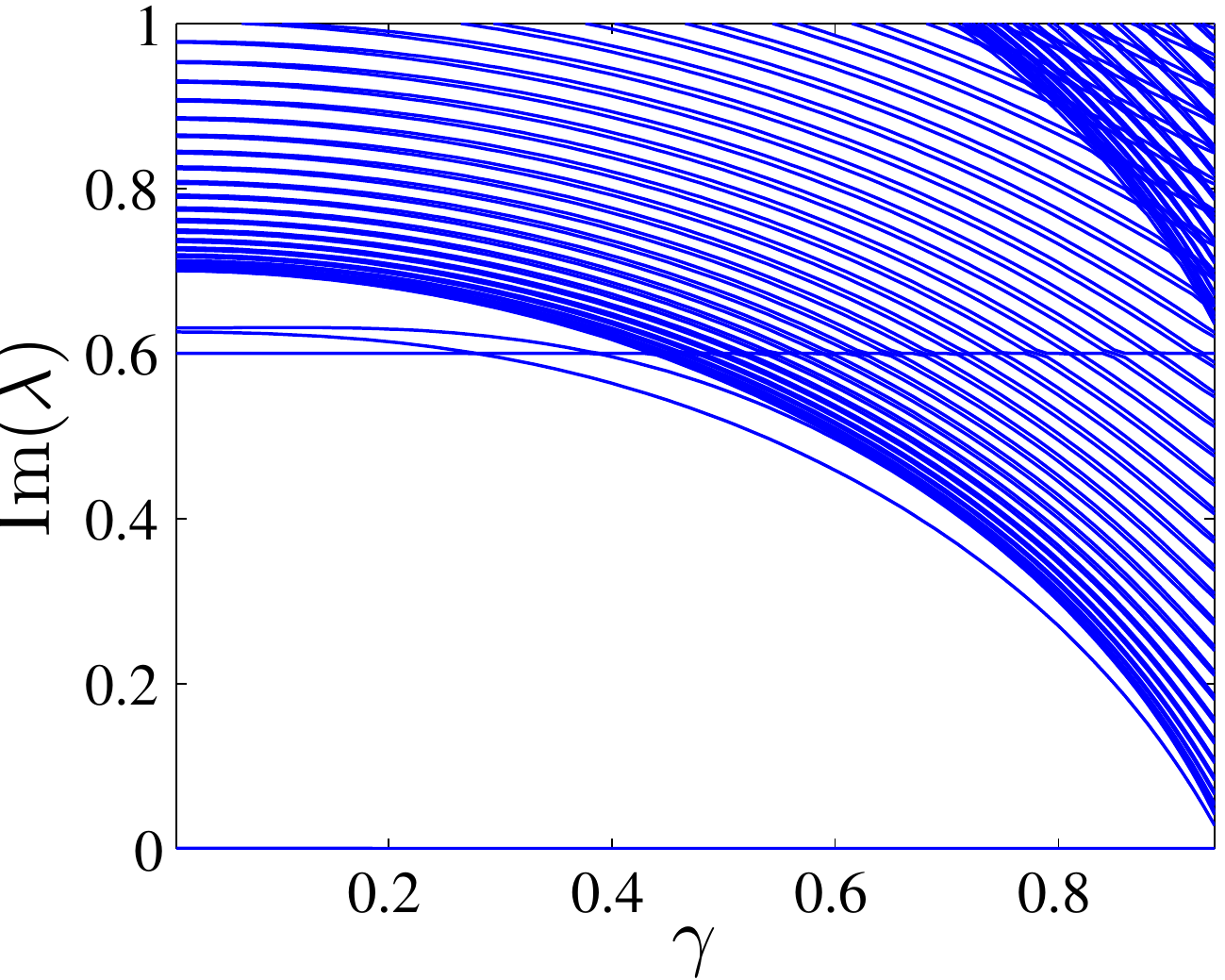} & \hfill
\includegraphics[width=.45\textwidth]{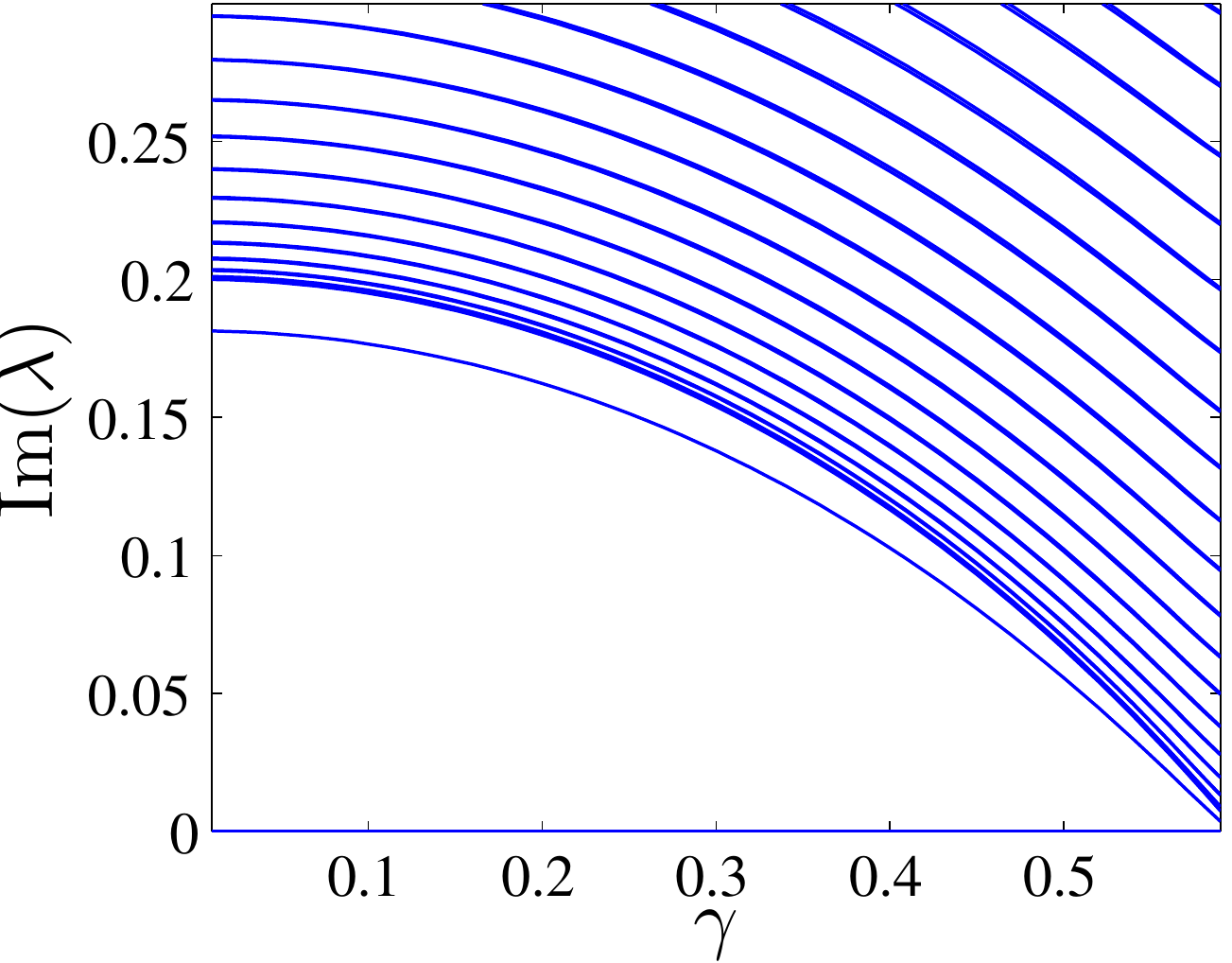} \\
\end{tabular}
\end{center}
\caption{Imaginary part of the eigenvalues dependence for {$\mathcal{P}\mathcal{T}$}-symmetric solitary waves with $\omega=0.3$ (left) and $\omega=0.8$ (right). Notice the existence on the left panel of a constant eigenvalue at $\lambda=2\omega{i}$. Notice also the approach of the eigenvlaues towards $0$, as per the discussed collision with the linear limit.}
\label{cuevas-fig36}
\end{figure}

The first numerical result found by studying the standing wave solution is the {$\mathcal{P}\mathcal{T}$} transition point. We have checked that this transition takes place when $\gamma=\gamma_\mathrm{PT}$ with $\gamma_\mathrm{PT}=\sqrt{1-\omega^2}$; notice that this is consonant with our analytical prediction from the previous section in the case of wavenumber $\kappa=0$. Fig.~\ref{cuevas-fig35} shows the profile of typical solitary waves with nonzero $\gamma$. Importantly, we have observed (see Fig.~\ref{cuevas-fig36}) that the relevant perturbation does not introduce instabilities to our system. The relevant spectrum features a zero eigenvalue of algebraic multiplicity four and geometric multiplicity two. This is present in the spectrum of the linearized equation due to the $\mathbf{U}(1)$ symmetry and due to the translational symmetry, which are both preserved when $\gamma\ne 0$, hence both the algebraic and geometric multiplicity of this eigenvalue are preserved for all values of $\gamma$, as is the presence of two generalized eigenvectors. The spectrum also features the ubiquitous eigenvalues $\lambda= \pm 2\omega{i}$ which is due to the $\mathbf{SU}(1,1)$-invariance, and is also preserved for any $\gamma$; the relevant eigenvalue, which persists under variations of $\gamma$, can be discerned in the left panel of Fig.~\ref{cuevas-fig36}. For the rest of the spectrum we note that, as discussed in Section \ref{sec:cuevas-stability_theory}, eigenvalues with nonzero real part can only be born in the embedded spectrum. Here, however, all the eigenvalues remain inside this interval for all $\gamma$, as illustrated in Fig.~\ref{cuevas-fig36}, solely tending towards $0$, as $\gamma$ approaches $\gamma_\mathrm{PT}$.

\begin{figure}
\begin{center}
\begin{tabular}{cc}
\includegraphics[width=.45\textwidth]{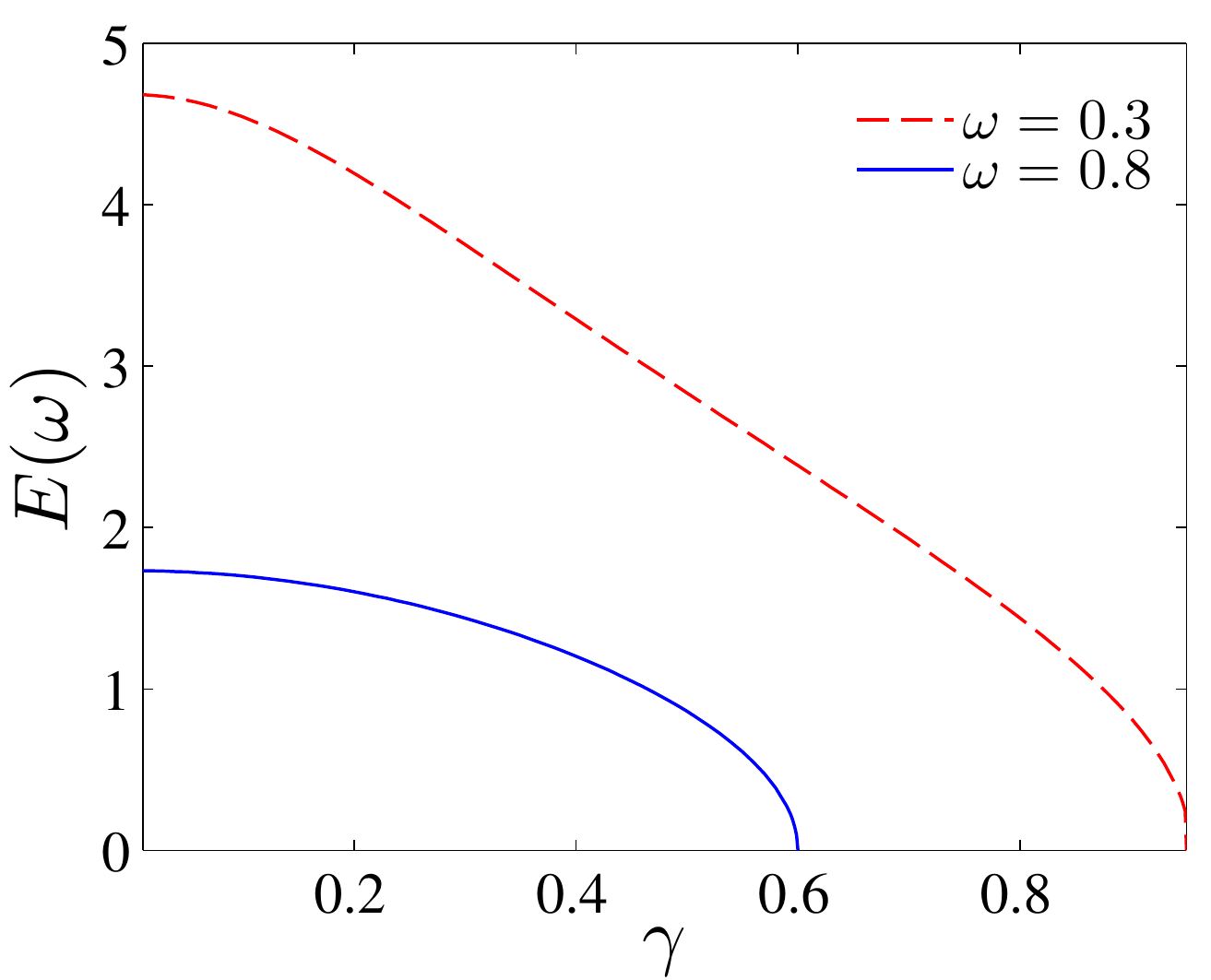} & \hfill
\includegraphics[width=.45\textwidth]{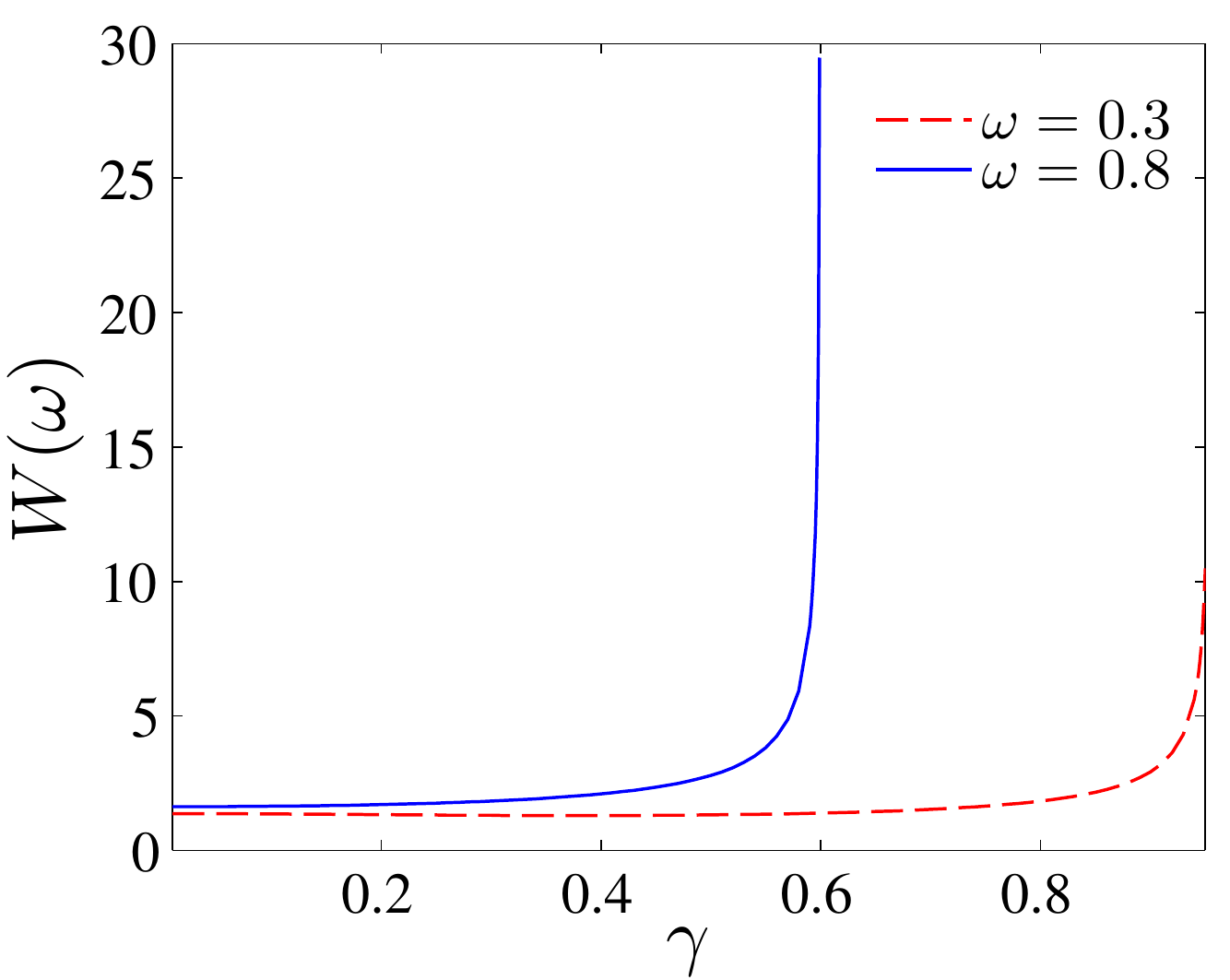} \\
\multicolumn{2}{c}{\includegraphics[width=.45\textwidth]{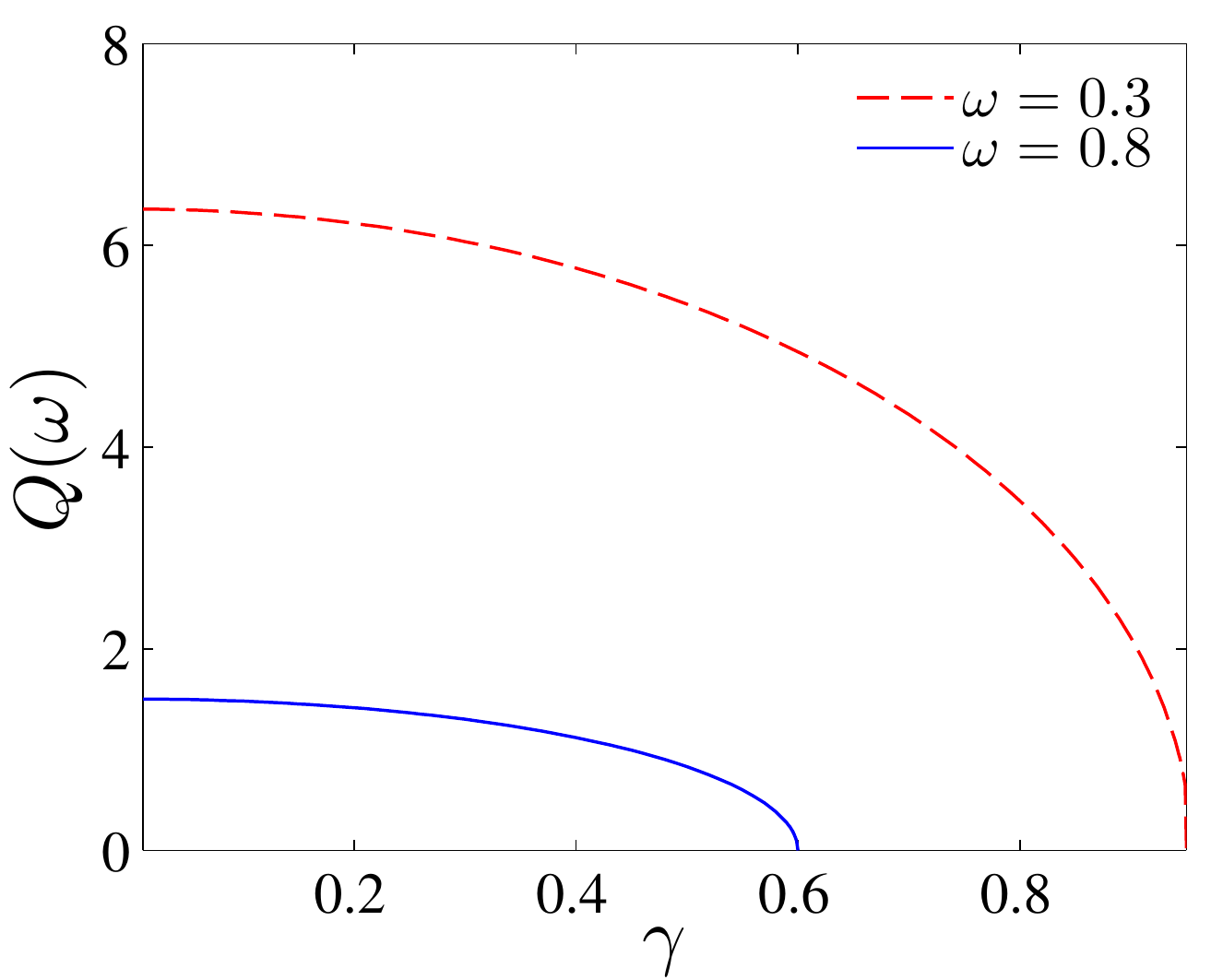}} \\
\end{tabular}
\end{center}
\caption{Energy, width, and charge dependence with respect to $\gamma$ of the
{$\mathcal{P}\mathcal{T}$}-symmetric solitary waves with $\omega=0.8$ and $\omega=0.3$.}
\label{cuevas-fig37}
\end{figure}

\begin{figure}
\begin{center}
\begin{tabular}{cc}
\includegraphics[width=.45\textwidth]{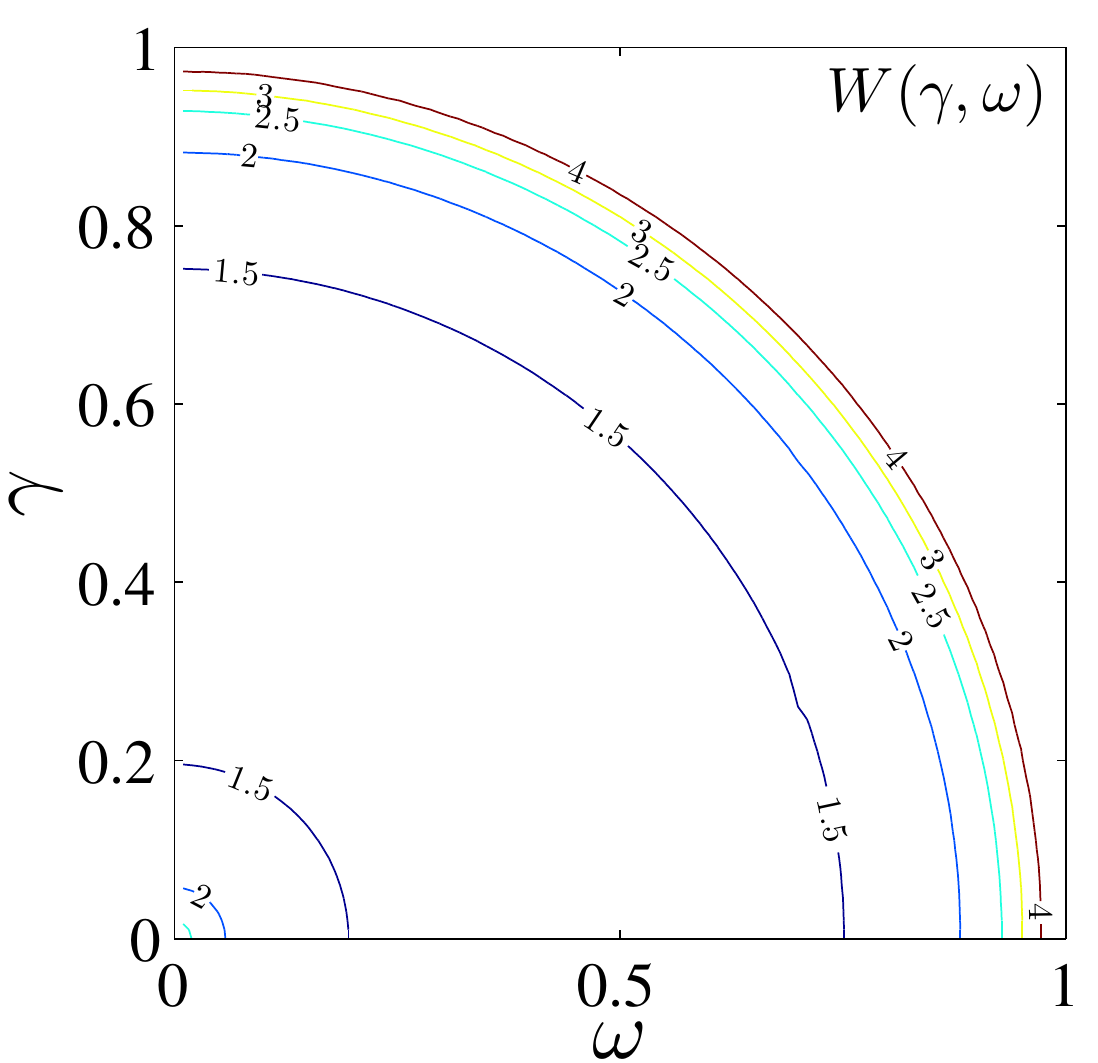} & \hfill
\includegraphics[width=.45\textwidth]{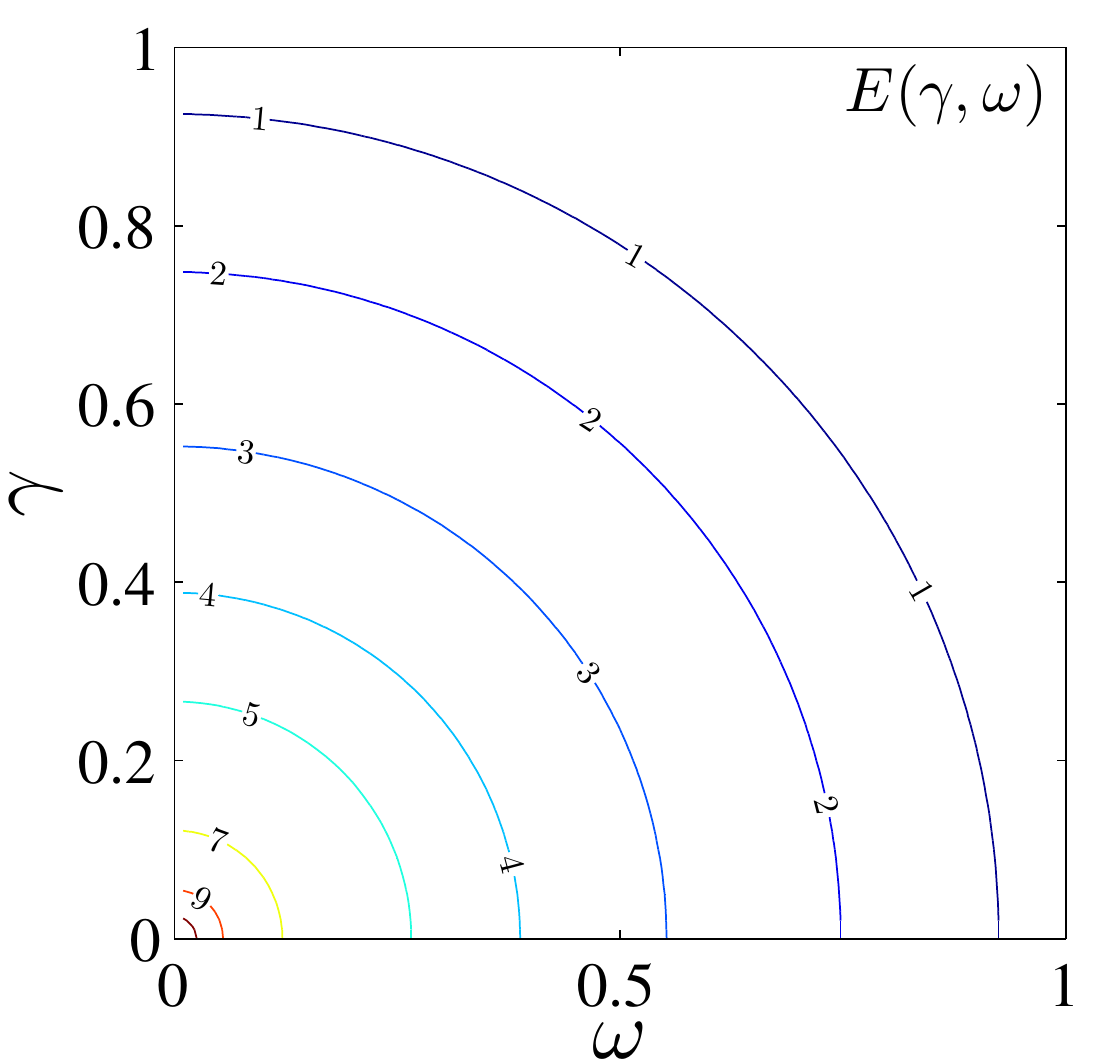} \\
\end{tabular}
\end{center}
\caption{Two-parameter diagram of the dependence of the width and energy of
{$\mathcal{P}\mathcal{T}$}-symmetric solitary waves on the frequency $\omega$ and gain-loss parameter $\gamma$.}
\label{cuevas-fig38}
\end{figure}

Notice that the {$\mathcal{P}\mathcal{T}$}-transition is caused by the nonlinear solutions colliding with (or degenerating into) linear modes. This fact can also be confirmed in the plots of Fig.~\ref{cuevas-fig37}, where the charge and energy tend to zero (while a width diagnostic~\footnote{Defined as $W=\frac{\int x^2\rho(x)\,dx}{\int \rho(x)\,dx}$.} of the solution diverges) when the transition point is reached (actually, we have not been able to reach this point exactly, as the solitary wave width increases drastically when approaching this point). It should be noted here that this is a distinct phenomenology in comparison to the NLS counterpart of the model. In the latter, typically at the {$\mathcal{P}\mathcal{T}$}-phase transition a stable (center) and an unstable (saddle) solution collide and disappear in a saddle-center bifurcation. Here, a fundamentally different scenario arises through the degeneration of the nonlinear modes into linear ones. In Fig.~\ref{cuevas-fig38}, we provide two-parameter diagrams of the relevant solutions as a function of the frequency $\omega$ and the gain-loss parameter $\gamma$. The dependencies strongly suggest a ``combined'' monoparametric dependence on $\gamma^2 + \omega^2$, although we have not been able to analytically identify solutions bearing this dependence.

\begin{figure}
\begin{center}
\begin{tabular}{cc}
\includegraphics[width=.45\textwidth]{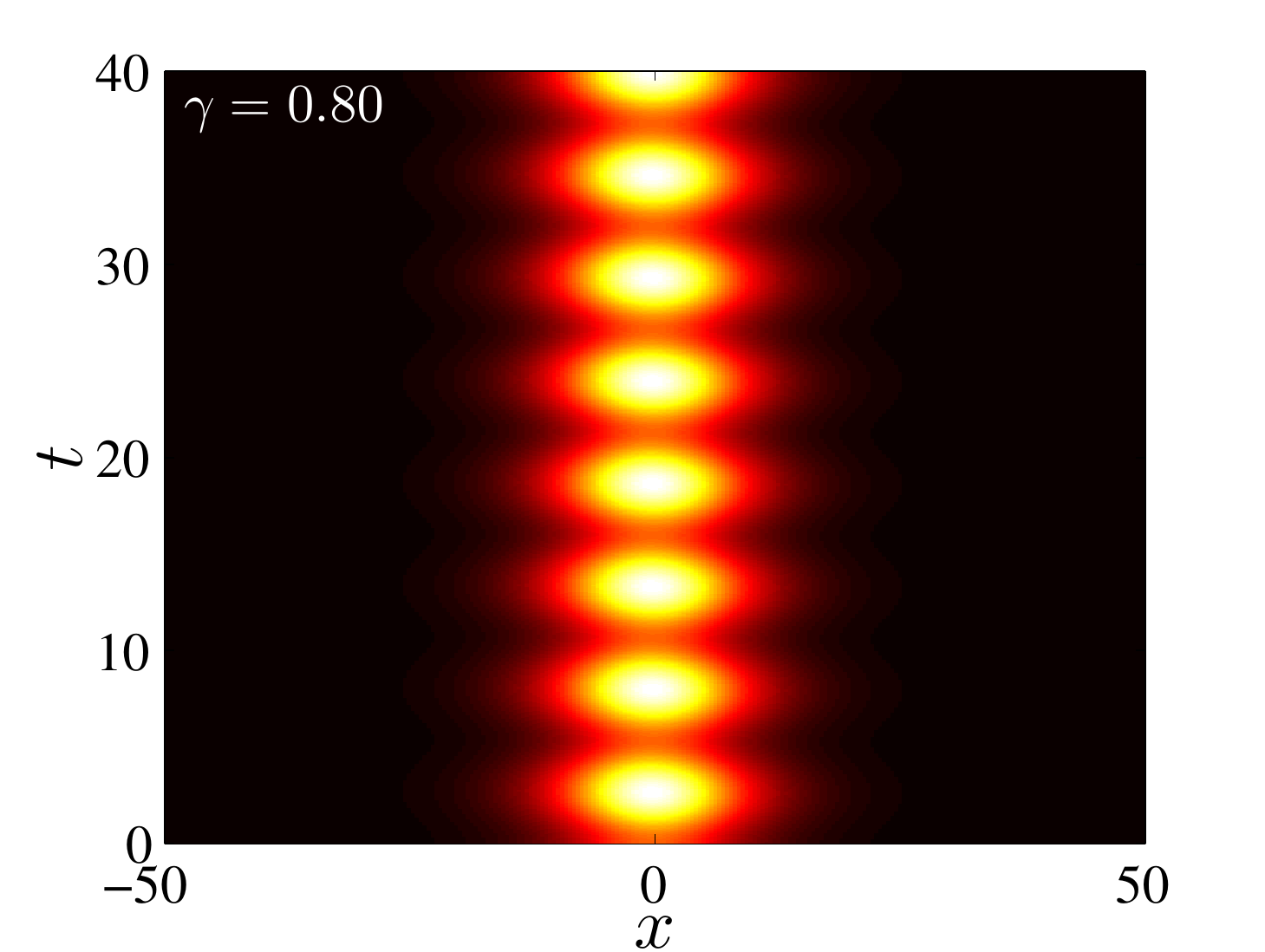} & \hfill
\includegraphics[width=.45\textwidth]{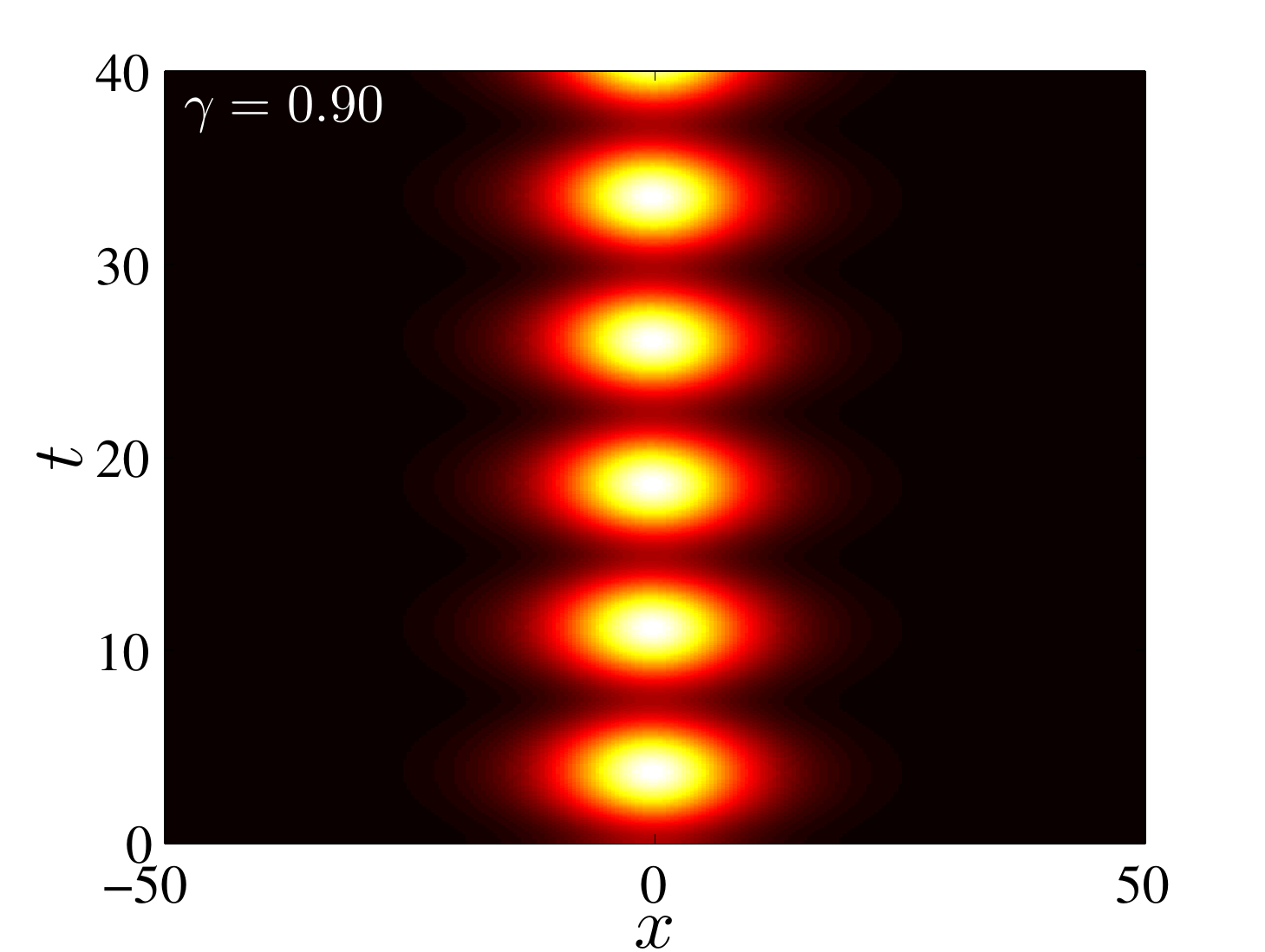} \\
\includegraphics[width=.45\textwidth]{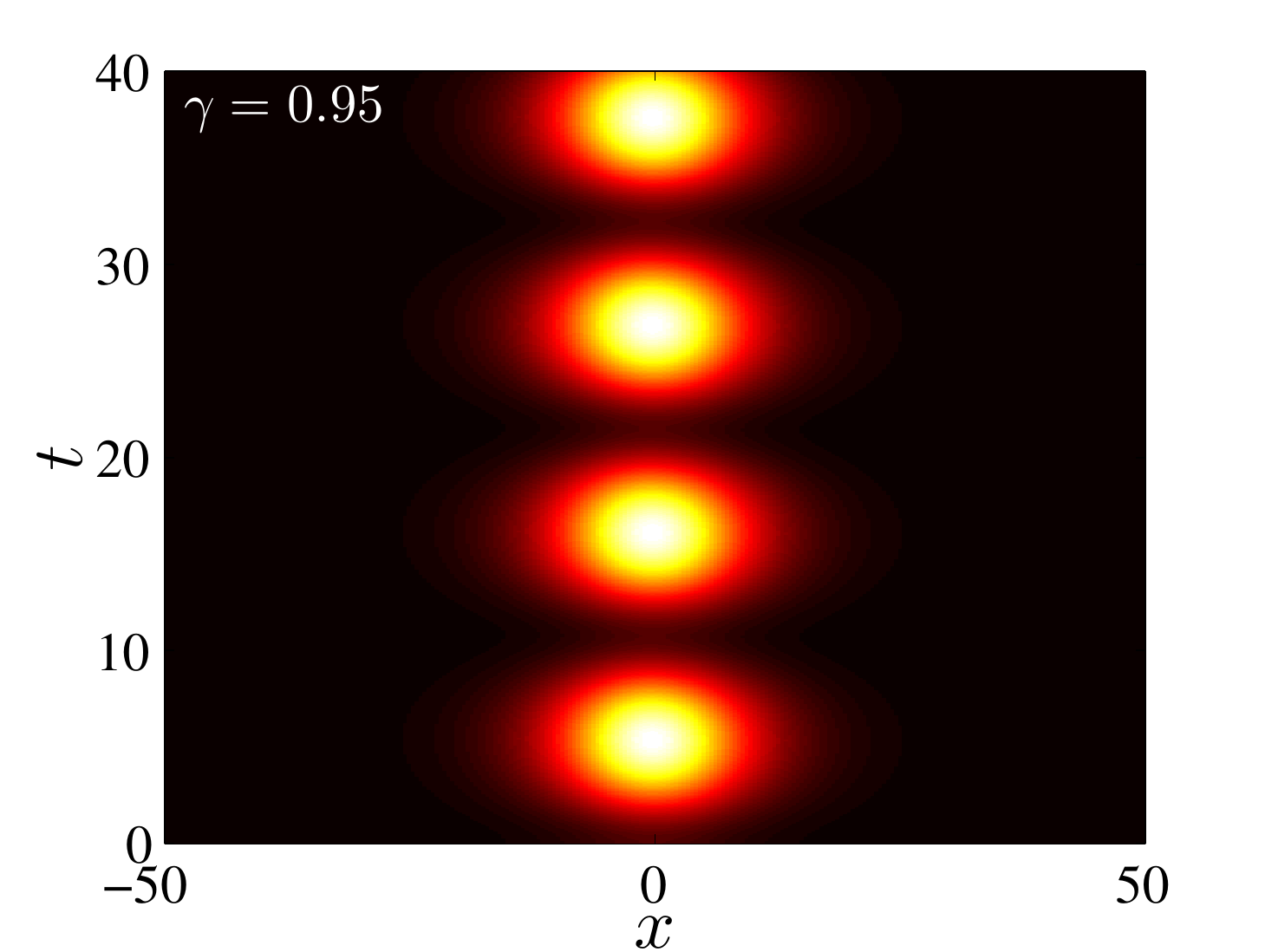} & \hfill
\includegraphics[width=.45\textwidth]{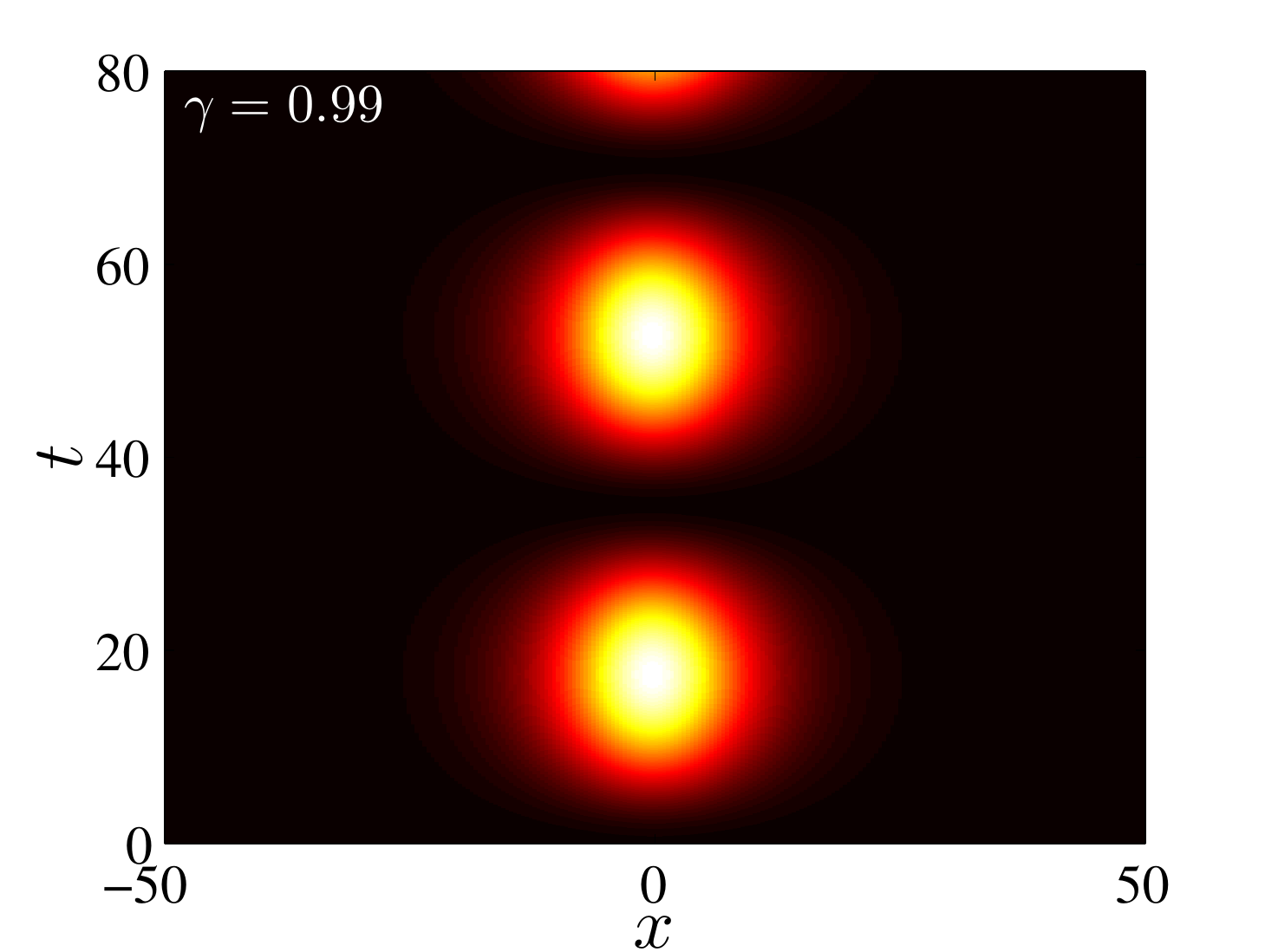} \\
\end{tabular}
\end{center}
\caption{Dynamics for the {$\mathcal{P}\mathcal{T}$}-symmetric NLD using as initial condition the solitary wave with $\omega=0.8$ and $\gamma=0.59$, but evolving Eq.~(\ref{eq:cuevas-SolerPT}) for the value of $\gamma$ shown in the respective panels. Each panel shows the space-time evolution of the solitary wave density.}
\label{cuevas-fig39}
\end{figure}

\begin{figure}
\begin{center}
\begin{tabular}{cc}
\includegraphics[width=.45\textwidth]{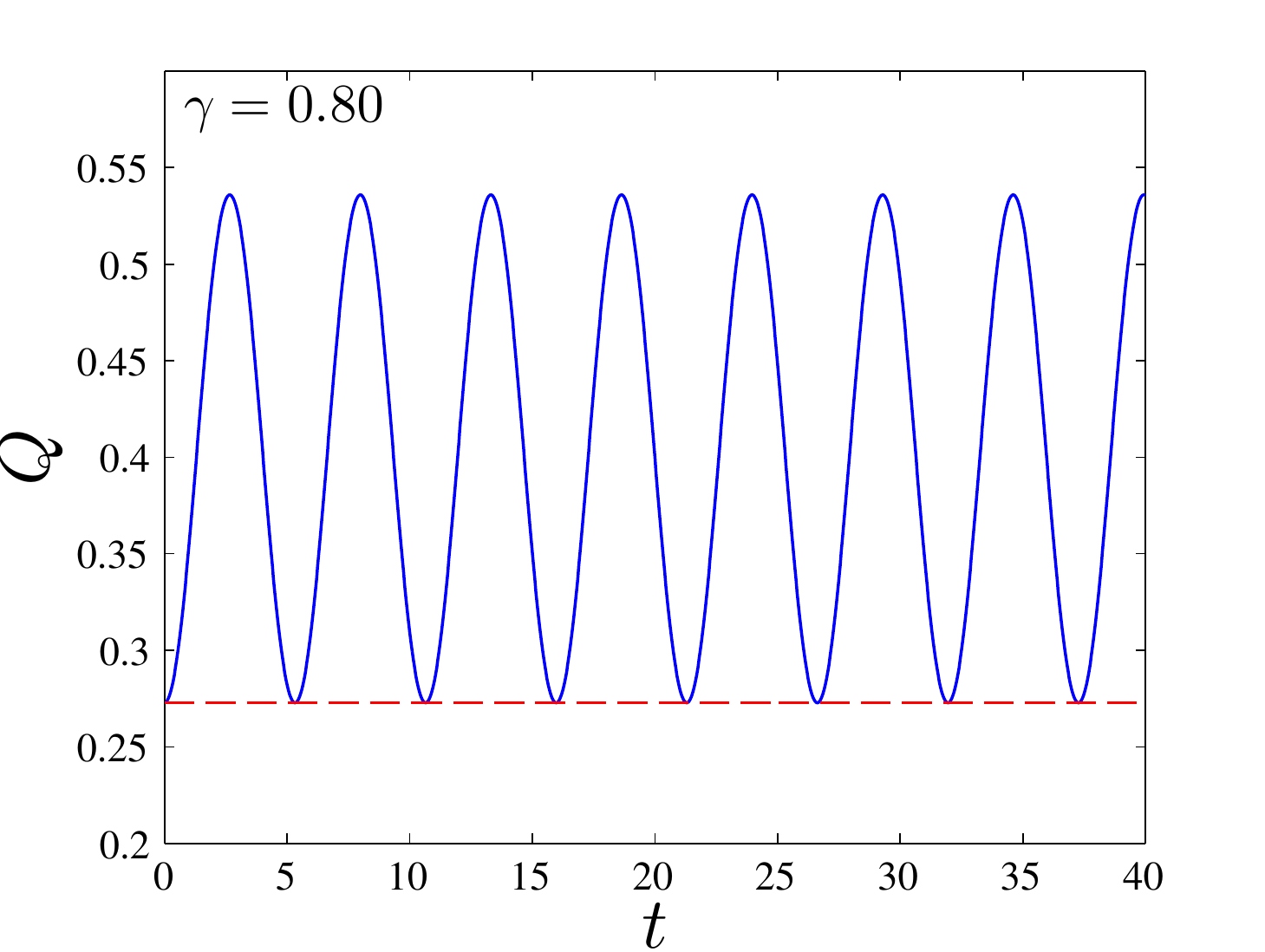} & \hfill
\includegraphics[width=.45\textwidth]{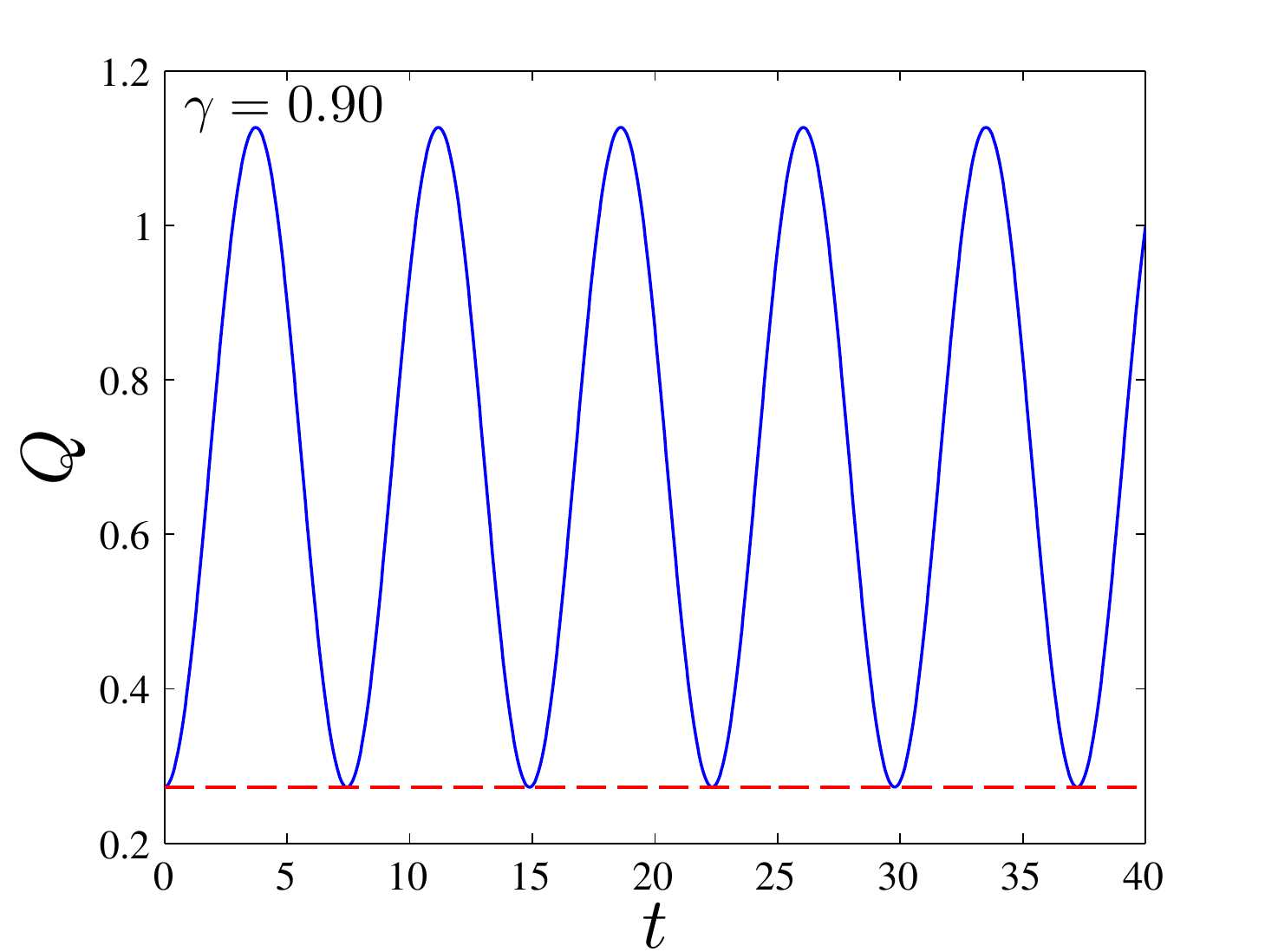} \\
\includegraphics[width=.45\textwidth]{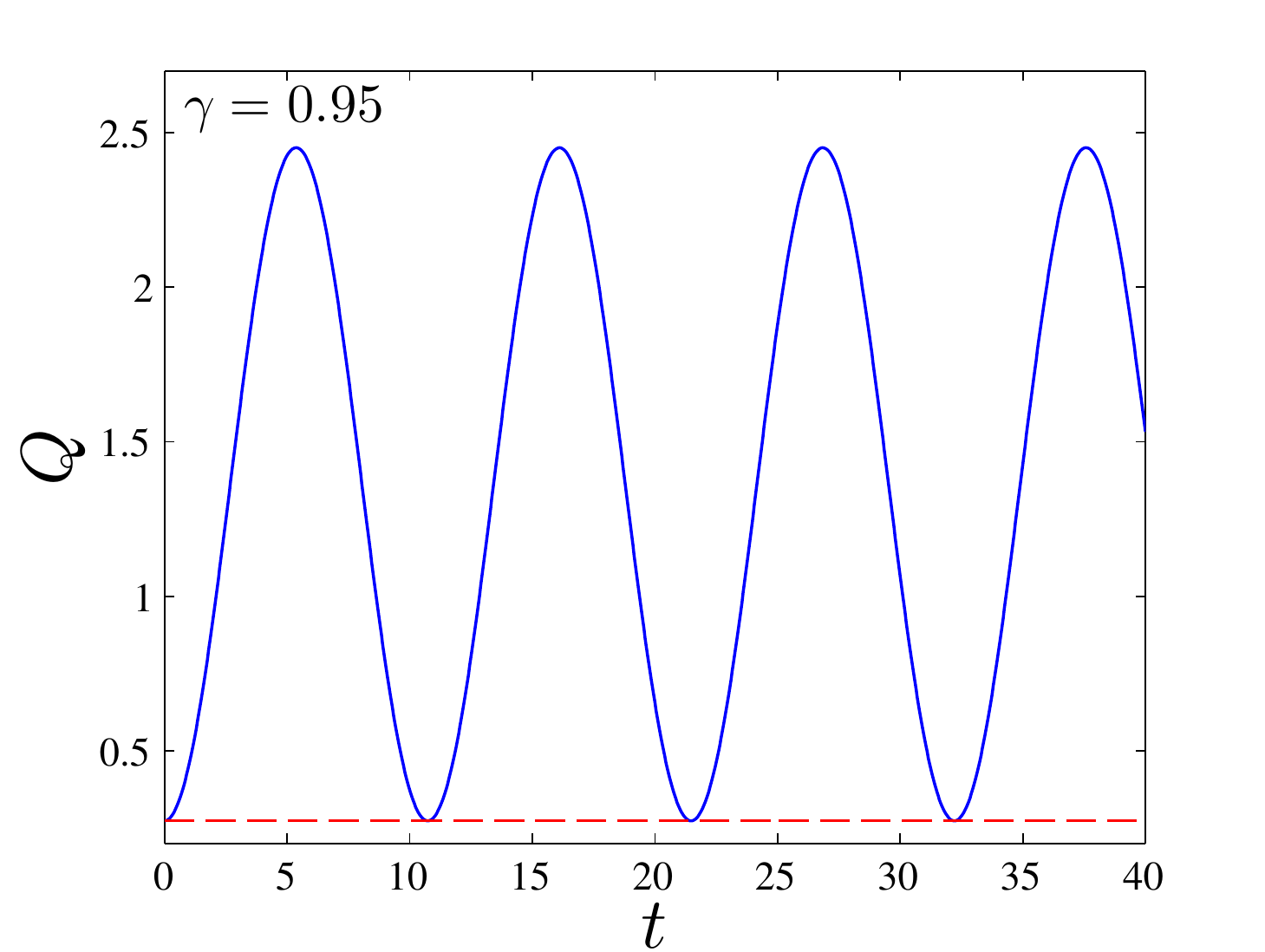} & \hfill
\includegraphics[width=.45\textwidth]{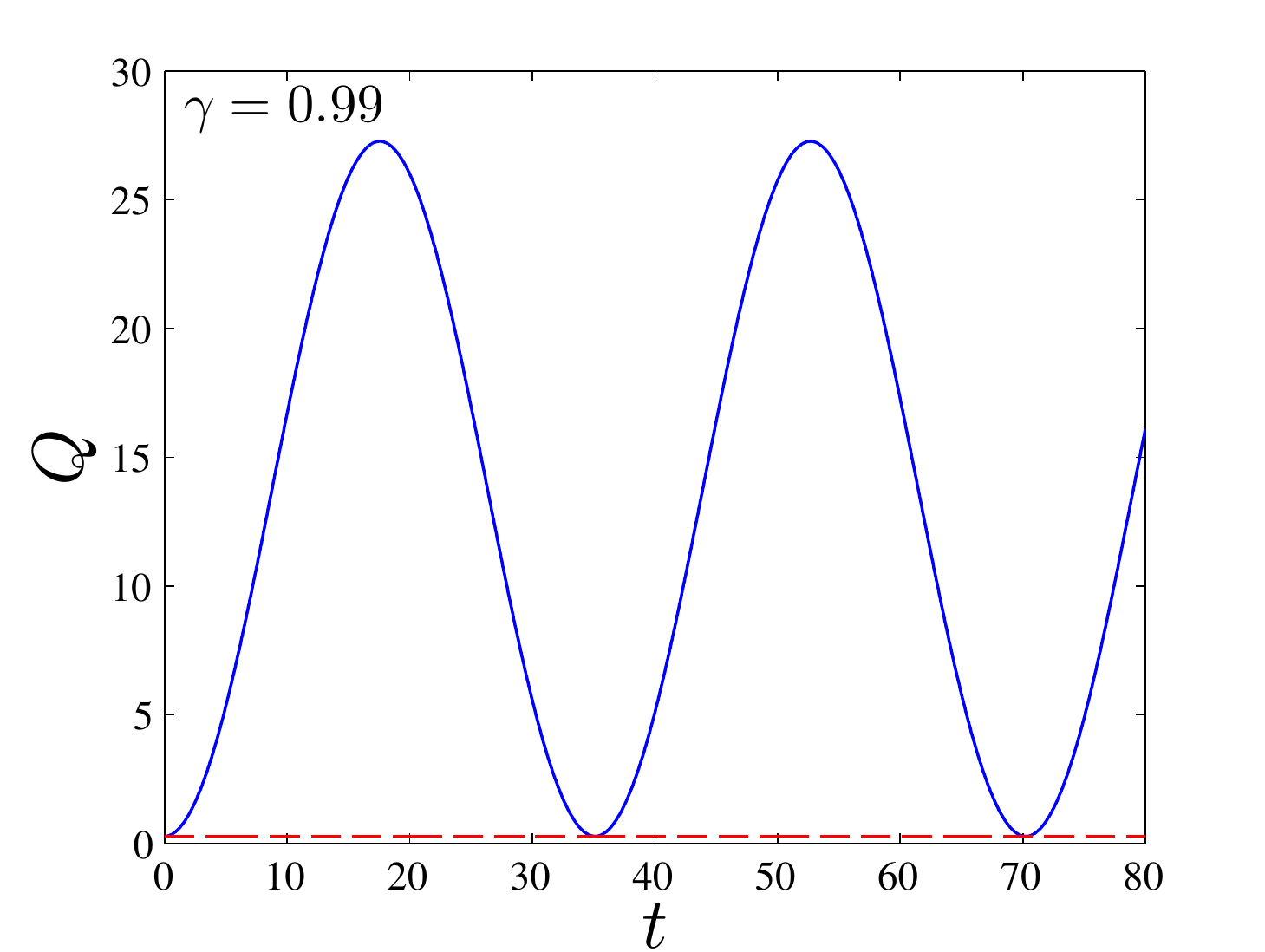} \\
\end{tabular}
\end{center}
\caption{Same as Fig.~\ref{cuevas-fig39} but showing the time evolution of the solitary wave charge. The red dashed line corresponds to the charge of the initial condition.}
\label{cuevas-fig40}
\end{figure}

\begin{figure}
\begin{center}
\begin{tabular}{cc}
\includegraphics[width=.45\textwidth]{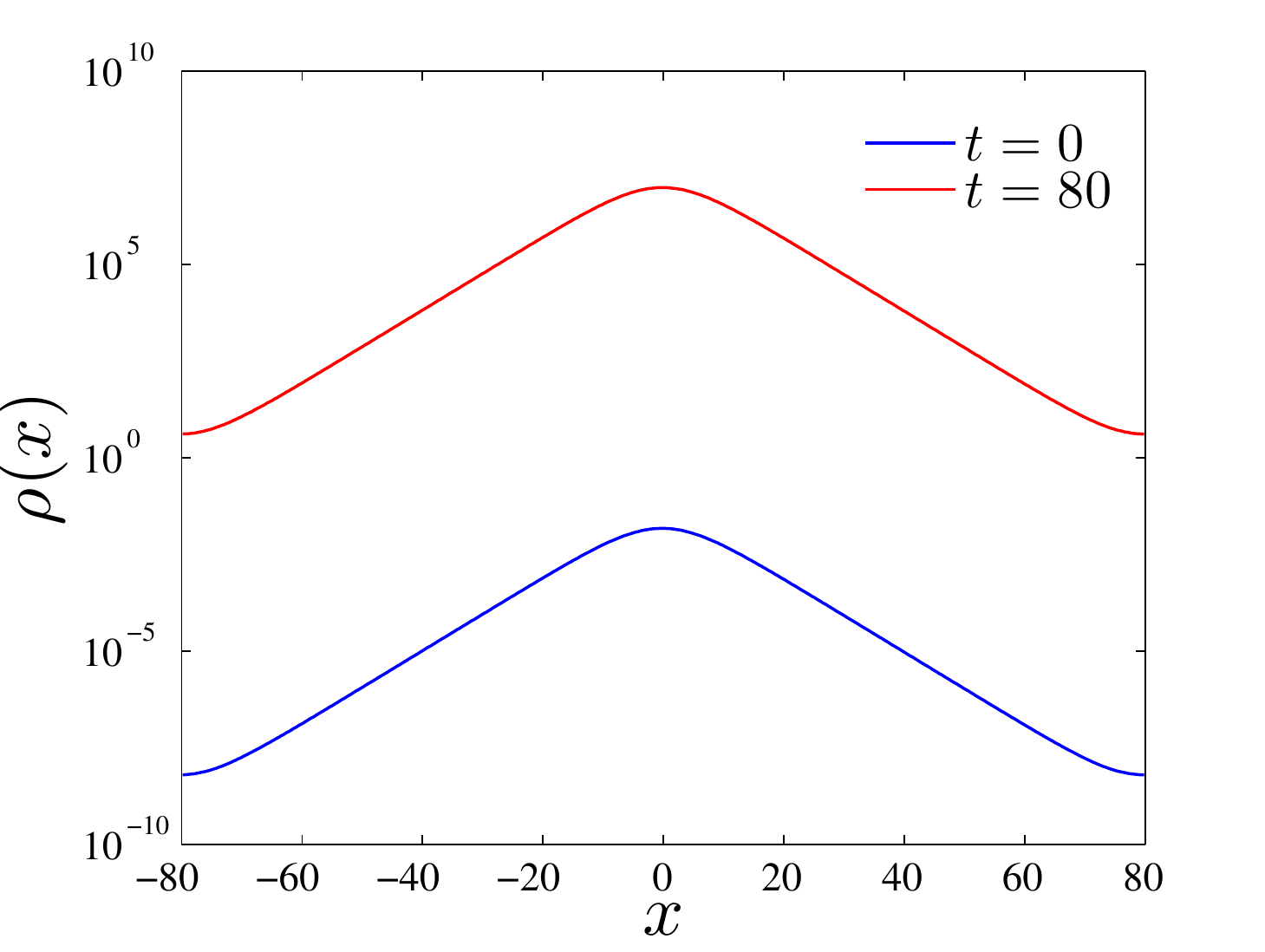} & \hfill
\includegraphics[width=.45\textwidth]{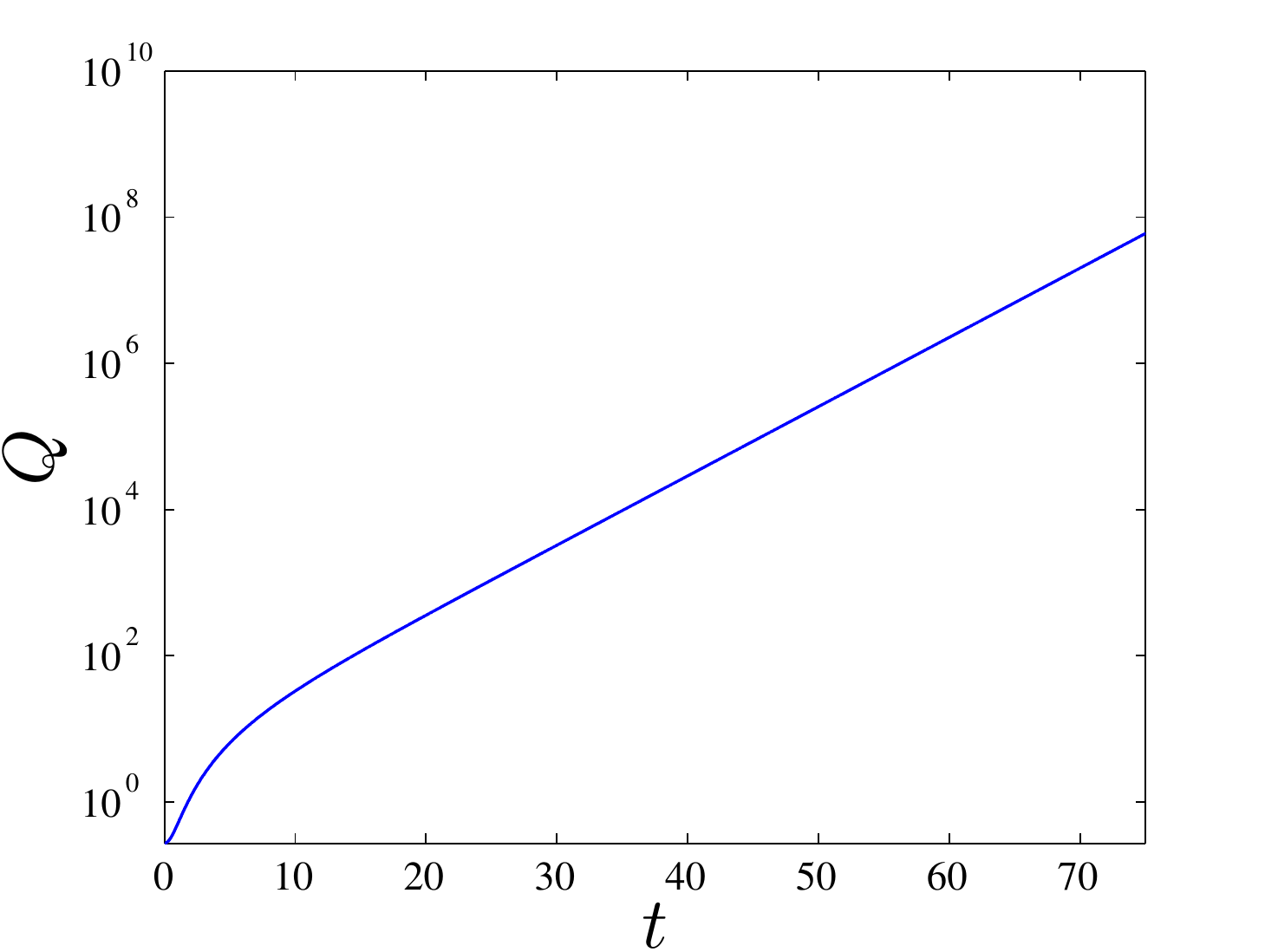} \\
\end{tabular}
\end{center}
\caption{Dynamics for the {$\mathcal{P}\mathcal{T}$}-symmetric NLD using as initial condition the solitary wave with $\omega=0.8$ and $\gamma=0.59$, but evolving Eq.~(\ref{eq:cuevas-SolerPT}) for the value $\gamma=1$. The left panel shows the charge density at different times. The right panel shows the total charge as a function of time. One can clearly observe the (spatially independent) exponential growth of the waveform.}
\label{cuevas-fig41}
\end{figure}

We now turn to the consideration of the dynamical evolution of solitary waves past the {$\mathcal{P}\mathcal{T}$}-transition point. As indicated above, given their generic stability for $\gamma < \gamma_\mathrm{PT}$, we do not consider the latter case. In the case of $\gamma > \gamma_\mathrm{PT}$, we have firstly taken as initial condition the solitary wave for $\omega=0.8$ and $\gamma=\gamma_0=0.59$ in the simulation with $\gamma=\gamma_s>\gamma_\mathrm{PT}=\sqrt{1-\omega^2}=0.6$. We observe that if $\gamma_s$ is close enough to $\gamma_\mathrm{PT}$ (i.e., for a ``shallow'' quench), the density oscillates with a frequency that decreases with $\gamma_s-\gamma_\mathrm{PT}$ (see Figs.~\ref{cuevas-fig39} and \ref{cuevas-fig40}). Notice that the charge of the new solitary wave is always higher than the charge of the initial one and that the maximum charge increases with $\gamma_s$. Interestingly, in all of these case examples we find that the ($\gamma$-independent) energy ($E=0.2738$ for $\omega=0.8$) is very well conserved as shown in \cite{cuevas-CKS+16}. When the maximum charge is above a threshold (this occurs for $\gamma_s\gtrsim0.995$, i.e., for a deep quench), the frequency of the new solitary wave tends to zero and the solution starts to grow indefinitely as shown in Fig.~\ref{cuevas-fig41}. If a smaller value of $\gamma_0$ is taken, the same phenomenology persists, but the indefinite growth emerges for a smaller value of $\gamma_s$. It was also confirmed in \cite{cuevas-CKS+16} that both the energy conservation law and the moment equation (\ref{eq:cuevas-chargederiv}) for the charge are satisfied in the dynamics of {$\mathcal{P}\mathcal{T}$}-symmetric solitary waves. The same is true for the case of Fig.~\ref{cuevas-fig41} where the charge grows exponentially (in the case shown in the figure, for which $\gamma_s=1$, approximately as $\exp(0.088t)$; although the characteristic growth rate depends on $\gamma_s$). Here, the solitary wave does not collapse, as its shape and width are preserved during the growth. Again, this type of growth appears to be very different than, say, the collapse in the Hamiltonian NLS model~\cite{cuevas-SS99}. In the latter, the width decreases and the amplitude increases, whereas here the entire solution grows without changing its spatial distribution.

Let us mention that the waveform with oscillating charge is fairly generic when the quench is not sufficiently deep to cause an exponential growth.
Remarkably, such solitary waves with oscillating charge can be obtained by performing an $\mathbf{SU}(1,1)$-transformation to a standard (one-frequency) solitary wave \cite{cuevas-Gal77,cuevas-DR79}.
This type of solution for a standing wave of frequency $\tilde{\omega}$ is, in fact, intrinsically connected to the invariance of the eigenvalue $2 \tilde{\omega}i$ (associated to eigenfrequency $2\tilde{\omega}$) in the spectrum. More specifically, these bi-frequency, oscillating charge coherent structures [which can be dubbed as $\mathbf{SU}(1,1)$-solitary waves] are of the form:
\begin{equation}\label{eq:cuevas-su11}
 \begin{split}
 \psi_1(t,x) =& \alpha_{-}\tilde{v}(x)e^{-i \tilde{\omega} t}-i\alpha_{+}\tilde{u}^*(x)e^{i \tilde{\omega} t}\,, \\
 \psi_2(t,x) =& \alpha_{-}\tilde{u}(x)e^{-i \tilde{\omega} t}-i\alpha_{+}\tilde{v}^*(x)e^{i \tilde{\omega} t}\,,
 \end{split}
\end{equation}
with
\begin{equation*}
 \alpha_{\pm}\in\mathbb{C},
 \qquad
 |\alpha_{-}|^2-|\alpha_{+}|^2=1 .
\end{equation*}

In this case, $\{\tilde{v}(x),\tilde{u}(x)\}$ is the standing wave solution with frequency $\tilde{\omega}$. Consequently, the charge oscillates with a frequency $2 \tilde{\omega}$ as long as $\gamma\neq0$. There is an $\mathbf{SU}(1,1)$-family of solutions for each value of $\gamma$ and $\tilde{\omega}$ which fulfills the same equations that the standing wave solutions satisfy. As a result, when $\gamma_s\neq\gamma_0$, an $\mathbf{SU}(1,1)$-solution with $\gamma=\gamma_s$ is apparently dynamically manifested. Since these periodic $\mathbf{SU}(1,1)$-solutions and the standing wave solutions only exist for $\gamma < m \equiv 1$, it is natural to expect that there are no nontrivial fixed points for the dynamics for $\gamma > 1$, hence giving rise to the observed growth dynamics.

\begin{figure}
\begin{center}
\begin{tabular}{cc}
\includegraphics[width=.45\textwidth]{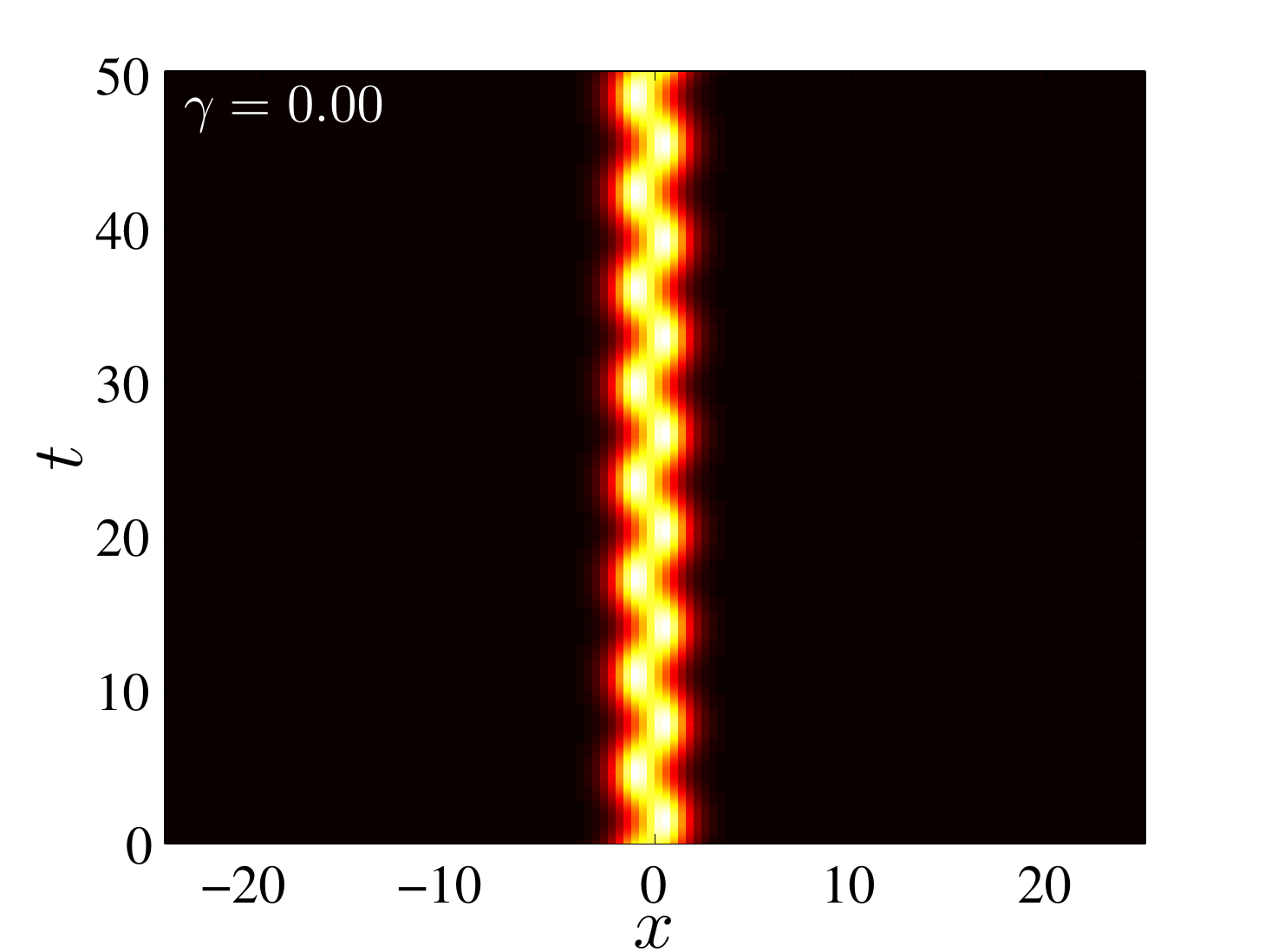} & \hfill
\includegraphics[width=.45\textwidth]{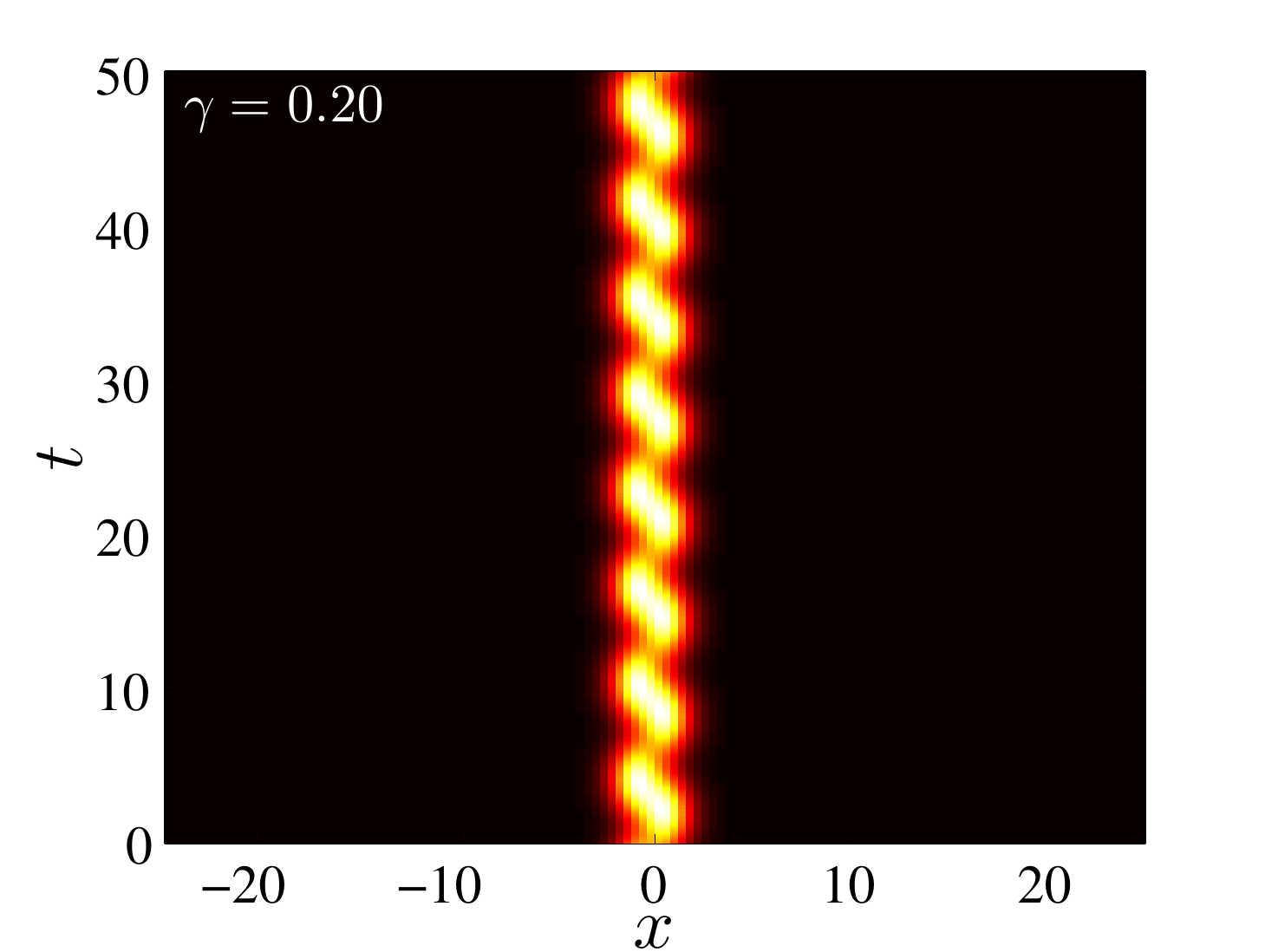} \\
\includegraphics[width=.45\textwidth]{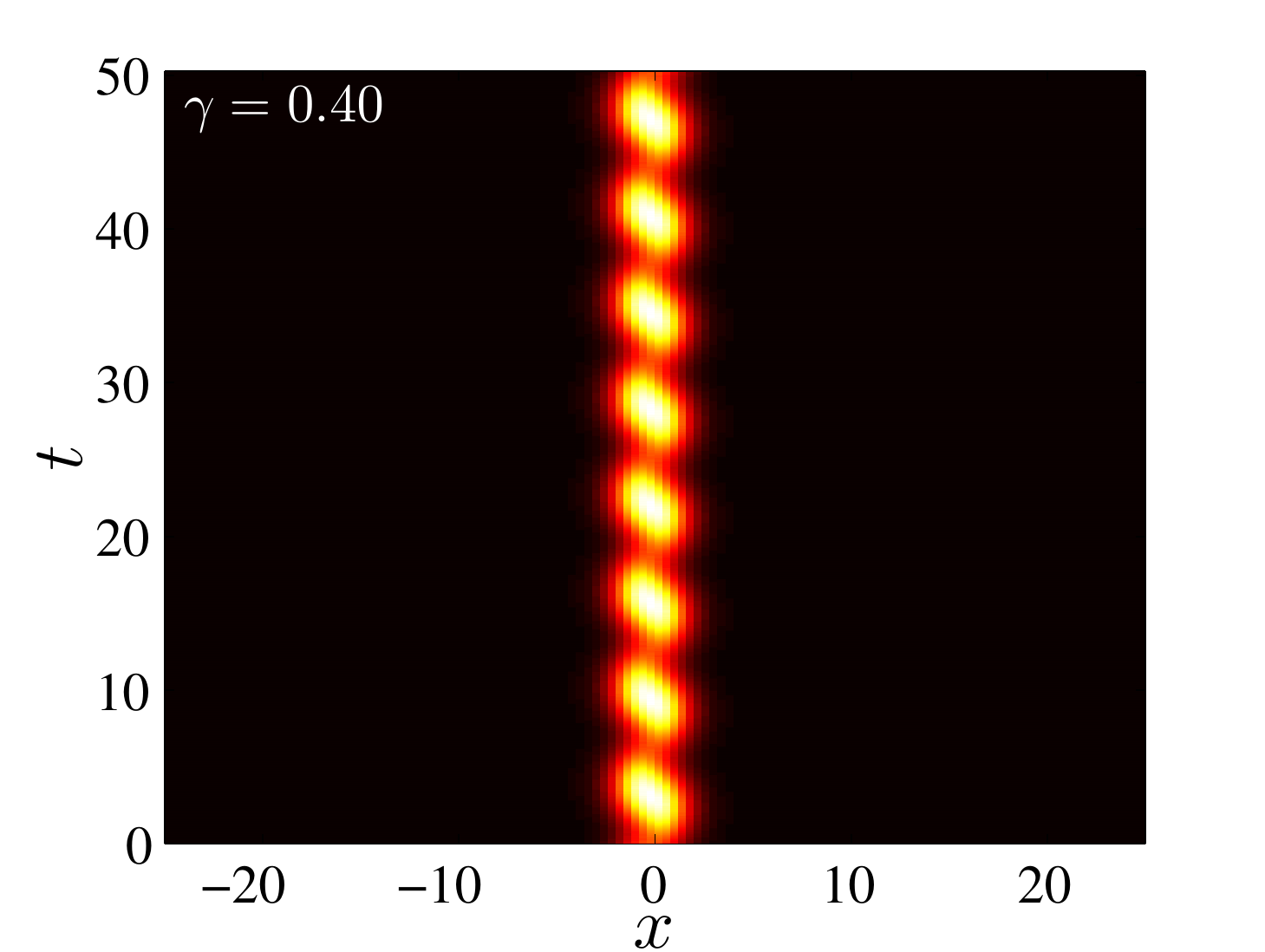} & \hfill
\includegraphics[width=.45\textwidth]{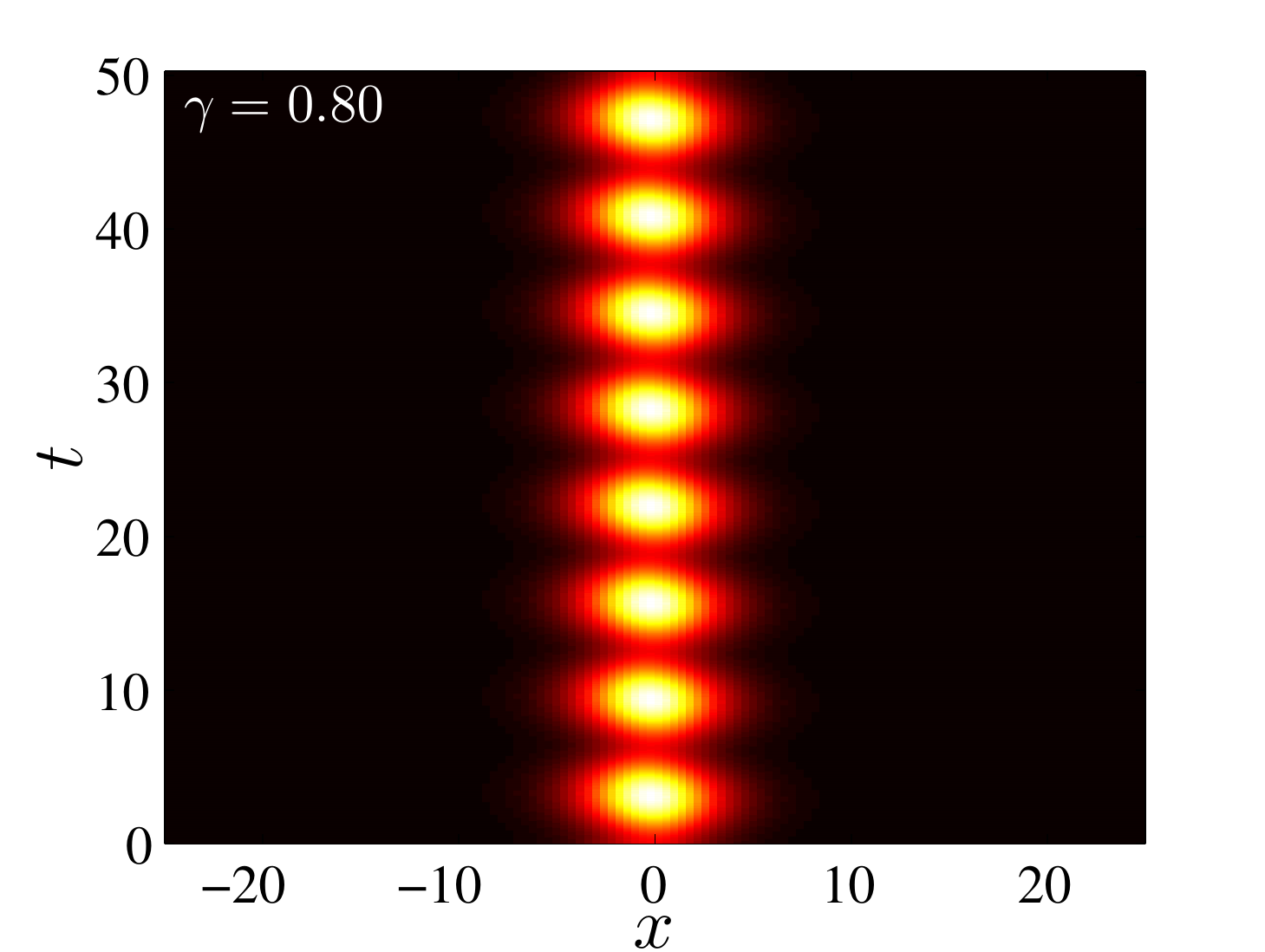} \\
\end{tabular}
\end{center}
\caption{Dynamics for $\mathbf{SU}(1,1)$-solitary waves with $\tilde{\omega}=0.5$, $\alpha_{-}=1.0500$, $\alpha_{+}=0.3202$, and the value of $\gamma$ indicated in each panel [see Eq.~(\ref{eq:cuevas-su11})]. Each panel shows the space-time evolution of the solution density.}
\label{cuevas-fig42}
\end{figure}

\begin{figure}
\begin{center}
\begin{tabular}{cc}
\includegraphics[width=.45\textwidth]{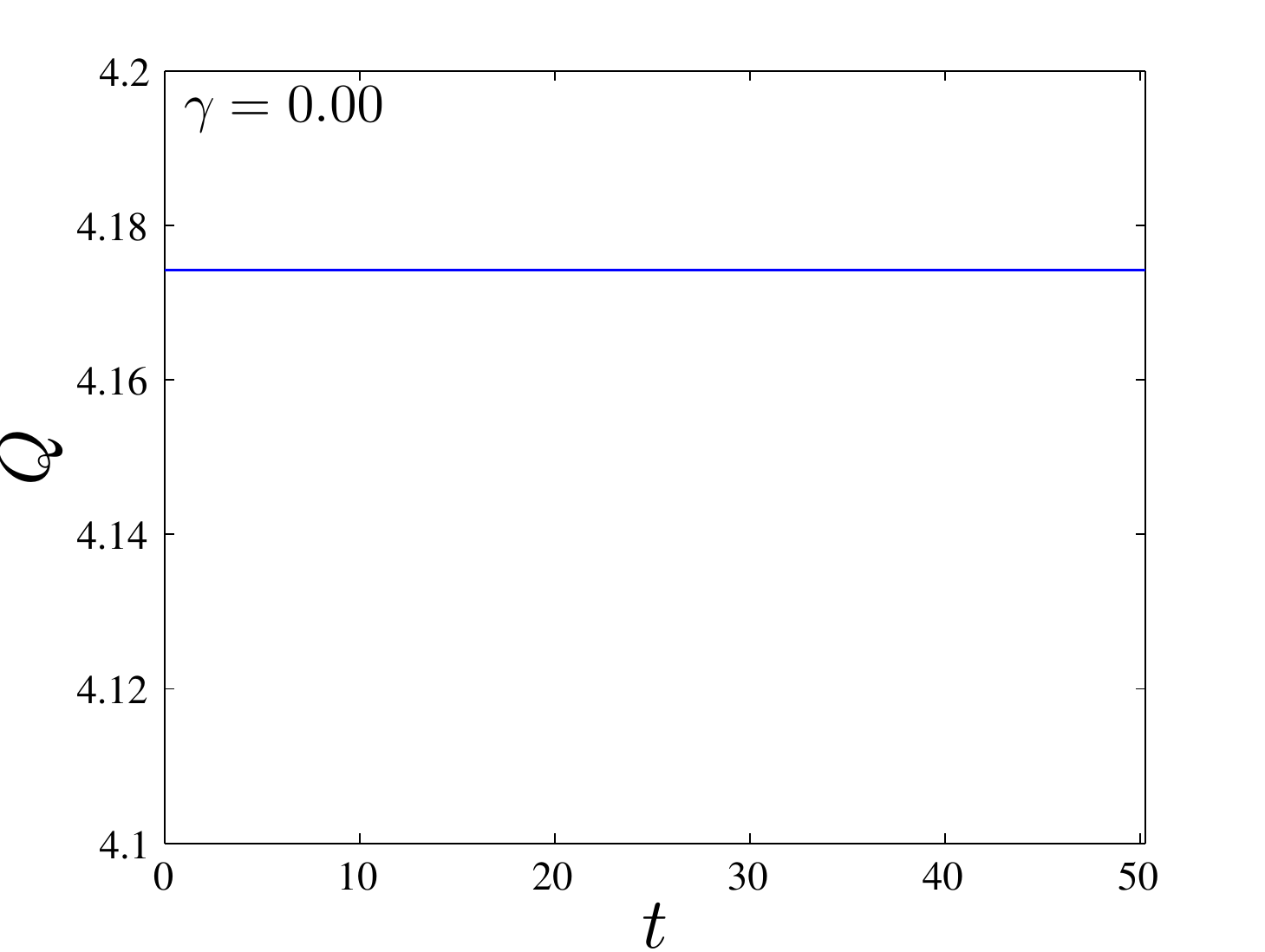} & \hfill
\includegraphics[width=.45\textwidth]{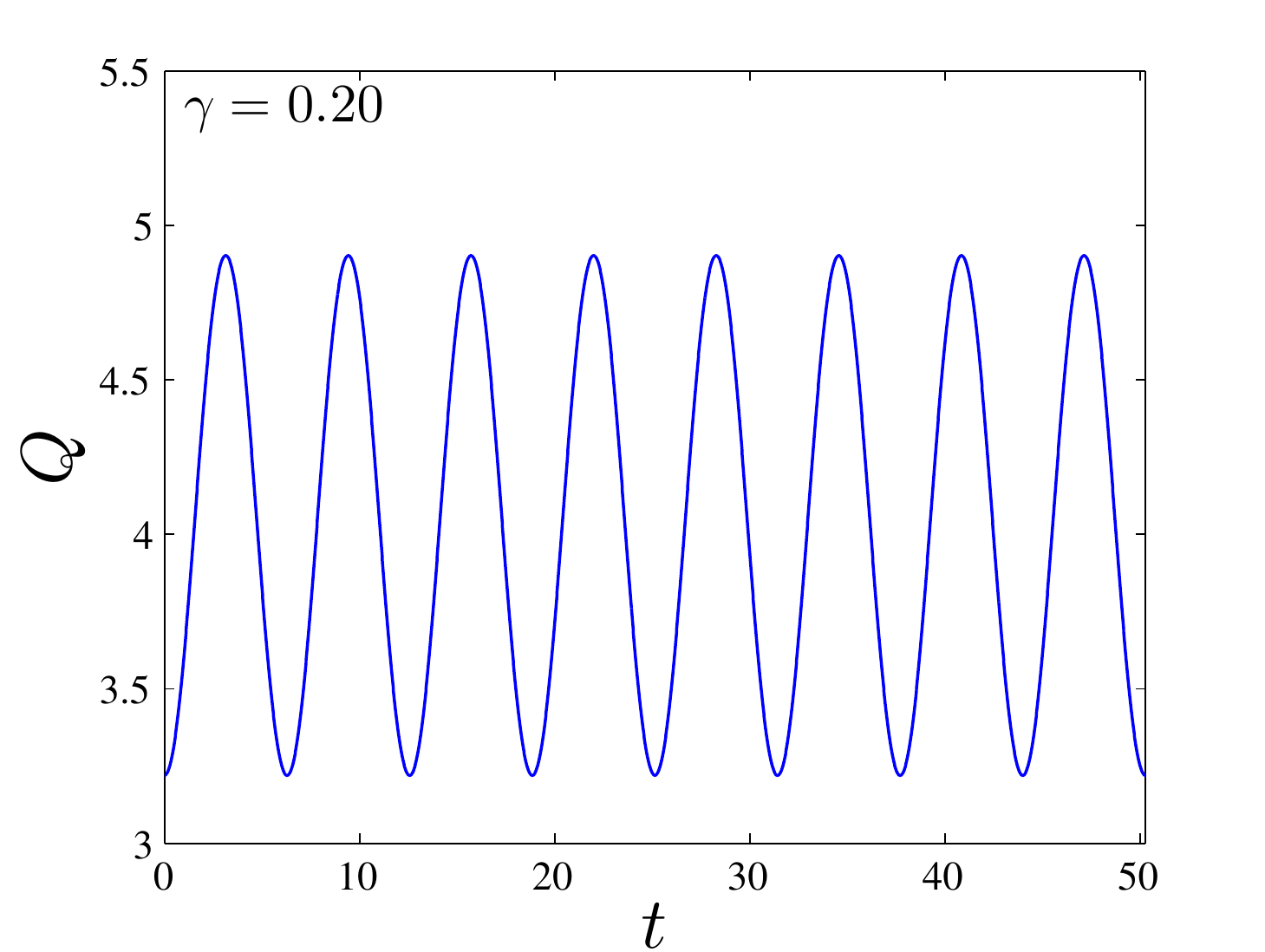} \\
\includegraphics[width=.45\textwidth]{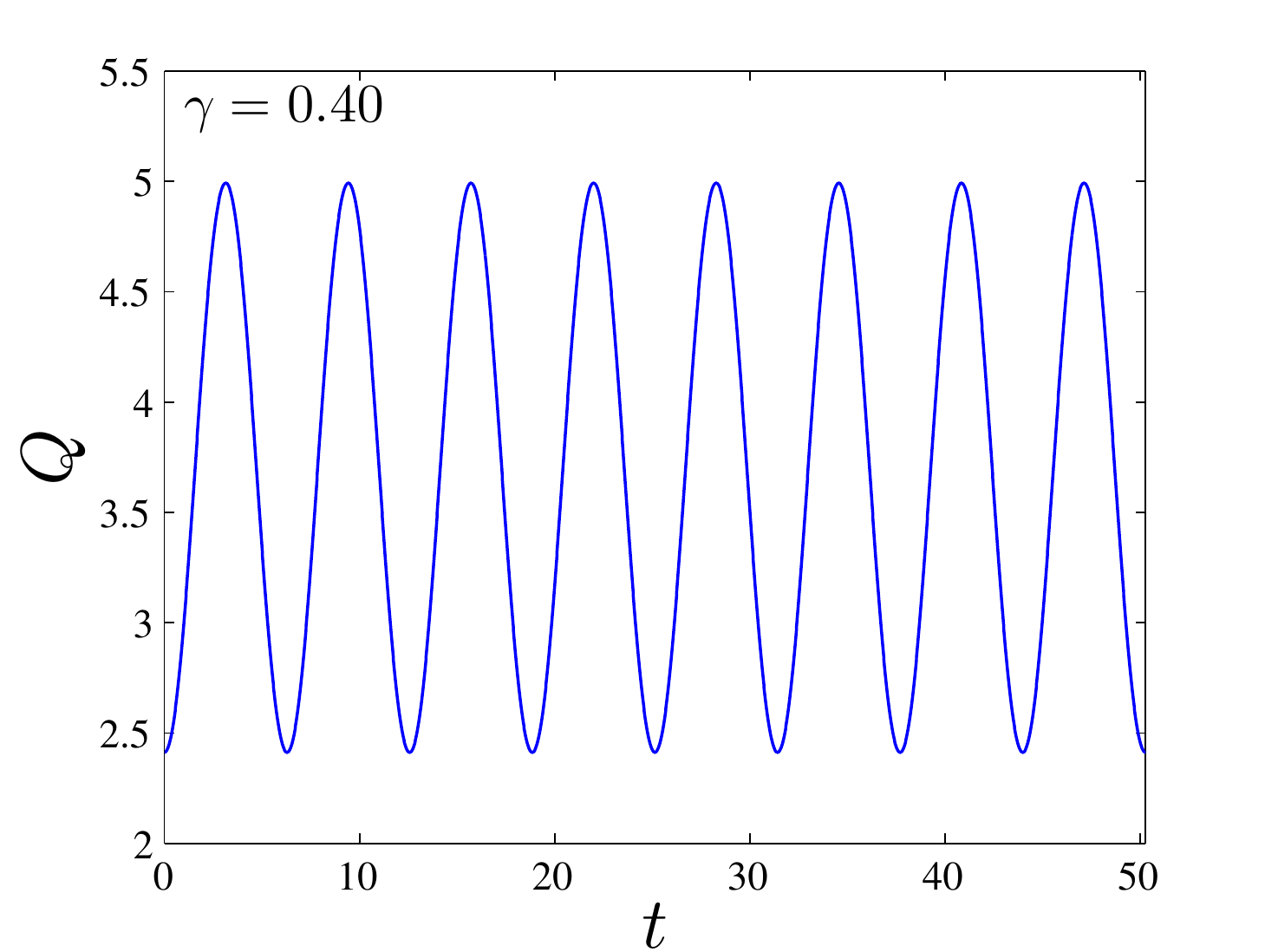} & \hfill
\includegraphics[width=.45\textwidth]{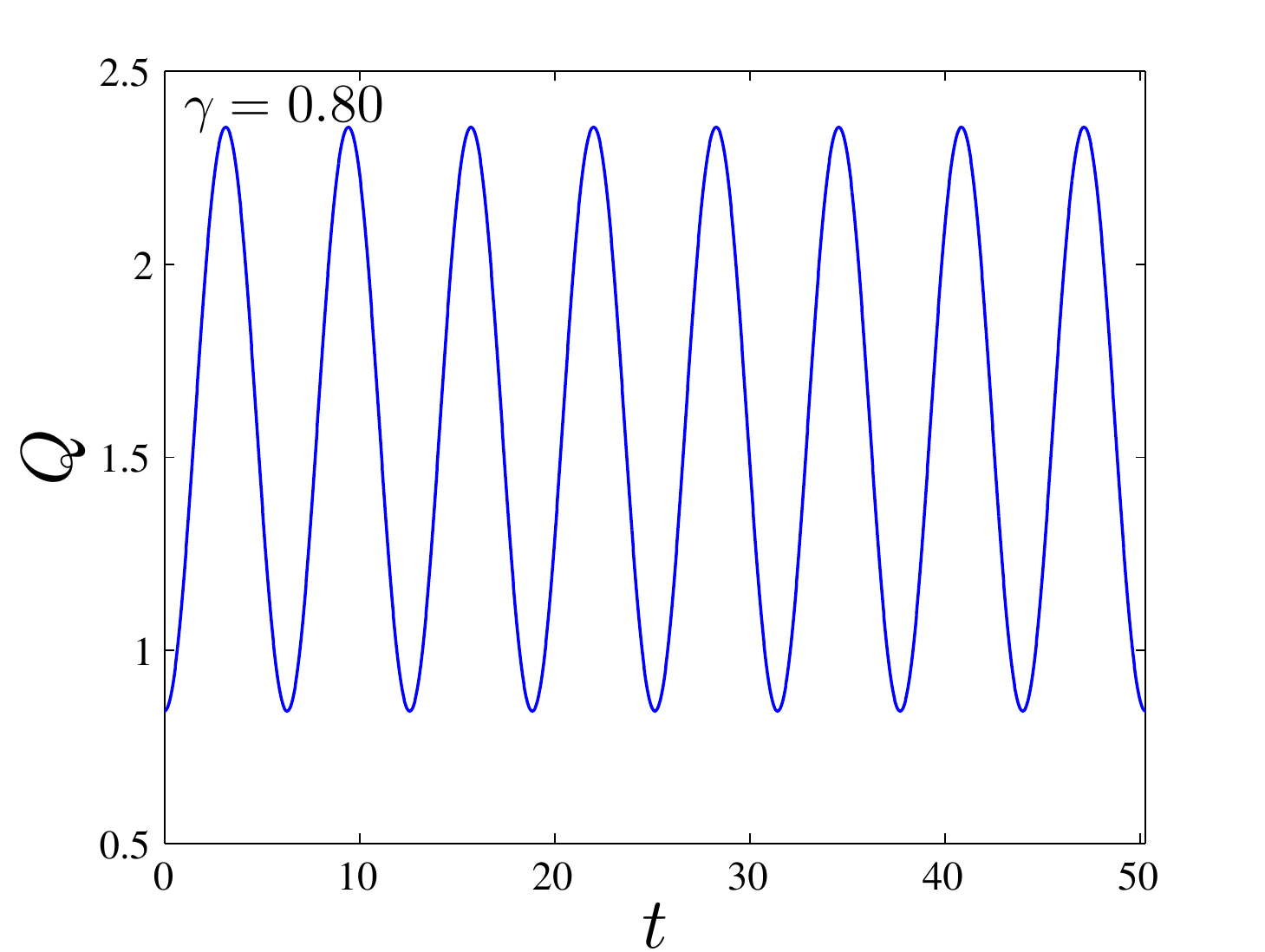} \\
\end{tabular}
\end{center}
\caption{Same as Fig.~\ref{cuevas-fig42} but showing the time evolution of the solution charge.}
\label{cuevas-fig43}
\end{figure}

We have confirmed that the dynamics observed, e.g., in Fig.~\ref{cuevas-fig39} corresponds to $\mathbf{SU}(1,1)$-solutions. For instance, for the solitary wave with $\gamma_0=0.59$, $\omega=0.8$, when initializing it for $\gamma_s=0.9$, it spontaneously gives rise to an oscillatory state of the above form of equation~(\ref{eq:cuevas-su11}) with $\tilde\omega=0.422$, $\alpha_{-}=1.0847$ and $\alpha_{+}=0.4201$. On the other hand, using a numerically exact (up to a prescribed tolerance) solution of our fixed point iteration scheme with a given frequency $\tilde{\omega}$ (for a desired $\gamma$), we can select values
of $\alpha_{-}$ and $\alpha_{+}$ and the exact form of Eq.~(\ref{eq:cuevas-su11}) in order to construct, at will, such bi-frequency $\mathbf{SU}(1,1)$-solutions.
An example of this form is shown in Figs.~\ref{cuevas-fig42} and \ref{cuevas-fig43} (even for $\gamma=0$) for $\tilde{\omega}=0.5$, $\alpha_{-}=1.0500$, and $\alpha_{+}=0.3202$.

\section{Summary and outlook}

In the present work, we have discussed some of the principal
properties of the nonlinear Dirac equation and its similarities,
as well as differences, in comparison to its extensively
studied cousin, namely the nonlinear Schr{\"o}dinger equation.
We have discussed nonlinear models that possess solitary wave solutions and
vortices, and
have placed particular emphasis on their spectral stability,
also mentioning the
orbital and asymptotic stability thereof and the corresponding
issues that arise. We have seen that especially in higher
dimensions the stability properties of solitary waves in the nonlinear Dirac
equation can be fundamentally different from the
NLS case, and may not feature collapse scenarios.
Moreover, solitary waves may be spectrally stable for suitable
parametric (i.e., frequency) regimes. For the three-dimensional
case, the stability properties are just starting to be explored
(in suitable subspaces), yet this problem is extremely interesting,
also due to its connections with the dynamics. In the context
of the latter, we explored some of the delicate features
that arise from different types of discretizations
(finite-difference, Fourier and Chebyshev spectral schemes)
and
the implications for the evolutionary dynamics. Generally,
we hope to have exposed some of the significant complications arising
in dynamically propagating such a system, especially when trying to
do so for long time scales.

   From every perspective that we can think of, nonlinear Dirac systems
pose significant
challenges ahead of us. From the
point of view of the mathematical analysis, understanding the spectral
properties observed herein and their dynamical implications is already
a formidable problem. Computing efficiently and systematically both
the solutions and their linearization eigenvalues emerges as a significant and
upcoming challenge. This is especially true in three spatial dimensions.
Devising numerical schemes -- possibly based on integrable
(semi-discrete or genuinely discrete) variants
of the model -- could prove to be of paramount importance towards future
robust computations of the dynamics. Finally, combining some of
the cutting edge themes in nonlinear waves (such as for instance
rogue waves~\cite{cuevas-KPS09}) with relevant scenarios involving
Dirac-type nonlinear models opens another highly promising vein
of research for future studies~\cite{cuevas-DWA15}.

\section*{Acknowledgments}

The research of Andrew Comech was carried out
at the Institute for Information Transmission Problems
of the Russian Academy of Sciences
at the expense of the Russian Foundation
for Sciences (project 14-50-00150).
J.C.-M. thanks financial support from MAT2016-79866-R project (AEI/FEDER, UE). P.G.K. gratefully acknowledges the support of NSF-PHY-1602994, the Alexander von Humboldt Foundation,
the Stavros Niarchos Foundation via the Greek Diaspora Fellowship Program, and the ERC under FP7, Marie Curie
Actions, People, International Research Staff Exchange Scheme (IRSES-605096).
This work was supported in part by the U.S. Department of Energy.

\bibliographystyle{spmpsci}

\end{document}